\documentclass[%
 reprint,
superscriptaddress,
 amsmath,amssymb,
 pra,
]{revtex4-2}

\DeclareMathOperator{\diag}{diag}
\DeclareMathOperator{\sgn}{sgn}

\usepackage{graphicx}
\usepackage{dcolumn}
\usepackage{multirow}
\usepackage{bm}
\usepackage[pdfpagemode=UseNone,pdfstartview=FitH,colorlinks=true,linkcolor=blue,citecolor=blue,urlcolor=blue]{hyperref}
\usepackage[all]{hypcap}

\def\({\left(}
\def\){\right)}

\def\bra#1{\mathinner{\langle{#1}|}}
\def\ket#1{\mathinner{|{#1}\rangle}}
\def\braket#1#2{\mathinner{\langle{#1}|#2 \rangle}}

\def\avg#1{\mathinner{\langle{#1} \rangle}}

\def\CZ{{\rm CZ}}

\def\T{{\rm T}}

\newcommand{\tr}{\operatorname{tr}}

\newtheorem{theorem}{Theorem}
\newtheorem{lemma}{Lemma}

\DeclareRobustCommand{\quickfig}[4]{
\begin{figure}
\begin{centering}
\includegraphics[width=#1]{#2}
\par\end{centering}
\caption{#3}
\label{#4}
\end{figure}
}

\newcommand{\xebfidelity}{\mathcal{F}_{\text{XEB}}}
\renewcommand{\thefigure}{S\arabic{figure}}
\begin{document}

\title{Supplementary information for\\ 
``Quantum supremacy using a programmable superconducting processor" \cite{arute2019}}

\author{Google AI Quantum and collaborators$^\dagger$}

\date{\today}


\maketitle

\tableofcontents

\section{\label{architecture}Device design and architecture}
The Sycamore device was designed with both the quantum supremacy experiment \cite{arute2019} and small noisy intermediate scale quantum (NISQ) applications in mind. The architecture is also suitable for initial experiments with quantum error correction based on the surface code.  While we are targeting 0.1\% error two-qubit gates for error correction, a quantum supremacy demonstration can be achieved with 0.3-0.6\% error rates. 

For decoherence-dominated errors, a 0.1\% error means a factor of about 1000 between coherence and gate times.  For example, a 25 $\mu$s coherence time implies a 25 ns gate.  A key design objective in our architecture is achieving short two-qubit gate time, leading to the choice of tunable transmon qubits with direct, tunable coupling.

A difficult challenge for achieving a high-performance two-qubit gate is designing a sufficiently strong coupling when the gate is active, which is needed for fast gates, while minimizing the coupling otherwise for low residual control errors. These two competing requirements are difficult to satisfy with a fixed-coupling architecture: our prior processors \cite{barends2014nine} used large qubit-qubit detuning ($\sim$1~ GHz) to turn off the effective interaction, requiring relatively high-amplitude precise flux pulses to tune the qubit frequencies to implement a CZ gate. In the Sycamore device, we use adjustable couplers \cite{charles_thesis} as a natural solution to this control problem, albeit at the cost of more wiring and control signals. This means that the qubits can idle at much smaller relative detuning. We chose a capacitor-coupled design \cite{charles_thesis, Yan-2018-MIT}, which is simpler to layout and scale, over the inductor-based coupler of previous gmon devices \cite{chen2014coupler, neill2018blueprint}. In Sycamore, the coupling $g$ is tunable from $5$ MHz to $-40$ MHz. The experiment uses `on' coupling of about $-20$ MHz.  

By needing only small frequency excursions to perform a two-qubit gate, the tunable qubit can be operated much closer to its maximum frequency, thus greatly reducing flux sensitivity and dephasing from $1/f$ flux noise.  Additionally, the coupling can be turned off during measurement, reducing the effect of measurement crosstalk, a phenomenon that has shown to be somewhat difficult to understand and minimize \cite{khezri2015transitions}.  
	
The interaction Hamiltonian of a system of on-resonance transmons with adjustable coupling (truncated to the qubit levels) has the following approximate form,
\begin{align}
H_{\mathrm{int}}(t) \approx 
\sum_{\left< i,j \right>} g_{ij}(t) \, 
(\sigma_i^+ \sigma_j^- + \sigma_i^- \sigma_j^+) +
\frac{g_{ij}^2(t)}{|\eta |}\, \sigma_i^z \sigma_j^z \, ,
\end{align}
where $g_{ij}$ is the nearest neighbor coupling, $\eta$ is the non-linearity of the qubits (roughly constant),  $i$ and $j$ index nearest-neighbor qubit pairs, and $\sigma^{\pm} = (\sigma^x \pm i \sigma^y)/2$.  We pulse the coupling in time to create coupling gates.

Our two-qubit gate can be understood using Cartan decomposition \cite{tucci507171cartan}, which enables an arbitrary two-qubit gate to be decomposed into four single-qubit gates around a central two-qubit gate that can be described by a unitary matrix describing only XX, YY and ZZ interactions, with 3 parameters indicating their strengths.  For the physical interaction describing our hardware, we see a swapping interaction between the $|01\rangle$ and $|10\rangle$ qubits states, corresponding to an XX+YY interaction.  Interaction of the qubit $|11\rangle$ state with the $|2\rangle$ states of the data transmons produce a phase shift of that state, corresponding to a ZZ interaction.  By changing the qubit frequencies and coupling strength we can vary the magnitude of these interactions, giving net control of 2 out of the 3 possible parameters for an arbitrary gate.  

\section{Fabrication and layout}

Our Sycamore quantum processor is configured as a diagonal array of qubits as seen in the schematic of Fig.~1 in the main text. The processor contains 142 transmon qubits, of which 54 qubits have individual microwave and frequency controls and are individually read out (referred to as qubits).  The remaining 88 transmons are operated as adjustable couplers remaining in their ground state during the algorithms (referred to as couplers).

The qubits consist of a DC SQUID sandwiched between two metal islands, operating in the transmon regime. An on-chip bias line is inductively coupled to the DC SQUID, which allows us to tune qubit frequency by applying control fluxes into the SQUID loop. For regular operations, we tune qubits through a small frequency range ($<$ 100 MHz). This corresponds to a relatively small control signal and makes qubit operation less sensitive to flux crosstalk. 

Each pair of nearest-neighbor qubits are coupled through two parallel channels: direct capacitive coupling and indirect coupling mediated by coupler \cite{charles_thesis, andrew_thesis, Yan-2018-MIT}. Both channels result in qubit-qubit coupling in the form of $\sigma_i^x \sigma_j^x + \sigma_i^y \sigma_j^y$ in the rotating frame, although with different signs. The indirect coupling is negative, given it is a second-order virtual process. The strength of the indirect coupling is adjusted by changing the coupler frequency with an additional on-chip bias line, giving a net zero qubit-qubit coupling at a specific flux bias. 

The Sycamore processor consists of two die that we fabricated on separate high resistivity silicon wafers. The fabrication process, using aluminum on silicon, requires a total of 14 lithography layers utilizing both optical and electron beam lithography. Crosstalk and dissipation are mitigated through ground plane shielding \cite{dunsworth2018method}.   After fabrication and die singulation, we use indium bump bonding \cite{rosenberg20173bumps, Foxen_2017} of the two separate dies to form the Sycamore processor. 

\begin{figure}[t!]
\includegraphics[width=\columnwidth]{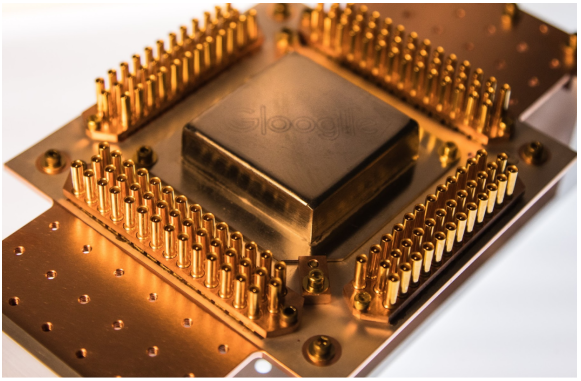}
\caption{\label{fig:package} \textbf{A photograph of a packaged Sycamore processor.} The processor is shielded from the electromagnetic environment by a mu-metal shield (middle) and a superconducting aluminum cap, inside the mu-metal shield. The processor control wires are routed, through PCB circuit board, to coaxial connectors shown around the edge.}
\end{figure}

The Sycamore processor is connected to a 3-layer Al-plated circuit board with Al wirebonds \cite{foxen2018high_speed}.  Each line is routed through a microwave connector to an individual coax cable.  We shield the processor from stray light using a superconducting Al lid with black coating, and from magnetic fields using a mu-metal shield as shown in Fig.~\ref{fig:package}.

\section{Qubit control and readout} \label{sec:control}

\subsection{Control}

Operating the device requires simultaneous synchronized control waveforms for each of the qubits and couplers.
We use 54 coherent microwave control signals for qubit XY rotations, 54 fast flux bias lines for qubit frequency tuning, and 88 fast flux biases for the adjustable couplers.
Dispersive readout requires an additional 9 microwave signals and phase sensitive receivers.
A schematic of the room temperature electronics is shown in Fig.~\ref{fig:sup_control:electronics}, and the cryogenic wiring is shown in Fig.~\ref{fig:sup_control:cryo_wiring}.

Waveform generation is based on a custom-built multi-channel digital to analog converter (DAC) module.
Each DAC module provides 8 DACs with 14-bit resolution and 1 GS/s sample rate.
Each DAC sample clock is synchronized to a global 10 MHz reference oscillator, and their trigger is connected by a daisy chain to synchronize all modules used in the experiment.
This set of DAC modules forms a $>$250-channel, phase-synchronous waveform generator.
We have measured 20 ps of jitter between channels.
The modules are mounted in 14-slot 6U rackmount chassis.
A single chassis, shown in FIG.~\ref{fig:sup_control:rack_photo}, can control approximately 15 qubits including their associated couplers and readout signals.
A total of 4 chassis are used to control the entire Sycamore chip.

The DAC outputs are used directly for fast flux biasing the qubits and couplers required for two-qubit gates.
Microwave control for single-qubit XY rotations and dispersive readout combine two DAC channels and a mixer module to form a microwave arbitrary waveform generator (Microwave AWG) via single-sideband upconversion in an IQ mixer as shown in Figure \ref{fig:sup_control:electronics}\,a.
The microwave AWG provides signals with arbitrary spectral content within $\pm$350 MHz of the local oscillator (LO).
A single LO signal is distributed to all IQ mixers so that all qubits' XY controls are phase coherent.
The mixer modules are mounted in the same chassis as the DAC modules.
Each mixer's I and Q port DC offsets are calibrated for minimum carrier leakage and the I and Q amplitudes and phases are calibrated to maximize image rejection.

Each DAC module contains an FPGA that provides a gigabit ethernet interface, SRAM to store waveform patterns, and sends the waveform data to the DAC module's 8 DACs.
To optimize the use of SRAM, the FPGA implements a simple jump table to allow reusing or repeating waveform segments.
A computer loads the desired waveforms and jump table onto each FPGA using a UDP-based protocol and then requests the first (master) FPGA to start.
The start pulse is passed down the daisy chain causing the remainder (slave) DACs and ADCs to start.

\quickfig{\columnwidth}{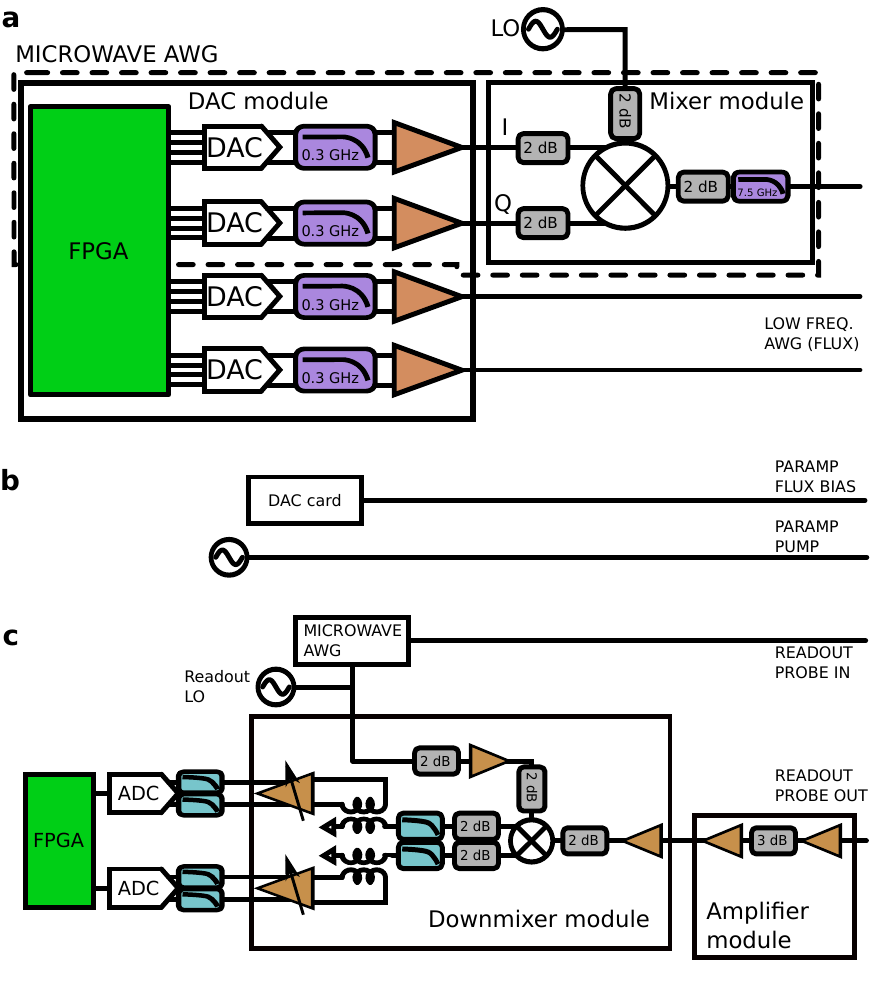}{\textbf{Control electronics.} \textbf{a,} The custom DAC module provides 8 DAC channels (4 shown). DACs are used individually for flux pulses or in pairs combined with a mixer module to comprise a microwave AWG channel (dashed box). \textbf{b,} A single DAC channel and a microwave source are used to bias and pump the parametric amplifier for readout. \textbf{c,} Readout pulses are generated by a microwave AWG. The reflected signal is amplified, mixed down to IF, and then digitized in a pair of ADCs. The digital samples are analyzed in the FPGA.}{fig:sup_control:electronics}

\quickfig{\columnwidth}{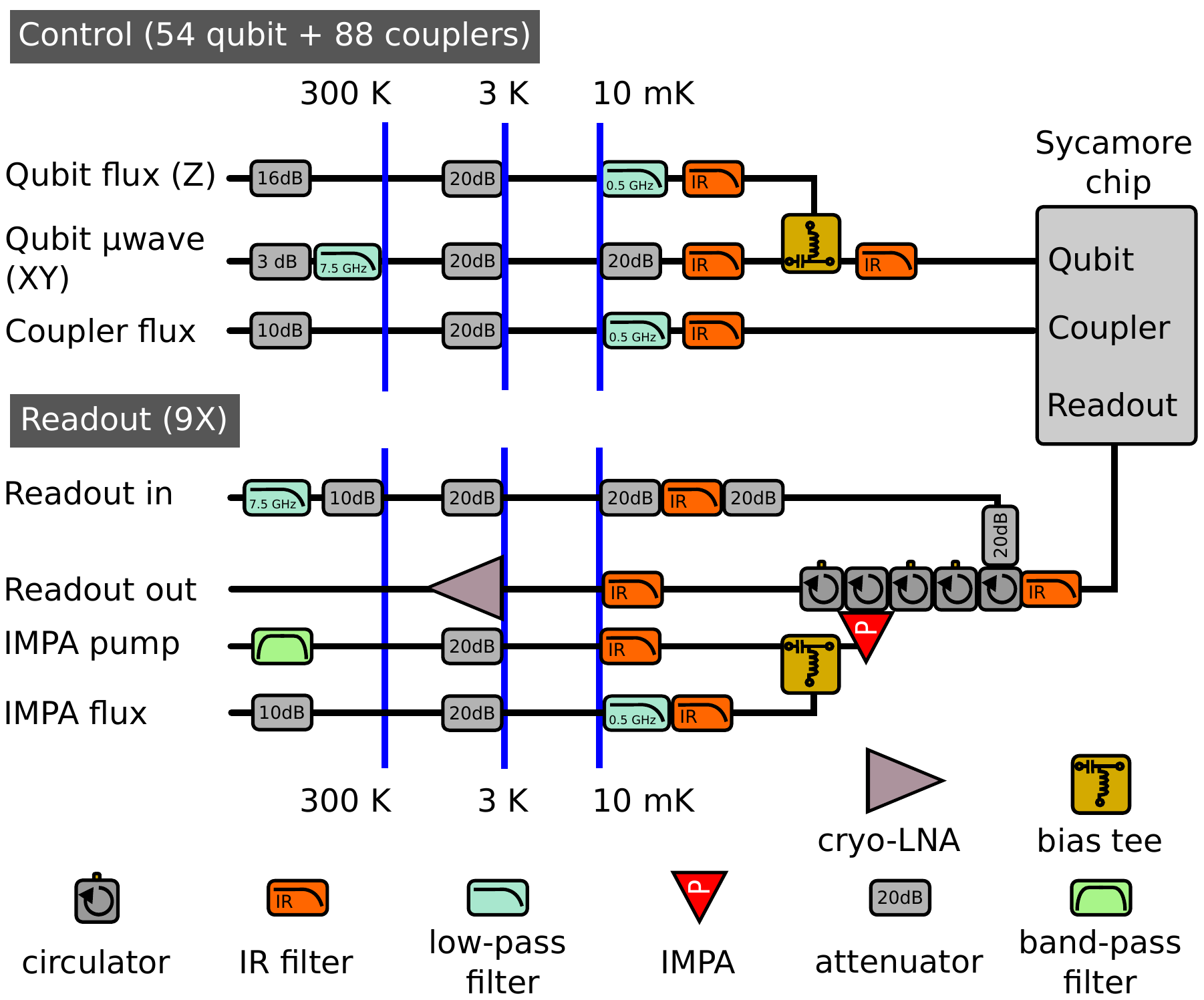}{\textbf{Cryogenic wiring.} Control and readout signals are carried to and from the Sycamore chip with a set of cables, filters, attenuators, and amplifiers. }{fig:sup_control:cryo_wiring}

\quickfig{\columnwidth}{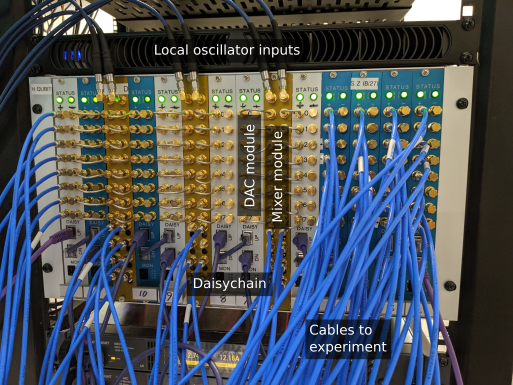}{\textbf{Electronics chassis.} Each chassis supports 14 DAC and/or mixer modules. Local oscillators are connected at the top of each mixer module. A set of daisychain cables connects from each ADC module to the next. Control signals exit the chassis through coaxial cables.}{fig:sup_control:rack_photo}

\subsection{Readout}

Qubit state measurement and readout (hereafter ``readout'') are done via the dispersive interaction between the qubit and a far-detuned harmonic resonator \cite{Blais2004CircuitQED, Gambetta2006, bultink2018general}.
A change in the qubit state from $\ket{0}$ to $\ket{1}$ causes a frequency shift of the resonator from $\omega_{\ket{0}}$ to $\omega_{\ket{1}}$.
A readout probe signal applied to the resonator at a frequency in between $\omega_{\ket{0}}$ and $\omega_{\ket{1}}$ reflects with a phase shift $\phi_{\ket{0}}$ or $\phi_{\ket{1}}$ that depends on the resonator frequency and therefore on the qubit state.
By detecting the phase of the reflected probe signal we infer the qubit state.
The readout probe signal is generated with the same microwave AWG as the XY control signals, but with a separate local oscillator, and is received and demodulated by the circuit shown in Figure \ref{fig:sup_control:electronics}\,c.

The readout probe intensity is typically set to populate the readout resonator with only a few photons to avoid readout-induced transitions in the qubit \cite{Sank2016Transitions}.
Detecting this weak signal at room temperature with conventional electronics requires 100 dB of amplification.
To limit the integration time to a small fraction of the qubit coherence time, the amplification chain must operate near the quantum noise limit \cite{clerk2010intro, caves1982Amplifiers}.

Inside the cryostat the signal is amplified by an impedance matched lumped element Josephson parametric amplifier (IMPA) \cite{Mutus2014IMPA} on the mixing chamber stage followed by a Low Noise Factory cryogenic HEMT amplifier at 3 K.
At room temperature the signal is further amplified before it is mixed down with an IQ mixer producing a pair of intermediate frequency (IF) signals $I(t)$ and $Q(t)$.
The IF signals are amplified by a pair of variable gain amplifiers to fine-tune their level, and then digitized by a pair of custom 1 GS/s, 8-bit analog to digital converters (ADC).
The digitized samples $I_n$ and $Q_n$  are processed in an FPGA which combines them into a complex phasor
\begin{equation*}
  z_n = I_n + iQ_n = E_n \exp(i (\omega n dt + \phi))
\end{equation*}
where $dt$ is the sample spacing, $\omega$ is the IF frequency, $\phi$ is the phase that depends on the qubit state, and $E_n$ is the envelope of the reflected readout signal.
The envelope is measured experimentally once and then used by the FPGA in subsequent experiments as the optimal demodulation window $w_n$ to extract the phase of the reflected readout signal \cite{ryan2015tomography, Jeffrey2014Fast}.
The FPGA multiplies $z_n$ by $w_n \exp(-i \omega n dt)$, and then sums over time to produce a final complex value $\exp(i \phi)$
\begin{equation*}
  \sum_{n=0}^{N-1} z_n w_n \exp(-i \omega n dt) \propto \exp(i \phi)
\end{equation*}
In the absence of noise, the final complex value would always be one of two possible values corresponding to the qubit states $\ket{0}$ and $\ket{1}$.
However, the noise leads to Gaussian distributions centered at those two points.
The size of the clouds is determined mostly by the noise of the IMPA and cryogenic HEMT amplifier, while the separation between the clouds' centers is determined by the resonator probe power and duration.
The signal to noise ratio of the measurement is determined by the clouds' separation and width \cite{Jeffrey2014Fast, Sank2015Noise}.

The 54 qubits are divided into nine frequency multiplexed readout groups of six qubits each.
Within a group, each qubit is coupled to its own readout resonator, but all six resonators are coupled to a shared bandpass Purcell filter \cite{Reed2010Purcell, Sete2015Purcell, Jeffrey2014Fast}.
All qubits in a group can be read-out simultaneously by frequency-domain multiplexing \cite{chen2012mux, barends2014nine} in which the total probe signal is a superposition of probe signals at each of the readout resonators’ frequencies.
The phase shifts of these superposed signals are independently recovered in the FPGA by demodulating the complex IQ phasor with each intermediate frequency.
In other words, we know what frequencies are in the superposed readout signal and we compute the Fourier coefficients at those frequencies to find the phase of each reflected frequency component.

\section{XEB theory}\label{sec:xeb_theory}

We use cross entropy benchmarking (XEB)~\cite{boixo2018characterizing,neill2018blueprint} to calibrate general single-
and two-qubit gates, and also to estimate the fidelity of random
quantum circuits with a large number of qubits. XEB is based on the
observation that the measurement probabilities of a random quantum
state have a similar pattern to laser ``speckles'', with
some bitstrings more probable than others \cite{wootters1990random, emerson2005}. The same holds for
the output state of random quantum circuits. As errors destroy the speckle
pattern, this is enough to estimate the rate of errors and fidelity in an experiment. Crucially, XEB does not require the reconstruction of
experimental output probabilities, which would need an exponential
number of measurements for increasing number of qubits. Rather, we use
numerical simulations to calculate the likelihood of a set of
bitstrings obtained in an experiment according to the ideal expected
probabilities. Below we describe the theory behind this technique in more
detail.

\subsection{XEB of a small number of qubits}\label{sec:sq}

We first consider the use of XEB to obtain the error rate for single- and two-qubit gates. As explained above, for a two-qubit XEB estimation we use sequences of cycles, each cycle
consisting of two sufficiently random single-qubit gates followed by the same two-qubit gate. 

 The  density operator of the system after application of a random circuit $U$ with $m$ cycles  can be written as a sum of two  parts
\begin{align}
\rho_U  = \varepsilon_m \ket{\psi_U}\bra{\psi_U}    + (1- \varepsilon_m)\chi_U\;,\quad D=2^n\;.\label{eq:xeb_ru0}
\end{align}
Here  $\ket{\psi_U}=U \ket{\psi_0}$ is the ideal output state and $\chi_U$ is an operator with unit trace that along with $  \varepsilon_m$ describes the effect of errors. For a depolarizing channel model  $\chi_U=I/D$ and $\varepsilon_m$ has the meaning of the depolarization fidelity after $m$ cycles. Nevertheless, in the case of small number of qubits, the part of the operator $\chi_U$ has nonzero matrix elements between the states with no error and the states with the error. However, if we undo the evolution of each random circuit and average over an ensemble of circuits such cross-terms are  averaged out and we expect
\begin{equation}
\overline{U^{\dagger}\chi_U U}= \frac I D \;.\label{eq:chiu_avg}
\end{equation}
Here and below we use the horizontal bar on the top to denote averaging over the ensemble of random circuits.
Because of this property it is possible to establish the connection between the quantity $\varepsilon_m$ and the depolarization fidelity after $m$ cycles.

From Eqs.~\eqref{eq:xeb_ru0} and \eqref{eq:chiu_avg} we get
\begin{align}
  \overline{U^{\dagger}\rho_U U} &= \overline{\varepsilon_m} \ket{\psi_0}\bra{\psi_0} + (1-\overline{\varepsilon_m}) \frac I D \;.\label{eq:dcm}
\end{align}
This is a depolarizing channel. From this and the exponential decay of fidelity we get 
\begin{equation}
\overline{\varepsilon_m}=p_c^m\;,\label{eq:Fd}
\end{equation}
connecting $\overline{\varepsilon_m}$ to the depolarization fidelity $p_c$   per cycle.  

The noise model (\ref{eq:xeb_ru0}) is very general in the context of random circuits. To
 provide some insight about the origin of this model we consider a specific case with pure systematic error in the two-qubit gate. In this case the resulting pure state after the application of the random circuit $\tilde U$ with the error can be expanded  into the direction of the ideal state vector and the orthogonal direction
 \begin{equation}
   \tilde U\ket{\psi_0}=\xi_m\, \ket{\psi_U} +\sqrt{1-|\xi_m|^2}\ket{\varphi_{\tilde U}}\;,\label{eq:Psi-er}
 \end{equation}
 where 
 \begin{equation}
 \braket{\psi_U}{\varphi_{\tilde U}}=0,\quad \braket{\varphi_{\tilde U}}{\varphi_{\tilde U}}=1\;.\label{eq:cond}
 \end{equation}
 
\begin{figure}
\includegraphics[scale=0.4]{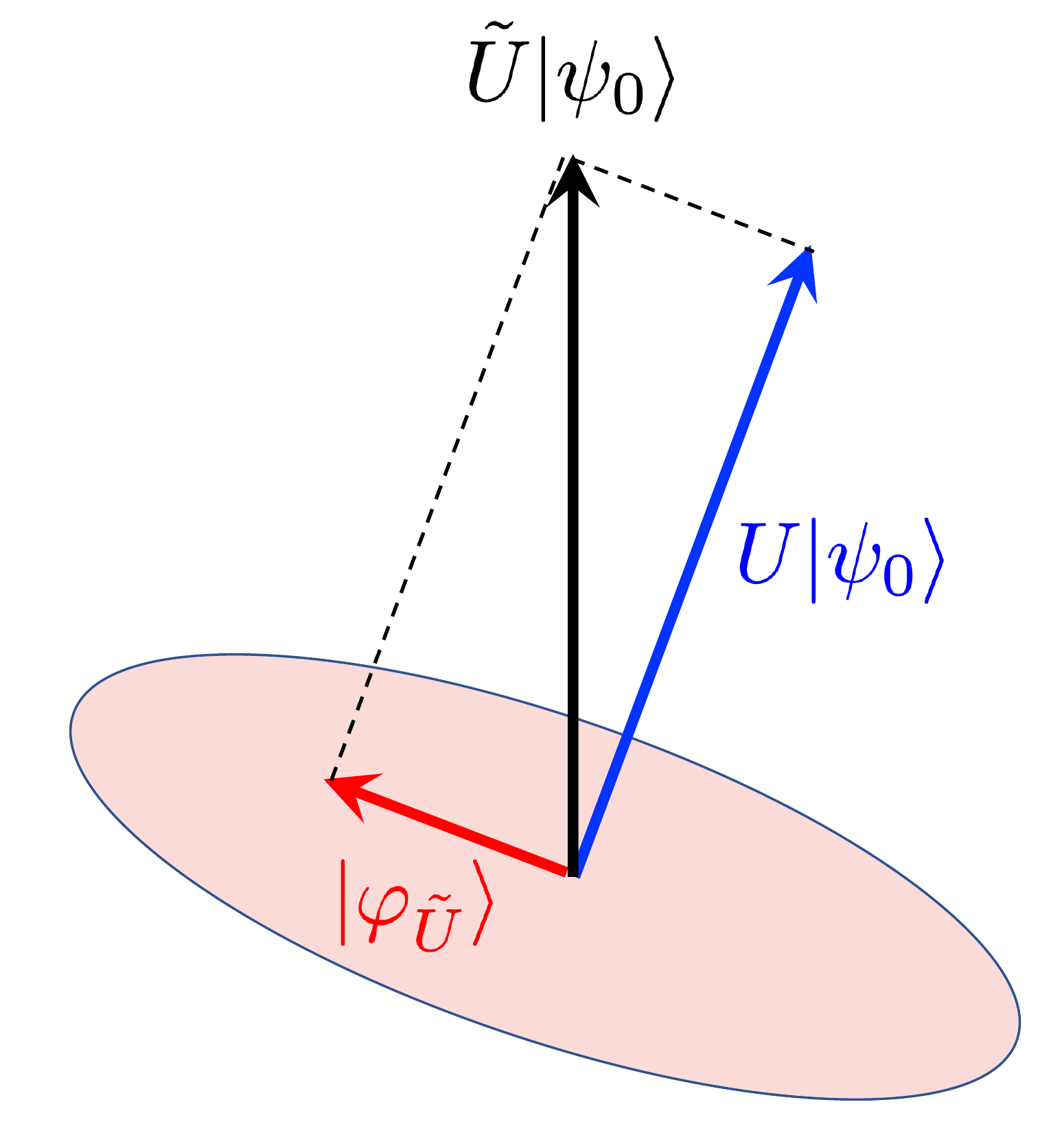}
\caption{\label{fig:vector-error}  \textbf{Cartoon: decomposition of the quantum state  into
the vector aligned with the ideal quantum state and its orthogonal complement}}
\end{figure}

 For the ensemble of random circuits $U$ the error vector  is distributed completely randomly in the  plane orthogonal to the ideal vector $U\ket{\psi_0}$ (see Fig.~\ref{fig:vector-error}).
This condition of orthogonality is the only constraint on the vector $\ket{\varphi_{\tilde U}}$ that involves  $\ket{\psi_U}$.
Therefore we expect
\begin{align}
  \overline{U^{\dagger}\ket{\varphi_{\tilde U}}\bra{\varphi_{\tilde U}} U} &= \frac 1 {D-1} (I - \ket{\psi_0}\bra{\psi_0})\;.\label{eq:xeb_uvu}
 \end{align}
 Also
 \begin{align}
   \overline{U^\dagger\left(\xi_m\sqrt{1-|\xi_m|^2}\, \ket{\psi_U}\bra{\varphi_{\tilde U}} +h.c\right)U} &= 0\;.
\end{align}
This gives  the connection between the error vector $\ket{\varphi_{\tilde U}}$ and the operator $\chi_U$ 

 \begin{multline}
 (1-\varepsilon_m)\chi_U-\frac{1-\varepsilon_m}{D} \ket{\psi_U}\bra{\psi_U} = 
 (1-|\xi_m|^2)\ket{\varphi_{\tilde U}}\bra{\varphi_{\tilde U}}\\+\left(\xi_m\sqrt{1-|\xi_m|^2}\, \ket{\psi_U}\bra{\varphi_{\tilde U}} +h.c\right)\;.\label{eq:conn}
 \end{multline}
The resulting equation 
 \begin{equation}
    \overline{|\xi_m|^2}=\overline{\varepsilon_m}+\frac{1-\overline{\varepsilon_m}}{D}\label{eq:xeb_xi_eps}
 \end{equation}
is to be expected, because $|\xi_m|^2$ is the average state fidelity while $\overline{\varepsilon_m}$ is the depolarization fidelity (see Sec. \ref{sec:computing_errors}). Note that Eqs.~\eqref{eq:xeb_uvu}--\eqref{eq:xeb_xi_eps} lead to Eq.~\eqref{eq:dcm}. This result can also be derived assuming that single qubit gates  form a 2-design in the Hilbert space of each qubit. 

\begin{figure}
\includegraphics[scale=0.45]{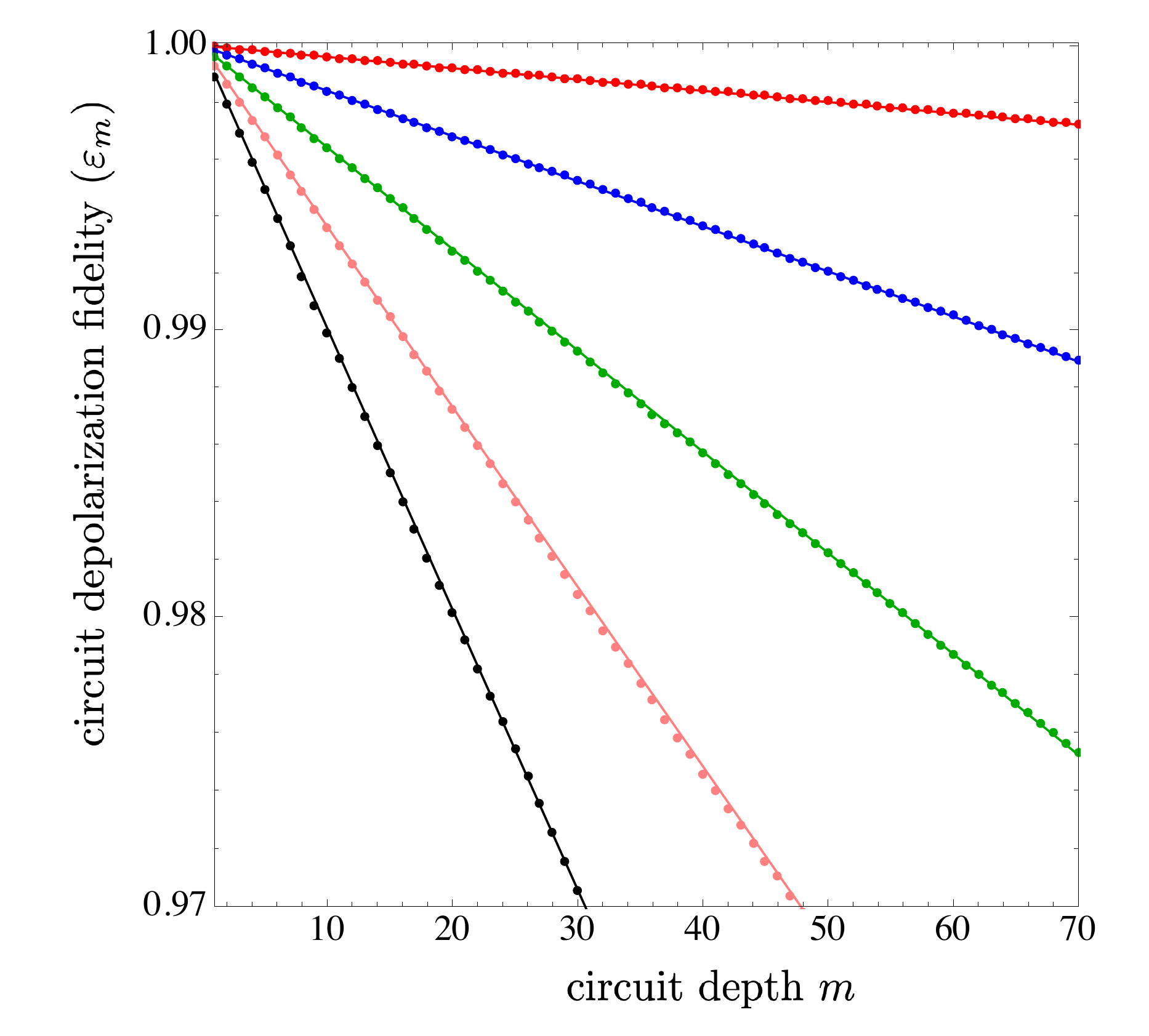}
\caption{\label{fig:syst-er}  \textbf{Plots of the  circuit depolarizing fidelity {\it vs} the circuit depth.}
Solid lines corresponds to the predictions from the depolarizing  channel model (\ref{eq:pc}) and
points correspond to $\overline{\varepsilon_m}$ (\ref{eq:pc-num}) obtained by  the averaging of the squared overlap over the ensemble of random circuits. Different colored  pots correspond to different values of the swap error
$\Delta\theta=$ 0.01(red), 0.02(blue), 0.03 (green), 0.04 (pink), 0.05 (black).}
\end{figure}

We demonstrate the above findings by numerically simulating  
the random circuits for 2 qubits that contains single qubit gates randomly sampled from 
Haar measure and ISWAP-like gate
\begin{equation}
 V(\theta) =  \left(
\begin{array}{cccc}
 1 & 0 & 0 & 0 \\
 0 & \cos\theta  & -i \sin\theta & 0 \\
 0 & -i \sin\theta & \cos\theta & 0 \\
 0 & 0 & 0 & 1 \\
\end{array}
\right)\;.
\end{equation}
The systematic error $\Delta\theta=\theta-\pi/2$ corresponds to the deviation of the swap angle from $\pi/2$.
Then assuming that the single qubit gates are error free the   depolarizing channel model gives the prediction for the depolarizing fidelity per cycle 
\begin{align}
    p_c &=\frac{| {\rm tr}(V(\theta)V^\dagger(\pi/2))|^2-1}{D^{2}-1}\nonumber\\
    &=\frac{1}{15} (8 \cos (\Delta \theta )+2 \cos (2 \Delta \theta )+5)\;.\label{eq:pc}
\end{align}
As shown in  Fig.~\ref{fig:syst-er} the depolarizing fidelity $p_c^m$ for the circuit of depth $m$ based on Eq.~(\ref{eq:pc})  closely  matches the corresponding quantity obtained by the averaging of the squared overlap over the ensemble of random circuits (cf. (\ref{eq:xeb_xi_eps})
\begin{equation}
 \overline{\varepsilon_m}=   \frac{D\,\overline{|\xi_{m}|^{2}}-1}{D-1}\;.\label{eq:pc-num}
\end{equation}

Returning to the generic case, property (\ref{eq:chiu_avg}) can be extended so that for any smooth function $f(u)$  the following relation holds
\begin{align}
\overline{\sum_{q\in\{0,1\}^n}f(p_s(q))\bra{q}\chi_U\ket{q}}&=\sum_{q\in\{0,1\}^n}\frac{\overline{f(p_s(q))}}{D}+\epsilon\;,\label{eq:corr}
\end{align}
where $\ket{q}$ is a computational basis state corresponding to bitstring $q$, and 
$p_s(q)=\bra{q} U \rho_0 U^{\dagger}\ket{q}$ is the simulated (computed) ideal probability of
$q$.    If the average is performed over a sample of random circuits of size $S$  then  the  correction is $\epsilon \in O(1/\sqrt S)$. We tested numerically for the case of $n=2$ that relation  (\ref{eq:corr}) holds even for purely systematic errors in the case of a sufficiently random set of single qubit gates. 

We now make the critical step of estimating the parameter $p_c^m$ from a set of
experimental realizations of random circuits with $m$ cycles. 
We map each measured bitstring $q$ with a function $f(p_s(q))$ and then average this function over the measured bitstrings. The standard XEB \cite{boixo2018characterizing, neill2018blueprint} uses the natural logarithm, $f(p_s(q))=\log (p_s(q))$. In the main text we use the linear version of XEB, for which $f(p_s(q))=D p_s(q) - 1$. Both these functions give higher values to bitstrings with higher  simulated probabilities. Another closely related choice is the Heavy Output Generation test~\cite{aaronson2017complexity}, for which $f$ is a step-function.

Under the model (\ref{eq:xeb_ru0}), in an experiment with ideal state preparation and measurement, we obtain the bitstring $q$ with
probability
\begin{align}
    p_c^m \;p_s(q) + (1-p_c^m)\bra{q}\chi_U\ket{q}\;,\label{eq:xeb_dn}
\end{align}For the linear XEB, the average value of $D p_s(q)
- 1$ when sampling with probabilities given by Eq.~\eqref{eq:xeb_dn} is
\begin{align}
  \overline{\avg{  D p_s(q) - 1}} = p_c^m\left(D \sum_q  \overline{ p_s(q)^2}  -1\right)\;.\label{eq:xeb_le}
\end{align}
Similarly to Eq.~(\ref{eq:corr}), the horizontal bar denotes averaging over the random circuits.

The sum on the right hand side of (\ref{eq:xeb_le}) goes over all bitstrings in the
computational basis, and  can be obtained with numerical
simulations.   It can also be found analytically assuming that the random circuit ensemble  approximates the Haar measure where for a given $q$ the quantity  $p_s(q)$ is distributed  with  the  beta distribution function $(D-1)(1-p_s)^{D-2}$. In this case the right hand side in  (\ref{eq:xeb_le}) equals $p_c^m(2D/(D+1)-1)$.

The experimental average on the left hand side of (\ref{eq:xeb_le}) can be
estimated with accuracy $1/\sqrt{S N_s}$ using $S$ random circuit realizations with
$N_s$ samples each
\begin{multline} \label{eq:xeb_exp}
 \frac{1}{S N_s} \sum_{j=1}^{S}\sum_{i=1}^{N_s} \left( D p_s^j(q_{i,j}) -1\right) \\= \overline{\avg{  D p_s(q) - 1}} +
 O\left( \frac{1} {\sqrt{S N_s}} \right)\;.
\end{multline}
This gives an estimate of $p_c^m$.

This estimate can be justified using Bayes rule.
The log-likelihood for a set of experimental measurements $\{q_{i,j}\}$ assuming that the experimental probabilities are given by Eq.~\eqref{eq:xeb_dn} is proportional to
\begin{align}
 \sum_{j=1}^{S} \sum_{i=1}^{N_s} \log\left(1 + p_c^m (D p_s^j(q_{i,j})-1)\right)\;,
\end{align}
where $p_s^j(q)$ is a simulated probability corresponding to the $j$-th circuit realization.
We want to maximize the log-likelihood as a function of $p_c^m$. Taking the derivative with respect to $p_c^m$ and equating to $0$ we obtain
\begin{align}
 \sum_{j=1}^{S}\sum_{i=1}^{N_s} {D p_s^j(q_{i,j})-1 \over 1 + p_c^m (D p_s^j(q_{i,j})-1)} = 0\;,
\end{align}
For $p_c^m \ll 1$ it is easy to solve this equation and obtain the estimate
\begin{align}
  p_c^m \simeq {\sum_{j=1}^{S}\sum_{i=1}^{N_s} \(D p_s^j(q_{i,j})-1\) \over \sum_{j=1}^{S}\sum_{i=1}^{N_s} \(D p_s^j(q_{i,j})-1\)^2} \simeq {\overline{\avg{  D p_s(q) - 1}} \over  D \sum_q \overline{p_s(q)^2}  -1}\;.
\end{align}

In the spirit of the XEB method, we can use other functions $f(p_s(q))$ to estimate $p_c^m$. One alternative is derived from the log-likelihood of a sample $\{q_{i,j}\}$ with respect to the simulated (computed) ideal probabilities
\begin{align}
  \log  \Pi_{j=1}^{S} \Pi_{i=1}^{N_s} p_s^j(q_{i,j}) =\sum_{j=1}^{S}\sum_{i=1}^{N_s} \log p_s^j(q_{i,j})\;,
\end{align}
which converges to the cross entropy between experimental probabilities and simulated probabilities. The experimental average of the function $f(p_s(q))= \log p_s(q)$ under the probabilities from Eq.~\eqref{eq:xeb_dn} with additional averaging over random circuits is
\begin{align}
  \overline{\avg{\log p_s(q)}} &\simeq p_c^m \left(\sum_q  \overline{(p_s(q)-1/D)\log p_s(q)}\right) \nonumber \\
  &+ \frac 1 D \sum_q \overline{\log p_s(q)}\;.\label{eq:xeb_xe}
\end{align}
As before, the sums on the right hand side  can be obtained with numerical simulations and the average value on the left hand side can be estimated  experimentally. This also gives an estimate of $p_c^m$.

Both Eq.~\eqref{eq:xeb_le} and Eq.~\eqref{eq:xeb_xe} give a linear
equation, from which we can obtain an estimate of the total
polarization $p_c^m$ for an experimental implementation of one quantum
circuit with $m$  cycles. We normally use mutiple circuits with the
same number of cycles $m$ to estimate $p_c^m$, which we can do using
the least squares method. Finally, we obtain an estimate of $p_c$ from a fit of the estimates $p_c^m$ as an exponential decay in $m$. This is standard in randomized
benchmarking~\cite{magesan_characterizing_2012,magesan_robust_2011}. One advantage of this method is
that it allows us to estimate the cycle polarization $p_c$ independently
of the
state preparation and measurement errors (SPAM).
See also below.

\subsection{XEB of a large number of qubits}\label{sec:xeb_large}

We now consider the case of a large number of qubits $n \gg 1$. We are typically interested in estimating the fidelity
$F$ of each of a set of circuits with a given number of qubits
and depth.
As above, we write the output of an approximate implementation of the random quantum circuit $U$ as
\begin{align}
  \rho_U = F \ket{\psi_U} \bra{\psi_U} + (1-F) \chi_U\;,\label{eq:xeb_ru}
\end{align}
where $\ket{\psi_U}$ is the ideal output and $F = \bra{\psi_U} \rho_U
\ket{\psi_U}$ is the fidelity. We do not necessarily assume $\chi_U = I /
D$, and we will ignore the small difference, of
order $2^{-n}$, $n \gg 1$, between the
fidelity $F$ and the depolarization  fidelity  $p$.

\begin{figure}
\includegraphics[width=\columnwidth]{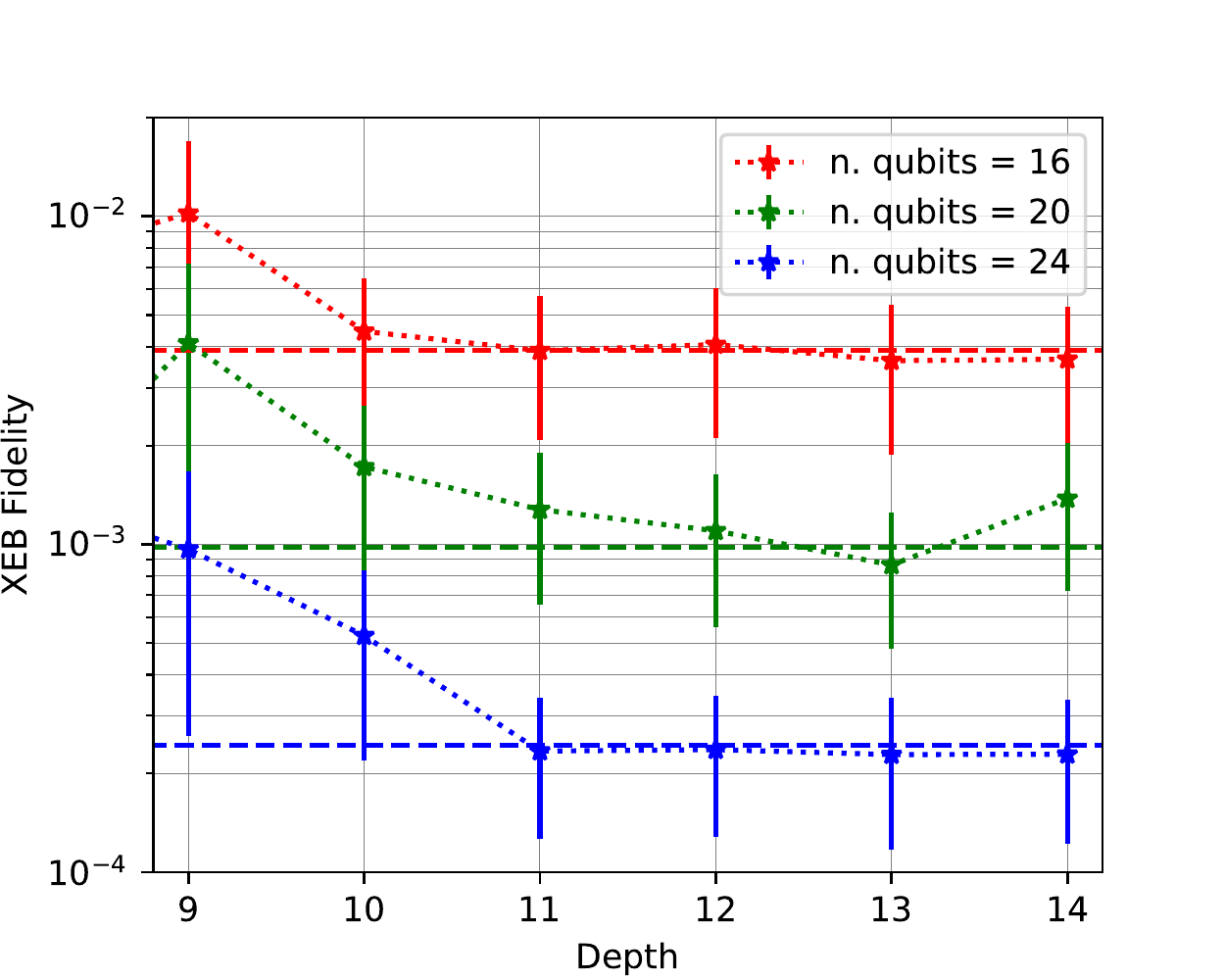}
\caption{\label{fig:xeb_puali_error} Absolute value of the XEB fidelity between a random quantum circuit and the same circuit with a single Pauli error. Markers show the median over all possible positions in the circuit for both bit-flip and phase-flip errors. Error bars correspond to the first and third quartile. The dashed lines are the $1/\sqrt{D}$ theory prediction.}
\end{figure}

As for the case of small number of qubits $n$, we map each output bitstring $q$ with a
function $f(p_s(q))$. Given that the values $\bra{q}
\chi_U\ket{q}$ resulting from errors are typically uncorrelated with the
chaotic ``speckles'' of $p_s(q)$, we make our main assumption
\begin{align}
  \sum_q   \bra{q} \chi_U\ket{q} f(p_s(q)) = \frac 1 D \sum_q
  f(p_s(q)) +\epsilon\;.\label{eq:xeb_cm}
\end{align}
This equation is trivial if we assume a depolarizing model, $\chi_U
= I/D$. More generally, it can be understood in the geometric context of concentration of
measure~\cite{Popescu2006entanglement,bremner2009random,Gross2009most,mcclean2018barren}
for high dimensional spaces, and from Levy's
lemma~\cite{ledoux2005concentration} we expect a typical statistical
fluctuation $\epsilon \in O(1/\sqrt
D)$ with $D=2^n$. We will only require $\epsilon \ll F$. We check
Eq.~\eqref{eq:xeb_cm} numerically for the output $\rho_e =
\ket{\psi_e}\bra{\psi_e}$ where $\ket{\psi_e}$ is the wave function
obtained after a single phase-flip or bit-flip
error is added somewhere in the circuit, see Fig.~\ref{fig:xeb_puali_error} and Ref.~\cite{boixo2018characterizing}. We
have also tested this assumption numerically comparing the fidelity
with the XEB estimate for a pure state $\sqrt{F}\ket{\psi_U} +
\sqrt{1-F}\ket{\psi_\perp}$, see also
Ref.~\cite{markov_quantum_2018} and Section \ref{sec:classical_sim}.

From Eqs.~\eqref{eq:xeb_ru} and \eqref{eq:xeb_cm} we obtain
Eq.~\eqref{eq:xeb_le} for linear XEB, $f(p_s(q)) = D p_s(q)
- 1$ ($\xebfidelity$ in the main text). We also obtain Eq.~\eqref{eq:xeb_xe} for XEB, $f(p_s(q))= \log
p_s(q)$, with  $p_c^m$ replaced by fidelity $F$. As before,
the sums on the right hand side  can be obtained with numerical simulations
and the average value on the left hand side can be estimated
experimentally with accuracy $1/\sqrt{N_s}$ using $N_s$ samples. This
gives an estimate of $F$.

\begin{figure}
\includegraphics[width=\columnwidth]{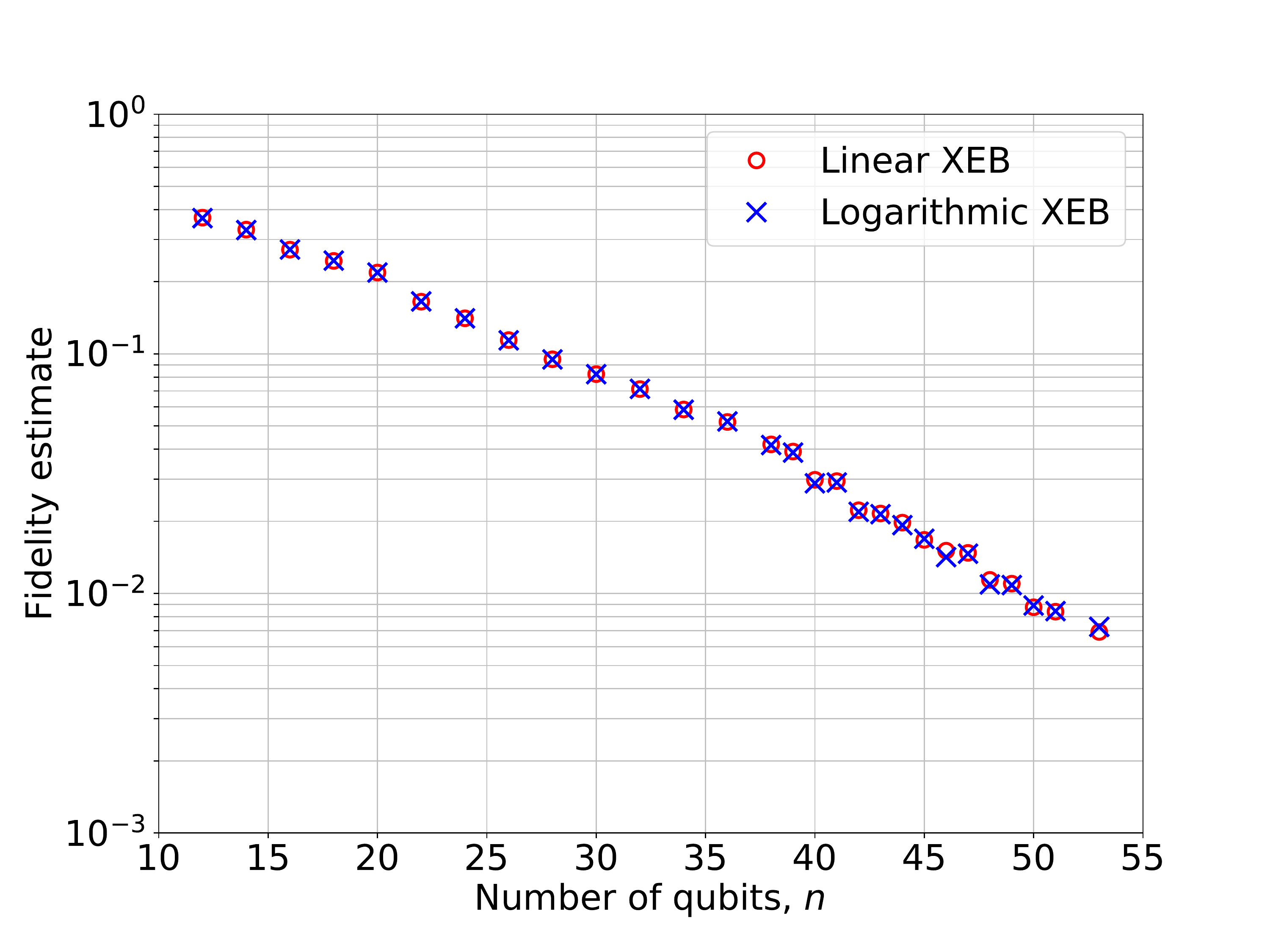}
\caption{\label{fig:linear_xeb_vs_log_xeb} Comparison of fidelity estimates obtained using linear XEB, Eq.~\eqref{eq:linear_xeb_fidelity} and logarithmic XEB, Eq.~\eqref{eq:log_xeb_fidelity} from bitstrings observed in our experiment using elided circuits (see Section \ref{subsubsection:gate_elision}). Standard deviation smaller than markers.}
\end{figure}

In practice,  circuits of enough depth (as in the
experiments reported here) exhibit the Porter-Thomas distribution for
the measurement probabilities $p=\{p_s(q)\}$, that is
\begin{align}\label{eq:porter_thomas}
  \Pr(p) = D e^{-D p}\;.
\end{align}
In this case the linear cross entropy Eq.~\eqref{eq:xeb_le} gives
\begin{align} \label{eq:linear_xeb_fidelity}
  F =   \avg{  D p_s(q) - 1} \;.
\end{align}
The standard deviation of the estimate of $F$ with $N_s$ samples
from the central limit theorem is $\sqrt{(1 + 2F -
  F^2)/N_s}$.
The cross entropy Eq.~\eqref{eq:xeb_xe} gives
\begin{align} \label{eq:log_xeb_fidelity}
  F =   \avg{  \log D p_s(q) } + \gamma \;,
\end{align}
where $\gamma$ is the Euler-Mascheroni constant $\approx$ 0.577. The standard deviation of the
estimate of $F$ with $N_s$ samples is
$\sqrt{(\pi^2/6 - F^2)/N_s}$. The logarithmic XEB has a smaller standard deviation for $F>0.32$ (it is the best estimate when $F\approx 1$), while for $F<0.32$ the linear XEB has a smaller standard deviation (it is the best estimate for $F \ll 1$, where it relates to the maximum likelihood estimator). See Fig.~\ref{fig:linear_xeb_vs_log_xeb} for comparison of the fidelity estimates produced by the linear and logarithmic XEB.

\begin{figure}
\includegraphics[width=\columnwidth]{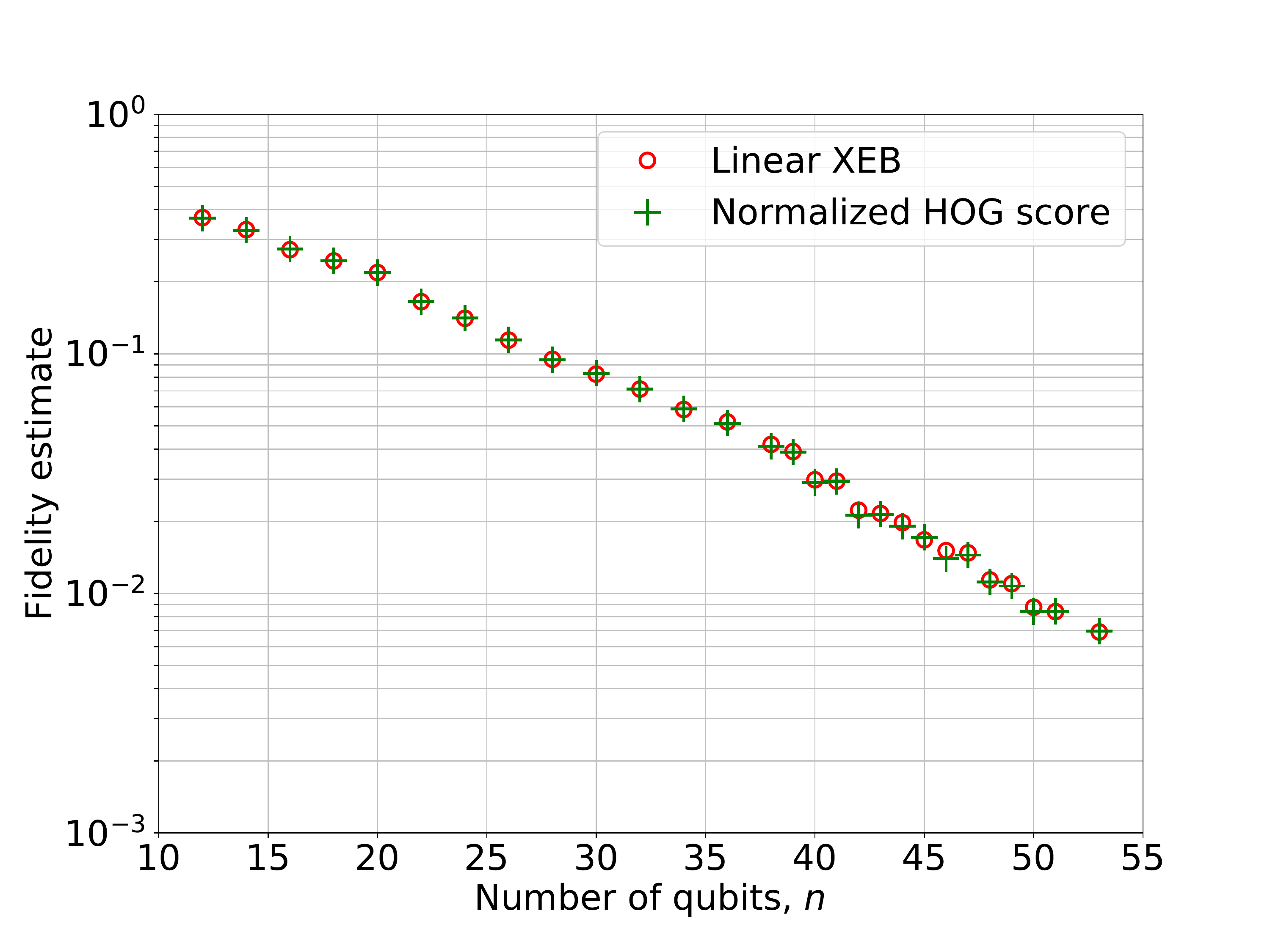}
\caption{\label{fig:linear_xeb_vs_hog} Comparison of fidelity estimates obtained using linear XEB, Eq.~\eqref{eq:linear_xeb_fidelity} and normalized HOG score, Eq.~\eqref{eq:hog_fidelity} from bitstrings observed in our experiment using elided circuits (see Section \ref{subsubsection:gate_elision}). Standard deviation smaller than markers.}
\end{figure}

We note in passing another example for an estimator of $F$ related to
the HOG test~\cite{aaronson2017complexity} which counts the number of measured bitstrings
with probabilities $p_s(q)$ greater than the median of the
probabilities. The function $f(p_s(q))$ in this case returns 1 for
$D p_s(q) \ge \log(2)$, and 0 in the other case. The fidelity estimator uses the following normalization
\begin{align} \label{eq:hog_fidelity}
  F =   \frac{1}{\log(2)} \avg{ 2 n_s(q) - 1} \;,
\end{align}
where $n_s(q)$ is defined to be 1 if $D p_s(q) \ge \log(2)$, and 0 otherwise. The standard deviation of this estimator is $\sqrt{ [\log ^{-2} (2)-F^2]/N_s}$, which is always larger than for the XEB. See Fig.~\ref{fig:linear_xeb_vs_hog} for comparison of the fidelity estimates produced by linear XEB and the HOG-based fidelity estimator. HOG test is also related to a definition of quantum volume~\cite{cross2018validating}. 

\subsection{Two limiting cases}\label{sec:limiting_cases}
Here, we consider two special cases of Eq.~\eqref{eq:linear_xeb_fidelity} and the formula (1) in the main paper \cite{arute2019}. First, suppose bitstrings $q_i$ are sampled from the uniform distribution. In this case the sampling probability is $1/D$ for every bitstring and $\xebfidelity = 0$. Therefore, if the qubits are in the maximally mixed state, the estimator yields zero fidelity, as expected.

Second, suppose that bitstrings are sampled from the theoretical output distribution of a random quantum circuit. Assume that the distribution has Porter-Thomas shape. By Eq.~\eqref{eq:porter_thomas}, the fraction of bitstrings with theoretical probability in $[p, p+dp]$ is

\begin{equation}
\Pr(p) \, dp = D e^{-Dp} dp
\end{equation}
and the total number of such bitstrings is

\begin{equation}
N(p) \, dp = D^2 e^{-Dp} \, dp.
\end{equation}
Therefore, the probability that a bitstring with probability in $[p, p+dp]$ is sampled equals

\begin{equation}
p \cdot N(p) \, dp = p D^2 e^{-Dp} \, dp = f(p) \, dp
\end{equation}
where $f(p)$ is the probability density function of the random variable defined as the ideal probability of a sampled bitstring, i.e. the random variable which is being averaged in the formula (1) of the main paper. Thus, the average probability of a sampled bitstring is

\begin{align}
\begin{split}
\langle p_s(q)\rangle = \int_0^1p f(p) \, dp = \int_0^1 p^2 D^2 e^{-Dp} \, dp \\
= \frac{2}{D}\left(1 - e^{-D} \left(\frac{D^2}{2} + D + 1\right)\right) \approx \frac{2}{D}.
\end{split}
\end{align}
Substituting into equation (1) in the main paper yields $\xebfidelity = 1$. The general case of a depolarizing error can be obtained from the two limiting cases by convex combination.

\subsection{Measurement errors}\label{sec:me}
We now consider how measurement errors affect the estimation of
fidelity.
Let us assume uncorrelated classical measurement errors, so that
if the ``actual'' measurement result of a qubit is 0, we can get 1 with
probability $e_{m0}$, and similarly with probability $e_{m1}$
we get 0 for actual result 1, i.e., $p(1|0)=e_{m0}$,
$p(0|0)=1-e_{m0}$, $p(0|1)=e_{m1}$, $p(1|1)=1-e_{m1}$. In
this case the probability to get measurement result $q=k_1 k_2 .. k_n$
for actual result $q'=k_1' k_2' .. k_n'$ is the product of the
corresponding factors. The probability of correct measurement result is then
 \begin{align}
 p_{\rm m} (q') &= (1-e_{m0})^{n-|q'|} (1-e_{m1})^{|q'|}  \nonumber \\&\approx (1-e_{m0})^{n/2} (1-e_{m1})^{n/2},
    \label{meas-good}
\end{align}
where $|q'|$ is the number of 1s (Hamming distance from 00..0) in
the initial bitstring $q'$, and in the second expression we approximated $|q'|$ with $n/2$ for large $n$.

Now let us make a natural assumption that if there was one or more
measurement errors, $q'\to q$, then the resulting ideal probability
$p_s(q)$ is uncorrelated with the
actual ideal probability $p_s(q')$. Using this assumption we
can write
\begin{align}\label{eq:au}
  F = F_U   p_{\rm m}
\end{align}
where  $F_U$ is the circuit fidelity and $F$ is the complete (effective) fidelity. The complete fidelity $F$ is estimated as
before. The measurement fidelity $p_{\rm m}$ can be
obtained independently. For instance, we can prepare a bistring $q$
and measure immediately to obtain the probability of a correct  measurement result for $q$. We obtain $p_{\rm m}$ by repeating this for
a set of random bitstrings. We can therefore obtain $F_U$ from
Eq.~\eqref{eq:au}. As explained above, fitting the depolarization
fidelity per cycle $p_c$ for different circuit depths $m$ is also a
method to separate measurement errors.

The state preparation errors can be treated similarly, assuming that a single error leads to uncorrelated resulting distribution $p_s(q)$, so that the measurement fidelity $p_{\rm m}$ in Eq.\ (\ref{eq:au}) is combined with a similar factor describing the state preparation fidelity.

\section{Quantifying errors}\label{sec:computing_errors}

An important test for this experiment is predicting XEB fidelity $\xebfidelity$ based on simpler measurements of single- and two-qubit errors.  Here we review how this is calculated, illustrating important principles with the example of a single qubit.  The general theory is described at the end of this section.  

First, we assume Pauli errors describe decoherence using a depolarizing model.  This model is used, for example, to compute thresholds and logical error rates for error correction.  The parameter describing decoherence in a single qubit is the Pauli error $e_P$, giving a probability $e_P/3$ for applying an erroneous X, Y, or Z gate to the qubit after the gate, corresponding to a bit and/or phase flip. 

Second, the depolarization model is assumed to describe the system state using simple classical probability.  The probability of no error for many qubits and many operations, corresponding to no change to the system state, is then found by simply multiplying the probability of no error for each qubit gate.  This is a good assumption for RB and XEB since a bit- or phase-flip error effectively decorrelates the state.  The depolarization model assumes that when there is an error with probability $e_d$, the system state randomly splits to all qubits states, which has Hilbert space dimension $D=2^n$.  This is described by a change in density matrix $\rho \rightarrow (1-e_d)\rho + e_d\times \openone/D$.  Note the depolarization term has a small possibility of the state resetting back to its original state.  For a single qubit where $D=2$, this can be described using a Pauli-error type model as a probability $e_d/4$ applying a I, X, Y, or Z gate.  Comparing to the Pauli model, the error probability thus needs to be rescaled by $e_d = e_P/(1-1/D^2)$.  This gives a net polarization $p$ of the qubit state due to many Pauli errors as 
\begin{equation}\label{eq:error_prod}
p=\prod_i \left[1-e_P(i)/\left(1-1/D^2\right)\right]. 
\end{equation} 

Third, the effect of this depolarization has to be accounted for considering the measured signal.   The measured signal for randomized benchmarking is given by $RB = p(1-1/D) + 1/D$, which can be understood in a physical argument that a complete randomization of the state has a 1/D chance to give the correct final state.  A cross-entropy benchmarking measurement gives $\xebfidelity = p$.  A measurement of $p$, which can have offsets and prefactors in these formulas, also includes other scaling factors coming from state preparation and measurement errors.  All of these scaling issues are circumvented by applying gates in a repeated number of cycles $m$ such that $p=p_c^m$.  A measurement of the signal versus $m$ can then directly pull out the fractional polarization change per cycle, $p_c$, independent of these scale factors.  

Fourth, from this polarization change we can then compute the Pauli error, which is the metric that should be reported since it is the fundamental error rate that is independent of $D$.  Unfortunately, a fidelity $1-e_P/(1+1/D)$ for RB is commonly reported, which has a $D$-dependent correction.  We recommend this practice be changed, but note that removing the $1/(1+1/D)$ factor decreases the reported fidelity value.  We also recommend reporting Pauli error, $e_P$ instead of entanglement fidelity $(1-e_P)$, since it is more intuitive to understand how close some quantity is to 0 than to 1. Table~\ref{table:error} summarizes the different error metrics and their relations.

\begin{table*}[t]
\caption{\label{table:error}
 A “Rosetta stone” translation between error metrics. In single- and two-qubit RB or XEB experiments, we measure the per-gate (or per-cycle) depolarization decay constant $p$. The second column shows conversions from this rate to the various error metrics. The last two columns are representative comparisons for 0.1\% Pauli error. 
}
\begin{ruledtabular}
\begin{tabular}{lcccc}
\textrm{Error metric}&
\textrm{Relation to depolarization decay constant $p$}&
\textrm{n=1 (D=2)}&
\textrm{n=2 (D=4)}\\
\colrule
Pauli error ($e_p$, $r_P$) ~\footnote{$1 - $ process fidelity, or $1 - $ entanglement fidelity}
 & $(1 - p) (1 - 1 / D^2)$ & 0.1\% & 0.1\% \\
Average error ($e_a$, $r$)
 & $(1 - p) (1 - 1 / D) $ & 0.067\% & 0.08\%\\
Depolarization error ($e_d$)
 & $1 - p$ & 0.133\% & 0.107\% \\
\end{tabular}
\end{ruledtabular}
\end{table*}

This general model can also account for non-depolarizing errors such as energy decay, since quantum states in an algorithm typically average over the entire Bloch sphere (as in XEB), or for example when the algorithm purposely inserts spin-echoes.  Thus the average effect of energy decay effectively randomizes the state in a way compatible with Pauli errors.  For a gate of length $t_g$ with a qubit decay time $T_1$, averaging over the Bloch sphere (2 poles and 4 equator positions) gives (to first order) an average error probability $e_a = t_g/3T_1$.  Using Table~\ref{table:error}, this converts to a Pauli error $e_P=t_g/2T_1$.

A detailed theory of the $D$ scaling factor is as follows.  In order to arrive at a first order estimate on how error rates accumulate on random quantum circuits, the errors can be modeled via the set of Kraus operators. The density matrix of the system $\rho$ after application of a gate is connected to the density matrix  $\rho_0$ before the gate as follows:
\begin{equation}
\rho=\Lambda(\rho_0)=\sum_{k=0}^{K} A_k \rho_0 A_k^\dagger,\quad \sum_{k} A_k^\dagger A_k=\openone.\label{eq:Kraus}
\end{equation}
For the  closed-system quantum evolution with unitary $U$ (no dephasing nor decay) the sum on the right hand side contains only one term with $k$=0 and  $A_0=U$. In general, Kraus operators describe the physical effects of many types of errors (control error, decoherence, etc.) that can explicitly depend on the gate.  Knowing the Kraus operators allows us to calculate the total  error budget as well as its individual components.

Conventionally, circuit fidelities are reported as a metric of its quality. To make a connection to physically observable quantities, the average fidelity can be expressed in terms of Kraus operators. In the absence of leakage errors and cross-talk the average fidelity equals
\begin{align}
F=1-\frac{e_P}{1+1/D},\quad e_P=1-\frac{1}{D^2}\sum_{k=0}^{K} |\tr(U A_k^\dagger)|^2\label{eq:FA}
\end{align}
where $D=2^n$ is the dimension of the Hilbert space and the quantity $e_P$ plays a role of a Pauli error probability in the depolarizing channel model (see below).
 
For random circuits the effects of errors can be described by a depolarizing channel model, with Kraus operators of the form
\begin{align}
A_{\bf k} &=\sqrt{\frac{ e_P}{D^2-1}} \,  P_{\bf k} U,\quad {\bf k} \neq {\bf 0},\label{eq:DPC}\\
A_{\bf 0} &=\sqrt{1-e_P} \,  P_{\bf 0} U,\nonumber\\
P_{\bf k} &=\sigma_{k_1}\otimes\sigma_{k_2}\ldots\otimes \sigma_{k_n}\nonumber
\end{align}
where $P_{\bf k}$ are strings of Pauli operators $\sigma_{k_j}$  for individual qubits for  $k_j=1,2,3$ and also identity matrices $\sigma_{0}$  in the qubit subspace for $k_j=0$.
This form assumes that individual Pauli errors all happen with the same probability $e_P$.

To make a connection to experimental measurements of the cross-entropy we substitute (\ref{eq:DPC}) into (\ref{eq:Kraus}) and obtain
\begin{multline}
\Lambda(\rho_0)=(1-e_P)U\rho_0 U^{-1}\\+\frac{e_P}{D-1/D}\left(\openone-\frac{ U \rho_0 U^{-1}}{D} \right).\label{eq:DP1}
\end{multline}
We compare this expression with the standard form of the depolarizing channel model 
\begin{align}
  \Lambda(\rho_0) = p U \rho_0 U^{-1} + (1-p) \frac \openone D\;,\label{eq:DP2}
\end{align}
expressed in terms of the depolarization fidelity parameter $p$. Note the difference between the expressions. On the one hand, in (\ref{eq:DP2}) the second term corresponds to full depolarization in all directions. On the other hand, in (\ref{eq:DP1}) the second term describes full depolarization in all directions except for the direction corresponding to the ideal quantum state.

From (\ref{eq:DP1}), (\ref{eq:DP2}) one can establish the connection between the Pauli error rate and depolarizing fidelity parameter $p$
\begin{equation}
e_P=(1-p)(1-1/D^2) 
\end{equation}

We note that the explicit assumption of connecting Pauli errors to depolarization is needed for the small $D$ case, typically for single- and two-qubit error measurements.  Once we have measured the Pauli errors, then only a simple probabilistic calculation is needed to compute $\xebfidelity$ in the large $D$ case.

\section{Metrology and calibration}\label{sec:calib_metro}

\subsection{Calibration overview}
Quantum computations are physically realized through the time-evolution of quantum systems steered by analog control signals. As quantum information is stored in continuous amplitudes and phases, these control signals must be carefully chosen to achieve the desired result. Calibration is the process of performing a series of experiments on the quantum system to learn optimal control parameters. 

Calibration is challenging for a number of reasons.  Analog control requires careful control-pulse shaping as any deviation from the ideal will introduce error. Qubits require individual calibration as variations in the control system and qubits necessitate different control parameters to hit target fidelities. Optimal control parameters can also drift in time, requiring calibrations to be revisited to maintain performance. Additionally, the full calibration procedure requires bootstrapping: using a series of control sequences with increasing complexity to determine circuit and control parameters to increasingly higher degrees of precision. Lastly, each qubit needs to perform a number of independent operations which are independently calibrated: single-qubit gates, two-qubit gates, and readout. 

Our Sycamore processor offers a high degree of programmability: we can dynamically change the frequency of each qubit, as well as the effective qubit-qubit coupling between nearest neighbor qubits. This tunability gives us the freedom to enact many different control strategies, as well as account for non-uniformities in the processor's parameters. However, these extra degrees of freedom are a double-edged sword. Additional control knobs always introduce a source of decoherence and control errors as well as an added burden on calibration. 

Our approach is to systematize and automate our calibration procedure as much as possible, thus abstracting complexity away. This automation allows us to turn calibration into a science, where we can compare calibration procedures to determine optimal strategies for time, performance, and reliability. By employing calibration science to study full-system performance with different control strategies, we have been able to improve full-system fidelities by over an order of magnitude from initial attempts while decreasing the calibration time and improving reliability. Lastly, we design our calibration to be done almost entirely at the single- or two-qubit level, rather than at the system level, in order to be as scalable as possible. 

\subsubsection{Device registry}
The device registry is a database of control variables and configuration information we use to control our quantum processors. The registry stores information such as operating frequencies, control biases, gate parameters such as duration, amplitude, parameterization of circuit models, etc. The goal of calibration is to experimentally determine and populate the registry with optimal control parameters. We typically store \textgreater100 parameters per qubit to achieve high fidelity across all of the various qubit operations. The large number of parameters and subtle interdependencies between them highlights the need for automated calibration. 

\subsubsection{Scheduling calibrations: ``Optimus"}
\begin{figure}
\includegraphics{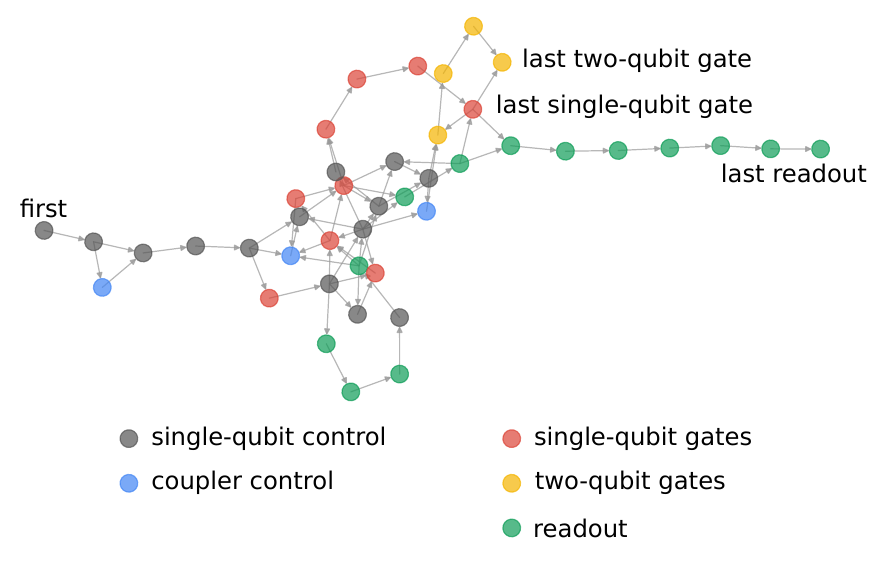}
\caption{\label{fig:optimus_graph} \textbf{Optimus calibration graph for Sycamore.} Calibration of physical qubits is a bootstrapping procedure between different pulse sequences or ``experiments" to extract control and system parameters. Initial experiments are coarse and have interplay between fundamental operations and elements such as single-qubit gates, readout, and the coupler. Final experiments involve precise metrology for each of the qubit operations: single-qubit gates, two-qubit gates, and readout. }
\end{figure}

We seek a strategy for identifying and maintaining optimal control parameters for a system of physical qubits given incomplete system information. To perform these tasks, we use the ``Optimus" formulation as in Ref~\cite{jkelly2018physical}, where each calibration is a node in a directed acyclic graph that updates one or more registry parameters, and the bootstrapping nature of calibration sequences is represented as directed edges between nodes. Now, calibrating a system of physical qubits becomes a well-defined graph traversal problem. The calibration graph used for the Sycamore device can be see in Figure~\ref{fig:optimus_graph}. This strategy is particularly useful for maintaining calibrations in the presence of drift, where we want to do the minimal amount of work to bring the system back in spec, and when extending the calibration procedure, as interdependencies are explicit. Typical timescales for bringup of a new Sycamore processor are approximately 36 hours upon first cooldown, and 4 hours per day thereafter for maintaining calibrations. These times are specific to current available technology, and can be significantly improved. 

\subsection{Calibration procedure}
\subsubsection{Device configuration}

\begin{figure}
\includegraphics{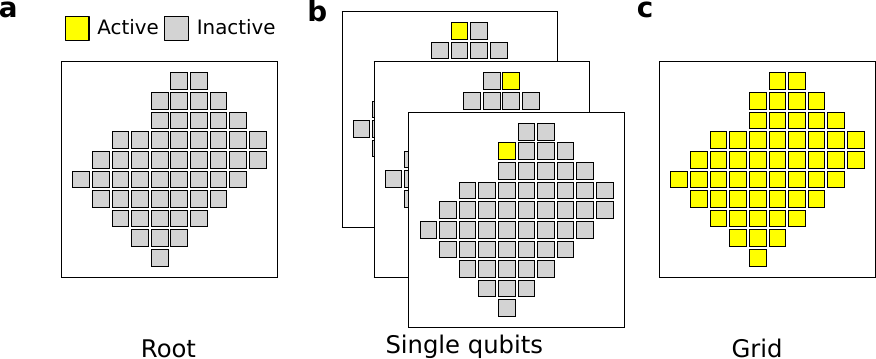}
\caption{\label{fig:device_configs}  \textbf{Configurations of the device over the course of calibration.} \textbf{a,} In the root configuration, we start with no knowledge of the system and measure basic device parameters. \textbf{b,} We create a single qubit configuration for each qubit, where all qubits except the qubit of interest are biased to near zero frequency. \textbf{c,} Using knowledge learned in the single qubit configurations, we build a grid of qubits. }
\end{figure}

Throughout the calibration procedure, the device registry may be configured in different states in order to calibrate certain parameters. We call these different states ``device configurations", and different kinds of configurations reflect our knowledge of the system at different points in the full calibration procedure. As illustrated in Figure~\ref{fig:device_configs}, the primary difference between the different configurations is the set of ``active" qubits, where active qubits are biased to an operating frequency between 5-7\,GHz, and ``inactive" qubits are biased near zero frequency. Following the outline above, we have three device configurations of interest:

\paragraph{Root config.}
The root configuration is the starting state of the system immediately after cool down and basic system verification. In this configuration, we calibrate coarse frequency vs bias curves for each readout resonator, qubit, and coupler.
\paragraph{Single qubit config.} 
After completing root calibrations, we now know how to bias each qubit to its minimum and maximum frequencies. We create one configuration of the device registry for each qubit, where the qubit of interest is biased in a useful region (5-7 GHz) and the remaining qubits are biased to their minimum frequencies in order to isolate the qubit of interest. In each of these configurations, we fine tune the bias vs frequency curves for the qubit and its associated couplers and resonators, and also measure $T_1$ as a function of frequency, necessary due to background TLS defects and modes.
\paragraph{Grid config.}
After completing calibrations in each isolated qubit configuration, we feed the information we learned into a frequency optimization procedure. The optimizer places the biases for each qubit and coupler in a user defined grid of any desired size up to the entire chip. We then proceed to calibrate high fidelity single qubit gates, two qubit gates, and readout.

\subsubsection{Root config: procedure}

We begin calibration with simple frequency-domain experiments to understand how each qubit and coupler responds to its flux bias line. 
\begin{itemize}
\item Calibrate each parametric amplifier (flux bias, pump frequency, pump power).
\item For each qubit, identify its readout resonator and measure the readout signal versus qubit bias (``Resonator Spectroscopy") \cite{Wallraff2004StrongCoupling}. Estimate the resonator and qubit frequency as a function of qubit bias.
\item For each coupler, place one of its qubits near maximum frequency and the other near minimum frequency, then measure the readout signal of the first qubit as a function of coupler bias. The readout signal changes significantly as the coupler frequency passes near the qubit frequency. Identify where the coupler is near its maximum frequency, so the qubit-qubit coupling is small (a few MHz) and relatively insensitive to coupler bias.
\end{itemize}

\subsubsection{Single-qubit config: procedure}
After setting the biases to isolate a single qubit, we follow the procedure outlined in \cite{chen2018metrology} which we will summarize here:
\begin{itemize}
\item Perform fixed microwave drive qubit spectroscopy while sweeping the qubit bias and detecting shifts in the resonator response, to find the bias that places the qubit at the desired resonant frequency.
\item Using the avoided level crossing identified in the root config, determine the operating bias to bring the qubit on resonance with its readout resonator to perform active ground state preparation. We use a 10~$\mu s$ pulse consistent with the readout resonator ringdown time.
\item Perform power Rabi oscillations to find the drive power that gives a $\pi$ pulse to populate the $|1\rangle$ state.
\item Optimize the readout frequency and power to maximize readout fidelity.
\item Fine tune parameters (qubit resonant frequency, drive power, drive detuning \cite{chen2016measuring}) for $\pi$ and $\pi/2$ pulses.
\item Calibrate the timing between the qubit microwave drive, qubit bias, and coupler bias.
\item Perform qubit spectroscopy as a function of qubit bias to fine tune the qubit bias vs frequency curves.
\item Measure $T_1$ vs.\ frequency by preparing the qubit in $|1\rangle$ then biasing the qubit to a variable frequency for a variable amount of time, and measuring the final population \cite{paultls}.
\item Measure the response of a qubit to a detuning pulse to calibrate the frequency-control transfer function \cite{kelly2014optimal, neill2018blueprint, chen2018metrology}.
\end{itemize}

With the single-qubits calibrated in isolation, we have a wealth of information on circuits parameters and coherence information for each qubit. We use this information as input to a frequency placement algorithm to identify optimal operating frequencies for when the full processor is in operation. 

\subsubsection{Optimizing qubit operating frequencies}
\label{sec:opt_freqs}

In our quantum processor architecture, we can independently tune each qubit's operating frequency. Since qubit performance varies strongly with frequency, selecting good operating frequencies is necessary to achieve high fidelity gates.  In arbitrary quantum algorithms, each qubit operates at three distinct types of frequencies: idle, interaction, and readout frequencies. Qubits idle and execute single-qubit gates at their respective idle frequencies. Qubit pairs execute two-qubit gates near their respective interaction frequencies. Finally, qubits are measured at their respective readout frequencies. In selecting operating frequencies, it is necessary to mitigate and make nontrivial tradeoffs between energy-relaxation, dephasing, leakage, and control imperfections. We solve and automate the frequency selection problem by abstracting it into an optimization problem. 

\begin{figure*}
\includegraphics{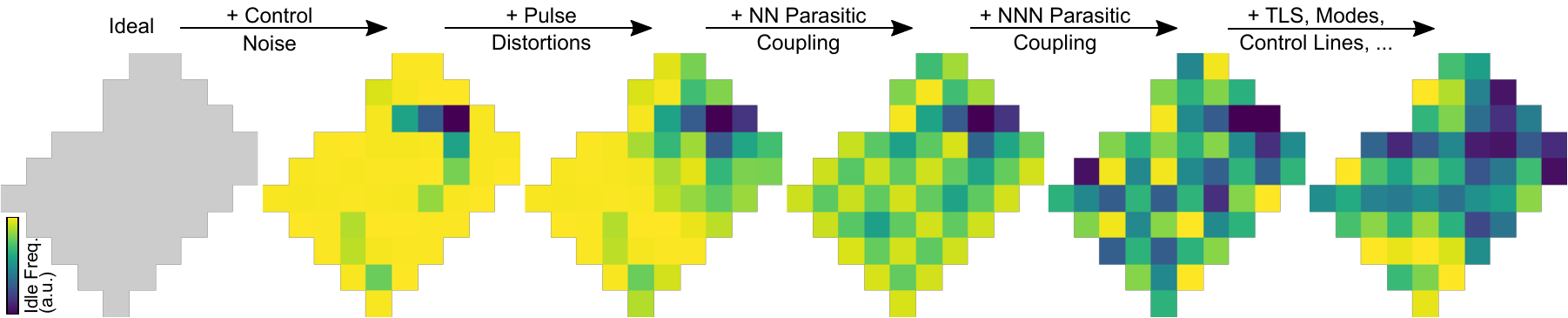}
\caption{\label{fig:snake}  \textbf{Idle frequency solutions found by our Snake optimizer with different error mechanisms enabled.} The optimizer makes increasingly complex tradeoffs as more error mechanisms are enabled. These tradeoffs manifest as a transition from a structured frequency configuration into an unstructured one. Similar tradeoffs are simultaneously made in optimizing interaction and readout frequencies. Optimized idle and interaction operating frequencies are shown in Figure \ref{fig:qubit_freqs} and optimized readout frequencies are shown in Figure \ref{fig:readout_freqs}. Color scales are chosen to maximize contrast. Grey indicates that there is no preference for any frequency.  }
\end{figure*}

\begin{figure}
\includegraphics{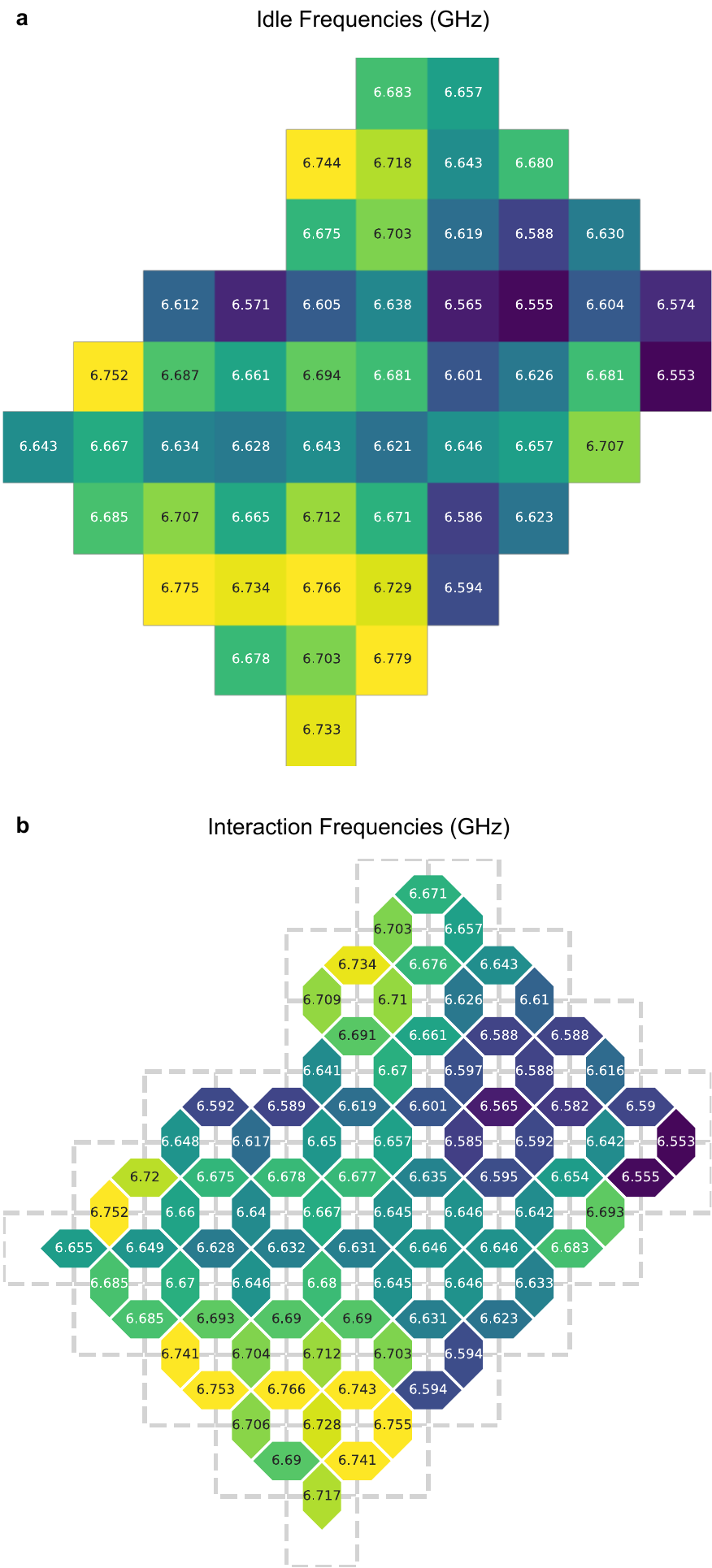}
\caption{\label{fig:qubit_freqs}  \textbf{Optimized idle and interaction frequencies found by our Snake optimizer.} \textbf{a,} Idle frequencies, \textbf{b,} interaction frequencies. Readout frequencies are shown in Figure \ref{fig:readout_freqs}. These solutions are sufficient for state-of-the-art system performance. See Figure \ref{fig:snake} to understand some of the tradeoffs that are made during optimization. Color scales are chosen to maximize contrast. }
\end{figure}

We construct a quantum-algorithm-dependent and gate-dependent optimization objective that maps operating frequencies onto a metric correlated with system error.  The error mechanisms embedded within the objective function are parasitic coupling between nearest-neighbor and next-nearest-neighbor qubits, spectrally-diffusing two-level-system (TLS) defects \cite{paultls}, spurious microwave modes, coupling to control lines and the readout resonator, frequency-control electronics noise, frequency-control pulse distortions, microwave-control pulse distortions, and microwave-carrier bleedthrough. Additional considerations in selecting readout frequencies are covered in Section \ref{sec:readout_cals}. The objective is constructed from experimental data and numerics, and the individual error mechanisms are weighted by coefficients determined either heuristically or through statistical learning. 

Minimizing the objective function is a complex combinatorial optimization problem. We characterize the complexity of the problem by the optimization dimension and search space. For a processor with $N$ qubits on a square lattice with nearest-neighbor coupling, there are $N$ idle, $N$ readout, and $\sim2N$ interaction frequencies to optimize. In an arbitrary quantum algorithm, all frequencies are potentially intertwined due to coupling between qubits. Therefore, the optimization dimension is $\sim4N$. The optimization search-space is constrained by qubits' circuit parameters and control-hardware specifications. Discretizing each qubit's operational range to 100 frequencies results in an optimization search space of $\sim100^{4N}$. This is much larger than the dimension of the Hilbert space of an $N$ qubit processor, which is $2^N$. 

Given the problem complexity, it is assumed that finding globally optimal operating frequencies is intractable.  However, we have empirically verified that locally optimal solutions are sufficient for state-of-the-art system performance. To find local optima, we developed the ``Snake" homebrew optimizer that combines quantum algorithm structure with physics intuition to exponentially reduce optimization complexity and take intelligent optimization steps. For the circuits used here, the optimizer exploits the time-interleaved structure of single-qubit gates, two-qubit gates, and readout. For our 53 qubit processor, it returns local optima in $\sim10$ seconds on a desktop. Because of its favorable scaling in runtime versus number of qubits, we believe the Snake optimizer is a viable long-term solution to the frequency selection problem. 

To illustrate how the Snake optimizer makes tradeoffs between error mechanisms, we plot idle frequency solutions with different error mechanisms enabled (Figure \ref{fig:snake}). Starting with an ideal processor with no error mechanisms enabled, there is no preference for any frequency configuration. Enabling frequency-control electronics noise, the optimizer pushes qubits towards their respective maximum frequencies, to minimize flux-noise susceptibility. Note that each qubit has a different maximum frequency due to fabrication variability. Enabling frequency-control pulse distortions forces a gradual transition between qubit frequencies to minimize two-qubit-gate frequency-sweep amplitudes. Enabling nearest-neighbor (NN) and next-nearest neighbor (NNN) parasitic coupling further lowers the degeneracy between qubit frequencies into a structure that resembles a multi-tiered checkerboard. Finally, enabling errors from TLS defects, spurious microwave modes, and all other known error mechanisms removes any obvious structure. A set of optimized idle and interaction frequencies is shown in Figure \ref{fig:qubit_freqs}, and readout frequencies are shown in Figure \ref{fig:readout_freqs}. 

\subsubsection{Grid config: procedure}
Calibrating a grid of qubits follows the same procedure as calibrating an isolated qubit with additional calibrations to turn off the qubit-qubit coupling. 
\begin{itemize}
\item Achieve basic state discrimination for each qubit at its desired frequency.
\item For each coupler, minimize the qubit-qubit coupling (note changing coupler biases affects qubit frequencies). For each case below, we choose the coupler bias minimizing the interaction.
\begin{itemize}
    \item For qubit pairs idling within 60 MHz of each other, use a resonant swapping experiment. We excite one qubit and apply flux pulses to nominally put the qubits on resonance and let the qubits interact over time \cite{andrew_thesis}.
    \item For qubit pair idling further apart, use a conditional phase experiment. We perform two Ramsey experiments on one qubit, where the other qubit is in the ground state and the excited state, to identify the state-dependent frequency shift of the first qubit.
\end{itemize}
\item Adjust the qubit biases to restore the desired qubit frequencies and proceed with qubit calibration as in the single-qubit configurations.
\item Calibrate the entangling gate.
\begin{itemize}
    \item Estimate the qubit pulse amplitudes to reach the desired interaction frequency with their frequency versus bias calibration.
    \item Fine-tune the qubit pulse amplitudes to reach resonance, compensating for pulse undershoot.
    \item Tune the coupler pulse amplitude to achieve a complete photon exchange.
\end{itemize}
\end{itemize}

In the next two sections, we describe in more detail the fine tuning required to achieve high fidelity two qubit gates and multiqubit readout.

\subsection{Two-qubit gate metrology}
High-fidelity two-qubit gates are very hard to achieve.  In an effort to make this easier, we design qubits with tunable frequencies and tunable interactions.  This added control allows for immense flexibility when implementing gates.  In the following subsections, we discuss a simple high-fidelity control and metrology strategy for two-qubit gates in our system. 

\subsubsection{The natural two-qubit gate for transmon qubits}
Consider two transmon qubits at different frequencies (say 6.0 and 6.1 GHz).  Here are two potential ways of generating a multi-qubit gate in this system.  If the qubits are tuned into resonance, then excitations swap back-and-forth and this interaction can be modeled as a partial-iSWAP gate \cite{bialczak2010quantum}.  If the qubits are detuned by an amount close to their nonlinearity, then the 11-state undergoes an evolution that can be modeled as a controlled-phase gate (assuming the population does not leak) \cite{martinis2014fast, dicarlo2009cz}.  In fact, any two-qubit control sequence that does not leak can be modeled as a partial-iSWAP followed by a controlled-phase gate.  

\begin{figure}
\includegraphics[width=3.2in]{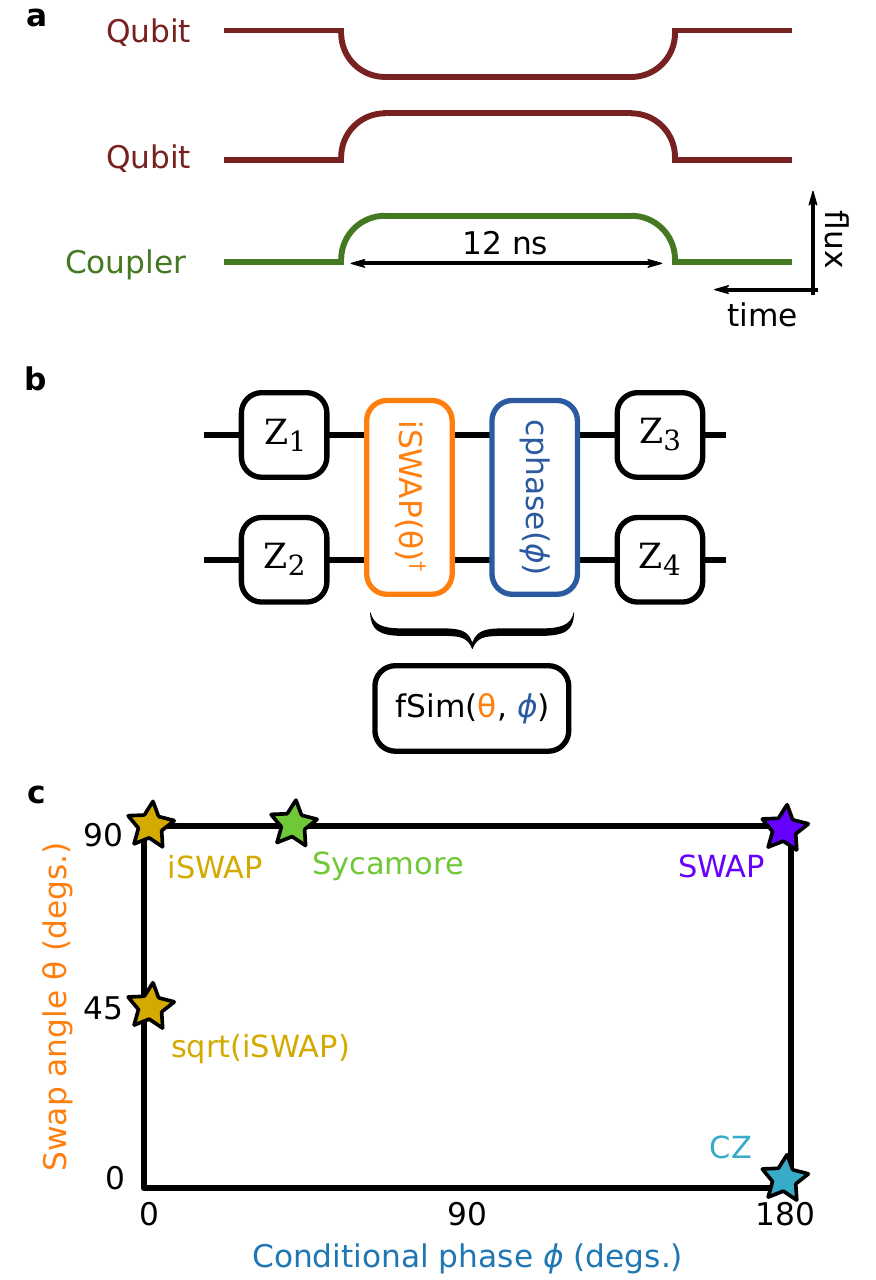}
\caption{\label{fig:two_qubit_gates}  \textbf{Two-qubit gate strategy.}  \textbf{a,} Control waveforms for two qubits and a coupler.  Each curve represents the control flux applied to the qubit's and coupler's SQUID loops as a function of time.  \textbf{b,} Generic circuit representation for an arbitrary two-qubit gate using flux pulses.  This family of gates have been named ``fSim'' gates, short for fermionic-simulation gates. Our definition of the fSim gate uses $\theta$ with the sign opposite to the common convention for the iSWAP gate. \textbf{c,} Control landscape for fSim gates as a function of the swap angle and conditional phase, up to single qubit rotations.  The coordinates of common entangling gates are marked along with the Sycamore gate fSim($\theta = 90^{\circ}, \phi = 30^{\circ}$). 
}

\end{figure}

A typical control sequence is shown Fig.~\ref{fig:two_qubit_gates}a.  Gate times of 12~ns are chosen to trade off decoherence (too slow) and leakage to higher states of the qubit (too fast).  Figure \ref{fig:two_qubit_gates}b depicts how this operation can be decomposed as a quantum circuit.  This circuit contains Z-rotations that result from the frequency excursions of the qubits, and can be expressed by the unitary:
\begin{equation}\label{eq:unitary_angles}
    \begin{bmatrix}
        1 & 0 & 0 & 0 \\
        0 & e^{i(\Delta_++\Delta_-)}\cos\theta & -ie^{i(\Delta_+-\Delta_{-,\mathrm{off}})}\sin\theta & 0 \\
        0 & -ie^{i(\Delta_++\Delta_{-,\mathrm{off}})}\sin\theta & e^{i(\Delta_+-\Delta_-)}\cos\theta & 0 \\
        0 & 0 & 0 & e^{i(2\Delta_+-\phi)}
    \end{bmatrix}.
\end{equation}
These gates have an efficient mapping to interacting fermions and have been coined `fSim' gates, short for fermionic simulation \cite{kivlichan2018quantum}.  The long-term goal is to implement the entire space of gates (shown in Fig.~\ref{fig:two_qubit_gates}c).

For quantum supremacy, the two-qubit gate of choice is the iSWAP gate. For example, CZ is less computationally expensive to simulate on a classical computer by a factor of two \cite{markov_quantum_2018, villalonga2019flexible}.  A dominant error-mechanism when trying to implement an iSWAP is a small conditional-phase that is generated by an interaction of the $|11\rangle$-state with higher states of the transmons ($|02\rangle$ and $|20\rangle$).  For this reason, the fSim gate with swap-angle $\theta \approxeq 90^{\circ}$ and conditional phase $\phi \approxeq 30^{\circ}$ has become the gate of choice in our supremacy experiment.  Note that small deviations from these angles are also viable quantum supremacy gates.  These gates result from the natural evolution of two qubits making them easy to calibrate, high intrinsic fidelity gates for quantum supremacy.

\subsubsection{Using cross entropy to learn a unitary model}
We have recently introduced cross-entropy as a fidelity metric for quantum supremacy experiments.   Cross-entropy benchmarking (XEB) was introduced as an analog to randomized benchmarking (RB) that can be used with any number of qubits and is independent of state-preparation and measurement errors \cite{neill2018blueprint, boixo2018characterizing}.

\begin{figure}
\includegraphics{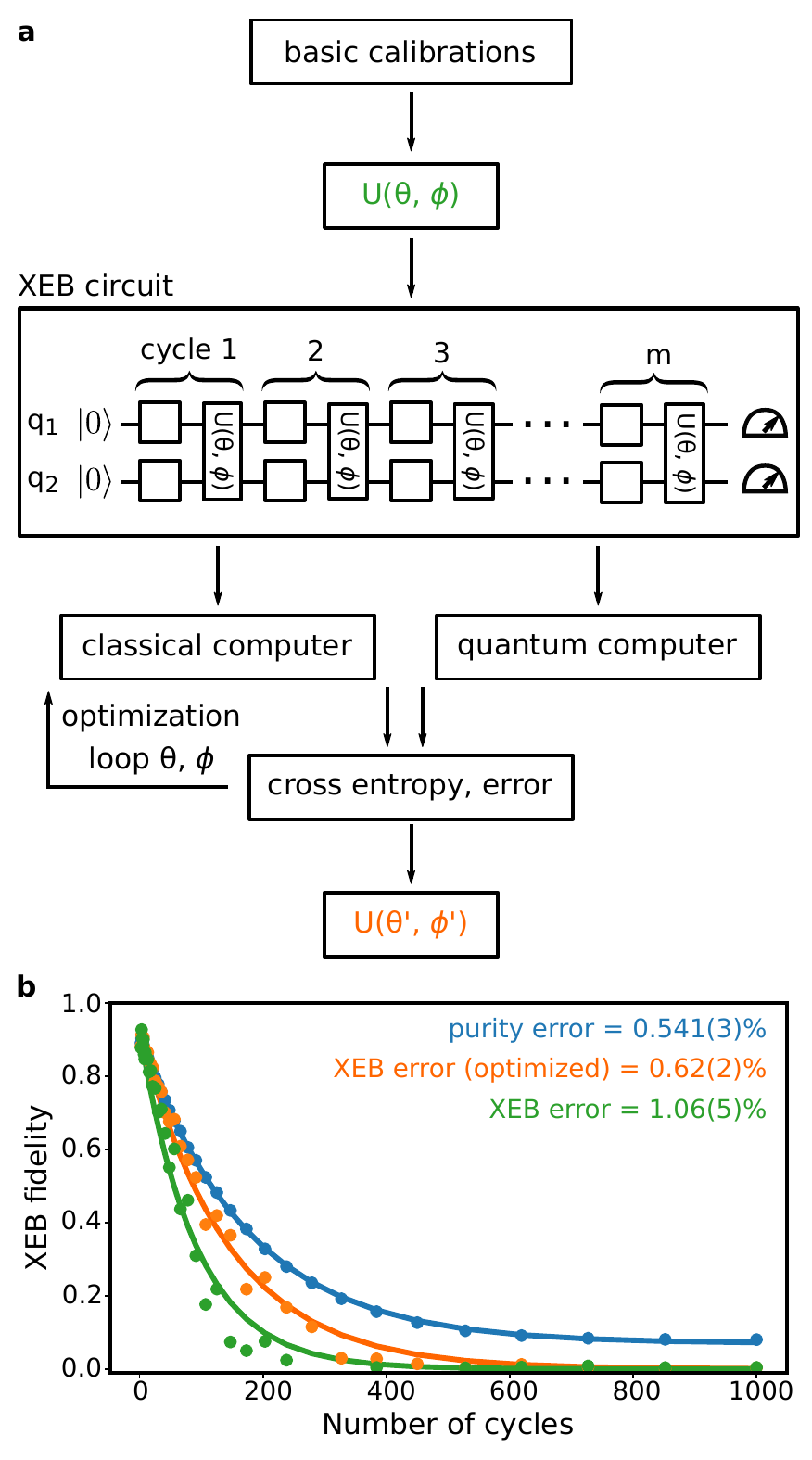}
\caption{\label{fig:xeb_learn_unitary}  \textbf{Using XEB to learn a unitary model.} \textbf{a,} Process flow diagram for using XEB to learn a unitary model. After running basic calibrations, we have an approximate model for our two-qubit gate. Using this gate, we construct a random circuit that is fed into both the quantum computer and a classical computer. The results of both outputs can be compared using cross-entropy.  Optimizing over the parameters in the two qubit model provide a high-fidelity representation of the two-qubit unitary. \textbf{b,} Data from a two-qubit XEB experiment. The two-qubit purity (blue) was measured tomographically and provides the coherence-limit of the operations. The decay of the XEB fidelity is shown in green and orange.  In orange, the parameters of a generic unitary model were optimized to determine a higher-fidelity representation of the unitary. All errors are quoted as Pauli errors.
}
\end{figure}

A distinct advantage of XEB is that the resulting data can be analyzed to find an optimal representation of a unitary; this process is outlined in Fig.~\ref{fig:xeb_learn_unitary}. The gate sequence for a two-qubit XEB experiment is shown in Fig.~\ref{fig:xeb_learn_unitary}a.  The sequence alternates between single-qubit gates on both qubits and a two-qubit gate between them.  At the end of the sequence, both qubits are measured and the probabilities of bitstrings (00, 01, 10, 11) are estimated.  This procedure is repeated for $\sim$10-20 instances of randomly selected single-qubit gates.  The measured probabilities can then be compared to the ideal probabilities using the expression for fidelity Eq.~(3) in Ref.~\cite{neill2018blueprint}.

The data from a two-qubit XEB experiment is shown in Fig.~\ref{fig:xeb_learn_unitary}b (green dots).  By performing additional sequences with tomography rotations prior to measurement, we can infer the decay of purity with increasing circuit depth (blue dots).  For two qubits, the decay of fidelity tells us the total error of our gates while the purity decay tells us the contribution from decoherence \textemdash the difference is control error.  Based on the data in green and blue, it appears that the total error is about half control and half decoherence.  

So far, we have established a generic unitary model (Fig.~\ref{fig:two_qubit_gates}b), a training dataset (Fig.~\ref{fig:xeb_learn_unitary}a), and a cost-function (Fig.~\ref{fig:xeb_learn_unitary}b).  These three ingredients form the foundation for using optimization techniques to improve fidelity.  Using a simple Nelder-Mead optimization protocol, we can maximize the XEB fidelity by varying the parameters of the unitary model.  The fidelity decay curve for the optimal unitary model are shown in Fig.~\ref{fig:xeb_learn_unitary}b (orange dots).  The optimized results are nearly coherence limited.  

\begin{figure}
\includegraphics{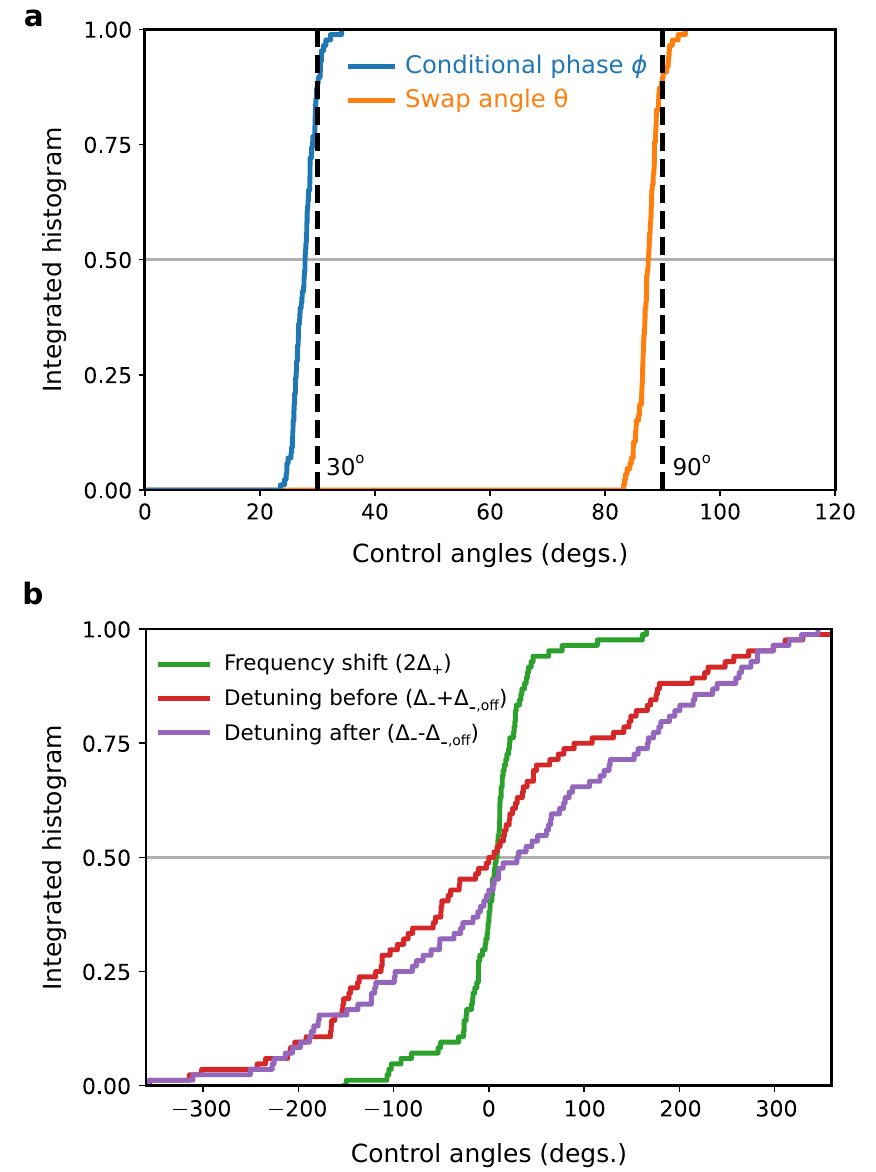}
\caption{\label{fig:params_control_model} \textbf{Parameters of the control model.}  A generic model for two-qubit gates using flux-control has five free parameters.  Using XEB we can measure these parameters with high fidelity.  \textbf{a,}  Integrated histogram (cumulative distribution) of the control parameters that determine the interaction between the qubits.  \textbf{b,}  An integrated histogram of the remaining three parameters that represent different flavors of single-qubit Z-rotations.  While the first two parameters (panel a) define the entangling gate, the final three parameters (panel b) are simply measured and then kept track of during an algorithm. Intuitively, these three angles correspond to a detuning before the swap, a detuning after the swap, and an overall frequency shift which commutes through the swap; these correspond to $\Delta_-+\Delta_{-,\mathrm{off}}$, $\Delta_--\Delta_{-,\mathrm{off}}$, and $2\Delta_+$ respectively in Eq.~(\ref{eq:unitary_angles}). Note that $\theta$ and $\phi$ angles are 360 degrees periodic and Z-rotation angles are 720 degree periodic.
}
\end{figure}

The optimal control-model parameters for all pairs are shown as integrated histograms in Fig.~\ref{fig:params_control_model}a,b.  Panel (a) shows the histograms for partial-iSWAP angles ($\sim$90 degrees)  and conditional phases ($\sim$30 degrees).  Panel (b) shows histograms for the various flavors of Z-rotations.  While conceptually there are four possible Z-rotations (see Fig.~\ref{fig:two_qubit_gates}b), only three of these rotations are needed to uniquely define the operation.  These three rotations can be thought of as the detuning of the qubits before the iSWAP, the detuning after the iSWAP, and an overall frequency shift of both qubits which commutes with the iSWAP.

\subsubsection{Comparison with randomized benchmarking}
\begin{figure}
\includegraphics[width=3.2in]{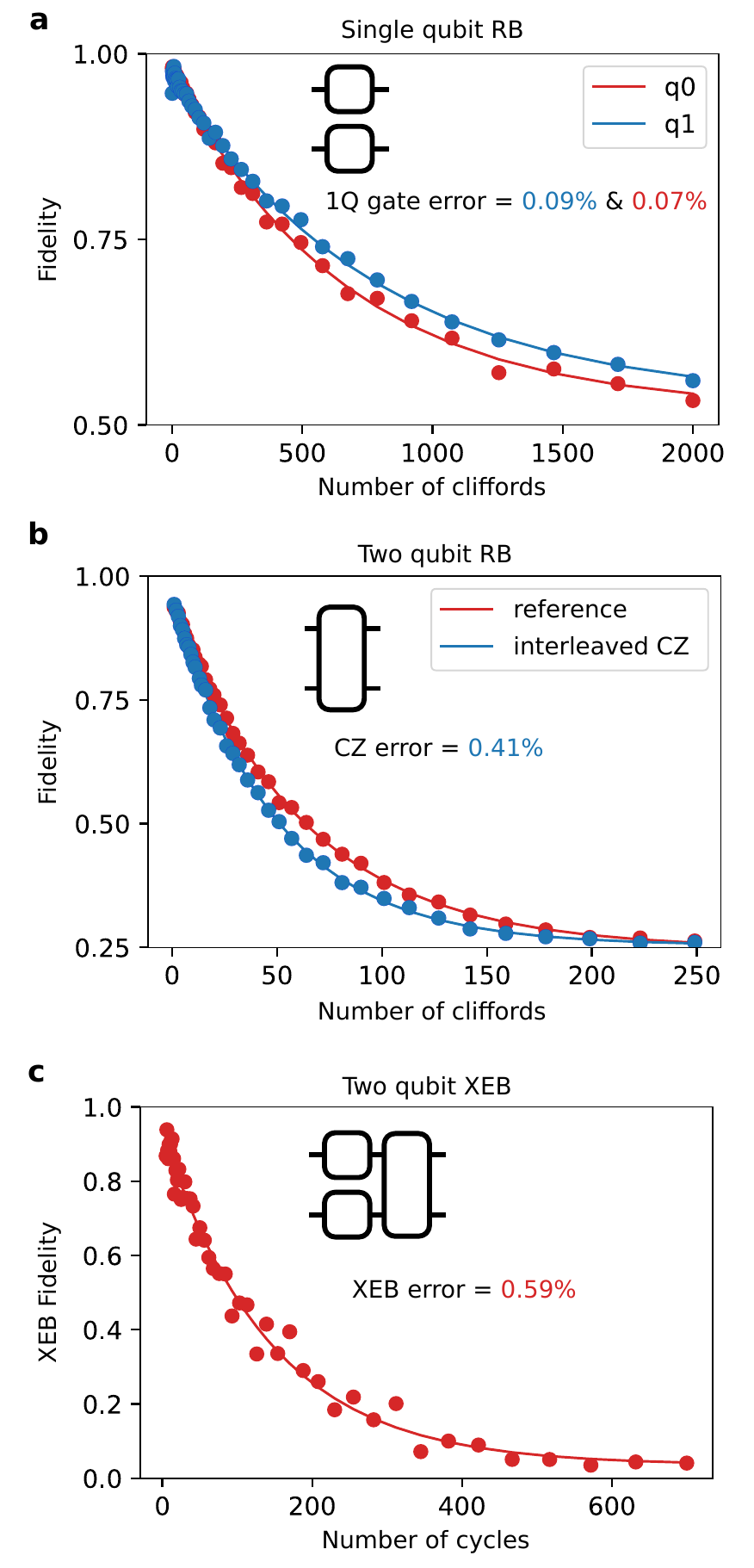}
\caption{\label{fig:compare_xeb_rb} \textbf{Sanity check:  XEB agrees with RB.}  \textbf{a,} Single-qubit randomized benchmarking (RB) data taken separately on two qubits.  \textbf{b,} Two-qubit randomized benchmarking data for a CZ on the same pair of qubits.  \textbf{c,} Two-qubit cross-entropy benchmarking (XEB) on the same pair of qubits.  The measured XEB error (0.59\% / cycle) agrees well with the prediction from single- and two-qubit RB (0.57\%).  All errors are quoted as Pauli errors.}
\end{figure}

In Fig.~\ref{fig:compare_xeb_rb} we show that two-qubit gate fidelity extracted using XEB agrees well with the fidelity as measured with RB, an important sanity check in validating XEB as a gate metrology tool.  In two-qubit XEB, we extract the error per cycle which consists of a single-qubit gate on each qubit and a two-qubit gate between them.  In Fig.~\ref{fig:compare_xeb_rb}a we show the individual RB decay curves for single-qubit gates.  In panel b, we show the RB decay curve for benchmarking a CZ gate.  Adding up the three errors from RB, we would expect an XEB cycle error of 0.57\%.  In panel c, we show the measured XEB decay curve which indicates a cycle error of 0.59\% \textemdash nearly identical to the value predicted by RB.

For single-qubit gate benchmarking on the Sycamore device used in this work (see Table \ref{tab:system_params}), we find that $\pi$ pulse fidelities are somewhat worse than $\pi/2$ pulse fidelities, which we attribute to reflections from the imperfect microwave environment. Because the XEB gateset we have used consists only of $\pi/2$ pulses, we find that the single-qubit gate errors extracted from conventional RB, which contains $\pi$ pulses, are somewhat higher than those extracted from single-qubit XEB. Using only $\pi/2$ pulses instead of $\pi$ pulses in single-qubit RB brings the extracted error close to that measured via XEB.

\subsubsection{Speckle purity benchmarking (SPB)}

It is experimentally useful to be able to extract state purity from XEB experiments in order to error-budget the contribution of decoherence. Conventionally, purity estimation can be done with state tomography, where the full density matrix $\rho$ is reconstructed and used to quantify the state purity. This involves expanding a single sequence into a collection of sequences each appended with single-qubit gates. Unfortunately, full tomographic reconstruction scales exponentially in the number of qubits, both for the number of sequences needed as well as the number of measurements needed per sequence. Here, we introduce an exponentially more efficient method to extract the state purity without additional sequences.

We use a re-scaled purity definition such that a fully-decohered state has a purity of 0, and a pure state has a purity of 1. We define
\begin{align}
  \label{eq:puritydef}
  \textrm{Purity} = \frac{D}{D-1}\left( \textrm{Tr}(\rho^2) - \frac{1}{D}\right),
\end{align}
which is consistent with what is defined in Ref.~\cite{jwallman2015estimating}. This can be understood as the squared length of the generalized Bloch vector in $D$ dimensions (for a qubit, $D=2$, this definition gives  $\langle X \rangle^2 + \langle Y \rangle^2 + \langle Z \rangle^2$).

\begin{figure}
\includegraphics[width=\columnwidth]{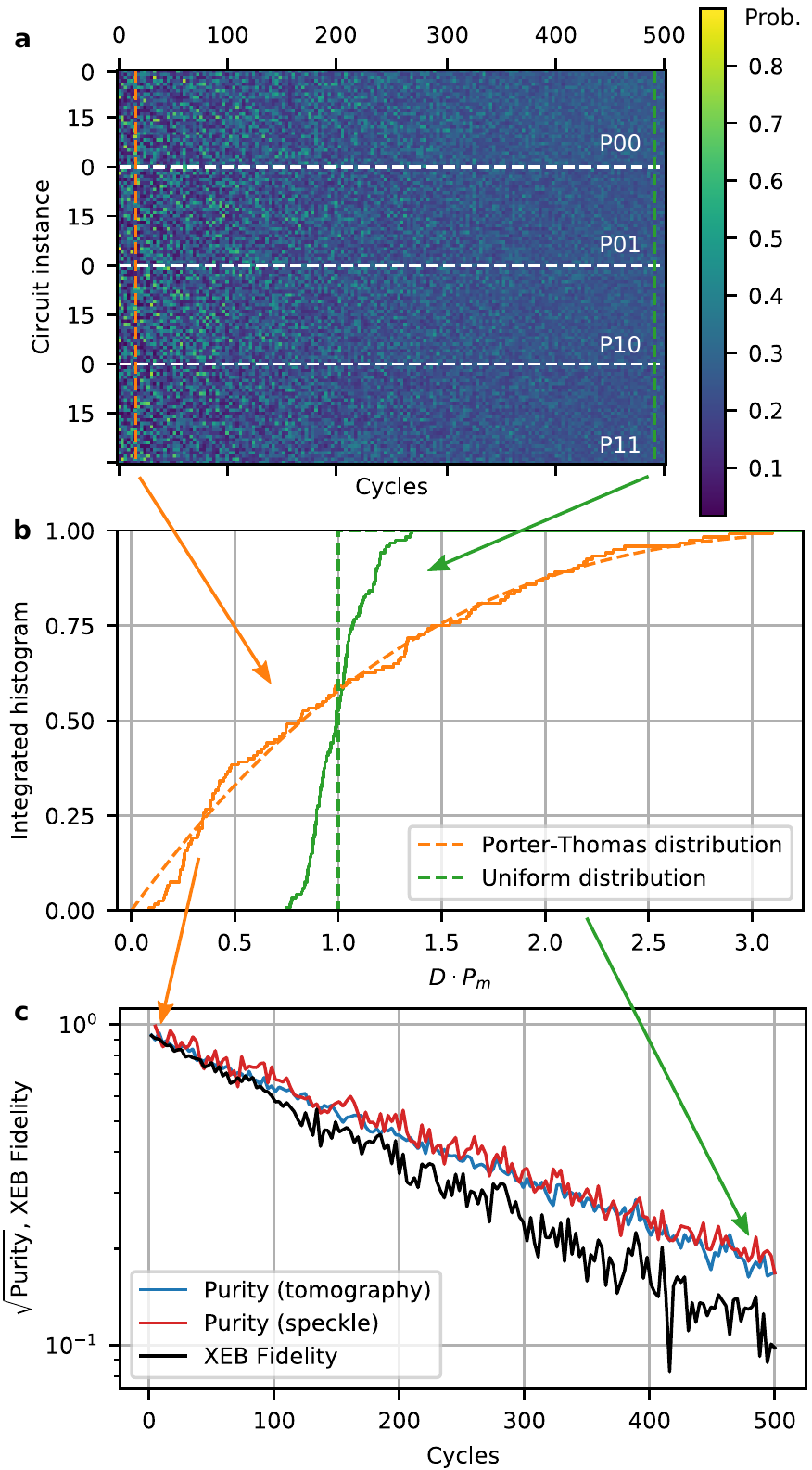}
\caption{\label{fig:speckle_xeb} \textbf{``Speckle" purity extracted from XEB.} \textbf{a,} Measured probabilities from XEB for a two-qubit system and 30 random circuits. Raw probabilities show a speckle pattern at low cycles (orange dashed) over circuit instance and probabilities ($|00\rangle$, $|01\rangle$, $|10\rangle$, $|11\rangle$). The speckle contrast decreases with cycles and thus decoherence (green dashed). \textbf{b,} Integrated histogram (cumulative distribution) of probabilities. The x-axis is scaled by the dimension $D=2^2$, so the uniform distribution is a step function at 1.0. At low cycles, the distribution is well-described by Porter-Thomas, and at high cycles, the distribution approaches the uniform distribution. \textbf{c,} We can directly relate the variance of the distribution to the average state purity. We fit an exponential to the square root of Purity. We compare this purity-derived number per-cycle$=0.00276$ to a similar number per-cycle=$0.00282$ derived from tomographic measure of purity, and see good agreement. The error of XEB, which also includes control errors, is slightly higher at error per-cycle=$0.00349$.}
\end{figure}

“Speckle” Purity Benchmarking (SPB) is the method of measuring the state purity from raw XEB data. Assuming the depolarizing-channel model with polarization parameter $p$, we can model the quantum state as 
\begin{align}
  \label{eq:polarizationmodel}
  \rho = p \, \ket{\psi} \bra{\psi} + (1-p)\, \frac{\openone}{D}.
\end{align}
Here, $p$ is the probability of a pure state $\ket{\psi}$ (which in this case is not necessarily known to us), while $1-p$ is the probability of being in the fully-decohered state ($\openone$ is the identity operator). For the state (\ref{eq:polarizationmodel}), from the definition (\ref{eq:puritydef}) it is easy to find the relation 
    \begin{align}
    \label{eq:purity-p-relation}
\textrm{Purity}=p^2. 
    \end{align}
We will now work out how to obtain $p^2$ from a distribution of measured probabilities $P_m$ of various bitstrings for a sequence, collected over many XEB sequences (Figs.\ \ref{fig:speckle_xeb}a and \ref{fig:speckle_xeb}b). 

First, we note that for $p=0$ the probabilities of all bitstrings are $1/D$, and the distribution is the $\delta$-function located at $1/D$ (the integrated histogram is then the step-function -- see Fig.\ \ref{fig:speckle_xeb}b). In contrast, if $p=1$, then the measured probabilities $P_m$ follow the $D$-dimensional Porter-Thomas distribution \cite{boixo2018characterizing}
    \begin{align}
    \label{eq:D-dim-Porter-Thomas}
    {\cal P}_{\rm PT}(P_m) = (D-1)(1-P_m)^{D-2},
    \end{align}
which has the same average $1/D$ and variance
    \begin{align}
     \label{eq:D-dim-Porter-Thomas-variance}
{\rm Var}_{\rm PT}(P_m)=\frac{D-1}{D^2(D+1)}.
    \end{align}

For the fully-decohered state all bitstrings have the same probability $1/D$, so in this case the variance of the distribution of probabilities is zero. For the state (\ref{eq:polarizationmodel}) with an arbitrary $p$, the histogram of probabilities $P_m$ will be described by the distribution (\ref{eq:D-dim-Porter-Thomas}) shrunk towards the average $1/D$ by the factor $p$. Consequently, the variance of the experimental probabilities will be $p^2$ times the Porter-Thomas variance (\ref{eq:D-dim-Porter-Thomas-variance}).

Thus, we can find $p^2$ by dividing the variance of experimentally measured probabilities $P_m$ by the Porter-Thomas variance (\ref{eq:D-dim-Porter-Thomas-variance}). Finally, using the relation (\ref{eq:purity-p-relation}) for the depolarization model (\ref{eq:polarizationmodel}), we can relate the variance of the experimental probabilities $P_m$ to the average state purity
\begin{align}
  \label{eq:specklypuritydefinition}
  \textrm{Purity} = \textrm{Var}(P_m) \, \frac{D^2 (D+1)}{D-1}.
\end{align}

With these convenient relations, we can directly compare the XEB fidelity $\xebfidelity=p$ to $\sqrt{\rm Purity}$ from SPB on the same scale, and check their dependence $p=p_c^m$ on the number of cycles $m$. Without systematic control errors, the XEB and SPB results should coincide. Experimentally, we always have control errors which lead us to incorrectly predict $\ket{\psi}$, so control errors give XEB a higher error than SPB. Thus, with a single XEB dataset we can extract the XEB error per-cycle, and the purity loss per-cycle with SPB. By subtracting these, we are left with the control error per-cycle. Thus, with a single experiment we can error budget total error into control error and decoherence error. 

These relationships can be seen experimentally in Figure \ref{fig:speckle_xeb}. Amazingly, computing the speckle purity can be done with no knowledge of the specific gate sequence performed; as long as the experiment introduces sufficient randomization of the Hilbert Space, Porter-Thomas statistics apply. Practically, SPB allows us to measure the state purity from raw XEB data with exponentially fewer number of pulse sequences as compared to full state tomography. This favorable scaling allows one to extend purity measurements to larger numbers of qubits. It is important to note that an exponential number of measurements are still required to fully characterize the probability distribution for a given sequence, as in tomography, so purity measurements of the full processor are impractical. 

\subsubsection{``Per-layer" parallel XEB}
To execute quantum circuits efficiently, it is helpful to run as many gates as possible in parallel. We wish to benchmark our entangling gates operating simultaneously. Resulting fidelities and optimized unitaries may differ from the isolated case, where we benchmark each pair individually, due to imperfections such as control crosstalk and stray qubit-qubit interactions. In the quantum supremacy algorithm, we partition the set of two-qubit gates into four layers, each of which can be executed in parallel. We then cycle through these layers interleaved with randomly chosen single-qubit gates (see Fig.~3a). However, it is intractable to directly use full-system XEB to benchmark our entangling gates for two reasons: we would simultaneously optimize over the unitary model parameters of every entangling gate, and the classical simulation would be exponentially expensive in system size.

\begin{figure*}[htbp]
\includegraphics{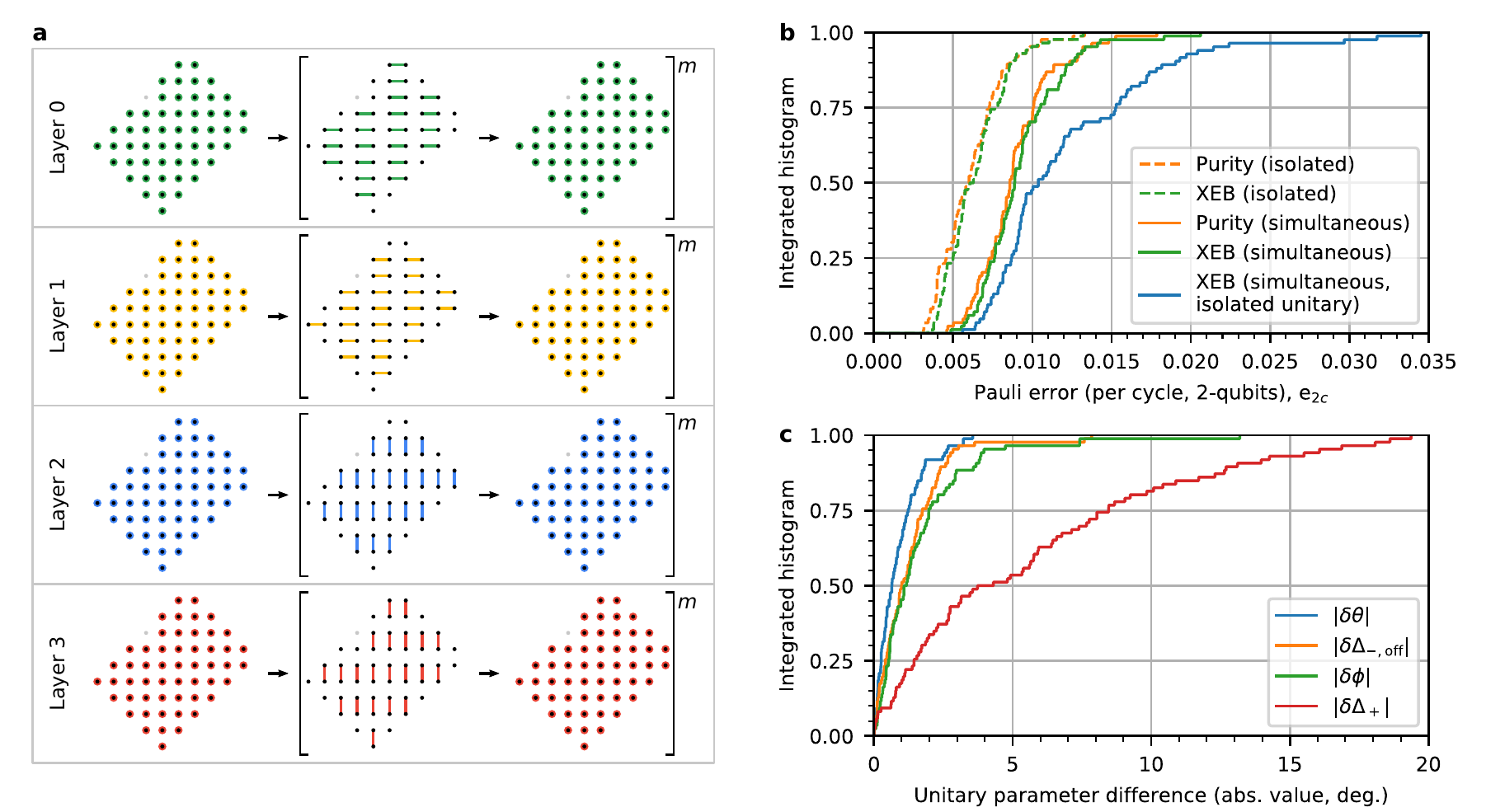}
\caption{\label{fig:per_layer_xeb} \textbf{Parallel XEB.} \textbf{a,} Schematics of four device-wide sequences, one for each entangler layer. Black points are active qubits, colored circles are single-qubit gates, and colored lines are two-qubit gates. We cycle between single- and two-qubit gates $m$ times. Compare to Fig.~3a, main text, where the layers are interleaved. \textbf{b,} Integrated histograms of Pauli error $e_{2c}$ (see Fig.~2a, main text). These include isolated results, where each entangler is measured in its own experiment, and simultaneous (parallel) results. Purity is “speckle” purity. \textbf{c,} Difference, $\delta$, in unitary model parameters (Eq.~\ref{eq:unitary_angles}) between the unitaries obtained in the isolated and simultaneous experiments. $\delta\Delta_-$ is not plotted because it has a negligible effect on the unitary when $\theta\approx90$ degrees.}
\end{figure*}

We solve this problem with ``per-layer" parallel XEB (see Ref.~\cite{erhard2019characterizing} for a related technique in the context of RB). Instead of alternating among the four layers of entanglers, where each qubit becomes entangled with each of its neighbors, we perform four separate experiments, one for each layer. The experiment sequences are illustrated in Fig.~\ref{fig:per_layer_xeb}a. For each layer, we construct parallel sequences where the layer is repeated with interleaved single-qubit gates; nominally, each qubit only interacts with one other. Following each parallel XEB sequence, we measure all the qubits and extract the equivalent XEB data for each pair. Every two-qubit gate can be characterized in these four experiments, regardless of system size. The optimization and classical simulation are also efficient, as each pair can be analyzed individually.

We present experimental results of ``per-layer" parallel XEB in Fig.~\ref{fig:per_layer_xeb}b-c. In Fig.~\ref{fig:per_layer_xeb}b, we compare the performance in the isolated and simultaneous (parallel) experiments. In both cases, the optimized XEB error is close to purity-limited. Simultaneous operation modestly increases the error, by roughly 0.003. This increase is primarily from purity error, which would arise from unintended interactions with other qubits, where coherent errors at the system scale manifest as incoherent errors when we focus on individual pairs. The unitaries we obtain in the simultaneous case differ slightly from the isolated case, which would arise from control crosstalk and unintended interactions. To quantify how these differences affect the gate error, we recalculate the error with the unitaries from the isolated optimization and the data from the simultaneous experiment, which increases the error. We also plot the distributions of the differences in unitary model parameters in Fig.~\ref{fig:per_layer_xeb}c. The dominant change is in $\Delta_+$, a single-qubit phase.

\subsection{Grid readout calibration}\label{sec:readout_cals}
\subsubsection{Choosing qubit frequencies for readout}
\begin{figure}[hbp]
\includegraphics[width=3.0in]{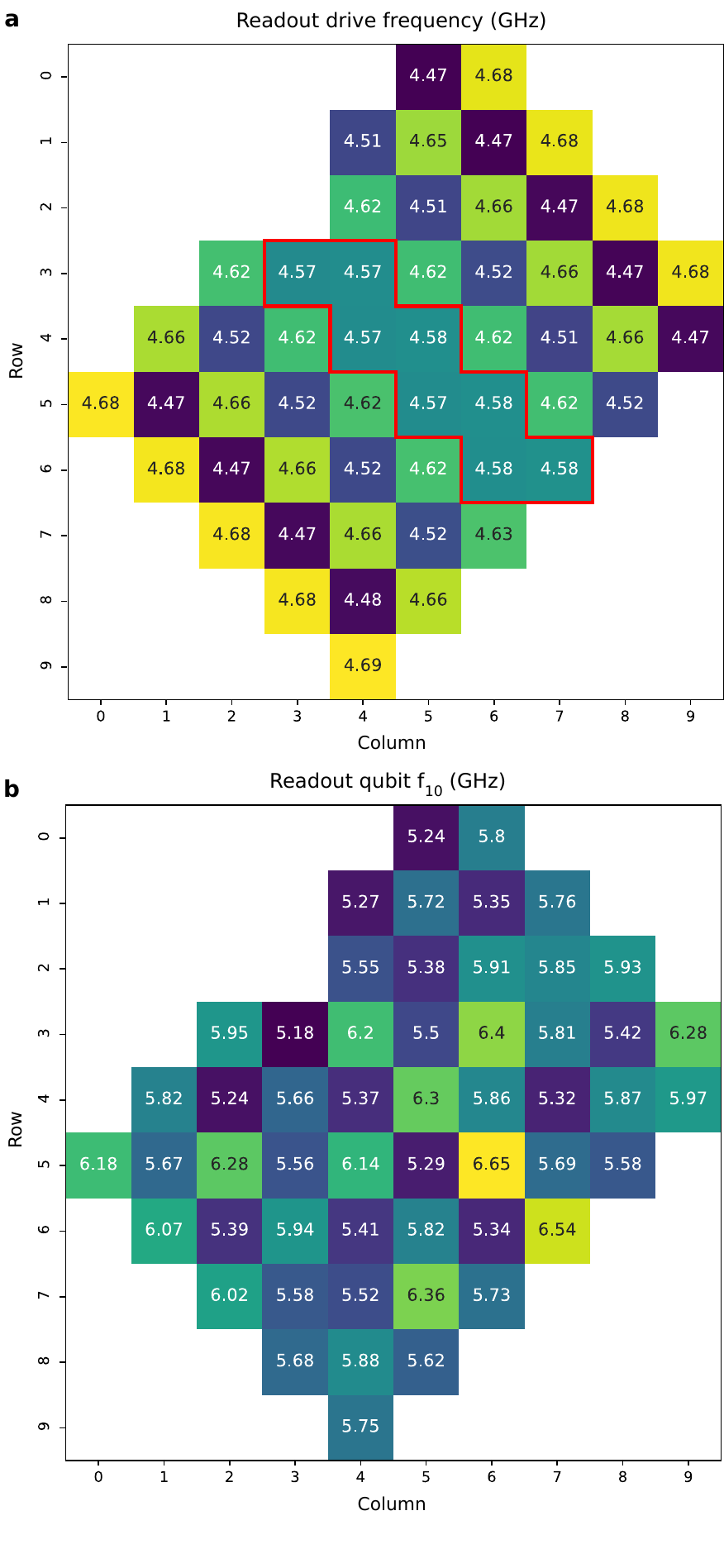}
\caption{\label{fig:readout_freqs} \textbf{a,} Drive frequencies for the readout resonators for each qubit. The red outline shows the area where we had to perform correlated discrimination because of unwanted cross-couplings between the resonators. \textbf{b,} Qubit frequencies during readout, found using a frequency optimization procedure.}
\end{figure}

The algorithm described in Section \ref{sec:opt_freqs} generally chooses qubit idling frequencies which are far detuned from the resonator to optimize for dephasing. However, these idling frequencies are not optimal for performing readout. To address this problem, we dynamically bias each qubit to a different frequency during the readout phase of the experiment. The qubit frequencies during readout are shown in Fig.~\ref{fig:readout_freqs} (compare to Fig.~\ref{fig:qubit_freqs}).

To choose the qubit frequencies for readout, we first measure readout fidelity as a function of qubit frequency and resonator drive frequency at a fixed resonator drive power, in each of the isolated single qubit configurations. This scan captures errors due to both non-optimal detuning between the qubit and resonator, as well as regions with low $T_1$ values due to TLSs. We then use the data for each qubit and a few constraints to optimize the placement of the qubit frequencies during readout, using the same optimization technique that was described in Section \ref{sec:opt_freqs}. We describe two of the important constraints and related error reduction techniques below.

First, because the coupling between qubits relies on a dispersive interaction with the coupler, the coupling would no longer be off when the qubits were detuned by a significant amount from their idling positions. Thus, we impose a constraint that qubits should not be placed near resonance during readout. Nevertheless, we found that for some pairs of qubits, we had to dynamically bias the coupler during readout to avoid any swapping transitions between the qubits during readout. This readout coupler bias is found by sweeping the coupler bias and maximizing the two-qubit readout fidelity.

Second, the pattern of the bare resonator frequencies on the chip as shown in Fig.~\ref{fig:readout_freqs} led to an unexpected problem. Pairs of readout resonators which were coupled to neighboring qubits and were also within a few MHz in frequency space were found to have non-negligible coupling. This coupling was strong enough to mediate swapping of photons from one resonator to the other. The pairs of qubits with similar resonator frequencies were all located in a diagonal chain bisecting the qubit grid, as shown by the red outline in Fig.~\ref{fig:readout_freqs}. To mitigate this problem, we arrange the qubit frequencies for these qubits so that the resonator eigenfrequencies are as far apart as possible. The resulting spectral separation is not quite enough to eliminate all deleterious effects, so in addition, we use correlated discrimination on the eight of the qubits in this chain. In other words, we use the results of all eight detector values to determine which one of $2^8 = 256$ states the eight qubits were in. All other qubits in the grid are discriminated as isolated qubits.

\subsubsection{Single qubit calibration}
After placing the qubit frequencies for readout, we calibrate and fine tune the readout parameters for each qubit. For each qubit, we use a 1~$\mu$s drive pulse and a 1~$\mu$s demodulation window. We summarize the procedure for choosing the remaining parameters as follows:
\begin{itemize}
    \item Choose the resonator drive frequency to maximize the separation between measurements performed with the qubit in either $|0\rangle$ and $|1\rangle$ \cite{bultink2018general}.
    \item Choose the resonator drive power to hit a target separation between $|0\rangle$ and $|1\rangle$, so that the error due to this separation is below a 0.3\% threshold. We do not choose the readout power to maximize the separation as doing so would saturate our amplifiers, and cause unwanted transitions of the qubit state \cite{johnson2012heralded, Jeffrey2014Fast, dan_thesis, Sank2016Transitions}.
    \item Find the optimal demodulation weight function by measuring the average detector voltage as a function of time during the course of the readout pulse \cite{bultink2018general, ryan2015tomography}.
    \item Finally, choose the discrimination line between the measurement results for $|0\rangle$ and $|1\rangle$, except as noted in the previous section where we need to apply correlated discrimination.
\end{itemize}

\begin{figure}[hbp]
\includegraphics[width=3.3in]{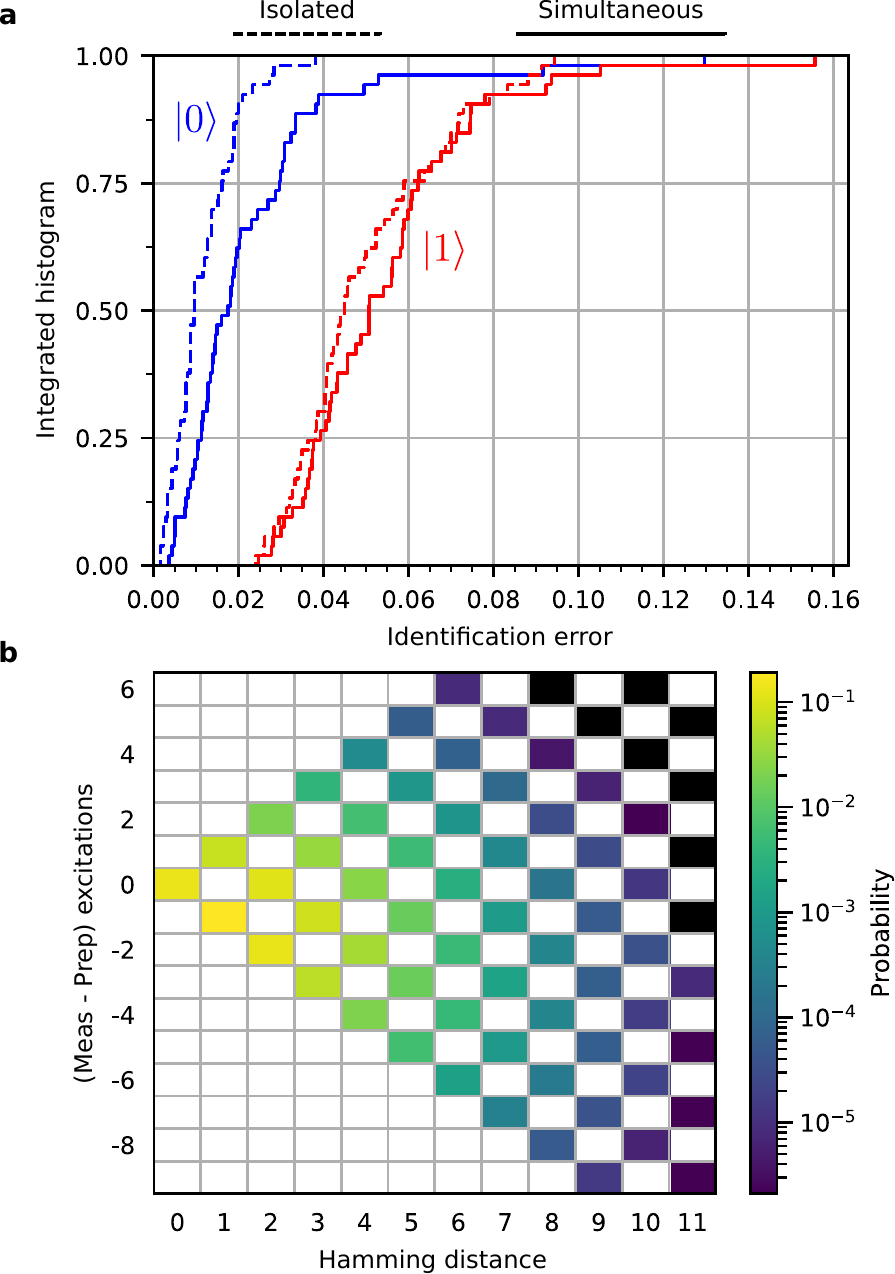}
\caption{\label{fig:mq_readout} \textbf{Readout errors.} \textbf{a,} Histogram of readout errors for each qubit when prepared in $|0\rangle$ or $|1\rangle$, and readout in isolation or simultaneously. \textbf{b,} Distribution of errors in multiqubit readout. The x-axis Hamming distance is the number of bits that are different between measured and prepared states, while the y-axis is the difference in the number of 1’s in the states. For example, if we prepare $|011\rangle$ and measure $|101\rangle$, the Hamming distance is 2 and the difference in the number of excitations is 0.}
\end{figure}

After completing these calibrations, we check each qubit's readout fidelity by preparing either $|0\rangle$ or $|1\rangle$ and reading the qubit out. We define the identification error to be the probability that the qubit was not measured in the state we intended to prepare. We achieve 0.97\% median identification error for the $|0\rangle$ state, and 4.5\% for $|1\rangle$, when each qubit is measured in isolation. The full distribution is shown in dashed lines in Fig.~\ref{fig:mq_readout}a. We conjecture that the error in $|0\rangle$ is due to thermal excitation during preparation or measurement, and that the error in $|1\rangle$ is due to energy relaxation during readout. 

\subsubsection{Characterizing multi-qubit readout}
To assess the fidelity of multi-qubit readout, we prepare and measure 150 random classical bitstring states with 53 qubits, with 3000 trials per state. We find that 13.6\% of all trials successfully identified the prepared state. We can decompose this overall fidelity in two ways.
First, we plot in solid lines in Fig.~\ref{fig:mq_readout} the errors for each qubit during simultaneous readout, averaged over the 150 random bitstrings. We find that the median errors increase from 0.97\% for $|0\rangle$ and 4.5\% for $|1\rangle$ in isolation, to 1.8\%  and 5.1\% for simultaneous readout. We do not yet understand the root causes of this increase in error. In addition, we show in Fig.~\ref{fig:mq_readout} the distribution of errors among the multiqubit results. We see that the most likely error is one lost excitation in the measured state.

\begin{table*}[htbp]
\caption{\label{tab:system_params} Aggregate system parameters}
\begin{ruledtabular}
\begin{tabular}{lccccr}
Parameter & Median & Mean & Stdev. & Units & Figure\\
\hline
Qubit maximum frequency & 6.924 & 6.933 & 0.114 & GHz & \ref{fig:qubit_freq_params}\\
Qubit idle frequency    & 6.661 & 6.660 & 0.057 & GHz & \ref{fig:qubit_freqs} \\
Qubit frequency at readout & 5.750 & 5.766 & 0.360 & GHz & \ref{fig:readout_freqs} \\
Readout drive frequency & 4.618 & 4.588 & 0.076 & GHz & \ref{fig:readout_freqs}\\
Qubit anharmonicity & -208.0 & -208.0 &  4.7 &  MHz & \ref{fig:qubit_freq_params}\\
Resonator linewidth $\kappa/2\pi$ & 0.64 & 0.69 & 0.23 & MHz & \ref{fig:qubit_freq_params} \\
Qubit-resonator coupling $g/2\pi$ & 72.3 & 72.1 &  2.8 & MHz & \ref{fig:qubit_freq_params}\\
$T_1$ at Idle Frequency & 15.54 & 16.04 & 4.00 & $\mu$s & \ref{fig:qubit_freq_params}\\
Readout error $|0\rangle$ isolated / simultaneous & 0.97 / 1.8 & 1.2 / 2.3 & 0.8 / 2.1 & \% & \ref{fig:mq_readout}\\
Readout error $|1\rangle$ isolated / simultaneous & 4.5 / 5.1 & 5.0 / 5.5 & 1.8 / 2.2 & \% & \ref{fig:mq_readout}\\
1Q RB\footnotemark[1] $e_1$ & 0.19 & 0.22 & 0.10 & \% & \ref{fig:single_qubit_gate_benchmarking}\\
1Q RB\footnotemark[1]  $e_1$ ($\pi/2$ gateset)& 0.15 & 0.16 & 0.06 & \% & \ref{fig:single_qubit_gate_benchmarking}\\
1Q RB\footnotemark[1] tomographic $e_1$ purity & 0.14 & 0.15 & 0.04 & \% & \ref{fig:single_qubit_gate_benchmarking}\\
1Q XEB $e_1$ isolated / simultaneous & 0.13 / 0.14 & 0.15 / 0.16 & 0.05 / 0.05 & \% & 3a (main) \ref{fig:single_qubit_gate_benchmarking}\\
1Q XEB $e_1$ purity isolated / simultaneous & 0.11 / 0.11 & 0.11 / 0.12 &  0.03 / 0.03 & \% & \ref{fig:single_qubit_gate_benchmarking}\\
2Q XEB $e_2$ isolated / simultaneous & 0.30 / 0.60 & 0.36 / 0.62 & 0.17 / 0.24 & \% & 3a (main) \\
2Q XEB $e_{2c}$ isolated / simultaneous & 0.64 / 0.89 & 0.65 / 0.93 & 0.20 / 0.26 & \% & 3a (main) \\ 
2Q XEB $e_{2c}$ purity isolated / simultaneous & 0.59 / 0.86 & 0.62 / 0.89 & 0.20 / 0.24 & \% & \ref{fig:per_layer_xeb}\\ 
Measurement $e_m$ isolated /  simultaneous & 2.83 / 3.50 & 3.05 / 3.77 & 1.09 / 1.61 & \% & 3a (main) \\

\end{tabular}
\end{ruledtabular}
\footnotetext[1]{RB data taken at a later date}
\end{table*}

\subsection{Summary of system parameters}
\begin{figure*}[htbp]
\centering
\includegraphics[width=2.0\columnwidth]{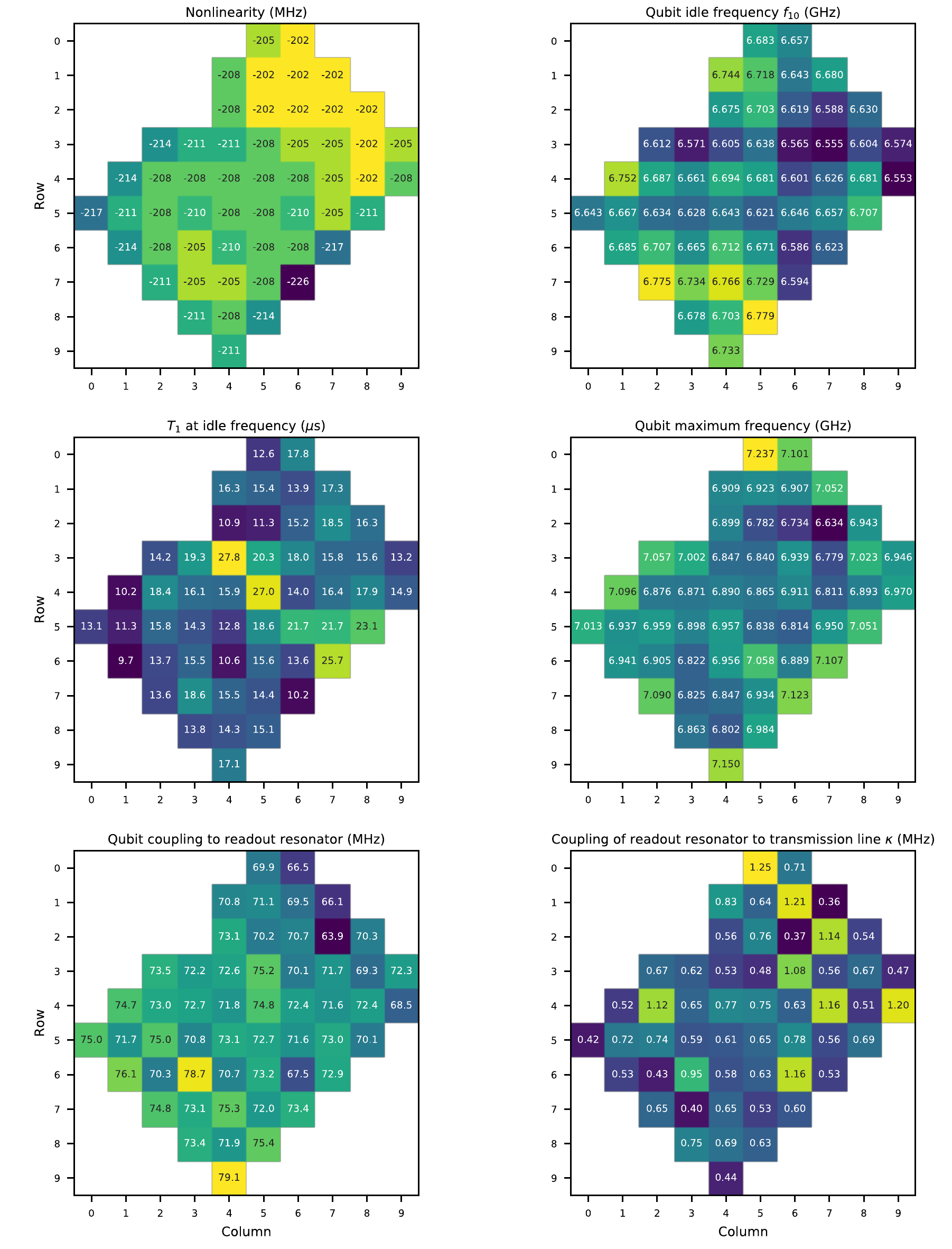}
\caption{\label{fig:qubit_freq_params} Typical distribution of single-qubit parameters over the Sycamore processor.}
\end{figure*}

\begin{figure*}[htbp]
\centering
\includegraphics[width=2.0\columnwidth]{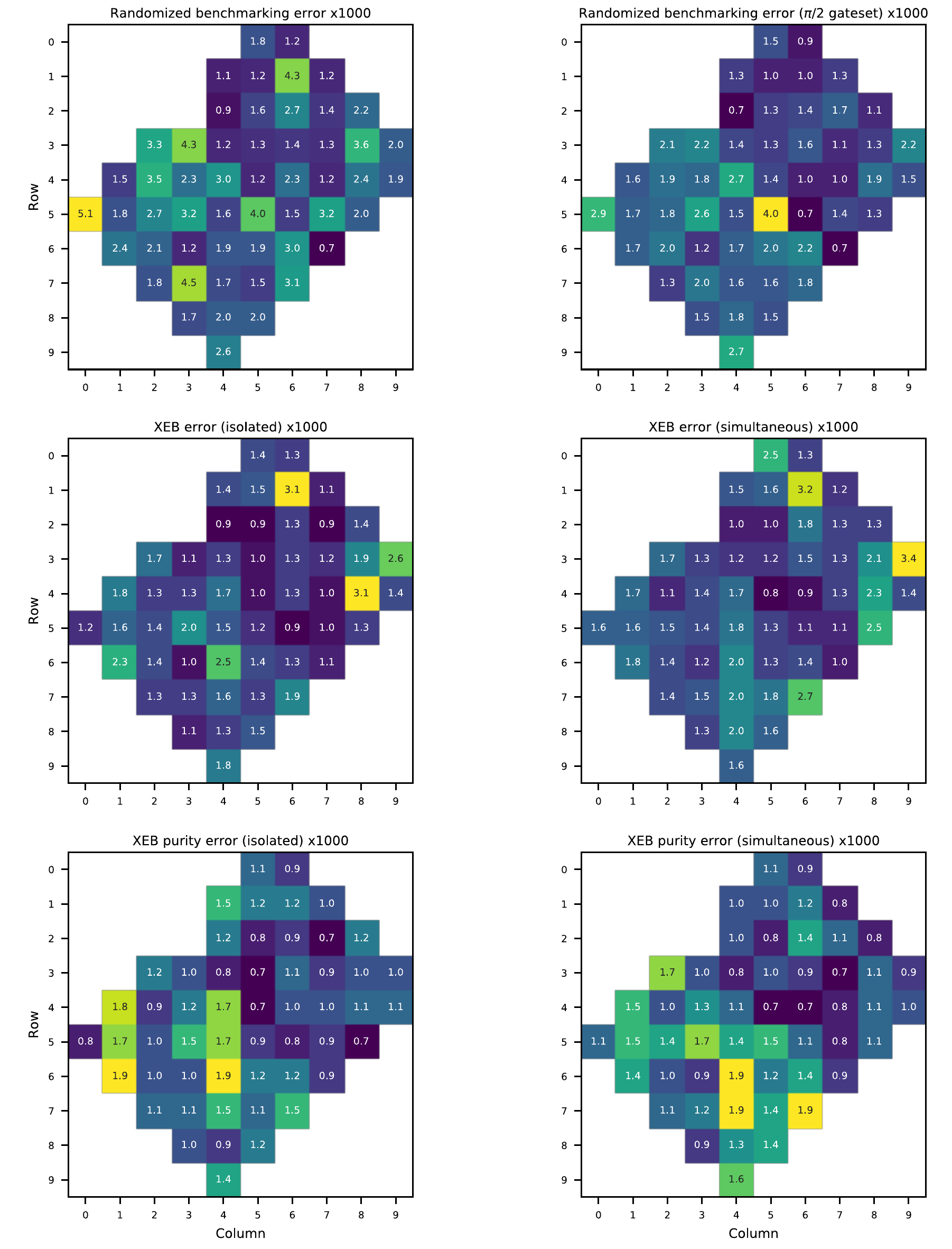}
\caption{\label{fig:single_qubit_gate_benchmarking} Typical distribution of single-qubit gate benchmarking errors over the Sycamore processor, for both isolated and simultaneous operation.}
\end{figure*}

\begin{figure*}[htbp]
\centering
\includegraphics[width=2.0\columnwidth]{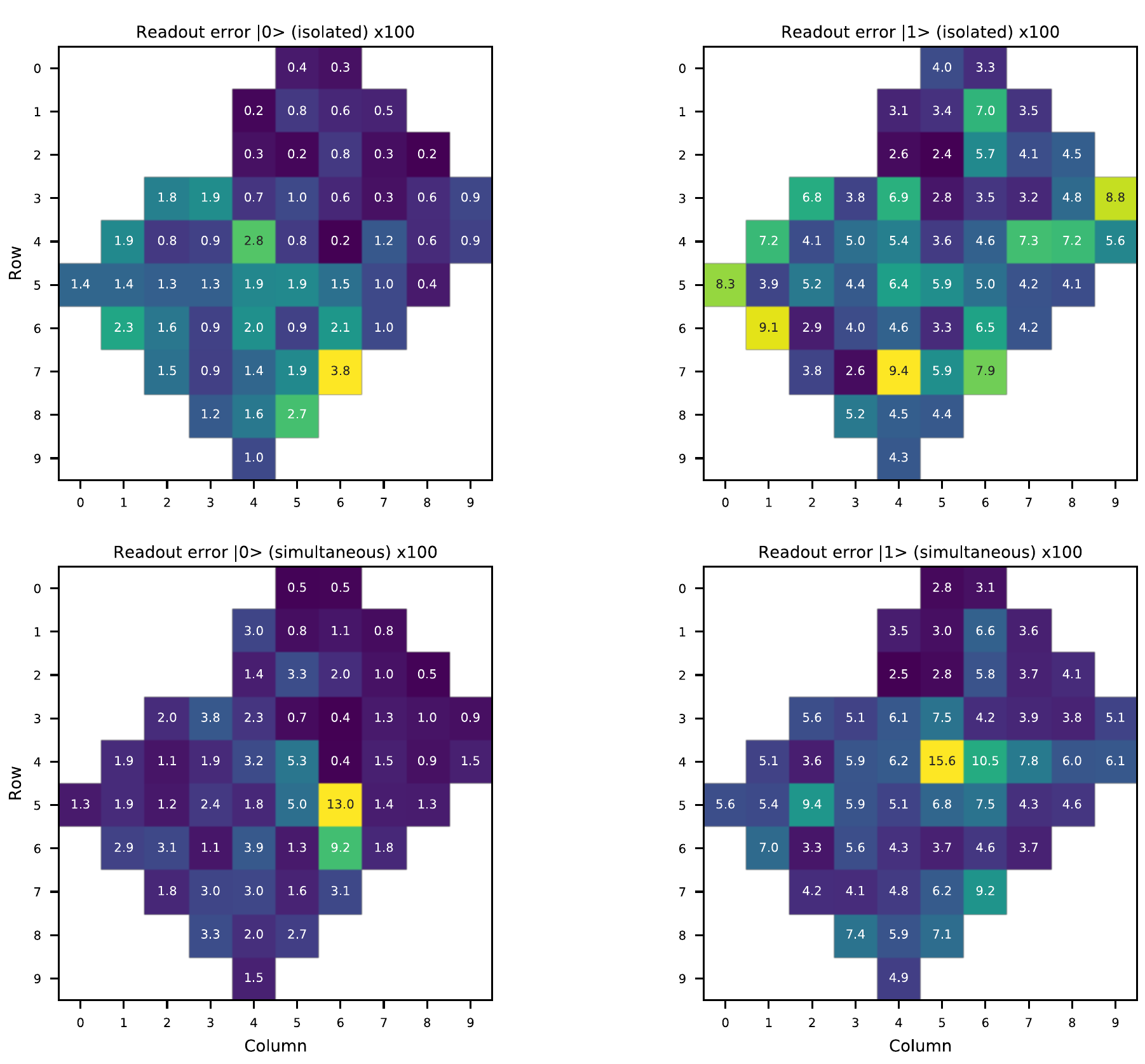}
\caption{\label{fig:qubit_readout_errors} Typical distribution of readout errors over the Sycamore processor, for both isolated and simultaneous operation.}
\end{figure*}

Table \ref{tab:system_params} reports aggregate values for qubit and pair parameters in our processor. A complete table of single-qubit parameter values by qubit is available in supporting online materials, Ref.~\cite{SOM_csv}, and illustrated in Figs.~\ref{fig:qubit_freq_params} through \ref{fig:qubit_readout_errors}. Single-qubit metrics represent a sample size of 53. Two-qubit metrics represent 86 pairs. 

\clearpage
\section{Quantum circuits}\label{sec:circuits}

\subsection{Background}
We sample the output of random quantum circuits (RQCs) with two use cases in mind: performing a computational task beyond the reach of state-of-the-art supercomputers (quantum supremacy); and estimating the experimental fidelity (performance evaluation).

In order for the RQCs to cover both use cases, we define a circuit family with a varying number of qubits $n$ and cycles $m$. Our quantum supremacy demonstration uses RQCs with a large number of qubits $n = 53$ and high depth $m = 20$. Large number of qubits hinders wave function (Schr\"odinger) simulation and high depth impedes tensor network (Feynman) simulation (see Sec. \ref{subsec:fa}). We find that the most competitive classical simulator for our hardest RQCs is the Schr\"odinger-Feynman algorithm (SFA, see Sec. \ref{subsec:sa_sfa}) which copes well with high depth circuits on many qubits.

SFA takes as input an $n$-qubit quantum circuit and a \textit{cut} which divides $n = n_1 + n_2$ qubits into two contiguous partitions with $n_1$ and $n_2$ qubits. The algorithm computes the output state as the sum over simulation paths formed as the product of the terms of the Schmidt decomposition of all cross-partition gates. By the distributive law there are $r^g$ such simulation paths for a circuit with $g$ cross-partition gates of Schmidt rank $r$. Consequently, the algorithm achieves runtime proportional to $(2^{n_1} + 2^{n_2}) r^g$. Circuit cuts with $n_1$, $n_2$ and $g$ that make the simulation task tractable are called \textit{promising cuts}. The most promising cut for our largest RQCs runs parallel to the shorter axis of the device starting in the vicinity of the broken qubit. The sum over the simulation paths can be interpreted as tensor contraction. In this view, the $r^g$ factor can be thought of as the \textit{bond dimension} associated with the circuit partitioning, i.e. the cardinality of the index set ranged over in the contraction corresponding to all cross-partition gates. SFA is described in more detail in \cite{markov_quantum_2018} and section \ref{sec:classical_sim}.

\subsection{Overview and technical requirements}
The two use cases for our RQCs give rise to a tension in technical requirements at the heart of quantum supremacy. On the one hand, supremacy RQC sampling should by definition be prohibitively hard to simulate classically. On the other hand, performance evaluation entails classical simulation of the RQCs. To resolve the conflict, we note that the fidelity of a RQC experiment depends primarily on the number and quality of the gates. By contrast, the simulation cost is highly sensitive to minor perturbations in the circuit. Consequently, experiment fidelity for RQCs that cannot be simulated directly may be approximated from the experiment fidelity of similar RQCs obtained as the result of transformations that reduce simulation cost without significantly affecting experiment fidelity (see Section \ref{sec:circuit_variants}).

Performance evaluation using XEB provides another design consideration. The procedure requires knowledge of the cross-entropy of the theoretical output distribution of the circuit. An analytical expression for this quantity has been derived in \cite{boixo2018characterizing} for circuits whose measurement probabilities approach the Porter-Thomas distribution. We find that our RQCs satisfy this assumption when the circuit depth is larger than 12, see Fig.~\ref{fig:distribution_of_Dp}a. Note that high circuit depth also increases the cost of classical simulation.

\subsection{Circuit structure}
A RQC with $n$ qubits generally utilizes qubits 1 through $n$ in the qubit order shown in Fig.~\ref{fig:qubit_order} with small deviations from this default qubit ordering in some circuits. The qubit order has been chosen to ensure that for most RQCs with fewer than 51 qubits, there is a partitioning of the qubits into two similarly sized blocks connected by only five couplers. The next larger RQC, with 51 qubits, has seven couplers along the most promising circuit cut. Since the cost of SFA grows exponentially in the number of gates across the partitions our circuit geometry leads to a steep increase in the simulation cost of 51-qubit RQCs relative to the circuits with fewer qubits. This creates a sizeable gap in the computational hardness between most of our evaluation circuits and the quantum supremacy circuits ($n = 53$).

\begin{figure*}
\includegraphics[scale=0.3]{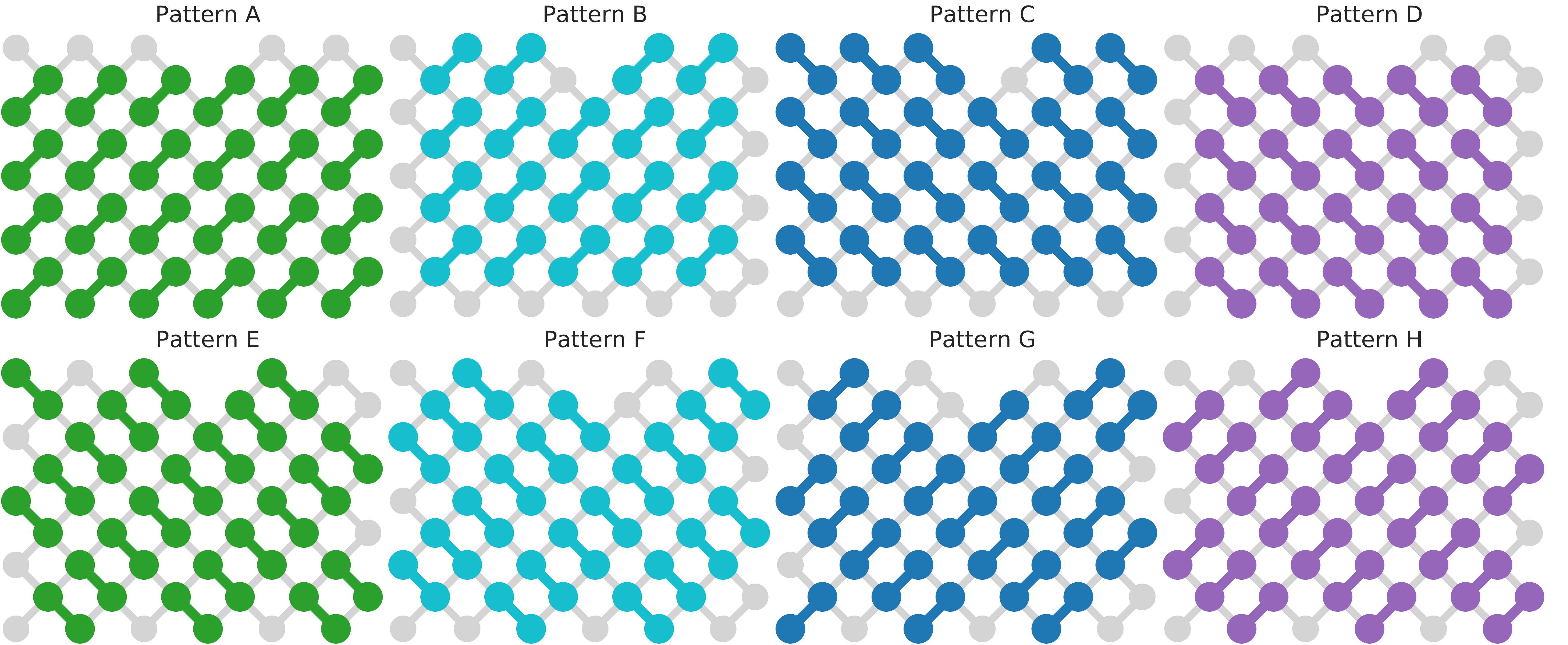}
\caption{\label{fig:gate_patterns} \textbf{Coupler activation patterns.} Coupler activation pattern determines which qubits are allowed to interact simultaneously in a cycle. Quantum supremacy RQCs utilize the staggered patterns shown in the top row in the sequence ABCDCDAB, repeated in subsequent cycles. Performance evaluation RQCs employ the patterns shown in the bottom row in the sequence EFGH, likewise repeated in subsequent cycles. The former sequence makes SFA simulation harder by facilitating prompt transfer of entanglement created at promising circuit cuts into the bulk of each circuit partition.}
\end{figure*}

In the time dimension, each RQC is a series of $m$ full cycles and one half cycle followed by measurement of all qubits. Every full cycle consists of two steps. In the first step, a single-qubit gate is applied to every qubit. In the second step, two-qubit gates are applied to pairs of qubits. Different qubit pairs are allowed to interact in different cycles. Specifically, in the supremacy RQCs we loop through the direct neighbors of every qubit over the eight-cycle sequence ABCDCDAB and in the evaluation RQCs we use the four-cycle sequence EFGH where A, B, ..., H are coupler activation patterns shown in Fig.~\ref{fig:gate_patterns}. The sequence is repeated in subsequent cycles. The cost of SFA simulation is highly sensitive to the specific sequence employed in a circuit, see \ref{sec:wedge_formation}. Border qubits have fewer than four neighbors and no gate is applied to them in some cycles. The half cycle preceding the measurement consists of the single-qubit gates only. The overall structure of our RQCs is shown in Fig.~3 of the main paper \cite{arute2019}.

\subsection{Randomness}
Single-qubit gates in every cycle are chosen randomly using a pseudo-random number generator (PRNG). The generator is initialized with a seed $s$ which is the third parameter for our family of RQCs. 
The single-qubit gate applied to a particular qubit in a given cycle depends only on $s$. Consequently, two RQCs with the same $s$ apply the same single-qubit gate to a given qubit in a given cycle as long as the qubit and the cycle belong in both RQCs as determined by their size $n$ and depth $m$ parameters. 

Conversely, the choice of single-qubit gates is the sole property of our RQCs that depends on $s$. In particular, the same two-qubit gate is applied to a given qubit pair in a given cycle by all RQCs that contain the pair and the cycle.

\subsection{Quantum gates}\label{subsec:quantum_gates}
In our experiment, we configure three single-qubit gates. Each one is a $\pi/2$-rotation around an axis lying on the equator of the Bloch sphere. Up to global phase, the gates are

\begin{align}
X^{1/2} \equiv R_X(\pi / 2) = \frac{1}{\sqrt{2}} \begin{bmatrix}
1  & -i \\
-i &  1 
\end{bmatrix}, \\
Y^{1/2} \equiv R_Y(\pi / 2) = \frac{1}{\sqrt{2}} \begin{bmatrix}
1 & -1 \\
1 &  1 
\end{bmatrix}, \\
W^{1/2} \equiv R_{X+Y}(\pi / 2) = \frac{1}{\sqrt{2}} \begin{bmatrix}
1  & -\sqrt{i} \\
\sqrt{-i} &  1
\end{bmatrix}
\end{align}
where $W = (X + Y)/\sqrt{2}$ and $\sqrt{\pm i}$ denotes the principal value of the square root. The first two belong to the single-qubit Clifford group, while $W^{1/2}$ is a non-Clifford gate. Single-qubit gates in the first cycle are chosen independently and uniformly at random from the set of the three gates above. In subsequent cycles, each single-qubit gate is chosen independently and uniformly at random from among the gates above except the gate applied to the qubit in the preceding cycle. This prevents simplifications of some simulation paths in SFA. Consequently, there are $3^n 2^{nm}$ possible random choices for a RQC with $n$ qubits and $m$ cycles.

Two-qubit gates in our RQCs are not randomized, but are determined by qubit pair and cycle number. The gates preserve the number of ground and excited states of the qubits which gives their matrices block diagonal structure with $1 \times 1$, $2 \times 2$ and $1 \times 1$ blocks. Therefore, up to global phase they belong to $U(1) \oplus U(2) \oplus U(1) / U(1)$ and thus can be described by five real parameters (see Fig.~\ref{fig:params_control_model}, and Eq.~\ref{eq:unitary_angles}). Each gate in this family can be decomposed into four Z-rotations described by three free parameters and the two-parameter fermionic simulation gate
\begin{equation} \label{eq:fsim}
{\rm fSim}(\theta, \phi) = \begin{bmatrix}
1 &             0  &              0 & 0 \\
0 &   \cos(\theta) & -i \sin(\theta) & 0 \\
0 & -i \sin(\theta) &   \cos(\theta) & 0 \\
0 &             0  &              0 & e^{-i \phi}
\end{bmatrix}
\end{equation}
which is the product of a fractional iSWAP and controlled phase gate (see Fig.~\ref{fig:two_qubit_gates}b).

In our experiment, we tune up the two-qubit gates close to $\theta \approx \pi/2$ and $\phi \approx \pi/6$ radians and then infer more accurate values of all five parameters for each qubit pair using XEB. Consequently, all five parameters of the two-qubit gate depend on the qubit pair. While inferred unitaries are suitable for RQC sampling, future applications of the Sycamore processor, for example, in quantum chemistry, will require precise targeting of the entangling parameters \cite{kivlichan2018quantum, babbush2018low}. The three parameters which control the Z-rotations implicit in the two-qubit gates can be canceled out with active Z-rotations turning an arbitrary five-parameter gate into pure $\rm{fSim} (\theta, \phi)$. In our RQCs, we have decided not to apply such correction gates. This choice affords us greater number of interactions within the available circuit depth budget and introduces additional implicit non-Clifford single-qubit gates into the RQCs.

The Z-rotations have two origins. First, they capture the phase shifts due to qubit frequency excursions during the two-qubit gate. Second, they account for phase changes due to different idle frequencies of the interacting qubits. The latter introduces dependency of the three parameters defining the Z-rotations on the time at which the gate is applied. By contrast, for a given qubit pair $\theta$ and $\phi$ do not depend on the cycle.

The $\rm{fSim} (\pi/2, \pi/6)$ gate is the product of a non-Clifford controlled phase gate and an iSWAP which is a two-qubit Clifford gate.

\subsection{Programmability and universality}

Programmability of Sycamore rests on our ability to tune up a variety of gate sets including sets that are universal for quantum computation. For example, the set of gates employed in our quantum supremacy demonstration is universal, as we show in this section.

The proof consists of two parts. First, we show that the {\rm CZ} gate can be obtained as a composition of two fSim gates and single-qubit rotations. Second, we outline how the well-known proof that the $H$ and $T$ gates are universal for {\rm SU(2)} \cite{Boykin1999OnUniversal} can be adapted for $X^{1/2}$ and $W^{1/2}$. The conclusion follows from the fact that the gate set consisting of the {\rm CZ} gate and {\rm SU(2)} is universal \cite{DiVincenzo1995Twobit}.

\subsubsection{Decomposition of {\rm CZ} into {\rm fSim} gates}

Here, we show how to decompose a controlled-phase gate into two {\rm fSim} gates and several single-qubit gates.  The {\rm fSim} gate is native to our hardware and can be decomposed into
\begin{align}\label{cz_to_syc:eq:Sycamore_a}
    {\rm fSim}(\theta, \phi) = e^{-i\theta(X\otimes X + Y\otimes Y)/2}\, e^{-i\phi(I-Z)\otimes (I-Z)/4}\,,
\end{align}
where the iSWAP angle $\theta \simeq \pi/2$ and the controlled-phase angle $\phi \simeq \pi/6$. The controlled-phase part can be further decomposed into
\begin{align}
&e^{-i\phi(I-Z)\otimes (I-Z)/4}\nonumber\\
&\quad = e^{-i\phi/4}\, e^{i\phi (Z\otimes I + I\otimes Z)/4}\, e^{-i\phi Z\otimes Z/4}\,.
\end{align}
To simplify notations, we introduce the two-qubit gate
\begin{align}\label{cz_to_syc:eq:Sycamore_b}
    \Upsilon(\theta, \phi)&= e^{-i\theta(X\otimes X + Y\otimes Y)/2}\, e^{-i\phi Z\otimes Z/4}\nonumber\\[2pt] 
    &= e^{i\phi/4}\, e^{-i\phi (Z\otimes I + I\otimes Z)/4}\: {\rm fSim}(\theta, \phi)
    \,,
\end{align}
which is equivalent to the {\rm fSim} gate up to single-qubit $Z$ rotations. The sign of $\theta$ in $\Upsilon(\theta, \phi)$ can be changed by the single-qubit transformation,
\begin{align}\label{cz_to_syc:eq:Sycamore_b}
  Z_1\, \Upsilon(\theta, \phi)\, Z_1  = \Upsilon(-\theta, \phi) \,,
\end{align}
where $Z_1 = Z\otimes I$ ($Z_2 = I\otimes Z$ works equally well).  

Multiplying two $\Upsilon$ gates with opposite values of $\theta$ on both sides the operator $X_1 = X\otimes I$, we have
\begin{align}
 \Upsilon(-\theta, \phi)\, X_1\, \Upsilon(\theta, \phi)&= e^{i\theta Y\otimes Y /2}\, X_1\,e^{-i\theta Y\otimes Y /2}\nonumber\\
 &= \cos\theta\, X_1 + \sin\theta\, Z\otimes Y\,.\label{cz_to_syc:eq:X_1_identity}
\end{align}
With the identity~\eqref{cz_to_syc:eq:X_1_identity}, we have
\begin{widetext}
\begin{align}
 \Upsilon(-\theta, \phi)\, e^{i\alpha X_1}\, \Upsilon(\theta, \phi)
 &= \cos\alpha\, \Big(\cos\frac{\phi}{2}\,I\otimes I -i\sin\frac{\phi}{2}\, Z\otimes Z\Big) + i\sin\alpha\, \Big(\cos\theta\, X\otimes I + \sin\theta\, Z\otimes Y\Big)\nonumber\\[3pt]
 &= \Big(\cos\alpha\cos\frac{\phi}{2}\, I + i\sin\alpha\cos\theta\, X\Big) \otimes I - i Z\otimes \Big(\cos\alpha\sin\frac{\phi}{2}\, Z - \sin\alpha\sin\theta\, Y\Big)\,,\label{cz_to_syc:eq:composite_gate}
\end{align}
\end{widetext}
where $0 \leq\alpha\leq \pi/2$ is to be determined.  We introduce the Schmidt operators
\begin{align}
    &\Gamma_1(\alpha) = \cos\alpha\cos(\phi/2)\, I + i\sin\alpha\cos\theta\, X\,,\\[3pt]
    &\Gamma_2(\alpha) = \cos\alpha\sin(\phi/2)\, Z - \sin\alpha\sin\theta\, Y\,,
\end{align}
and the unitary~\eqref{cz_to_syc:eq:composite_gate} takes the simple form
\begin{align}\label{cz_to_syc:eq:composite_gate_schmidt}
  \Upsilon(-\theta, \phi)\, e^{i\alpha X_1}\, \Upsilon(\theta, \phi) =  \Gamma_1\otimes I -iZ\otimes \Gamma_2\,.
\end{align}
The Schmidt rank of this unitary is two.  Therefore, it is equivalent to a controlled-phase gate (also with Schmidt rank two) up to some single-qubit unitaries.  The two non-zero Schmidt coefficients of the unitary~\eqref{cz_to_syc:eq:composite_gate} are equal to the operator norms of $\Gamma_{1,\,2}$.  

The target controlled-phase gate that we want to decompose into the {\rm fSim} gate is
\begin{align}
\diag\big(1,\, 1,\, 1,\, e^{-i\delta}\big) = e^{-i\delta(I-Z)\otimes (I-Z)/4} \,,
\end{align}
where $0\leq\delta\leq 2\pi$.  It has two non-zero Schmidt coefficients  $\cos(\delta/4)$ and $\sin(\delta/4)$.  For example, we set the operator norm of $\Gamma_2$ to be equal to the second Schmidt coefficient of the target unitary
\begin{align}
\big\lvert\Gamma_2(\alpha)\big\rvert &= \sqrt{\big(\cos\alpha\sin(\phi/2)\big)^2 + \big(\sin\alpha\sin\theta\big)^2}\nonumber\\[3pt]
&= \sin(\delta/4)\,,\label{cz_to_syc:eq:second_sch_coeff}
\end{align}
and the parameter $\alpha$ can be determined
\begin{align}\label{cz_to_syc:eq:alpha}
\sin\alpha = \sqrt{\frac{\sin(\delta/4)^2 - \sin(\phi/2)^2}{\sin(\theta)^2-\sin(\phi/2)^2}}\;.
\end{align}
This equation has a solution if and only if one of the following two conditions is satisfied
\begin{align}
\lvert\sin\theta\rvert \leq \sin(\delta/4) \leq \lvert\sin(\phi/2)\rvert\,,\\[3pt]
\lvert\sin(\phi/2)\rvert\leq \sin(\delta/4)  \leq \lvert\sin\theta\rvert \,.
\end{align}
A large set of controlled-phase gates can be implemented with the typical values of $\theta$ and $\phi$ of the {\rm fSim} gate, except for those that are very close to the identity. 

To fix the local basis of the first qubit in Eq.~\eqref{cz_to_syc:eq:composite_gate}, we introduce two $X$ rotations of the same angle
\begin{gather}
 e^{-i\xi X/2}\, \Gamma_1(\alpha)\, e^{-i\xi X/2} = \cos(\delta/4)\,I\,,\\[2pt]
  e^{-i\xi X/2}\, Z\:  e^{-i\xi X/2}=  Z\,,
\end{gather}
where the angle $\xi$ is 
\begin{align}
 \xi = \arctan \bigg(\frac{\tan\alpha \cos\theta}{\cos(\phi/2)}\bigg) + \frac{\pi}{2}\, \Big(1-\sgn\big(\cos(\phi/2)\big)\Big)\,.
\end{align}
To fix the local basis of the second qubit in Eq.~\eqref{cz_to_syc:eq:composite_gate}, we introduce two $X$ rotations of opposite angles
\begin{align}
 e^{i\eta X/2}\, \Gamma_2(\alpha)\, e^{-i\eta X/2}= \sin(\delta/4)\,Z\,,
\end{align}
where the angle $\eta$ is
\begin{align}
\eta = \arctan \bigg(\frac{\tan\alpha \sin\theta}{\sin(\phi/2)}\bigg) + \frac{\pi}{2}\, \Big(1-\sgn\big(\sin(\phi/2)\big)\Big)\,.
\end{align}
Applying these local $X$ rotations before and after the gate sequence in Eq.~\eqref{cz_to_syc:eq:composite_gate}, we have
\begin{align}
 & e^{-i(\xi X_1 - \eta X_2)/2}\,\Upsilon(-\theta, \phi)\, e^{i\alpha X_1}\, \Upsilon(\theta, \phi)\, e^{-i(\xi X_1 + \eta X_2)/2}\nonumber\\[2pt]
 & \quad= \cos(\delta/4)\: I \otimes I - i\sin(\delta/4)\, Z\otimes Z\,,\label{cz_to_syc:eq:final_decomposition}
\end{align}
which is the desired controlled-phase gate up to some single-qubit $Z$ rotations.  

The target controlled-phase gate equals to the \CZ\ gate for $\delta = \pi$.  We numerically checked that the decomposition~\eqref{cz_to_syc:eq:final_decomposition} yields the \CZ\ gate for all 86 {\rm fSim} gates (with different values of $\theta$ and $\phi$) in our device.

\subsubsection{Universality for {\rm SU(2)}}

Here, we show how the argument for the well-known result that the $H$ and $T$ gates are universal for {\rm SU}(2) \cite{Boykin1999OnUniversal} can be adapted for the $X^{1/2}$ and $W^{1/2}$ gates. At the core of the argument lies the observation that $T \equiv R_Z(\pi / 4)$ followed by $HTH \equiv R_X(\pi / 4)$ is a single-qubit rotation by angle $\alpha$ which is an irrational multiple of $\pi$. Specifically, $\alpha$ is such that

\begin{equation}
\cos\frac{\alpha}{2} = \cos^2\frac{\pi}{8} = \frac{1}{2}\left(1 + \frac{1}{\sqrt{2}}\right).
\end{equation}
By Theorem B.1 in Appendix B of \cite{Boykin1999OnUniversal}, $\alpha/\pi$ is irrational because the monic minimal polynomial with rational coefficients of $e^{i\alpha}$

\begin{equation}\label{eq:minimal_polynomial}
x^4 + x^3 + \frac{1}{4}x^2 + x + 1
\end{equation}
is not cyclotomic (since not all its coefficients are integers).

Similarly, $W^{1/2} \equiv R_{X+Y}(\pi / 2)$ followed by $X^{1/2} \equiv R_X(\pi / 2)$ is a single-qubit rotation by angle $\beta$ such that

\begin{equation}
\cos\frac{\beta}{2} = \cos^2\frac{\pi}{4} - \frac{1}{\sqrt{2}}\sin^2\frac{\pi}{4} = \frac{1}{2}\left(1 - \frac{1}{\sqrt{2}}\right).
\end{equation}
The monic minimal polynomial with rational coefficients of $e^{i\beta}$ is \eqref{eq:minimal_polynomial}, the same as that of $e^{i\alpha}$. Therefore, $\beta$ is also an irrational multiple of $\pi$. The rest of the universality argument for $H$ and $T$ also applies in the case of $X^{1/2}$ and $W^{1/2}$.

\subsection{Circuit variants}
\label{sec:circuit_variants}
Since XEB entails classical simulation, it is hard or impossible to use it to estimate experimental fidelity of circuits which are hard or impossible to simulate classically. As described above, we designed our RQCs to ensure that an effective partitioning for SFA exists for circuits with fewer than 51 qubits. This gives rise to a significant gap in the cost of classical simulation between quantum supremacy circuits and most of our performance evaluation circuits. This gap facilitates performance evaluation of the Sycamore processor near the quantum supremacy frontier. In practice, however, we would like greater control over the simulation hardness, for two reasons. First, performance evaluation is still very costly for large $n$ approaching the supremacy frontier. Second, we would like to be able to estimate the fidelity of supremacy RQCs more directly, even though classical simulation of this case is unfeasible by definition.

In order to achieve more fine-grained control over the cost of classical simulation of our RQCs, we exploit the fact that the experimental fidelity depends primarily on the number and quality of the gates while the simulation cost is highly sensitive to the structure of the quantum circuit. Therefore, we approximate the experimental fidelity of RQCs which are hard or impossible to simulate from the fidelity of similar RQCs obtained as the result of transformations that reduce simulation cost without significantly affecting experimental fidelity.

We employ two such transformations. Each decreases simulation cost by reducing the bond dimension of promising circuit cuts. The first one removes some or all cross-partition gates. We say that the removed gates have been \textit{elided} and term the transformation \textit{gate elision}. The second transformation changes the sequence of coupler activation patterns shown in Fig.~\ref{fig:gate_patterns} to enable the formation of \textit{wedges} which reduce the bond dimension by slowing the spread of entanglement generated at the circuit cut.

The two transformations complete the description of RQCs used in our experiment. Consequently, each RQC is uniquely determined by five parameters: number of qubits $n$, number of cycles $m$, PRNG seed $s$, number of elided gates and the sequence of coupler activation patterns.

\begin{table}
\centering
\begin{tabular}{|c|c|c|}
\hline
\textbf{Circuit variant} & \textbf{Gates elided} & \textbf{Sequence of patterns} \tabularnewline
\hline
\hline
non-simplifiable full & none & ABCDCDAB \tabularnewline
non-simplifiable elided & some & ABCDCDAB \tabularnewline
non-simplifiable patch & all & ABCDCDAB \tabularnewline
simplifiable full      & none & EFGH \tabularnewline
simplifiable elided    & some & EFGH \tabularnewline
simplifiable patch     & all  & EFGH \tabularnewline
\hline
\end{tabular}
\caption{\label{table:rqc_variants} \textbf{Circuit variants.} Six variants of RQCs employed in quantum supremacy demonstration (non-simplifiable full) and performance evaluation (remaining five variants) classified by transformations applied in order to control the cost of classical simulation. The eight coupler activation patterns A, B, ..., H are shown in Fig.~\ref{fig:gate_patterns}.}
\end{table}
 
\subsubsection{Gate elision}
\label{subsubsection:gate_elision}
The most straightforward way to reduce the cost of classical simulation of a RQC is to remove a number of cross-partition gates across the most promising circuit cut. In order to enable independent propagation by the SFA of the wave function of each circuit partition for the first few cycles, the gates are elided beginning with the initial cycle. Each elided gate reduces the bond dimension of the partitioning by a factor of two or four, see Section \ref{sec:classical_sim}.

We refer to RQCs with a small number of elided gates as \textit{elided circuits}. A particularly dramatic speedup is possible when all two-qubit gates across the partitions are elided leading to two disconnected circuits running in parallel. We refer to such disconnected RQCs as \textit{patch circuits}. Base RQCs in which no gates have been elided are referred to as \textit{full circuits}.

If the error probability of the elided two-qubit gate is similar to the error probability of the two-qubit identity gate which it is replaced with, the circuit resulting from gate elision exhibits fidelity that is similar to the fidelity of the original circuit. This assumption holds when the two-qubit gate errors are dominated by the same decoherence processes that govern the single-qubit gate errors such as finite $T_1$ and $T_2$. Indeed, for circuit sizes where XEB on full circuits is possible, we have observed good agreement between fidelity estimates produced for patch, elided and full circuits. For harder circuits, we have observed good agreement between fidelity estimates for patch and elided circuits. See Section \ref{sec:large_scale_xeb} for detailed discussion of these results.

\begin{table*}
\centering
\begin{tabular}{|c|c|c|c|c|c|}
\hline 
\textbf{Circuit variant} & \textbf{$n$} & \textbf{$m$} & \textbf{Single-qubit gates} & \textbf{All two-qubit gates} & \textbf{Cross-partition two-qubit gates} \tabularnewline
\hline 
\hline 
non-simplifiable full & 53 & 20 & 1113 & 430 & 35 \tabularnewline
non-simplifiable elided & 53 & 20 & 1113 & 408 & 13 \tabularnewline
non-simplifiable patch & 53 & 20 & 1113 & 395 & 0 \tabularnewline
simplifiable full & 38 & 14 & 570 & 210 & 18 \tabularnewline
simplifiable elided & 38 & 14 & 570 & 204 & 12 \tabularnewline
simplifiable patch & 38 & 14 & 570 & 192 & 0 \tabularnewline
\hline 
\end{tabular}
\caption{\label{table:rqc_params} \textbf{Gate counts.} Number of gates in selected random quantum circuits employed for quantum supremacy demonstration and performance evaluation of the Sycamore processor.} 
\end{table*}

\subsubsection{Wedge formation}
\label{sec:wedge_formation}
\begin{figure}
\includegraphics[scale=0.67]{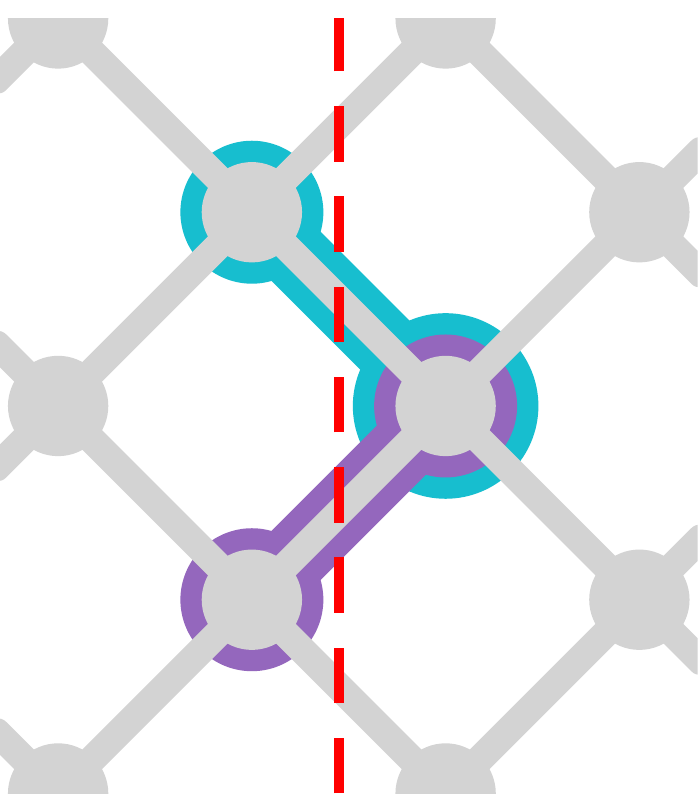}
\caption{\label{fig:wedge} \textbf{Cross-partition wedge.} Two consecutive cross-partition gates which share a qubit form a \textit{wedge}, as illustrated here with gates highlighted in turquoise and magenta. Schmidt rank of a single two-qubit gate is at most four. Schmidt rank of a wedge is also at most four. Therefore, generally wedges are not efficient at increasing entanglement across partitions and can be simulated efficiently by the SFA.}
\end{figure}

The most competitive algorithm for our hardest circuits, SFA (see Sec. \ref{subsec:sa_sfa}) scales proportionally to the bond dimension of the circuit partitioning which is equal to the product of Schmidt rank of all cross-partition gates (see Sec. \ref{subsec:sim_target_fidelity}). The Schmidt decomposition of most two-qubit gates in our RQCs consists of four terms (a few gates can be replaced with simpler gates with Schmidt rank of two, see Section \ref{sec:classical_sim}). Therefore most cross-partition gates contribute a factor of four to the bond dimension of the partitioning. However, when two consecutive cross-partition gates share a qubit forming a \textit{wedge} as shown in Fig.~\ref{fig:wedge}, the Schmidt decomposition of the resulting three-qubit unitary also has only four terms. In other words, the second cross-partition gate does not generally produce substantial new entanglement (as quantified by the Schmidt rank) among the partitions in excess of the entanglement produced by the first gate. Consequently, every wedge reduces the bond dimension of the partitioning by a factor of four.

The eight-cycle sequence ABCDCDAB and the four constituent coupler activation patterns A, B, C and D shown in Fig.~\ref{fig:gate_patterns} have been designed to prevent formation of wedges across promising circuit cuts. In other words, the sequence ensures that entanglement created in a given cycle by cross-partition gates is transferred into the bulk of each partition in the following cycle. 

On the other hand, the four-cycle sequence EFGH enables formation of wedges and thus efficient simulation of RQCs using SFA. We employ the latter sequence in most evaluation circuits and use the former eight-cycle sequence for the quantum supremacy circuits and largest evaluation circuits, see Table \ref{table:rqc_variants}.

\section{Large scale XEB results}\label{sec:large_scale_xeb}

In Section \ref{sec:calib_metro}, we have detailed the device calibration processes used for individual components such as qubits, couplers, and coupled pairs of qubits. We have also introduced cross-entropy benchmarking (XEB) as a method that allows us to evaluate the performance of a quantum system. In this section, we describe how we use a few circuit variations to benchmark our Sycamore processor at a larger scale. In particular, we present a modular version of XEB with ``patch circuits" that does not require exponential classical computation resources for estimating XEB fidelities $\xebfidelity$ of larger systems. We also describe the effect of choice of unitary model on large-scale $\xebfidelity$, as well as how we use patch circuits to monitor the stability of the full system.

\subsection{Limitations of full circuits}

\begin{figure}
\includegraphics{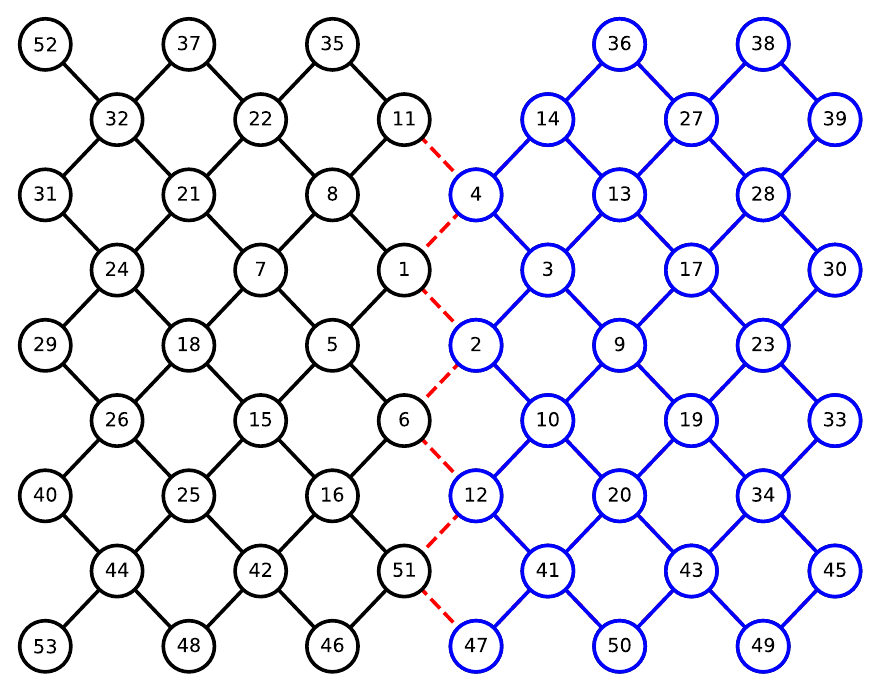}
\caption{\label{fig:qubit_order} \textbf{Qubit ordering for large-scale XEB experiments.} Illustration of the order in which qubits are added for large-scale experiments. The partition between left (black) and right (blue) qubits along the boundary (dashed red lines) is used in patch and elided circuits, as explained below.}
\end{figure}

We first discuss what we refer to as ``full circuits", where for a given set of qubits, all possible two-qubit gates participate in the circuit. With full circuits, we benchmarked the system as a function of size, where as discussed below the classical resources and techniques used to compute the $\xebfidelity$ is a function of the number of qubits. The order in which each qubit was added is labeled in Fig.~\ref{fig:qubit_order}. The rationale behind this ordering is explained in Section \ref{sec:circuits}. At each system size, we executed 10 randomly generated circuit instances on the processor and sampled output bitstrings 500k times for each circuit (unless otherwise specified). To minimize potential instance-to-instance fluctuations, we chose the gate sequences in a persistent, ``stable" manner: using a known seed for a random number generator, for each circuit, each time a new qubit is added, we maintain the same gateset for all the ``existing" qubits and new gates are only introduced to qubits and pairs associated with the added qubit (see Section \ref{sec:circuits} for details).

Once a sufficient number of bitstrings are collected, $\xebfidelity$ can be calculated for each system size, following the method described in Section \ref{sec:xeb_theory}. As the system size increases, the computational complexity of XEB analysis grows exponentially, which can be qualitatively divided into three regimes. For system size from 12 to 37 qubits, XEB analysis was carried out by evolving the full quantum state (Schr\"odinger method) on a high-performance server (88 hyper-threads, 1.5TB memory in our case) using the ``qsim" program. At 38 qubits we used a n1-ultramem-160 VM in Google's cloud (160 hyperthreads, 3.8TB memory). Above 38 qubits, Google's large-scale cluster computing became necessary, and in addition a hybrid Schr\"{o}dinger-Feynman approach, the ``qsimh" program, was used to improve the efficiency: in this case, we break the system up into two patches, where each patch can be efficiently computed via the Schr\"{o}dinger method and then connected by a Feynman path-integral approach (see Section \ref{sec:classical_sim} for more details). Finally we used a Schr\"odinger algorithm in the J\"{u}lich supercomputer for some circuits up to 43 qubits. 

In order to reduce the computational cost, we introduce two modified circuit types in the following sections. By using slightly simplified gate sequences, these two methods can provide good approximate predictions of system performance all the way out to the ``quantum supremacy" regime.

\subsection{Patch circuits: a quick performance indicator for large systems}

\begin{figure}
\includegraphics{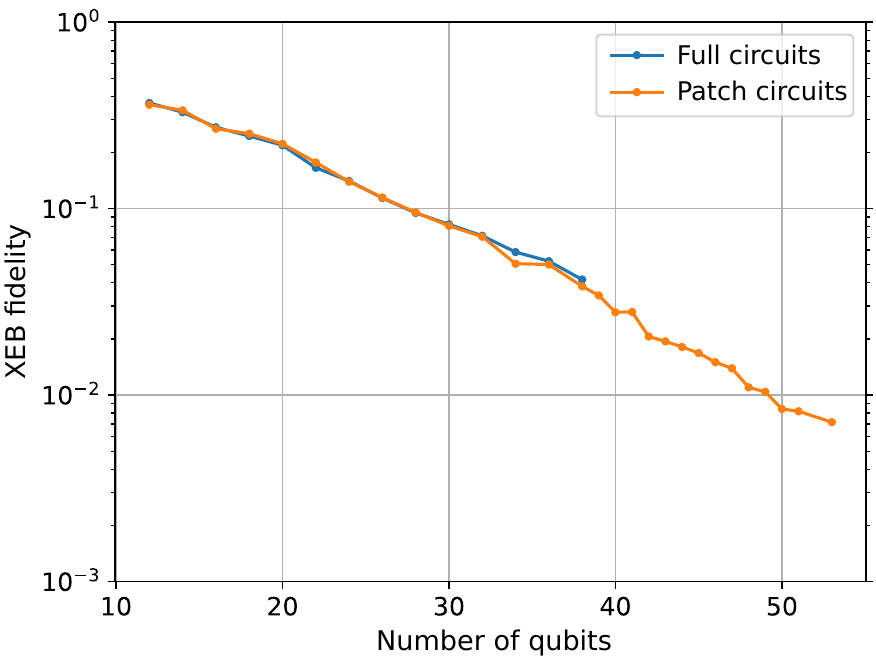}
\caption{\label{fig:patch_vs_full} \textbf{Comparison between XEB with patch circuits and full circuits.} Full vs.~patch circuit benchmarking up to 38 qubits with 14 cycles, showing close agreement to within the intrinsic fluctuations of the system. We plot the results for patch circuits out to 53 qubits.}
\end{figure}

The simplest approach to large-scale performance estimation is referred to as ``patch circuits," which predicts the performance of the full system by multiplying together the fidelities of non-interacting subsystems, or ``patches". In this work, we use two such subsystems, where each patch is roughly half the size of the full system. The two subsystems are run simultaneously, so that effects such as gate and measurement crosstalk between patches are included, but the two patches are analyzed separately when computing the fidelity. The two patches are defined by the gates removed along their boundary, as illustrated in Fig.~\ref{fig:qubit_order}. For sufficiently large systems, these removed two-qubit gates represent a small portion of the whole circuit. As a consequence, $\xebfidelity$ of the full system can be estimated as the product of the fidelities of the two subsystems; compared with full circuits, the main missing factor is the absence of entanglement between the two patches.

We evaluate the efficacy of using patch circuits by comparing it against full circuits with the same set of qubits. The experimental results can be seen in Fig.~4a (main text), where we show fidelities measured by these two methods for systems from 12 qubits to 53 qubits, in an interleaved fashion. We re-plot this data here in Fig.~\ref{fig:patch_vs_full} as well. As expected, the fidelities obtained via patch XEB show a consistent exponential decay (up to fluctuations arising from qubit-dependent gate fidelities and a small amount of system fluctuations) as a function of system size. For every system size investigated, we found that patch and full XEB provide fidelities that are in good agreement with each other, with a typical deviation of $\sim$5\% of the fidelity itself (we attribute the worst-case disagreement of 10\% at 34 qubits due to a temporary system fluctuation in between the two datasets, which was also seen in interleaved measurement fidelity data). Theoretically, one would expect patch circuits to result in $\sim10$\% higher fidelity than full circuits due to the slightly reduced gate count. We find that patch circuits perform slightly worse than expected, which we believe is due to the fact that the two-qubit gate unitaries are optimized for full operation and not patch operation. In any case, agreement between patch and full circuits shows that patch circuits can be a good estimator for full circuits, which is quite remarkable given the drastic difference in entanglement generated by the two methods. These results give us a good preview of the system performance in all three regimes discussed earlier.

The advantage of using patch circuits lies in its exponentially reduced computational cost, as it only requires calculating $\xebfidelity$ of subsystems at half the full size (or less if a larger number of smaller patches is used). This allows for quick estimates of large-scale system performance on a day-to-day basis, including for system and circuit sizes in the ``quantum supremacy" regime. As a consequence, we typically use patch circuits as a quick system performance indicator, which we use for rapid turnarounds between system calibration and performance evaluation, as well as for monitoring full system stability (see Section \ref{subsec:stability}). We also note that patch circuits can be used well beyond 50 qubits, and in fact can be extended to arbitrary numbers of qubits while keeping the analysis time at most linear in the number of qubits (or even constant if the patches can be analyzed in parallel), assuming that the patch size stays roughly constant and more non-interacting patches are added as the number of qubits grows.

\subsection{Elided circuits: a more rigorous performance estimator for large systems}
\label{subsec:performance_estimator_with_elided_circuits}

\begin{figure}
\includegraphics{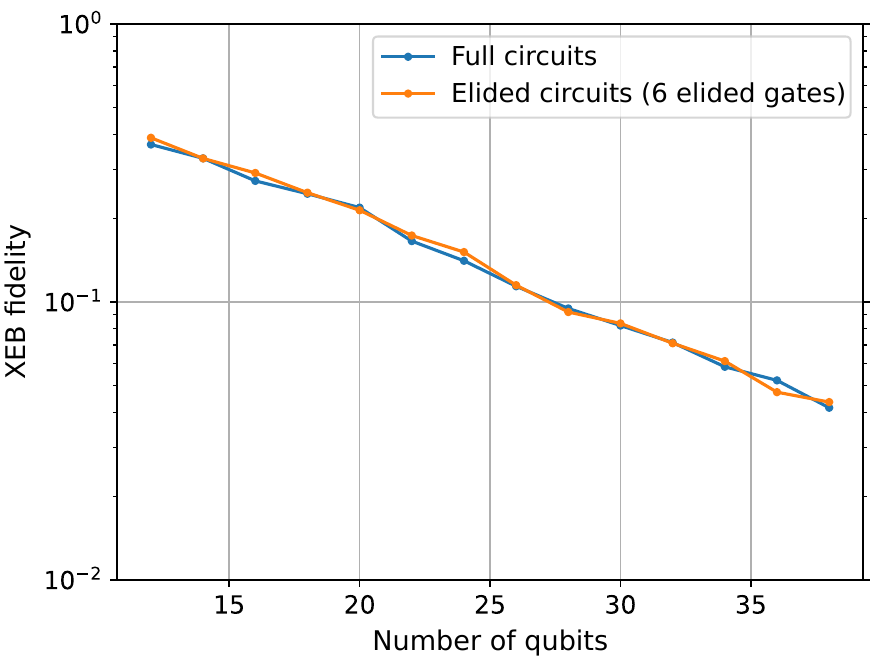}
\caption{\label{fig:elided_vs_full} {\bf Comparison between XEB with elided circuits and full circuits.} Full vs. elided circuit benchmarking up to 38 qubits at 14 cycles, showing close agreement to within the intrinsic fluctuations of the system.}
\end{figure}

For a more rigorous prediction of full $\xebfidelity$, we introduce a more sophisticated approach referred to as ``elided circuits".  Similar to patch circuits, we partition a given set of qubits into two subsets separated by a boundary, but elide (remove) only a fraction of the two-qubit gates along this boundary during a few early cycles of the sequence (more specifically, we elide the earliest gates in time, meaning early layers will have none of their gates along the boundary while later layers will have all of their usual gates across the boundary). Accordingly, the two subsets of qubits are no longer isolated from each other and we cannot simply compute their fidelities separately and multiply. Rather, we must still compute the evolution of the full system. Given that a sufficient number of gates are elided, we can take advantage of the ``weak link" between patches with a hybrid analysis technique: we compute each patch via the Schr\"odinger method and then connect them with a Feynman path-integral approach (see Section \ref{sec:classical_sim} for more details on this ``qsimh" program).

Compared with patch circuits, elided circuits more closely approach a description of the full system performance under a full circuit: in addition to capturing issues such as control and readout crosstalk, elided circuits allow entanglement to form between the two weakly connected subsystems. It covers essentially all the possible processes that occur in the full  circuit, and therefore can be used to predict system performance at a dramatically reduced computational cost, albeit significantly costlier than patch circuits.

In order to validate the use of elided circuits as a system performance estimator, we evaluated its accuracy via a direct comparison with full circuits.  In Fig.~\ref{fig:elided_vs_full} we show two sets of fidelities from interleaved full and elided circuit experiments. For every system size investigated, using elided circuits yields a fidelity value that is in good agreement with the one obtained with the corresponding full circuits. The average ratio of elided circuit fidelity to full circuit fidelity over all verification circuits was found to be 1.01, with a standard deviation of $5\%$, dominated by system fluctuations. It is this agreement that certifies elided circuits as a precise predictor for full circuits (within a systematic relative uncertainty of $5\%$), which we rely on to extrapolate the system performance in the regimes where full circuit analysis is too expensive to perform (i.e., Fig.~4b of the main text).

Compared with full circuits, elided circuits can result in a reduced amount of quantum entanglement in the system. The amount of reduced entanglement can be bounded from above by counting the number of iSWAP gates across the boundary: one iSWAP gate generates at most two units of bipartite entanglements (ebits). This upper bound translates directly into the exponential cost of a Schr\"odinger-Feynman simulation. For elided circuits with 50 qubits and 14 cycles, the full circuit has approximately 25 ebits of entanglement, while with 6 elisions the elided circuit has at most 12 ebits entanglement between the two patches. For the 53-qubit elided circuits used in the main paper \cite{arute2019}, there were enough iSWAPs across the boundary that the amount of entanglement between patches for full vs.~elided circuits should be close, giving us even more confidence in using elided circuits to predict the fidelity of the circuit used to claim quantum supremacy.

\subsection{Choice of unitary model for two-qubit entangling gates}

\begin{figure}
\includegraphics{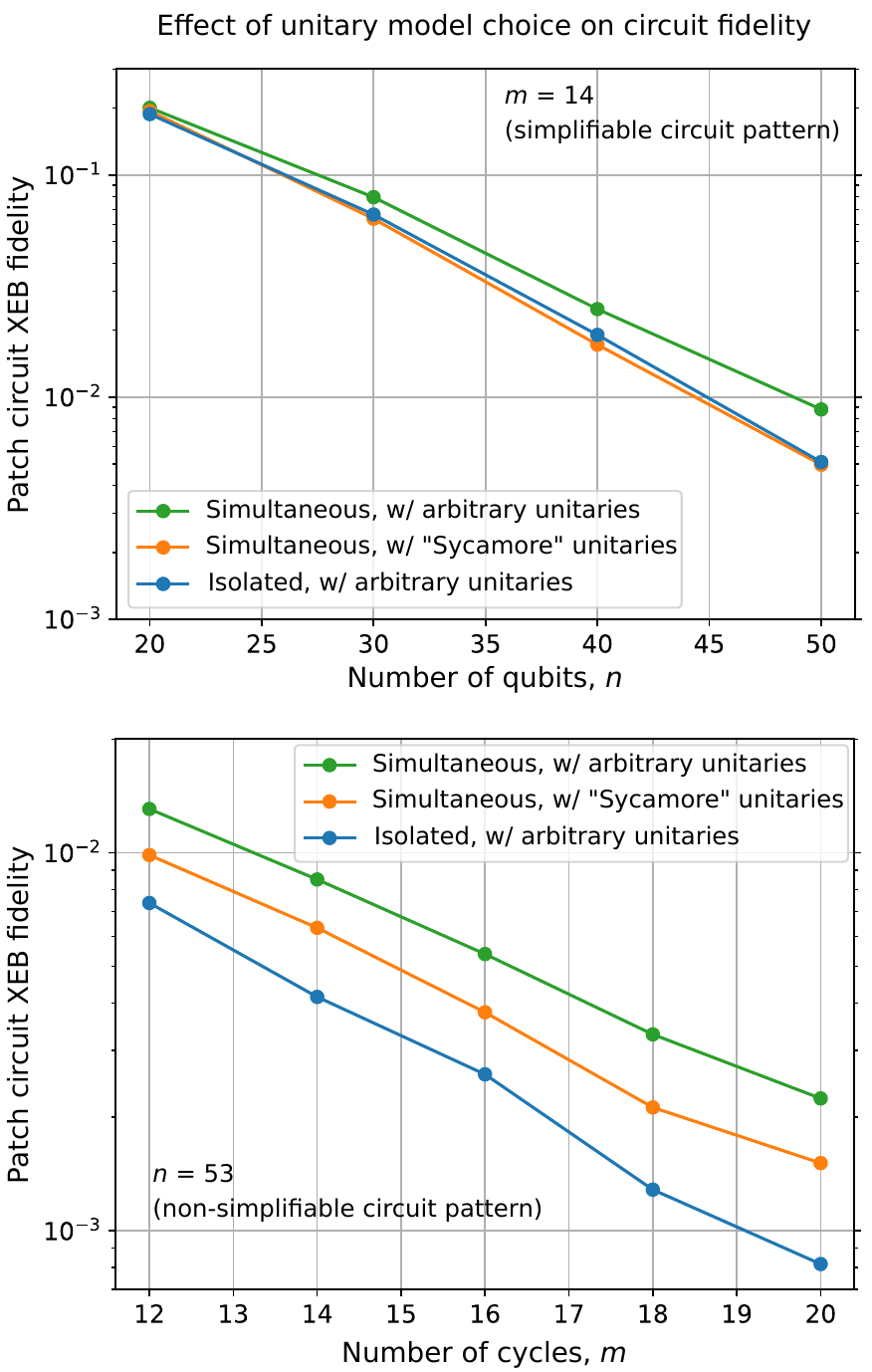}
\caption{\label{fig:choice_of_unitary_model} \textbf{Effect of unitary model on full system fidelity.} \textbf{a,} Patch circuit fidelity versus number of qubits and choice of unitary model. \textbf{b,} Same but versus number of cycles and for the non-simplifiable supremacy circuits. Blue: patch XEB fidelities using the unitaries deduced from the best-fit fSim unitary from isolated pair XEB. Green: patch XEB fidelities using the unitaries deduced from the best-fit fSim unitary from per-layer simultaneous pair XEB. Orange: patch XEB fidelities using the unitaries deduced from the best-fit ``Sycamore unitary" ($\theta = \pi/2$, $\phi = \pi/6$) from per-layer simultaneous pair XEB. As expected, the best fidelities arise from fitting to the most general unitary in parallel operation, although the fidelities are high enough to achieve quantum supremacy with the Sycamore unitary model as well.}
\end{figure}

In Section \ref{sec:calib_metro}, we discussed how the two-qubit gate unitaries can be measured by two different approaches: isolated two-qubit XEB and per-layer simultaneous two-qubit XEB. These two methods resulted in two different unitary models when deducing the best-fit unitary. Since we must specify the two-qubit gate unitary matrices in order to compute $\xebfidelity$ of the larger system, a natural question is which unitary model should be used. To address this question, we point out that full XEB on the large system occurs in repeated cycles, where during each two-qubit gate layer, all the two-qubit gates in the same orientation take place at the same time (see Fig.~3 in the main text). As a consequence, the two-qubit gate layers during simultaneous pair XEB in Fig.~\ref{fig:per_layer_xeb} emulate the corresponding layer when running full XEB on a large system. Accordingly, learning the unitaries in parallel operation captures any small coherent modifications induced by the simultaneous application of the other two-qubit gates, such as flux control crosstalk and dispersive shifts from stray interactions. This is evident from the fact that by re-learning the two-qubit unitary parameters, the errors extracted from simultaneous pair XEB become purity-limited (see Fig.~\ref{fig:per_layer_xeb}). This correspondence assures us that unitary parameters extracted from simultaneous pair XEB provides a more accurate description of the full system when full XEB is performed. 

In Fig.~\ref{fig:choice_of_unitary_model}, we show patch circuit fidelities at different system sizes, where the fidelity is evaluated using three different unitary models: the best-fit unitaries from isolated pair XEB, the best-fit unitaries from simultaneous pair XEB, and the best-fit ``Sycamore" unitaries from simultaneous pair XEB. The Sycamore unitaries are the unitaries obtained when keeping the swap angle fixed at $\theta = \pi/2$ and conditional phase fixed at $\phi = \pi/6$ for all qubits, and then fitting only for two single-qubit phase terms. For the purpose of benchmarking the system fidelity for the operations we performed, we have focused on using unitaries learned from simultaneous pair XEB, which provide the most accurate description of the system. The validity of this approach is experimentally verified\textemdash for the same gate sequences, using the simultaneous pair XEB unitaries leads to the best full-system fidelity values at every system size. This is direct evidence that the unitaries learned from simultaneous pair XEB form a more accurate description of the system than those from isolated pair XEB.

On the other hand, in order to be useful for generic quantum algorithms, it will be desirable to use calibrated gatesets that are independent of the specific gate sequences used.  For this purpose, it is important to check the circuit fidelity under the other two unitary models, where the two-qubit gate unitaries were calibrated in more generic settings. One can see that fidelities calculated from these two unitary models still demonstrate nearly as good performance despite the addition of small coherent control errors. They differ from the fidelities using the simultaneous pair XEB unitaries by less than a factor of 2 at 50 qubits (fidelity goes from $9\times10^{-3}$ to $5\times10^{-3}$ at 50 qubits). This is remarkable since it suggests going from a 2-qubit setting to 50-qubit setting, our full system calibration precision degrades only by a factor of $<2$ despite the system size increasing by a factor of 25. This high precision in gate calibration gives us confidence to use our processors in NISQ algorithms.

\subsection{Understanding system performance: error model prediction}\label{sec:error_model}

\begin{figure}
\includegraphics{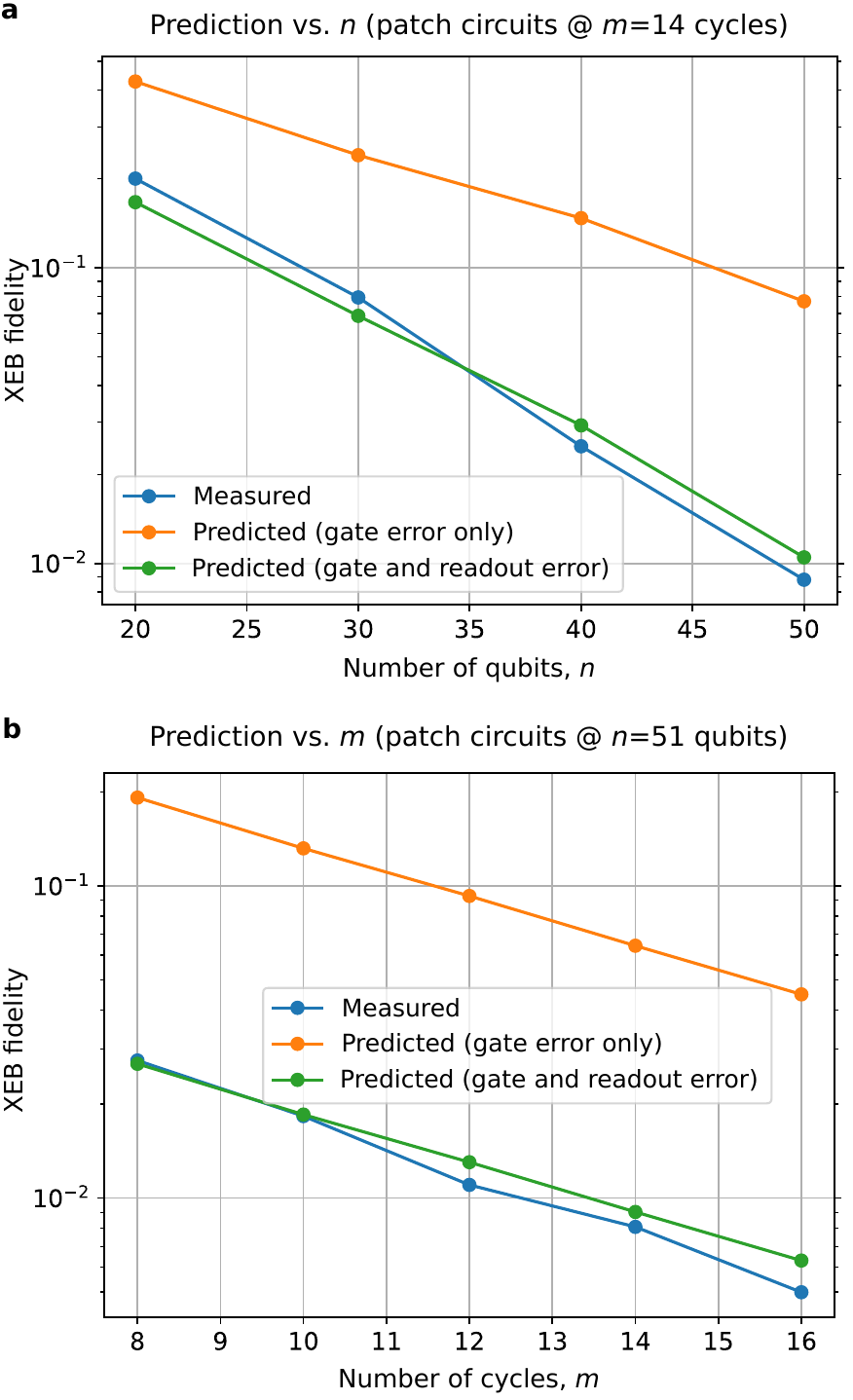}
\caption{\label{fig:predictions} \textbf{Predicted vs.~measured large-scale XEB fidelity.} \textbf{a,} Data and two predictions for 14-cycle patch circuits vs.~number of qubits. Predictions are based on the product of single- and two-qubit gate entanglement fidelities under simultaneous operation. Blue curve contains measured fidelities. Orange is the prediction based only on gate errors during parallel operation, but without taking measurement error into account. Green is the same but multiplied by the measured readout fidelities. \textbf{b,} Same as the first panel, but vs.~number of cycles at a fixed number of qubits $n = 51$. Again, the prediction from simultaneous gate fidelities and measurement fidelity is a good prediction of the actual system performance.}
\end{figure}

In this section, we perform additional analysis to compare the measured fidelities to that predicted from the constituent gate and measurement errors.

The most commonly used error model in quantum computing theory is the digital error model. Analogous to the independent noise model in classical information theory, the digital error model is based on the assumption that there are no space and time correlations between errors of quantum gates \cite{shor1995scheme, knill2008randomized, boixo2018characterizing}. If this assumption is valid, it should be possible to construct the fidelity  of a large quantum system from the fidelities of its constituent parts: single- and two-qubit gates, and measurement. It is important to point out that the gate fidelity metric that should be used here is the entanglement fidelity, $1 - e_P$ (see Section \ref{sec:computing_errors} for more details).
This is the correct quantity to describe the fidelity of quantum operations since, in contrast to other metrics such as the commonly used average fidelity, it is independent of the dimension of the Hilbert space.

In Fig.~\ref{fig:predictions}, we show fidelities as a function of both system size and number of cycles (circuit depth), measured with patch circuits. In each plot, we compare the measured fidelities to the predicted fidelities, which are calculated from a simple multiplication of individual gate entanglement fidelities as measured during simultaneous operation, along with the measurement fidelities obtained during simultaneous measurement. We note that the measured readout fidelities actually also automatically include the effect of state preparation errors as well. More explicitly, if a circuit contains the set of single-qubit gates $G_1$, the set of two-qubit gates $G_2$, and the set of qubits $Q$, then we approximate the fidelity $F$ as 
\begin{equation}
F = \prod_{g \in G_1} (1 - e_g) \prod_{g \in G_2} (1 - e_g) \prod_{q \in Q} (1 - {e_q}),
\end{equation}
where $e_g$ are the individual gate Pauli errors and $e_q$ are the state preparation and measurement errors of individual qubits. It is evident that there is a good agreement between the measured and predicted fidelities, with deviations of up to only ~10-20\%. Given that the sequence here involves tens of qubits and $\sim1000$ quantum gates, this level of agreement provides strong evidence to the validity of the digital error model.

This conclusion can be further strengthened by the close agreement between the fidelities of full circuits, patch circuits, and elided circuits.  Even though these three methods differ only slightly in the gate sequence, they can result in systems with drastically different levels of computational complexity and entanglement between subsystems. The agreement between the fidelities measured by these different methods, as well as the agreement with the predicted fidelity from individual gates, gives compelling evidence confirming the assumptions made by the digital error model. Moreover, these assumptions remain valid even in the presence of quantum entanglement. 

The validation of the digital error model has crucial consequences, in particular for quantum error correction. The absence of space or time correlations in quantum noise has been a commonly assumed property in quantum error correction since the very first paper on the topic \cite{shor1995scheme}. Our data is evidence that such a property is achievable with existing quantum processors. 

\subsection{Distribution of bitstring probabilities}
\label{subsec:xeb_distribution_and_ks_test}

In Section~\ref{sec:xeb_theory}, we motivate two different estimates for fidelity $F$, one based on the cross entropy, Eq.~\eqref{eq:log_xeb_fidelity}, and the other based on linear cross entropy, Eq.~\eqref{eq:linear_xeb_fidelity}.  In this section, we examine the probabilities of sampled bitstrings and compare them against theoretical distributions. We use bitstring samples from non-supremacy region to demonstrate the analysis methodology, then apply it to the sample in the supremacy region.

The theoretical PDF for the bitstring probability $p$ with linear XEB is
$$
P_l(x|F) = (F x + (1 - F)) e^{-x}
$$
where $x\equiv Dp$ is the probability $p$ scaled by the Hilbert space dimension $D$, and $F$ is the linear cross entropy fidelity. The PDF for $\log p$ is
$$
P_c(x|F) = (1 + F(e^x - 1)) e^{x - e^x}
$$
where $x\equiv \log(Dp)$ and $F$ is the cross entropy fidelity.

From a set of bitstrings $\{q_i\}$, the fidelity is estimated from the ideal probabilities $\{p_i = p_s(q_i)\}$ as
\begin{eqnarray}
\hat F_l &=& \avg{D p} - 1, \\
\hat F_c &=& \avg{\log(Dp)} + \gamma,
\end{eqnarray}
where $\gamma$ is the Euler-Mascheroni constant, see Sec.~\ref{sec:xeb_large}. 

\begin{figure} \label{fig:experimental_distribution_of_Dp}
\includegraphics[width=\columnwidth]{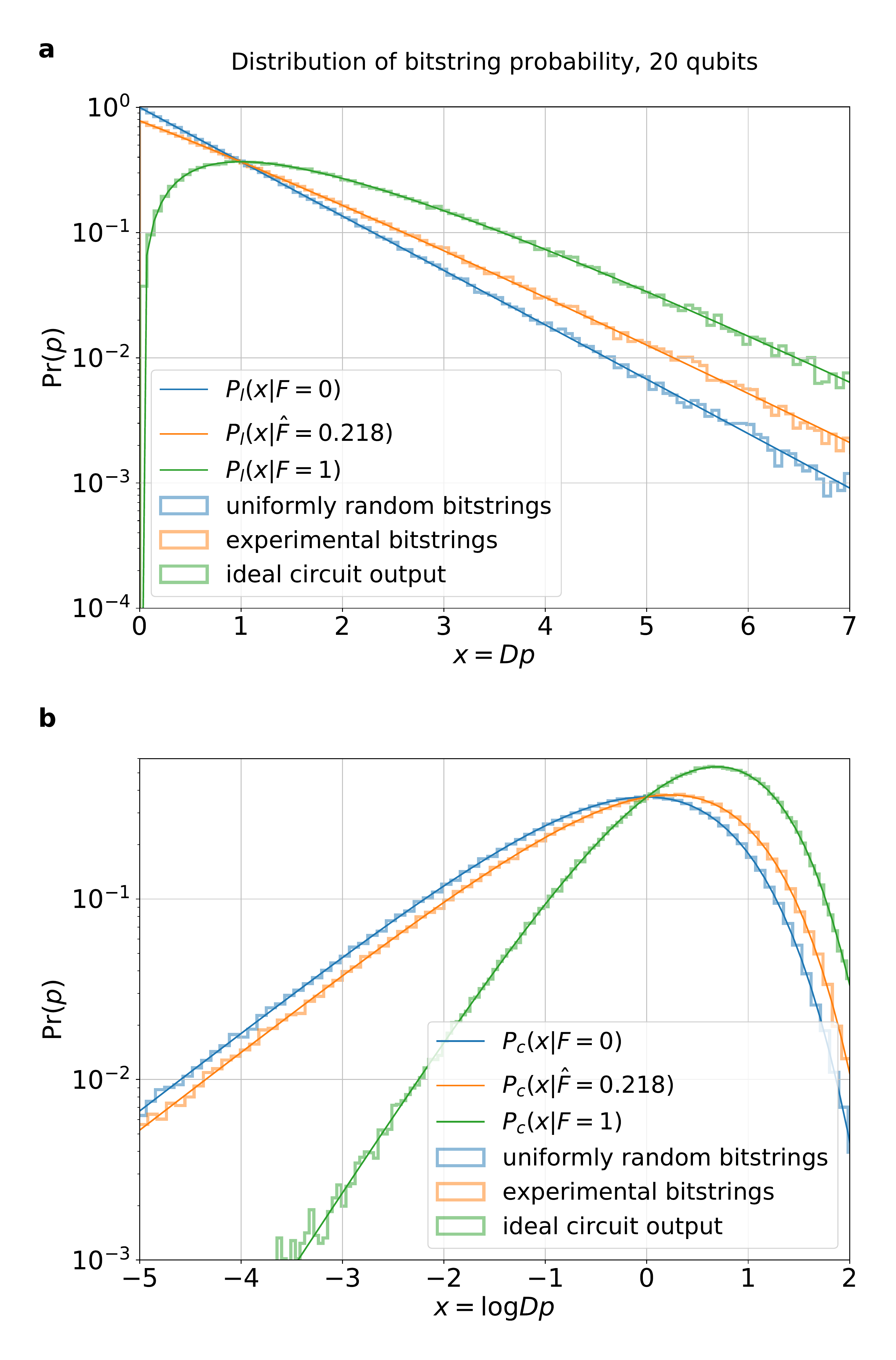}
\caption{{\bf Histograms of ideal probabilities.} The ideal probability $p$ is calculated from the final state amplitudes of a (20-qubit 14-cycle) random circuit. The blue, orange, and green histogram is the ideal probabilities of bitstrings sampled uniformly at random, from the experiment, and ideal output, respectively.
{\bf a,} The distribution of $D p$ and theoretical curves $P_l(x|F_l)$ normalized to histogram counts for $F_l = 0, \hat F_l, 1$, respectively. {\bf b,} The distribution of $\log(Dp)$ and theoretical curves $P_c(x|F_c)$ for $F_c = 0, \hat F_c, 1$, respectively.}
\end{figure}

Figure~\ref{fig:experimental_distribution_of_Dp} shows the distribution of $\{p_i\}$ from 0.5 million bitstrings obtained in an experiment with a 20-qubit 14-cycle random quantum circuit. For comparison, we produce 0.5 million bitstrings sampled uniformly at random and 0.5 million bitstrings sampled from the output distribution of the ideal circuit and show them in the same figure. The theoretical distribution curves are also shown, where the fidelity estimated from data is fed into the curve $P_l(x|\hat F)$ and $P_c(x|\hat F)$.

\begin{figure} \label{fig:kolmogorov_distribution}
\includegraphics[width=\columnwidth]{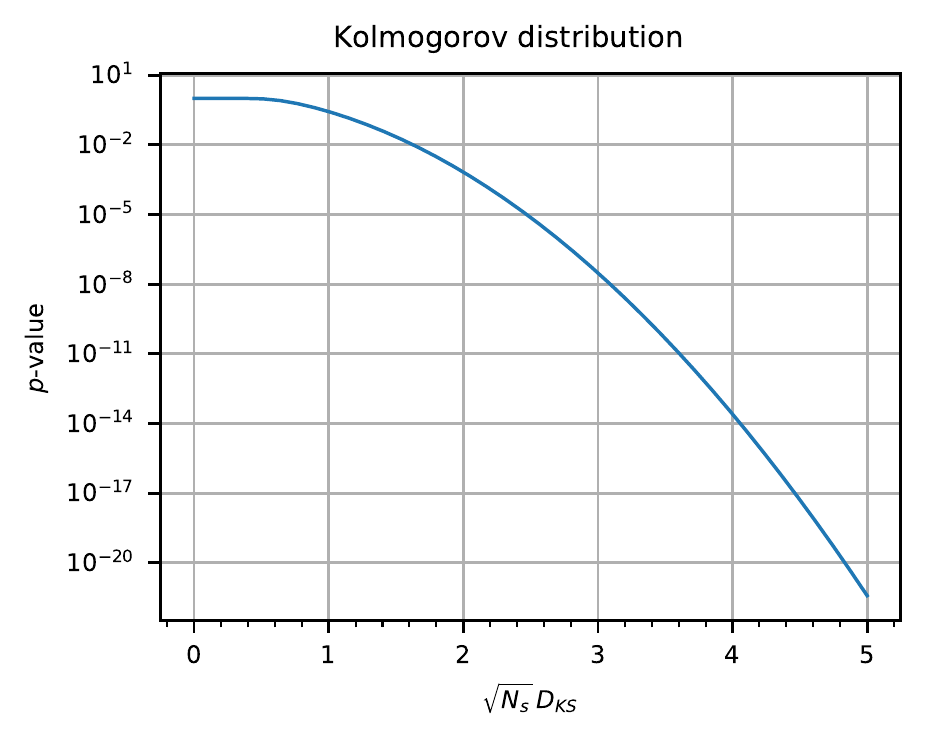}
\caption{{\bf The Kolmogorov distribution function.} This function is used to compute $p$-value from a given $D_{KS}$ and number of samples $N_s$.}
\end{figure}

\begin{figure} \label{fig:ks_test_d_and_p_values_20qubits}
\includegraphics[width=\columnwidth]{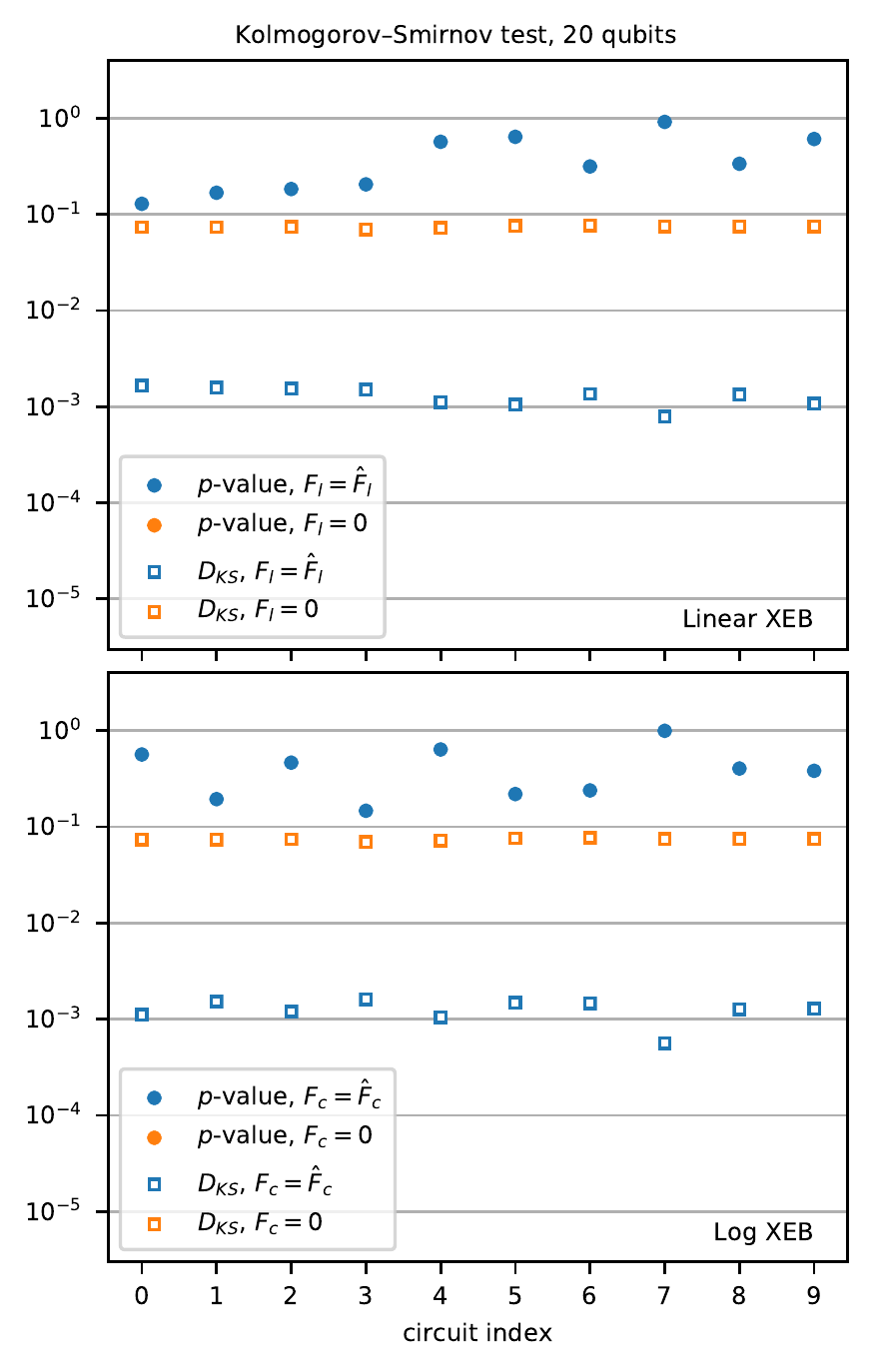}
\caption{{\bf The Kolmogorov-Smirnov test results for each of 10 circuits for a (20-qubit 14-cycle) random circuit.}
See text for the definition of $D_{KS}$ and $p$-value. The upper plot is for linear XEB, and the lower one is for log XEB.}
\end{figure}

We see good agreements between experiment and theory. To quantify the agreements, we use the Kolmogorov-Smirnov test~\cite{kolmogorov-smirnov-test} to characterize the goodness of fit of data $\{p_i\}$ to theoretical PDFs. First we compute the Kolmogorov-Smirnov statistics $D_{KS}$, that is, the distance between data and theory as the supremum of point-wise distances between the empirical cumulative distribution function of data $\text{ECDF}(p)$ and the theoretical cumulative distribution function $\text{CDF}(p)$:
$$
D_{KS} = \sup_{i} |\text{ECDF}(p_i) - \text{CDF}(p_i)|.
$$
We then convert the distance $D_{KS}$ to a $p$-value using the Kolmogorov distribution shown in Fig.~\ref{fig:kolmogorov_distribution}. The $p$-value is used for rejecting the null hypothesis that the data $\{p_i\}$ is consistent with the theoretical distribution. The whole Kolmogorov-Smirnov test is done using the scipy package~\cite{scipy} and checked against R package ks.test~\cite{R_package}. Both packages produce consistent results.

We test the ideal probabilities of bitstrings observed in the experiment $\{p_i\}$ against 2 theoretical distributions, one with estimated fidelity $F = \hat F$ and one with fidelity $F = 0$. The Kolmogorov-Smirnov statistics $D_{KS}$ and the $p$-value of every circuit are shown in figure~\ref{fig:ks_test_d_and_p_values_20qubits}. Note that the $p$-values for $F=0$ are not shown because they are $\ll 10^{-20}$ due to the large $D_{KS} \approx 0.07$ with $N_s = 5\times 10^5$ points in the sample. That is evident from reading off Fig.~\ref{fig:kolmogorov_distribution}.

We reject the null hypothesis that the experimental bitstrings are consistent with the uniform random distribution with very high confidence for this (20-qubit 14-cycle) random circuit.

Now we turn our attention to the supremacy circuits.

We use random circuits with gate elisions for checking the distributions because it is exponentially expensive to calculate the ideal theoretical probability of a bitstring without gate elisions. The effect on fidelity from gate elisions is well understood, see Sec.~\ref{subsec:performance_estimator_with_elided_circuits}. The gate elisions are chosen to minimize the effect while making the classical estimation feasible, see Sec.~\ref{subsubsection:gate_elision}. We sample $N_s = 3\times 10^6$ bitstrings $\{q_i | i = 1 ... N_s\}$ from each of 10 (53-qubit 20-cycle) random circuits, and compute the theoretical ideal probabilities of each bitstring $\{p_i | i = 1 ... N_s\}$.

\begin{figure} \label{fig:distribution_of_Dp}
\includegraphics[width=\columnwidth]{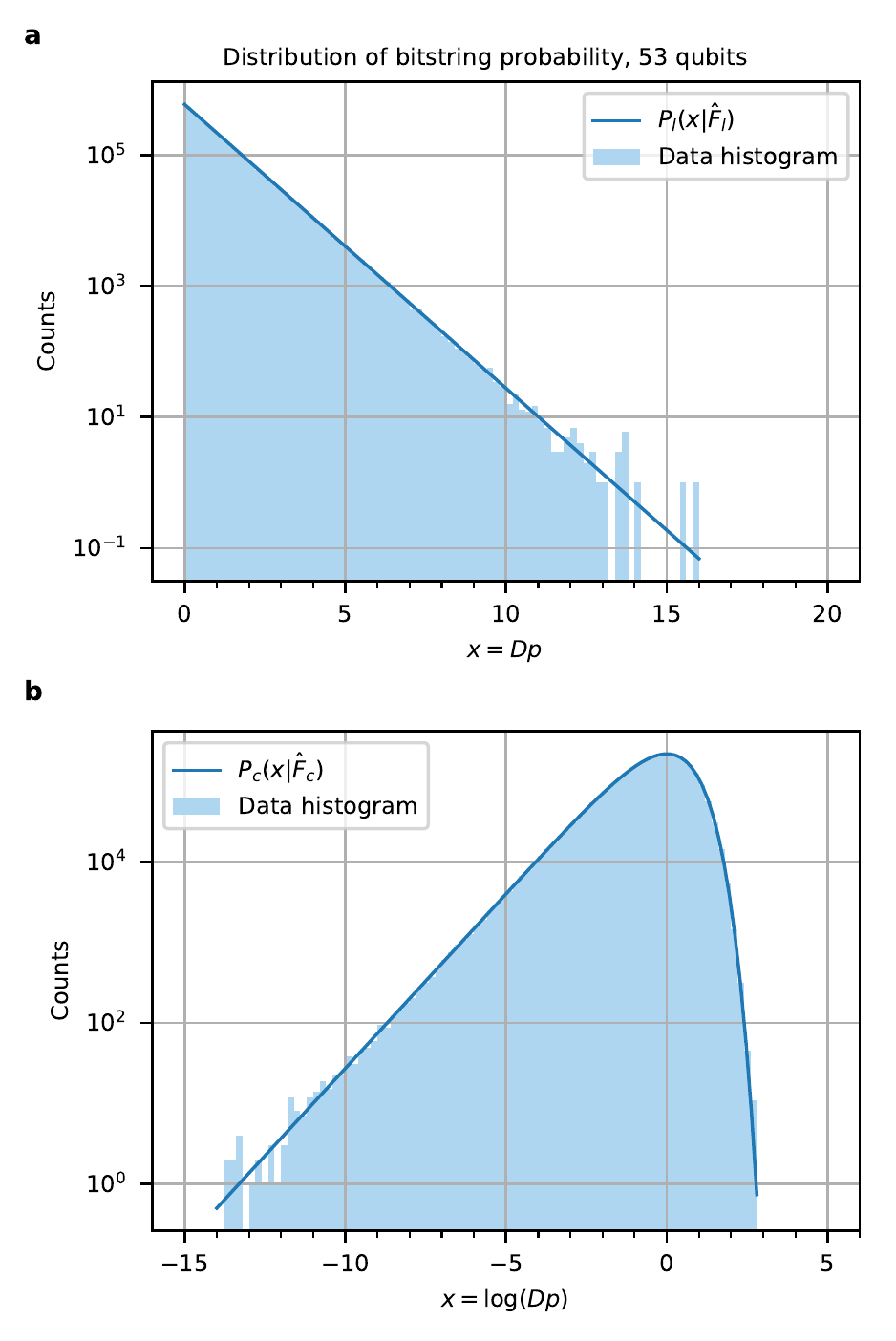}
\caption{{\bf Distribution of bitstring probabilities from a 53-qubit 20-cycle circuit.} We calculate the theoretical probabilities of experimentally-observed bitstrings. {\bf a,} The distribution of $D p$ and the theoretical curve $P_l(x|\hat F_l)$ normalized to histogram counts. {\bf b,} The distribution of $\log(Dp)$ with theoretical curve $P_c(x|\hat F_c)$.}
\end{figure}

The distributions of $D p$ and $\log(Dp)$ from one such circuit along with the corresponding theoretical curves are shown in Fig.~\ref{fig:distribution_of_Dp}.

\begin{figure} \label{fig:ks_test_d_and_p_values_53qubits}
\includegraphics[width=\columnwidth]{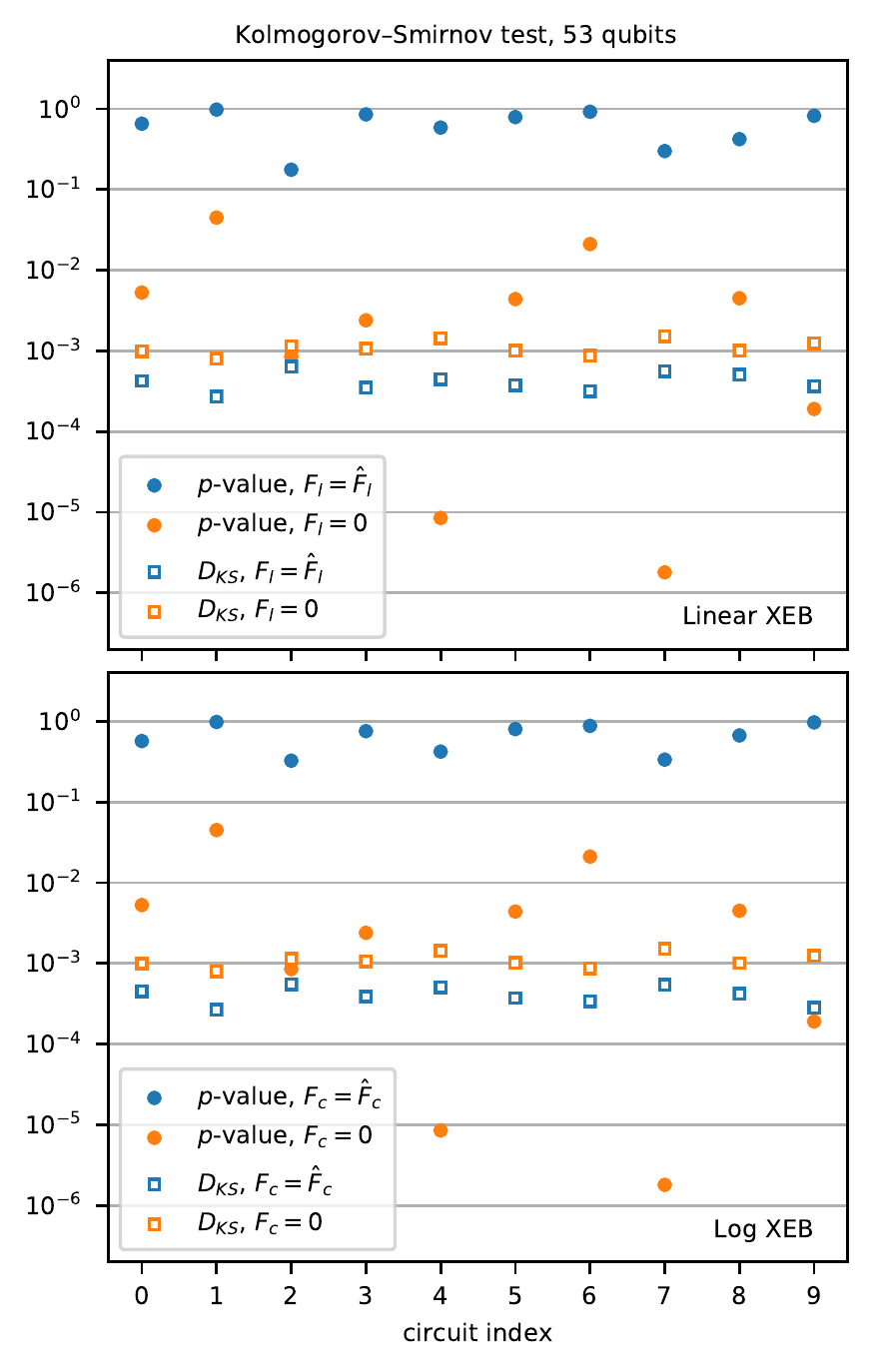}
\caption{{\bf The Kolmogorov-Smirnov test results for random circuits with 53 qubits.}
The upper plot is for linear XEB, and the lower one is for log XEB.}
\end{figure}

We again use the Kolmogorov-Smirnov test to characterize the goodness of fit of data $\{p_i\}$ to theoretical PDFs with estimated fidelity $F=\hat F$ and zero fidelity $F = 0$. The Kolmogorov-Smirnov statistics $D_{KS}$ and the $p$-value of every circuit are shown in figure~\ref{fig:ks_test_d_and_p_values_53qubits}.

The $p$-value for the null hypothesis of zero fidelity is generally small for every circuit, with a maximum of 0.045 for circuit number 1. We say that the null hypothesis of zero fidelity is rejected better than a 95\% confidence level for each circuit. On the other hand, the $p$-value of null hypothesis of estimated fidelity $\hat F$ is generally large. The $p$-value is between 0.18 and 0.98 for linear XEB, and between 0.33 and 0.98 for log XEB. That indicates that the empirical cumulative distribution functions $\text{ECDF}(p_i)$  from data is quite consistent with the theoretical $\text{CDF}(p_i|\hat F)$.

As will be seen in Fig.~\ref{fig:elided_stat_error} in section~\ref{subsec:xeb_stat_uncertainty} below, the fidelity of individual circuits are consistent with each other within the statistical uncertainties. Therefore it makes sense to do a Kolmogorov-Smirnov test on all samples combined, containing 30 million bitstrings. The estimated fidelities from the combined sample are $\hat F_l = 2.24\times 10^{-3}$ and $\hat F_c = 2.34\times 10^{-3}$, respectively. The $D_{KS}$ and $p$-values are listed in table~\ref{tab:ks_test_for_combined_samples}. The $p$-value for the null hypothesis of $F=0$ is very small: $p$-value = $3\times 10^{-24}$ from scipy, and $p$-value $< 2.2\times 10^{-16}$ from R. We note the more conservative value in the table. The null hypothesis of $F=0$ is rejected with much higher confidence levels than individual circuits.

\begin{table}[] \label{tab:ks_test_for_combined_samples}
\renewcommand{\arraystretch}{1.3}
\begin{tabular}{|l|c|c|c|c|} \hline
  & \multicolumn{2}{c|}{$D_{KS}$} & \multicolumn{2}{c|}{$p$-value} \\ \hline
           & $F=\hat F$ & $F=0$  & $F=\hat F$ & $F=0$ \\ \hline
Linear XEB & $1.3\times 10^{-4}$  & $9.6\times 10^{-4}$ & 0.66 & $< 2.2\times 10^{-16}$ \\
Log XEB & $9.5\times 10^{-5}$  & $9.6\times 10^{-4}$  & 0.95 & $< 2.2\times 10^{-16}$ \\ \hline
\end{tabular}
\caption{{\bf The Kolmogorov-Smirnov test results on combined samples}.}
\end{table}

\subsection{Statistical uncertainties of XEB measurements}
\label{subsec:xeb_stat_uncertainty}

In this section we check the statistical uncertainties of our fidelity estimates against theoretical predictions.

The statistical uncertainties of $\hat{F}_l$ and $\hat{F}_c$ are estimated from data using the
standard error-on-mean formula as
\begin{eqnarray*} \label{eq:fidelity_stat_uncertainty}
  \hat \sigma_{F_l} &=&  D \sqrt{\mbox{Var}(p) / N_s}, \\
  \hat \sigma_{F_c} &=&  \sqrt{\mbox{Var}(\log p) / N_s},
\end{eqnarray*}
where $\mbox{Var}(x)$ is the variance estimator of sample $\{x_i\}$. Because the distribution of $p$ and $\log p$ have finite variances both experimentally and theoretically, we can use the bootstrap procedure~\cite{bootstrapping} to verify the estimate of statistical uncertainties.

\begin{figure} \label{fig:bootstrap_fidelity}
\includegraphics[width=\columnwidth]{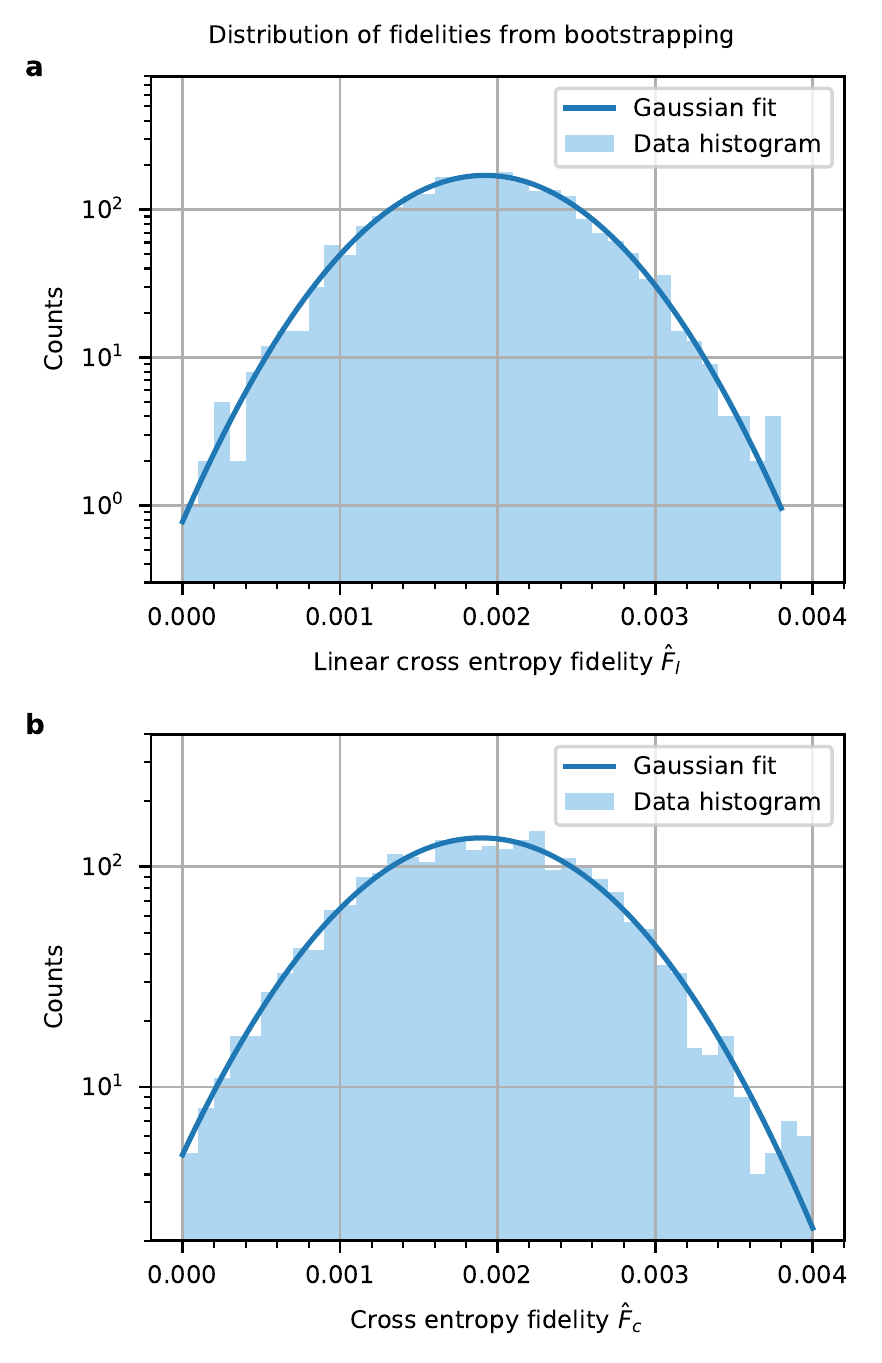}
\caption{Distribution of fidelity from 4000 bootstrap samples. {\bf a,} The distribution of bootstrap $\hat{F_l}$. The theoretical curve is a Gaussian fit normalized to histogram counts. {\bf b,} The distribution of bootstrap $\hat{F_c}$, with Gaussian fit.}
\end{figure}

The fidelity distribution from 4000 bootstrap samples are shown in Fig.~\ref{fig:bootstrap_fidelity}. The distribution of $\hat{F}_l$ and $\hat{F}_c$ are each fit to a Gaussian distribution function using maximum likelihood.

The Kolmogorov-Smirnov test on the Gaussian fit produces $p$-values of 0.99 and 0.41 for
$\hat{F}_l$ and $\hat{F}_c$ bootstrap distributions, respectively. It indicates that the central limit theorem is at work and the distributions are consistent with Gaussian distributions.

The estimated statistical uncertainty, the standard deviation of the bootstrap distribution, and the $\sigma$ parameter of the Gaussian fit are compared against each other to verify that the statistical uncertainty estimate is minimally biased. For the example circuit used in the figures, the three parameters are 5.78, 5.78, 5.78 ($\times 10^{-3}$) for $\hat{\sigma}_{F_l}$, respectively. The same parameters for $\hat{\sigma}_{F_c}$ are 7.40, 7.46, 7.46 ($\times 10^{-3}$). The relative differences are less than 1\%, consistent with the expected agreement of parameters for 4000 bootstrap samples.

We repeat the bootstrap procedure on all ten 53-qubit 20-cycle circuits with 2500 bootstrap resamples. The statistical uncertainty estimates are all within 3.1\% of the bootstrap standard deviation.

The combined linear cross entropy fidelity and statistical uncertainty of 10 random circuits is calculated using inverse-variance weighting to be $\hat{F}_l = (2.24\pm 0.18) \times 10^{-3}$. The theoretical prediction of the statistical uncertainty, $\sqrt{(1 + 2 F - F^2) / N_s}$, is $1.8 \times 10^{-4}$, which agrees with the experimental estimate. As a comparison, the combined cross entropy fidelity is $\hat{F}_c = (2.34\pm 0.23)\times 10^{-3}$. The theoretical prediction of statistical uncertainty, $\sqrt{(\pi^2/6 - F^2) / N_s}$, is $2.3 \times 10^{-4}$, which agrees with the experimental estimate as well. Thus, the cross entropy fidelity and linear cross entropy fidelity estimators produce consistent results.  Furthermore,  the statistical uncertainty of the linear cross entropy estimator is smaller, as expected from its theoretical formula.

\begin{figure*}
\includegraphics[width=2.0\columnwidth]{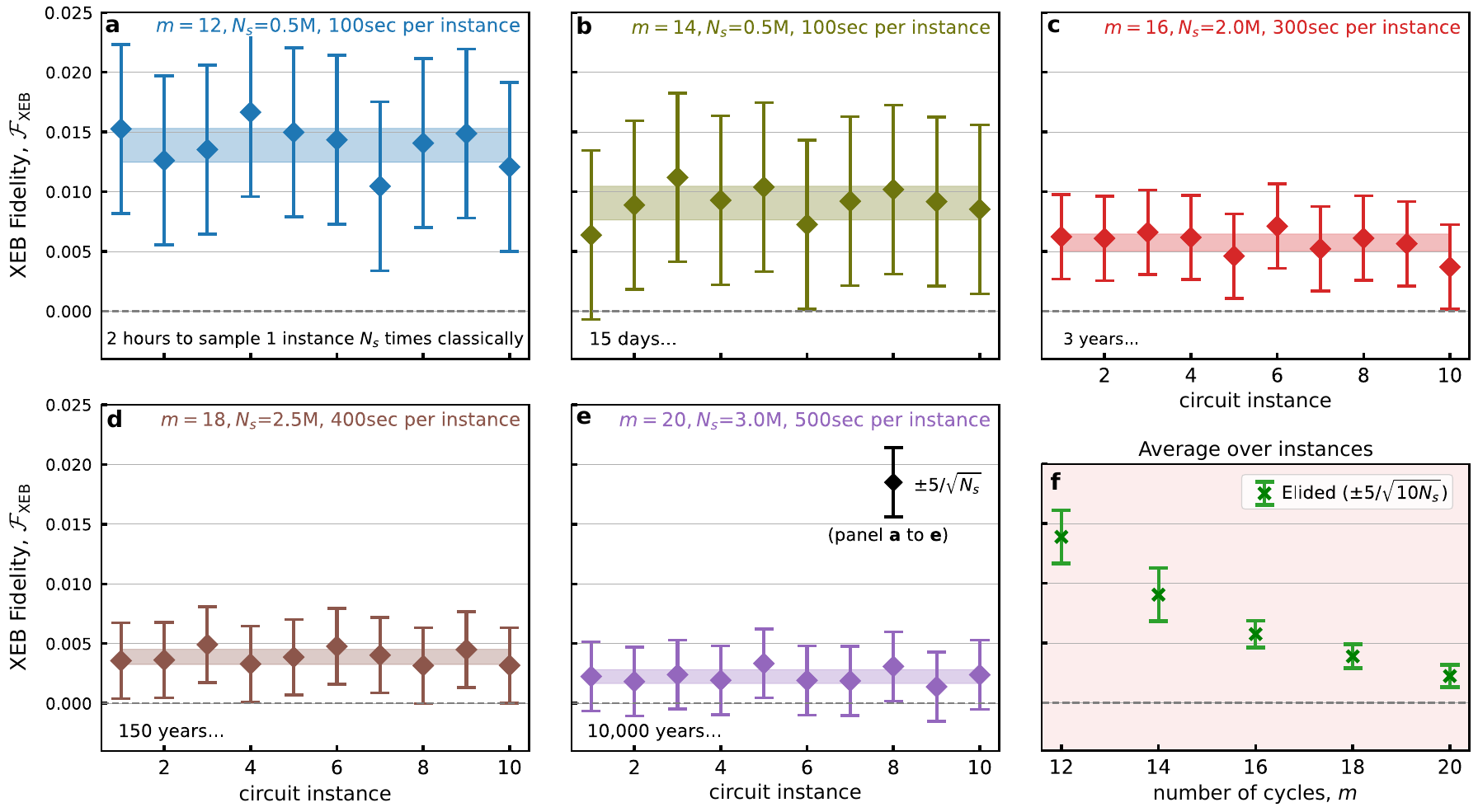}
\caption{\label{fig:elided_stat_error} Per-instance elided circuit fidelities and statistical uncertainties. XEB fidelities of all 10 elided circuit instances for each circuit depth from Fig.~4b of the main text. \textbf{a} to \textbf{e}, Here, each panel corresponds to a single circuit depth $m$. In these panels, $\pm 5\sigma$ statistical error bars, where $\sigma = 1/\sqrt{N_s}$, are shown for each of the individual circuit instance fidelities. Also shown is a band corresponding to $\pm\sigma$ for a single instance, but about the mean fidelity of the 10 instances, showing that the variations between circuits can be explained by statistical fluctuations from the finite number of samples. \textbf{f}, Fidelity averaged over all 10 circuits along with $\pm 5\sigma$ error bars are shown (the same quantity is plotted in Fig.~4b of the main text but on a log scale), where in this case $\sigma = 1/\sqrt{10 N_s}$. 
Here, for all circuit depths, the mean fidelity is more than $5\sigma$ above 0.001.}
\end{figure*}

In Fig.~\ref{fig:elided_stat_error}, we also show the linear XEB fidelities and $5\sigma$ statistical uncertainties of all 10 elided circuit instances for each circuit depth from Fig.~4b of the main text. Variations between the fidelities of different circuit instances are consistent with the expected statistical noise due to the finite number of samples. In the last panel, we also show the smaller statistical uncertainties of the fidelity averaged over the 10 circuit instances for each depth.

\subsection{System stability and systematic uncertainties}\label{subsec:stability}

\begin{figure}
\includegraphics[width=\columnwidth]{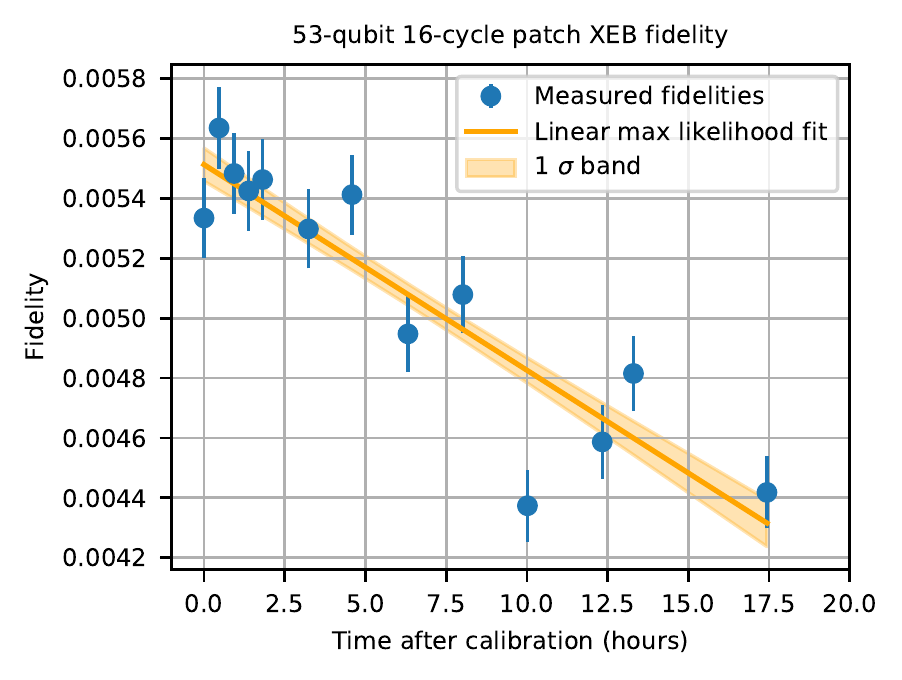}

\caption{\label{fig:patch_xeb_stability} Stability of repeated 53-qubit 16-cycle patch circuit benchmarking over 17.4 hours, without any system recalibration. Statistical error bars from the finite bitstring sample number are included. The intrinsic system fluctuations are likely dominated by a small number of TLSs moderately coupled to a few qubits at their idling and/or readout biases.}
\end{figure}
In addition to statistical errors, XEB fidelity is also subject to systematic drift as the system performance may fluctuate and/or degrade over time. To quantify these mechanisms, we performed a patch circuits time stability measurement on 53 qubits using a circuit of 16 cycles and 1 million bitstrings for 17.4 hours after calibration. In between these measurements, we measured the fidelity of other 53-qubit circuits with 16 to 20 cycles. The analyzed results are shown in Fig.~\ref{fig:patch_xeb_stability}. The statistical uncertainties of the fidelities are estimated to be $1.29\times 10^{-4}$, as indicated by the error bars.

We repeated the stability measurements twice, with different circuits and on different days. Fig.~\ref{fig:patch_xeb_stability} shows the one that exhibits greater degradation as a conservative estimate of the effect. The measurement indicates a degradation of fidelity within the range of time. A linear fit with $F = p_0 + p_1 t$ results in estimated parameters $\hat{p}_0 = (5.51\pm0.055) \times 10^{-3}$, $\hat{p}_1 = (-6.87\pm0.64) \times 10^{-5}$, and a correlation coefficient of $\hat{p}_0$ and $\hat{p}_1$, $\rho$, to be -0.76. The $\chi^2$ per degree of freedom is $26.3/11$.

The $p$-value for the $\chi^2$ for 11 degrees of freedom is 0.0058, indicating that it is not a very good fit.
Because the correctness of the estimates of statistical uncertainties has been verified in Section~\ref{subsec:xeb_stat_uncertainty},
this is attributed to systematic fluctuation in addition to degradation.
It is supported by the larger variance of fidelity than the $1\sigma$ band in Fig.~\ref{fig:patch_xeb_stability}.

The $1\sigma$ band depends on the statistical uncertainties of fidelities and the variance of time on the $x$-axis,
but is independent of the variance of fidelity. To take the variance of fidelity into account,
we use the variance of the residuals of the linear fit as an estimator of the variance of fidelity.
The standard deviation of residuals is estimated to be $1.84\times 10^{-4}$,
which is added to $\sigma_{p_0}$ in quadrature to be the total $\sigma_{p_0}$.
The estimate is total $\sigma_{p_0} = 1.92\times 10^{-4}$, 3.5 times larger than the
statistical-only $\sigma_{p_0}$ of $5.5\times 10^{-5}$.

The uncertainty on a fidelity measured at time $t$ can be estimated by the standard error propagation, assuming that $t$ is uncorrelated with either $p_0$ or $p_1$.
\begin{equation} \label{eq:drift_sigma_fidelity}
\sigma_F = \left[ \sigma_{p_0}^2 + 2 t \sigma_{p_0} \sigma_{p_1} \rho +
    \sigma_{p_1}^2 t^2 \right]^{1/2}
\end{equation}
The value of $\sigma_F$ as well as the ratio $\sigma_F/F$ in the range of measured fidelities monotonically decreases.
We take $\max(\sigma_F / F)$ as the estimate of relative systematic uncertainty for fidelities measured
in the same run. The value is found to be 4.4\% and is used in subsequent analysis.

\begin{figure*}[htbp]
\centering
\includegraphics[width=2.0\columnwidth]{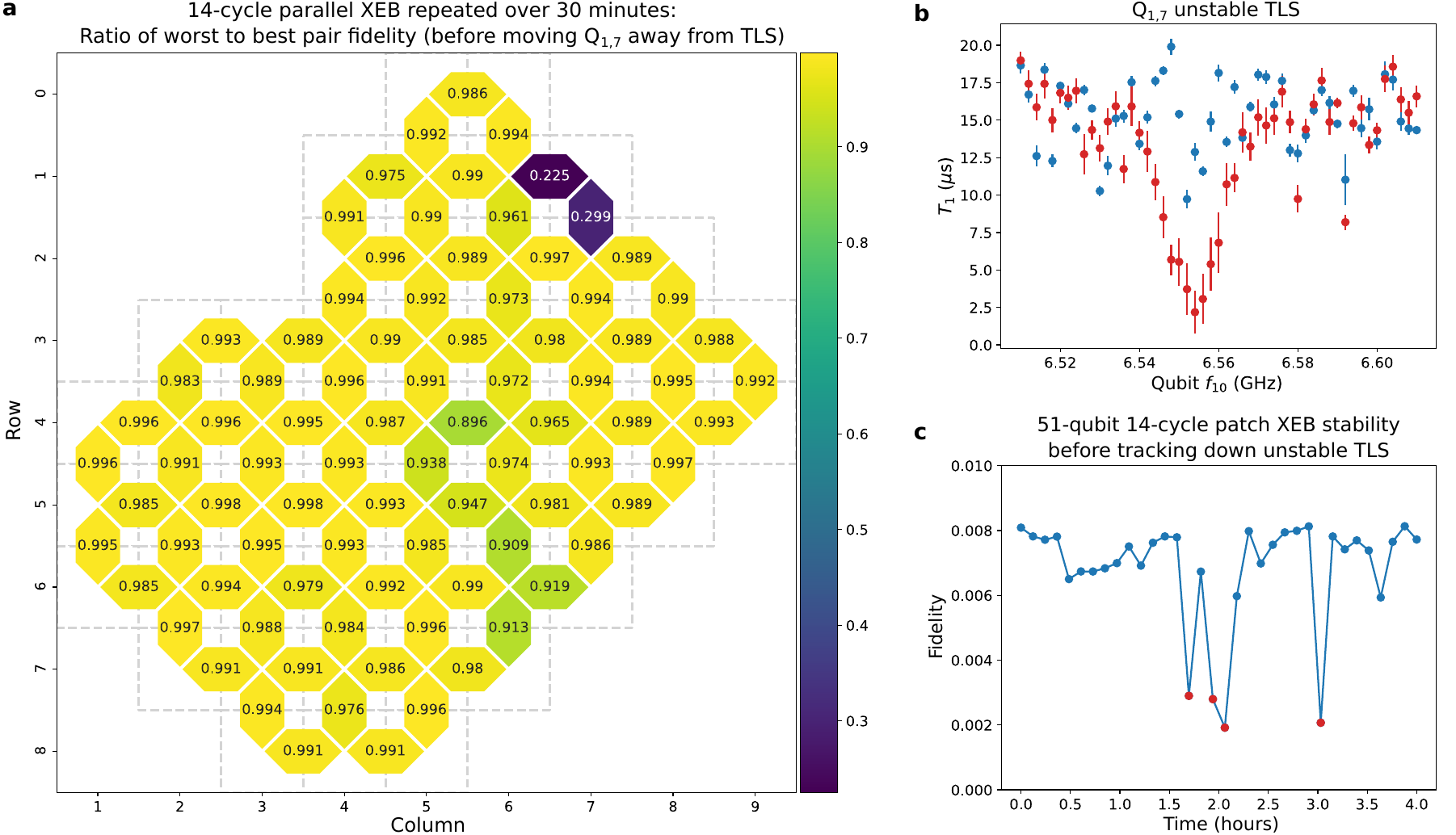}
\caption{\label{fig:unstable_tls_stability} \textbf{Identifying sources of fluctuations with repetitive per-layer simultaneous pair XEB.}
\textbf{a,} Per-pair ratio of worst fidelity to best fidelity measured via per-layer simultaneous pair XEB at a depth of 14 cycles over the course of 30 minutes. During this time, fluctuations were dominated by a single TLS. \textbf{b,} Measured qubit $T_1$ vs.~$f_{10}$ for Q\textsubscript{1,7} at two different times a few minutes apart (red vs.~blue points), showing an unstable TLS that was dominating the fluctuations in full system fidelity seen in \textbf{c,}. Moving Q\textsubscript{1,7} far from this TLS led to the stability seen in Fig.~\ref{fig:patch_xeb_stability}.}
\end{figure*}

The physical origin of the observed system fluctuations can be attributed to many possible channels: $1/f$ flux noise, qubit $T_1$ fluctuations, control signal drift, etc.  We speculate that the dominant mechanism is the moderate interaction between a small number of TLS’s and a few qubits at their idling and/or readout biases. In Fig.~\ref{fig:unstable_tls_stability}a, we show the result of measuring per-layer simultaneous pair XEB at a fixed depth of 14 cycles repeatedly over time. The quantity plotted is the ratio of the worst pair fidelity to best fidelity observed over the course of 30 minutes. This type of repetitive measurement allows us to pinpoint which pairs dominate the fluctuations in full system fidelity. Note that because we used fidelity at a fixed cycle depth rather than the one extracted from the exponential decay, these numbers contain the effect of fluctuating measurement fidelity as well. 

As shown in Fig.~\ref{fig:unstable_tls_stability}a, the depth-14 fidelity of most pairs fluctuates downward by only $\sim$1\% at depth 14, which translates to either a $\sim$1\% fluctuation in measurement fidelity for a pair, or a $\sim$0.08\% fluctuation in the two-qubit gate fidelity for a pair. Before finding the unstable TLS defect in Fig.~\ref{fig:unstable_tls_stability}b, a single qubit dominated the fluctuations in full system fidelity seen in Fig.~\ref{fig:unstable_tls_stability}c. After we moved this problematic qubit far from the fluctuating TLS, the fluctuations in fidelity during the actual quantum supremacy experiment (Fig.~\ref{fig:patch_xeb_stability}) were dominated by a handful of pairs containing qubits in the ``degenerate" readout region (described in section \ref{sec:calib_metro}). For these qubits, due to constraints from readout crosstalk we had little freedom in what readout detunings we could choose, and so the best we could do was to put some qubits near defects or transmon-resonator transition modes during readout. We speculate that this is where the remaining dominant fluctuations originate.

\subsection{The fidelity result and the null hypothesis on quantum supremacy}\label{subsec:final_fidelity}

\begin{figure}
\includegraphics[width=\columnwidth]{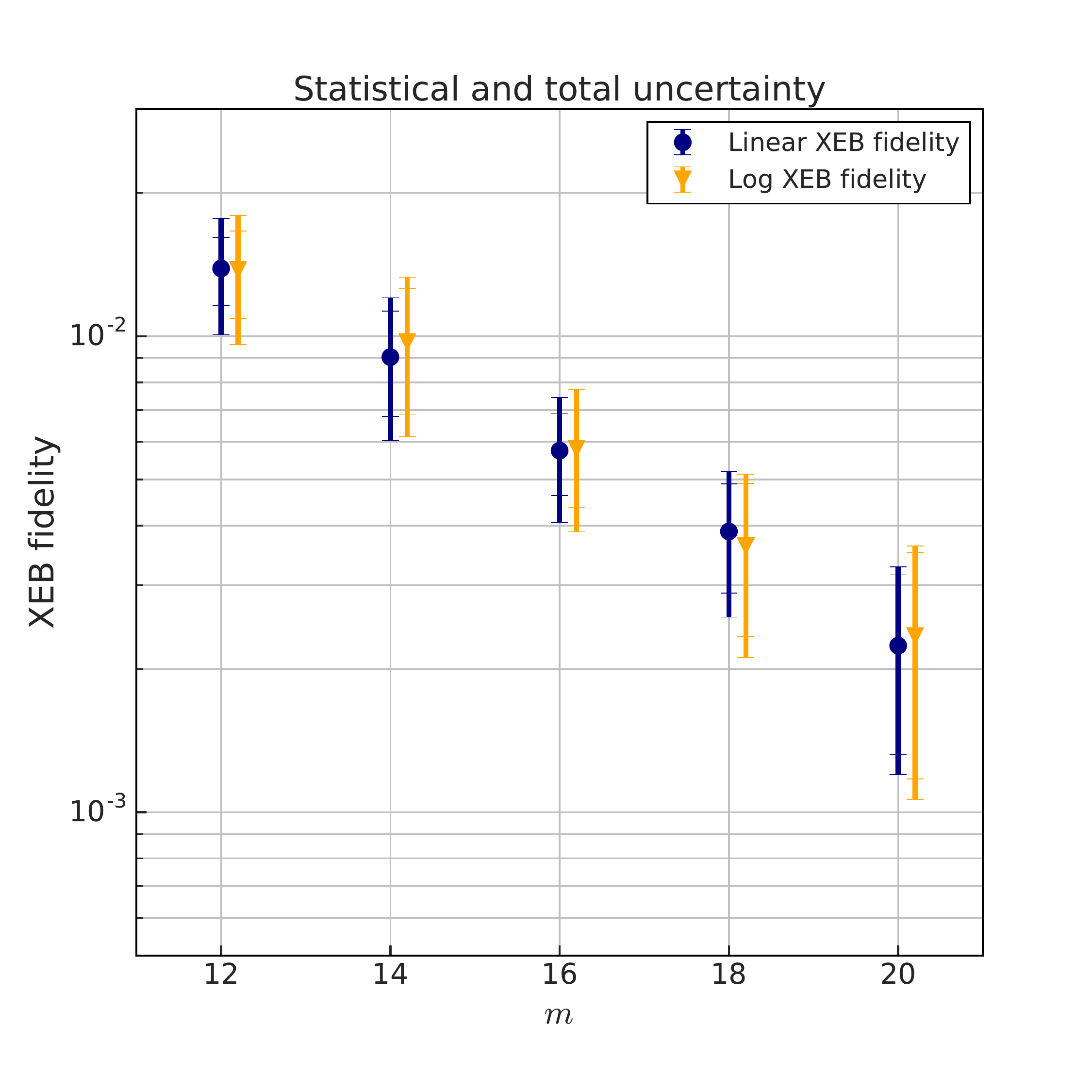}
\caption{\label{fig:stat_and_syst_error} Statistical and total uncertainty of fidelity estimates produced by the linear XEB and logarithmic XEB from ten random quantum circuits with 53 qubits and $m$ cycles. The inner error bars represent the statistical uncertainty discussed in section~\ref{subsec:xeb_stat_uncertainty}. The outer error bars represent the total uncertainty discussed in section~\ref{subsec:stability}.}
\end{figure}

We use the mean fidelity of ten 53-qubit 20-cycle circuits as the final benchmark of the system. In section~\ref{subsec:xeb_stat_uncertainty} we estimated the fidelity and statistical uncertainty to be $(2.24\pm 0.18) \times 10^{-3}$ using the linear cross entropy. In section~\ref{subsec:stability} we estimated the relative systematic uncertainty due to drift to be 4.4\%. Combining these 2 estimations we arrive at the final fidelity as $(2.24 \pm 0.10(\mbox{syst.}) \pm 0.18 (\mbox{stat.})) \times 10^{-3}$. Fidelity estimates with statistical and systematic uncertainty for other quantum circuits are shown in Figure~\ref{fig:stat_and_syst_error}.

As we show in section~\ref{sec:classical_sim}, a noisy sampling of a random quantum circuit at fidelity $F = 10^{-3}$ requires 5000 years with a classical computer with CPU power equivalent to 1 million cores, and it scales linearly with fidelity $F$. It takes a quantum computer less than an hour to complete the
same noisy sampling. Therefore we form the null hypothesis that the fidelity of the quantum computer is $F \le 10^{-3}$, and the alternative hypothesis that $F > 10^{-3}$. If the alternative hypothesis is true, we can say that a classical computer can not perform the same noisy sampling task as the quantum computer.

The total uncertainty on fidelity is estimated with addition in quadrature of systematic uncertainty and statistical uncertainty. The mean fidelity of 10 random
circuits with 53 qubits and 20 cycles is $(2.24 \pm 0.21) \times 10^{-3}$. The
null hypothesis is therefore rejected with a significance of $6\sigma$.

While our analysis of the uncertainty in $\xebfidelity$ was computed from both statistical and systematic errors, some care should be taken in the consideration of systematic errors as they pertain to the claim of quantum supremacy. Systematic errors should be included if we wish to use the XEB fidelity value, for example comparing fidelities of patch, elided and full circuits. However for quantum supremacy, a false claim would arise if $\xebfidelity$ was zero, but we obtained a non-zero value because of a fluctuation.  Systematic fluctuations produce a change in magnitude of XEB, as seen in the data in this section, which is thus a multiplicative-type error that does not change the XEB fidelity value when it is zero.  A false positive is only produced by a additive-type statistical fluctuations and thus it is the only mechanism that should be considered when computing the uncertainty. Therefore, the $6\sigma$ significance of our claim should be considered as conservative.

Some skeptics have warned that a quantum computer may not be possible \cite{Dyakonov18,Kalai19}, for example due to the fragility of quantum information at large qubit number and exponentially large Hilbert space.  The demonstration here of quantum behavior at $~10^{16}$ Hilbert space is strong confirmation that nothing unusual or unexpected happens to our current understanding of quantum mechanics at this scale.

\section{Sensitivity of XEB to errors}\label{sec:xeb_sens}

\begin{figure*}
\includegraphics[scale=0.4]{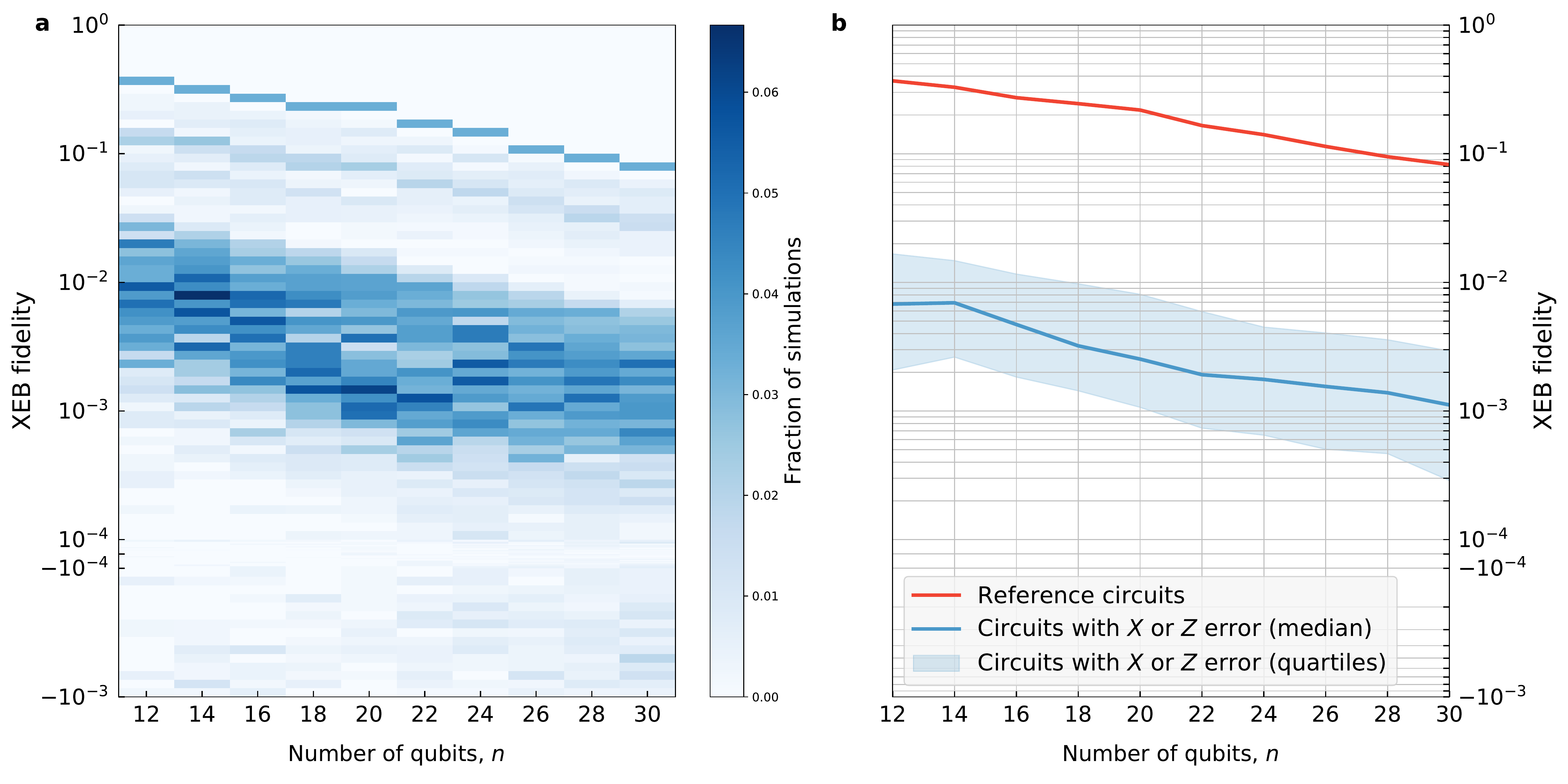}
\caption{\label{fig:discrete_error} \textbf{Impact of one single-qubit Pauli error on fidelity estimate from XEB.} \textbf{a,} Distributions of fidelity estimates from XEB using measured bitstrings and quantum circuits with one bit-flip or one phase-flip error. For each $n$, shades of blue represent the normalized histogram of the estimates obtained for the error gate placed at different circuit locations. The highest fidelity estimates correspond to phase-flip errors immediately preceding measurement and are equal to the fidelity estimates from XEB using error-free circuits. \textbf{b,} Quartiles of the distributions shown in \textbf{a} (blue) compared to the fidelity estimates from XEB using measured bitstrings and unmodified quantum circuits (red). Both plots use linear scale between $10^{-4}$ and $-10^{-4}$ and logarithmic scale everywhere else.}
\end{figure*}

\begin{figure}
\includegraphics[scale=0.55]{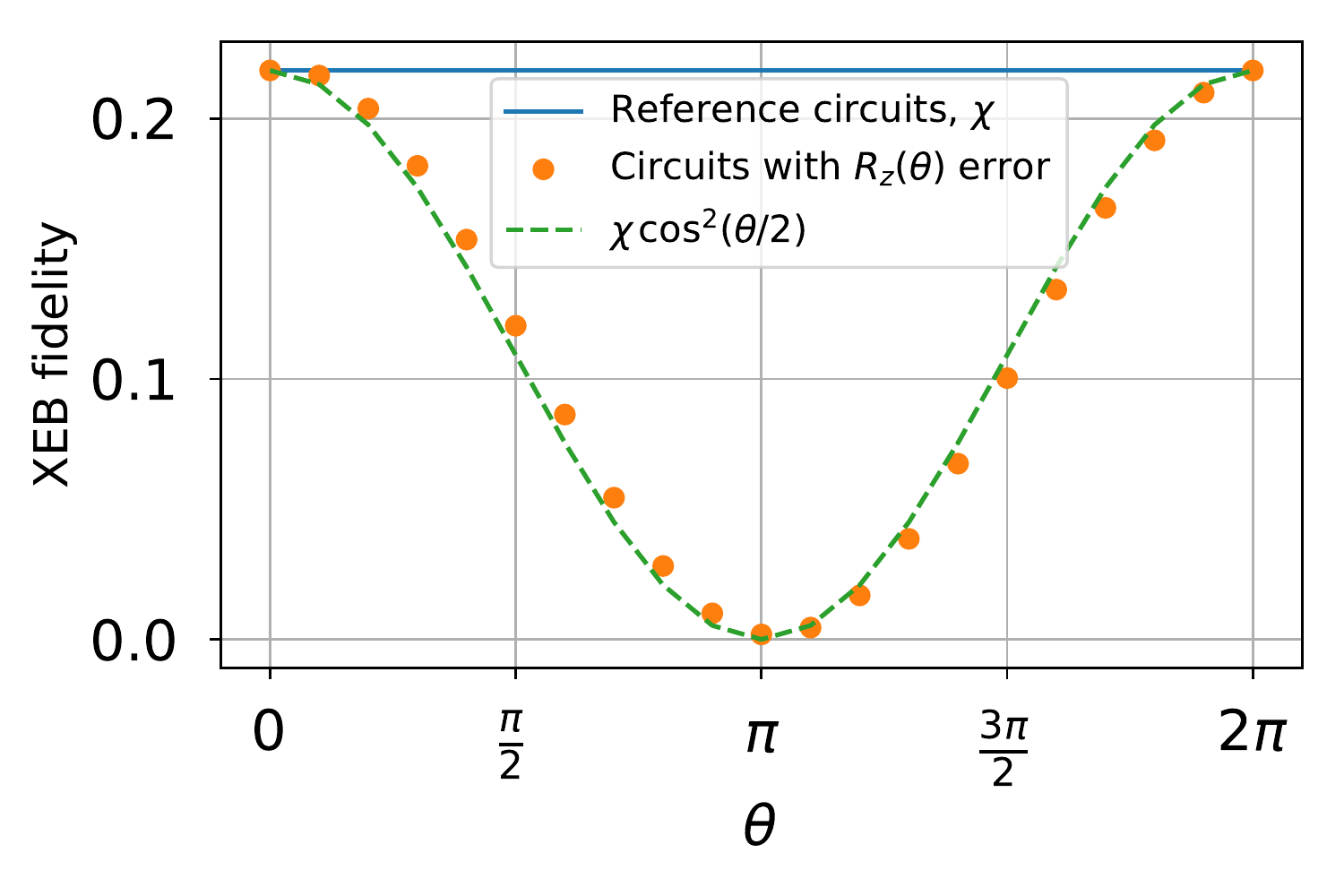}
\caption{\label{fig:continuous_error} \textbf{Impact of the $R_z(\theta)$ error on XEB.} Fidelity estimates computed by XEB from measured bitstrings and circuits with $n=20$ qubits and $m=14$ cycles modified to include $R_z(\theta)$ error applied in 10th cycle to one of the qubits as a function of $\theta$ (orange dots). Also shown is XEB fidelity computed using the same bitstrings and unmodified circuits (blue solid line) and a simple model which predicts the effect of the error (green dashed line).}
\end{figure}

An important requirement for a procedure used to evaluate quantum processors, such as XEB, is sensitivity to errors. Qubit amplitudes are complex variables and therefore quantum errors are inherently continuous. Nevertheless, they can be given a discrete description, for example in the form of a finite set of Pauli operators. The digital error model is used for instance in quantum error correction where errors are discretized by syndrome extraction. In this section we examine the impact of both discrete and continuous errors on the fidelity estimate obtained from the XEB algorithm.

The XEB procedure uses a set of random quantum circuits $\mathcal{U} = \{U_1, \dots, U_S\}$ with $n$ qubits and $m$ cycles. Every circuit is executed $N_s$ times on the quantum processor under test. Each execution of the circuit $U_j$ applies the quantum operation $\Lambda_j$, which is an imperfect realization of $U_j$, to the input state $\ket{0}\bra{0}$. The result of the experiment is a set $\mathcal{B}$ of $S N_s$ bitstrings $q_{i,j}$ sampled from the distributions $p_e(q_{i,j}) = \bra{q_{i,j}} \rho_j \ket{q_{i,j}}$ where $\rho_j = \Lambda_j(\ket{0}\bra{0})$ is the output state in the experiments with circuit $U_j$. For each bitstring $q_{i,j}$, a simulator computes the ideal probability $p_s(q_{i,j}) = |\braket{q_{i,j}}{\psi_j}|^2$ where $\ket{\psi_j} = U_j\ket{0}$ is the ideal output state of the circuit $U_j$. Finally, XEB uses Eq.~\eqref{eq:linear_xeb_fidelity} or \eqref{eq:log_xeb_fidelity} to compute an estimate $\xebfidelity(\mathcal{B}, \mathcal{U})$ of fidelity $F(\ket{\psi_j}\bra{\psi_j}, \rho_j) = \bra{\psi_j}\rho_j\ket{\psi_j}$ averaged over circuits $\mathcal{U}$. The result quantifies how well the quantum processor is able to realize quantum circuits of size $n$ and depth $m$. See section \ref{sec:xeb_theory} for more details on XEB.

The estimate $\xebfidelity(\mathcal{B}, \mathcal{U})$ is a function of bistrings $\mathcal{B}$ obtained in experiment and of the set of quantum circuits $\mathcal{U}$ used to compute ideal probabilities. This enables a test of the sensitivity of the method to errors by replacing the error-free reference circuits $\mathcal{U} = \{U_1, \dots, U_S\}$ with circuits $\mathcal{U}_E = \{U_{1,E}, \dots, U_{S,E}\}$ where $U_{j,E}$ is the quantum circuit obtained from $U_j$ by the insertion at a particular location in the circuit of a gate $E$ representing the error. We identify errors inserted at different circuit locations that lead to the same output distribution since XEB cannot differentiate between them.

We first consider the impact of a discrete single-qubit Pauli error $E$ placed in a random location in the circuit. In Fig.~\ref{fig:discrete_error} we plot $\xebfidelity(\mathcal{B}, \mathcal{U}_E)$ where $\mathcal{B}$ are bitstrings observed in our experiment and $\mathcal{U}_E$ are quantum circuits modified by the insertion of an additional $X$ or $Z$ gate following an existing single-qubit gate. Each fidelity estimate corresponds to a different circuit location where the error gate has been inserted. For every $n$, the highest fidelity values correspond to the insertion of the $Z$ gate in the final cycle of the circuit. They have no impact on measurements and thus are equivalent to absence of error. The corresponding fidelity estimates match the estimates for the unmodified circuits.

The probability of only seeing the error $E$ is approximately $q = e p$ where $e$ is the probability of $E$ arising at the particular circuit location and $p$ is the probability that no other error occurs. The fraction $q$ of executions realize circuit $U_{j,E} \in \mathcal{U}_E$ yielding bitstrings $\mathcal{B}_E$ while the remaining fraction $1 - q$ yield bitstrings $\mathcal{B}_*$. XEB averages over circuit executions, so

\begin{multline}
\xebfidelity(\mathcal{B}, \mathcal{U}_E) = \\ q \, \xebfidelity(\mathcal{B}_E, \mathcal{U}_E) + (1 - q) \, \xebfidelity(\mathcal{B}_*, \mathcal{U}_E).
\end{multline}
Since bitstrings $\mathcal{B}_E$ originated in a perfect realization of $\mathcal{U}_E$ we have $\xebfidelity(\mathcal{B}_E, \mathcal{U}_E) \simeq 1$ with high probability. Also, assuming the circuits randomize the output quantum state sufficiently, we have $\xebfidelity(\mathcal{B}_*, \mathcal{U}_E) \simeq 1/\sqrt{D}$, where $D = 2^n$, see Eq.~\eqref{eq:xeb_cm} and Fig.~\ref{fig:xeb_puali_error}. Therefore, for large $n$

\begin{equation}
\xebfidelity(\mathcal{B}, \mathcal{U}_E) \simeq q + \frac{1 - q}{\sqrt{D}} \simeq q
\end{equation}
with high probability.

Now, the probability $p$ that no error other than $E$ occurs is approximately equal to the experimental fidelity $F$ which is approximated by $\xebfidelity(\mathcal{B}, \mathcal{U})$, so

\begin{equation} \label{eq:xeb_proportionality}
\xebfidelity(\mathcal{B}, \mathcal{U}_E) \simeq e \, \xebfidelity(\mathcal{B}, \mathcal{U})
\end{equation}
which means that XEB result obtained using circuits modified to include $E$ is approximately proportional to the XEB result obtained using the error-free reference circuits. Moreover, the ratio of the two XEB results is approximately equal to the probability of $E$.

The data in Fig.~\ref{fig:discrete_error} agrees with the approximate proportionality in Eq.~\eqref{eq:xeb_proportionality} and allows us to estimate the median probability of a Pauli error. Based on the drop in XEB fidelity estimate by a factor of almost 100 due to the insertion of one single-qubit Pauli error into the circuit, the probability is on the order of $1\%$. While more work on the gate failure model needs to be done to correctly relate Sycamore gate error rates to the probability of specific Pauli errors, we already see that $e$ has the same order of magnitude as our per cycle and per qubit error given by $e_{2c}/2 \simeq 0.5\%$, see Table \ref{tab:system_params}. A possible resolution of the factor of two discrepancy may lie in the fact that more than one gate failure can manifest itself as a particular Pauli error $E$ in a particular circuit location.

Lastly, we consider the impact of continuous errors on XEB result. Fig.~\ref{fig:continuous_error} shows the fidelity estimate obtained from XEB using bitstrings observed in our experiment and quantum circuits modified to include a single rotation $R_Z(\theta)$. The middle point of the plot is equal to the fidelity estimate obtained for one of the discrete errors in Fig.~\ref{fig:discrete_error} whereas the leftmost and rightmost points correspond to the fidelity estimate obtained from XEB using the error-free reference circuit.

The analysis above illustrates how questions about the behavior and performance of quantum processors can be formulated in terms of modifications to the reference quantum circuits and how XEB can help investigate these questions. While XEB has proven itself a powerful tool for calibration and performance evaluation (see sections \ref{sec:calib_metro} and \ref{sec:large_scale_xeb}), more work is required to assess its efficacy as a diagnostic tool for quantum processors.
\section{Classical simulations}\label{sec:classical_sim}

\subsection{Local Schr\"{o}dinger and Schr\"{o}dinger-Feynman simulators}
\label{subsec:sa_sfa}
We have developed two quantum circuit simulators: qsim and qsimh. The first simulator, qsim, is a Schr\"{o}dinger full state vector simulator. It computes all $2^n$ amplitudes, where $n$ is the number of qubits. Essentially, the simulator performs matrix-vector multiplications repeatedly. One matrix-vector multiplication corresponds to applying one gate. For a 2-qubit gate acting on qubits {\tt q1} and {\tt q2} ({\tt q1} $<$ {\tt q2}), it can be depicted schematically by the following pseudocode.

\begin{verbatim}
#iterate over all values of qubits q > q2
for (int i = 0; i < 2^n; i += 2 * 2^q2) {
  #iterate values for q1 < q < q2
  for (int j = 0; j < 2^q2; j += 2 * 2^q1) {
    #iterate values for q < q1
    for (int k = 0; k < 2^q1; k += 1) {
      #apply gate for fixed values 
      #for all q not in [q1,q2]
      int l = i + j + k;

      float v0[4]; #gate input
      float v1[4]; #gate output

      #copy input
      v0[0] = v[l];
      v0[1] = v[l + 2^q1];
      v0[2] = v[l + 2^q2];
      v0[3] = v[l + 2^q1 + 2^q2];

      #apply gate
      for (r = 0; r < 4; r += 1) {
        v1[r] = 0;
        for (s = 0; s < 4; s += 1) {
          v1[r] += U[r][s] * v0[s];
        }
      }

      #copy output
      v[l] = v1[0];
      v[l + 2^q1] = v1[1];
      v[l + 2^q2] = v1[2];
      v[l + 2^q1 + 2^q2] = v1[3];
    }
  }
}
\end{verbatim}
Here {\tt U} is a 4x4 gate matrix and {\tt v} is the full state vector. To make the simulator faster, we use gate fusion \cite{smelyanskiy2016qhipster}, single precision arithmetic, AVX/FMA instructions for vectorization, and OpenMP for multi-threading. We are able to simulate 38-qubit circuits on a single Google cloud node that has 3844 GB memory and four CPUs with 20 cores each (n1-ultramem-160). The run times for different circuit sizes at depth 14 are listed in Table \ref{table:qsimruntimes}.

\begin{table}
\begin{tabular}{| r | r |}
\hline
qubits, $n$ & run time in seconds \\
\hline
32 & 111 \\
\hline
34 & 473 \\
\hline
36 & 1954 \\
\hline
38 & 8213 \\
\hline
\end{tabular}
\label{table:qsimruntimes}
\caption{Circuit simulation run times using qsim on a single Google cloud node (n1-ultramem-160).}
\end{table}

The second simulator, qsimh, is a hybrid Schr\"{o}dinger-Feynman algorithm (SFA) simulator \cite{markov_quantum_2018}. We cut the lattice into two parts and use the Schmidt decomposition for the 2-qubit gates on the cut. If the Schmidt rank of each gate is $r$ and the number of gates on the cut is $g$ then there are $r^g$ paths, corresponding to all the possible choices of Schmidt terms for each 2-qubit gate across the cut. To obtain fidelity equal to unity, we need to simulate all the $r^g$ paths and sum the results. The total run time is proportional to $(2^{n_1}+2^{n_2})r^g$, where $n_1$ and $n_2$ are the qubit numbers in the first and second parts. Each part is simulated by qsim using the Schr\"odinger algorithm. Path simulations are independent of each other and can be trivially parallelized to run on supercomputers or in data centers. Note that one can run simulations with fidelity $F < 1$ just by summing over a fraction $F$ of all the paths (see Ref.~\cite{markov_quantum_2018} and Sec.~\ref{subsec:sim_target_fidelity}). In order to speed up the computation, we save a copy of the state after the first $p$ 2-qubit gates across the cut, so the remaining $r^{g-p}$ paths can be computed without re-starting the simulation from the beginning. We call the specific choice of Schmidt terms for the first $p$ gates in the cut a {\it prefix}.

\subsection{Feynman simulator}
\label{subsec:fa}

\begin{table*}
\centering
\begin{tabular}{|c|c|c|c|c|c|c|c|c|c|c|c|}
\cline{7-8} \cline{9-10}
\multicolumn{1}{c}{} & \multicolumn{1}{c}{} & \multicolumn{1}{c}{} & \multicolumn{1}{c}{} & \multicolumn{1}{c}{} &  & \multicolumn{2}{c|}{\textbf{PFlop/s*}} & \multicolumn{2}{c|}{\textbf{efficiency (\%)}} & \multicolumn{1}{c}{} & \multicolumn{1}{c}{}\tabularnewline
\hline 
\textbf{qubits} & \textbf{cycles} & \textbf{$\xebfidelity$} (\%) & \textbf{$N_s$} & \textbf{nodes} & \textbf{runtime} & \textbf{peak} & \textbf{sust.} & \textbf{peak} & \textbf{sust.} & \textbf{power (MW)} & \textbf{energy (MWh)}\tabularnewline

\hline 
\hline 
\multirow{7}{*}{53} & \multirow{3}{*}{12} & 0.5 & 1M & \multirow{7}{*}{4550} & 1.29 hours & \multirow{3}{*}{235.2} & \multirow{3}{*}{111.7} & \multirow{3}{*}{57.4} & \multirow{3}{*}{27.3} & \multirow{3}{*}{5.73} & 8.21 \tabularnewline
\cline{3-4} \cline{6-6} \cline{12-12}
 &  & 1.4 & 0.5M &  & 1.81 hours** &  &  &  &  &  & 11.2** \tabularnewline
\cline{3-4} \cline{6-6} \cline{12-12}
 &  & 1.4 & 3M &  & 10.8 hours** &  &  &  &  &  & 62.7** \tabularnewline
\cline{2-2} \cline{3-4} \cline{6-6} \cline{7-11} \cline{12-12}
 & \multirow{4}{*}{14} & $2.22\times 10^{-6}$ & 1M &  & 0.72 hours & \multirow{4}{*}{347.5} & \multirow{4}{*}{252.3} & \multirow{4}{*}{84.8} & \multirow{4}{*}{61.6} & \multirow{4}{*}{7.25} & 6.11 \tabularnewline
\cline{3-4} \cline{6-6} \cline{12-12}
 &  & 0.5 & 1M &  & 67.7 days** &  &  &  &  &  & $1.18 \times 10^{4}$** \tabularnewline
\cline{3-4} \cline{6-6} \cline{12-12}
 &  & 1.0 & 0.5M &  & 67.7 days** &  &  &  &  &  & $1.18 \times 10^{4}$** \tabularnewline
\cline{3-4} \cline{6-6} \cline{12-12}
 &  & 1.0 & 3M &  & 1.11 years** &  &  &  &  &  & $7.07 \times 10^{4}$** \tabularnewline
\hline 
\end{tabular}
\caption{\label{table:qflex_runs} \textbf{Runtimes, efficiency and energy consumption for the simulation of random circuit sampling of $N_s$ bitstrings from Sycamore with fidelity $\mathcal{F}$ using qFlex on Summit}. Simulations used 4550 nodes out of 4608, which represents about $99\%$ of Summit. Single batches of $64$ amplitudes were computed on each MPI task using a socket with three GPUs (two sockets per node); given that one of the 9100 MPI tasks acts as master, 9099 batches of amplitudes were computed. For the circuit with 12 cycles, $144/256$ paths for these batches were computed in 1.29 hours, which leads to the sampling of about 1M bitstrings with fidelity $\mathcal{F}\approx 0.5\%$ (see Ref.~\cite{villalonga2019flexible} for details on the sampling procedure); runtimes and energy consumption for other sample sizes and fidelities are extrapolated linearly in $N_s$ and $\mathcal{F}$ from this run. At 14 cycles, $128/524288$ paths were computed in 0.72 hours, which leads to the sampling of about 1M bitstrings with fidelity $2.22\times 10^{-6}$. In this case, one would need to consider 288101 paths on all 9099 batches in order to sample about 1M (0.5M) bitstrings with fidelity $\mathcal{F}\approx 0.5\%$ (1.0\%). By extrapolation, we estimate that such computations would take 1625 hours (68 days). For $N_s=$3M bitstrings and $\mathcal{F}\approx1.0\%$, extrapolation gives us an estimated runtime of 1.1 years. Performance is higher for the simulation with 14 cycles, due to higher arithmetic intensity tensor contractions. Power consumption is also larger in this case. Job, MPI, and TAL-SH library initialization and shutdown times, as well as initial and final IO times are not considered in the runtime, but they are in the total energy consumption. *Single precision. **Extrapolated from the simulation with a fractional fidelity.}
\end{table*}

\begin{figure*}
\includegraphics[width=0.66\columnwidth]{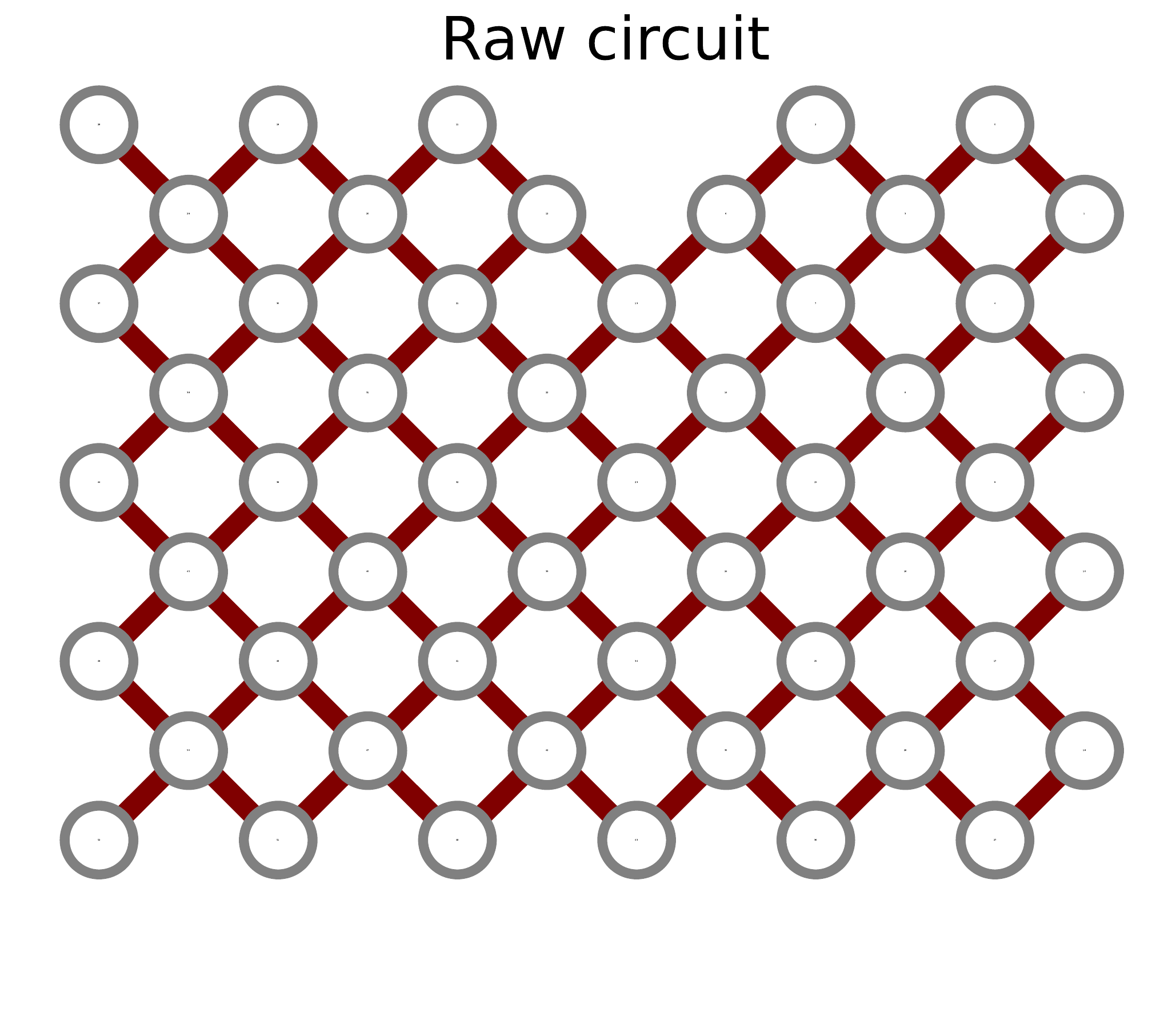}
\includegraphics[width=0.66\columnwidth]{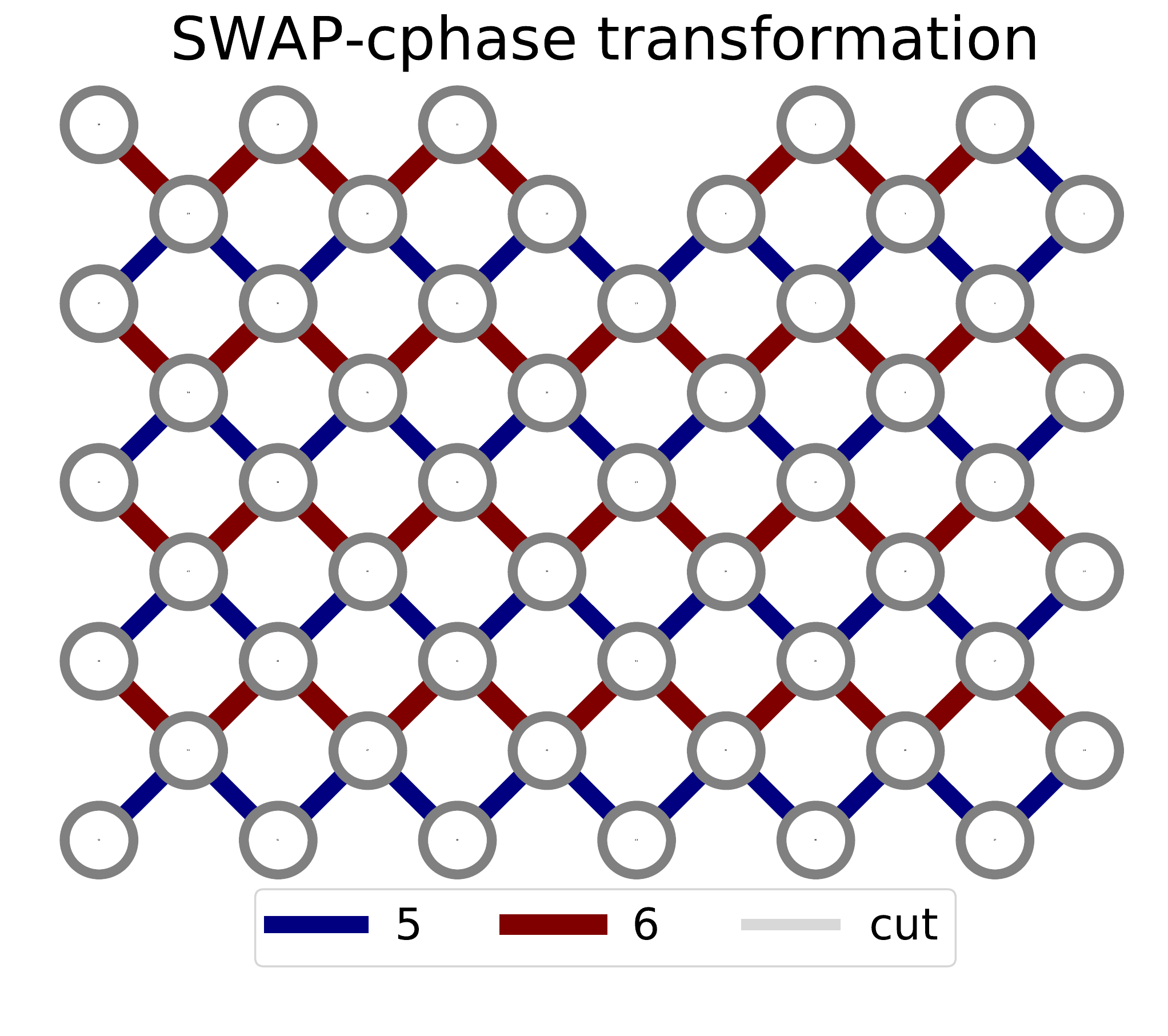}
\includegraphics[width=0.66\columnwidth]{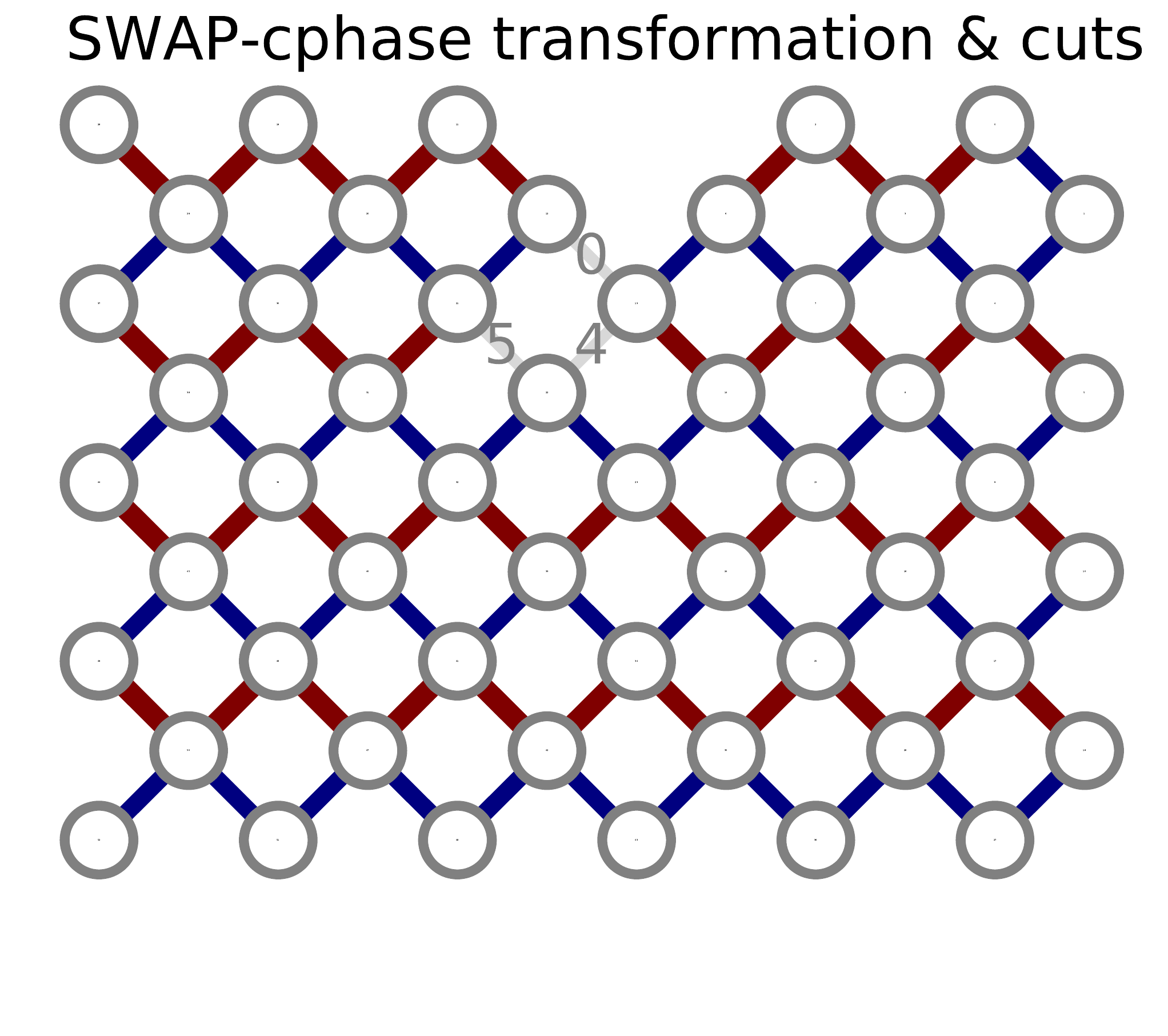}
\includegraphics[width=0.66\columnwidth]{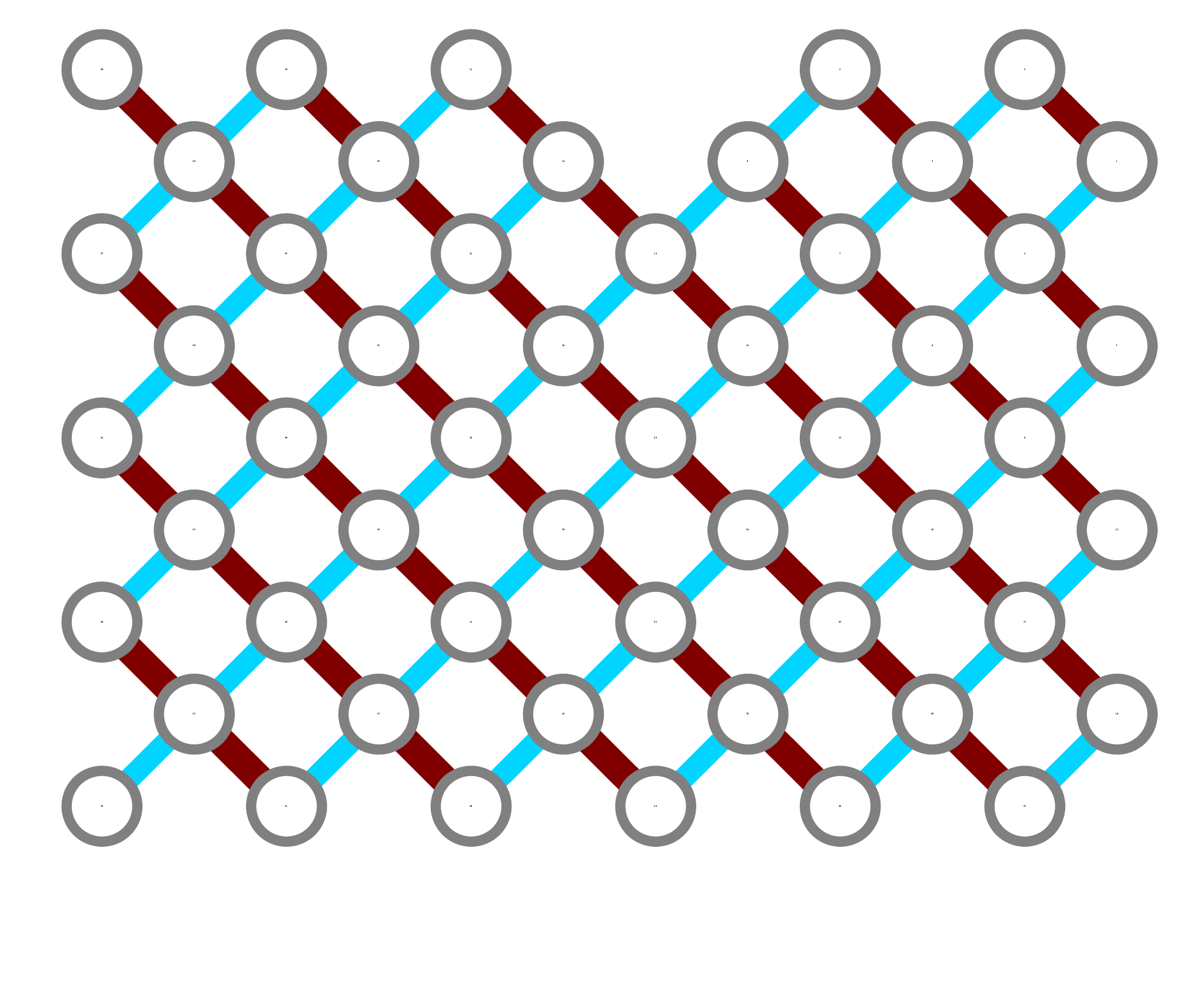}
\includegraphics[width=0.66\columnwidth]{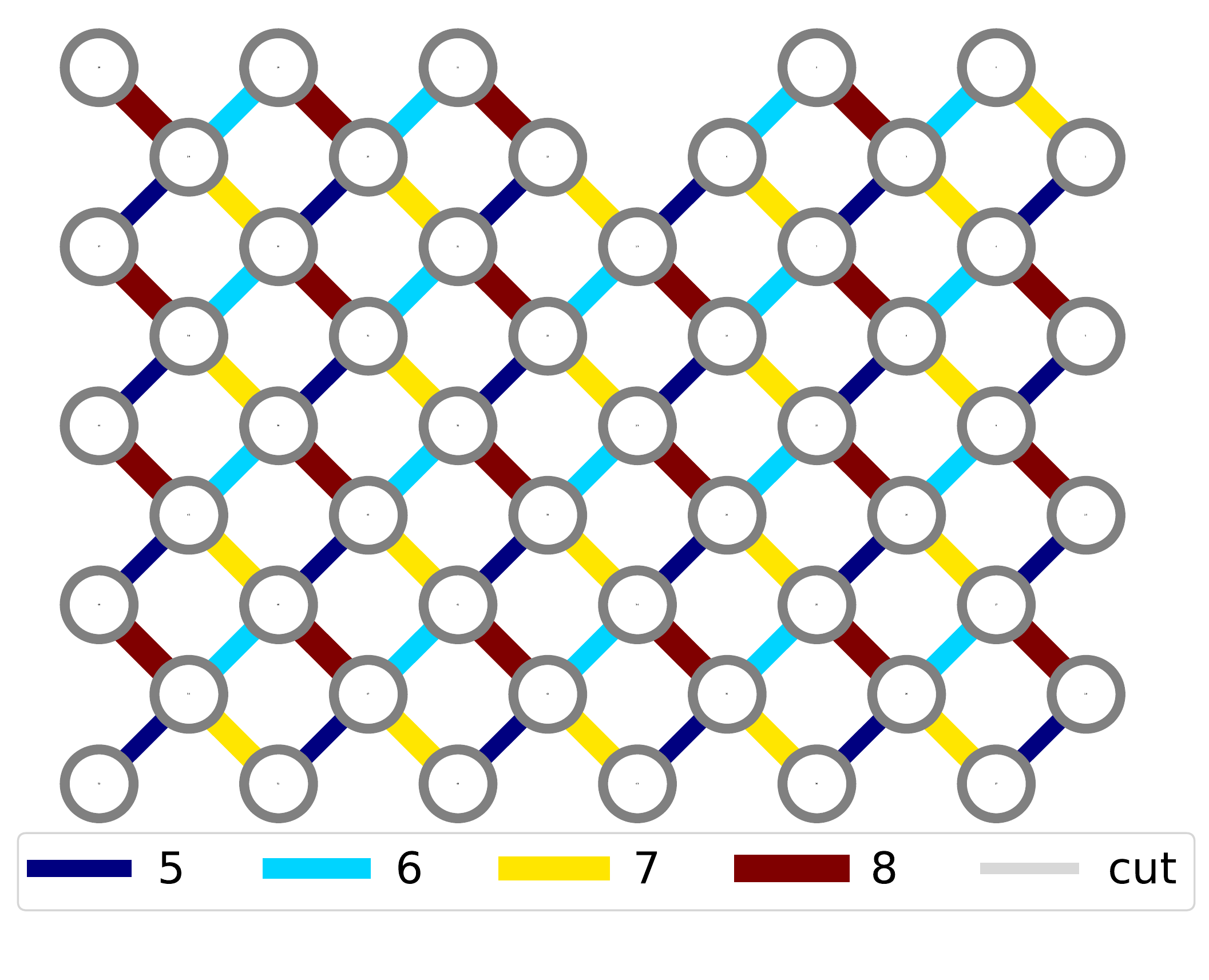}
\includegraphics[width=0.66\columnwidth]{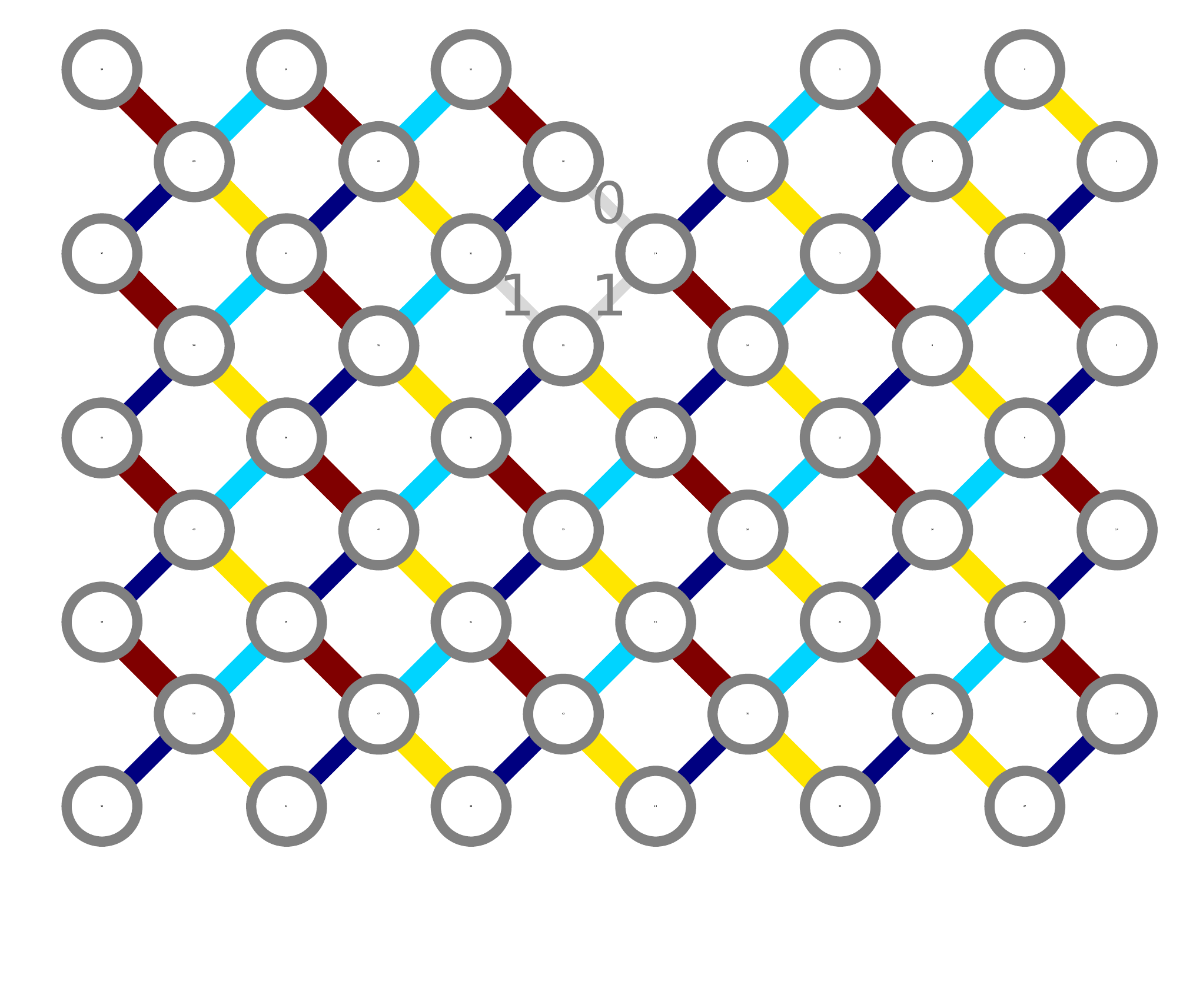}
\caption{\label{fig:qflex_bond_dimensions} \textbf{Logarithm base 2 of the bond (index) dimensions of the tensor network to contract for the simulation of sampling from Sycamore with 12 cycles (top) and 14 cycles (bottom) using qFlex}. The left plots represent the tensor network given by the circuit. The middle plots represent the tensor network obtained from a circuit where fSim gates have been transformed, when possible (see main text). The right plots represent the tensor network after the gate transformations and cuts (gray bonds) have been applied; the $\log_2$ of the bond dimensions of the indexes cut are written explicitly. For 12 cycles, there are $2^5\times2^1\times2^2 = 2^8 = 256$ cut instances (paths); for 14 cycles, there are $2^7\times2^7\times2^5 = 2^{19} = 524288$ cut instances.}
\end{figure*}

\begin{figure*}
\includegraphics[width=0.49\columnwidth]{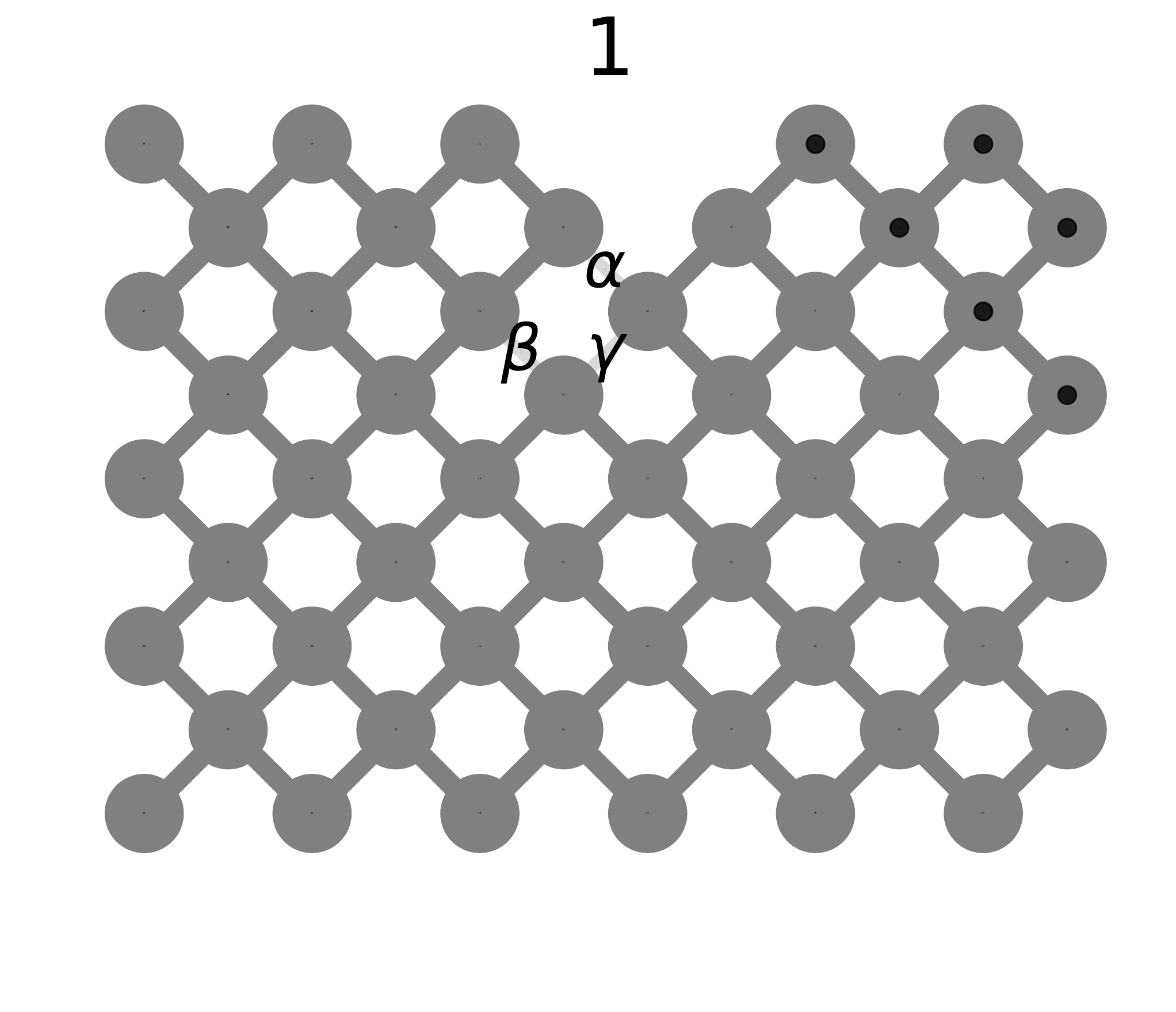}
\includegraphics[width=0.49\columnwidth]{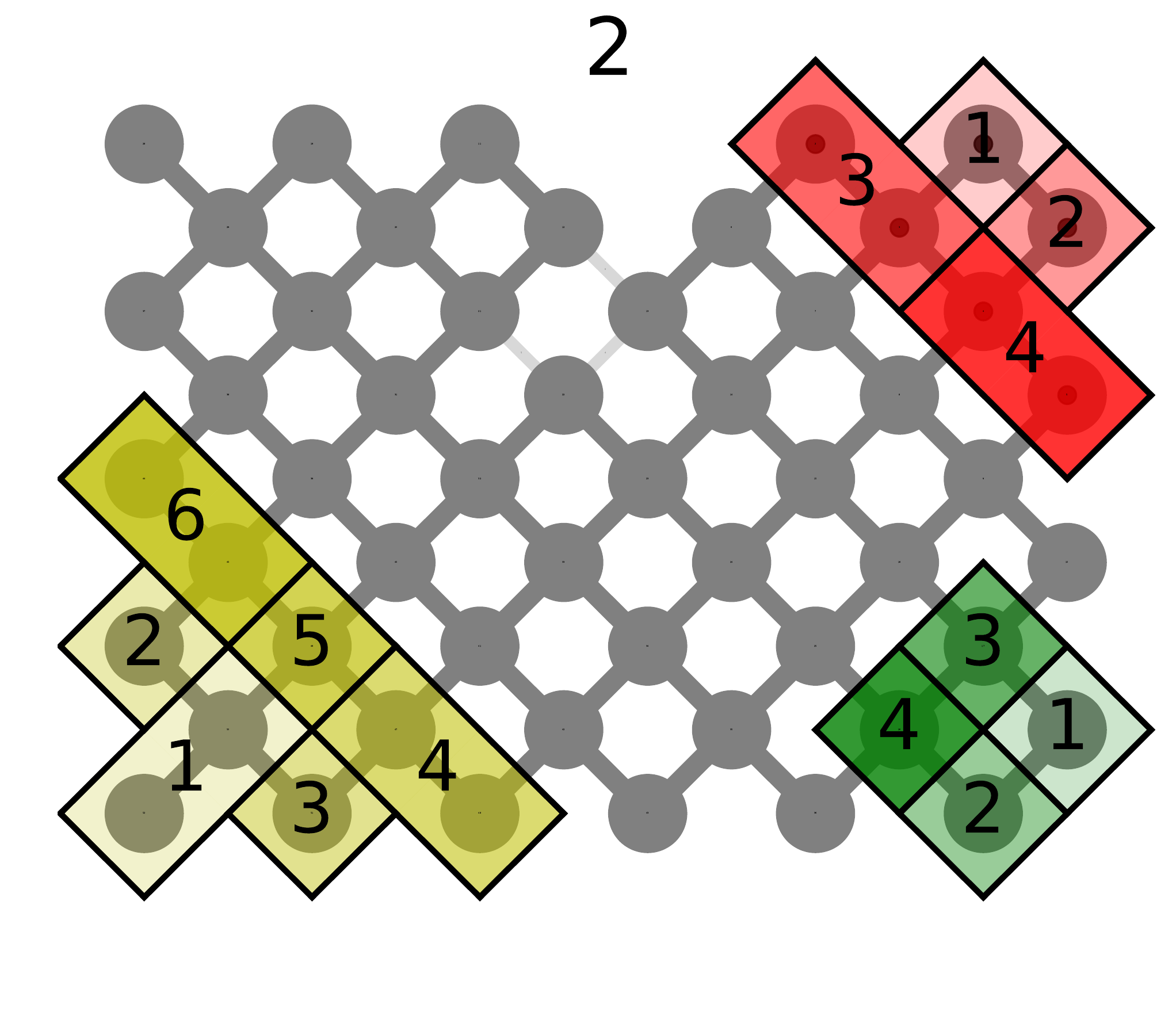}
\includegraphics[width=0.49\columnwidth]{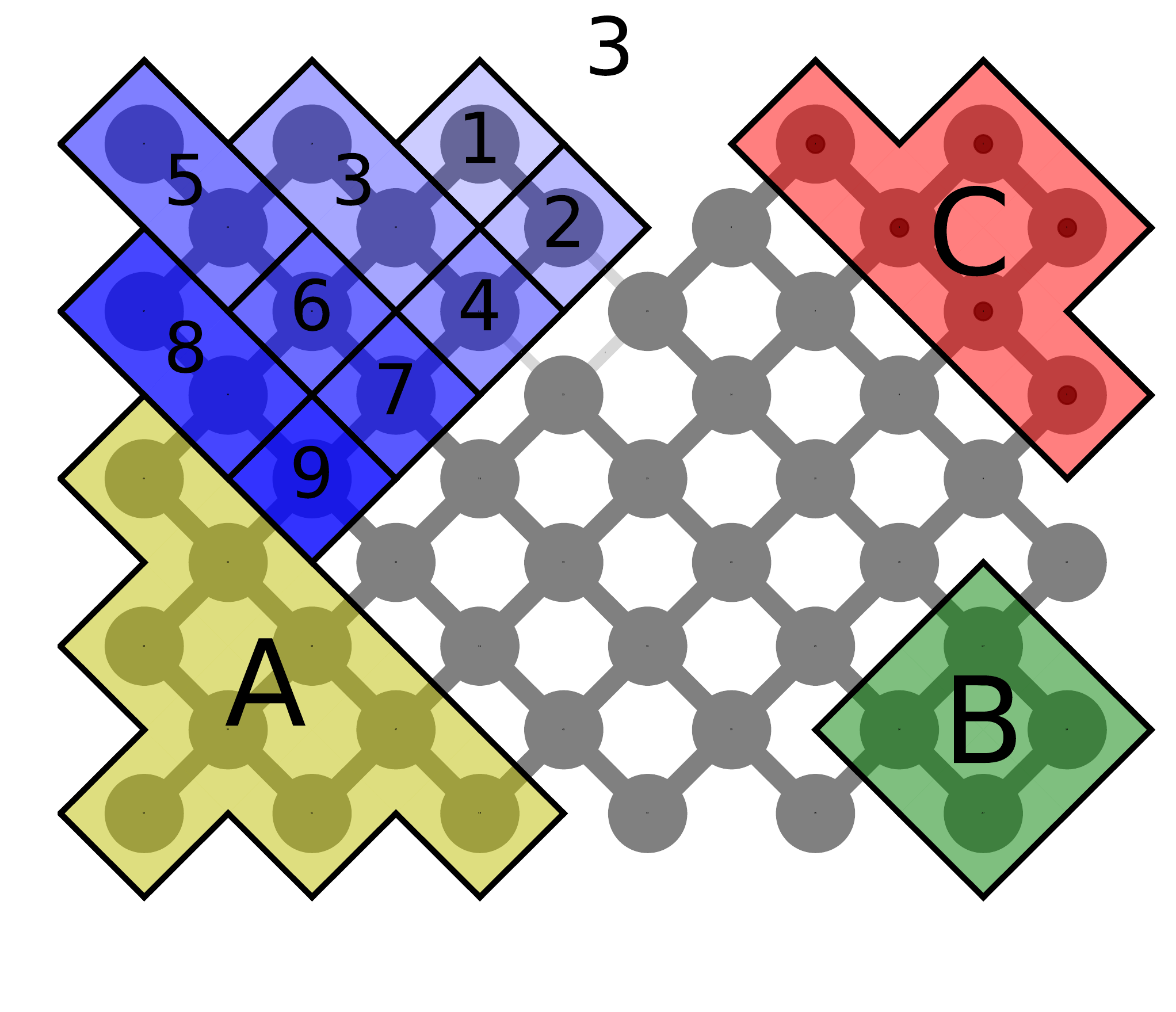}
\includegraphics[width=0.49\columnwidth]{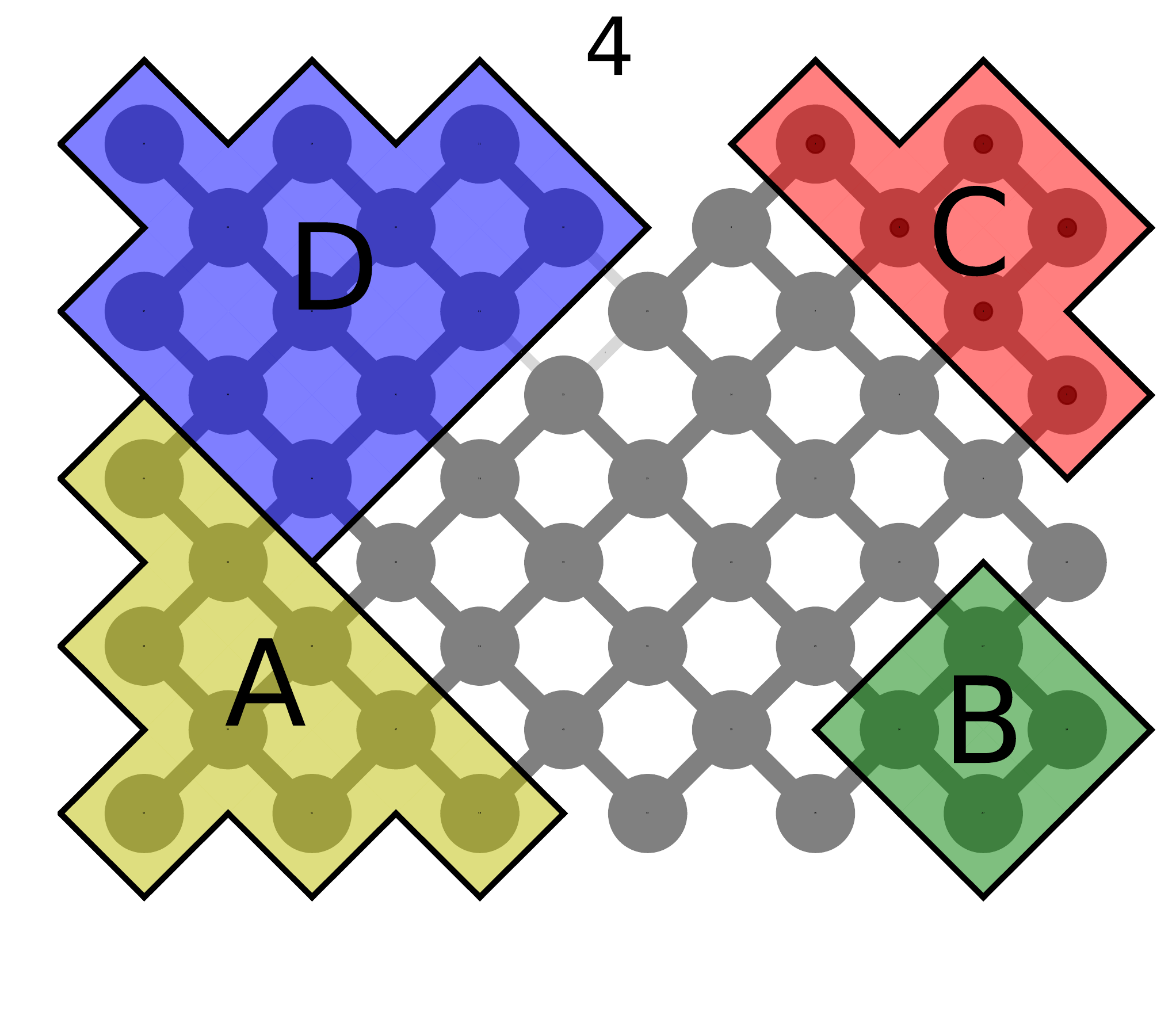}
\includegraphics[width=0.49\columnwidth]{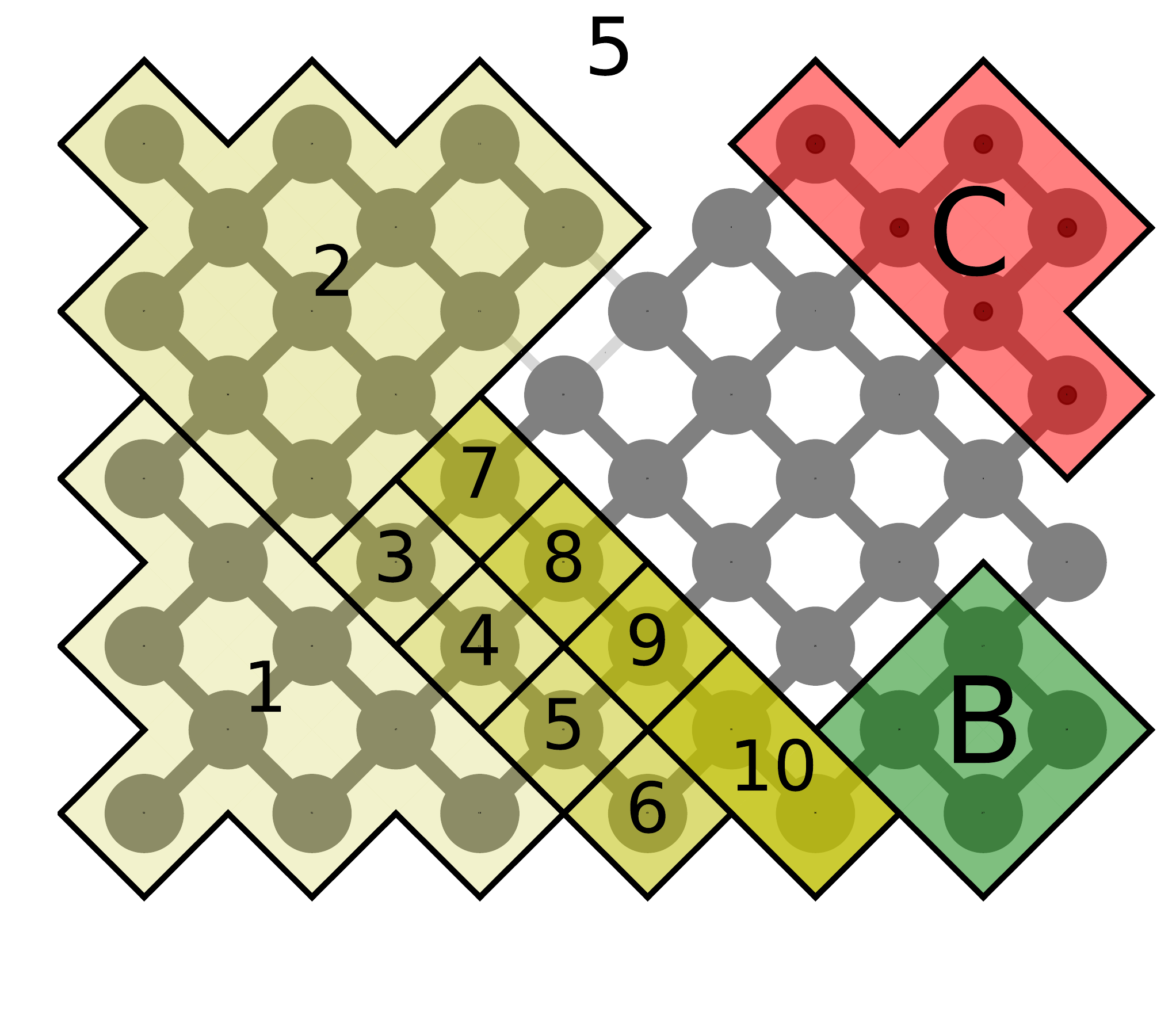}
\includegraphics[width=0.49\columnwidth]{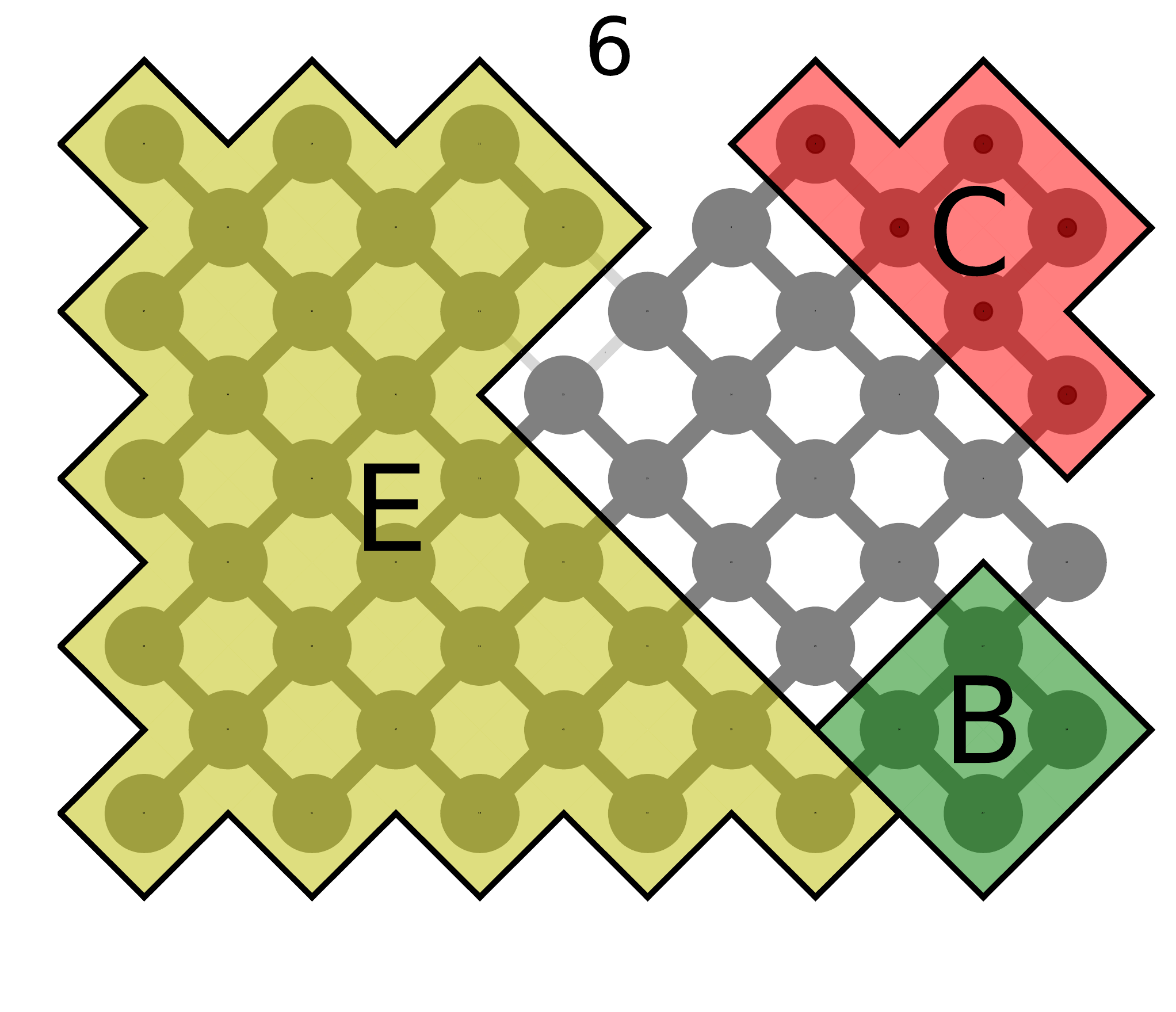}
\includegraphics[width=0.49\columnwidth]{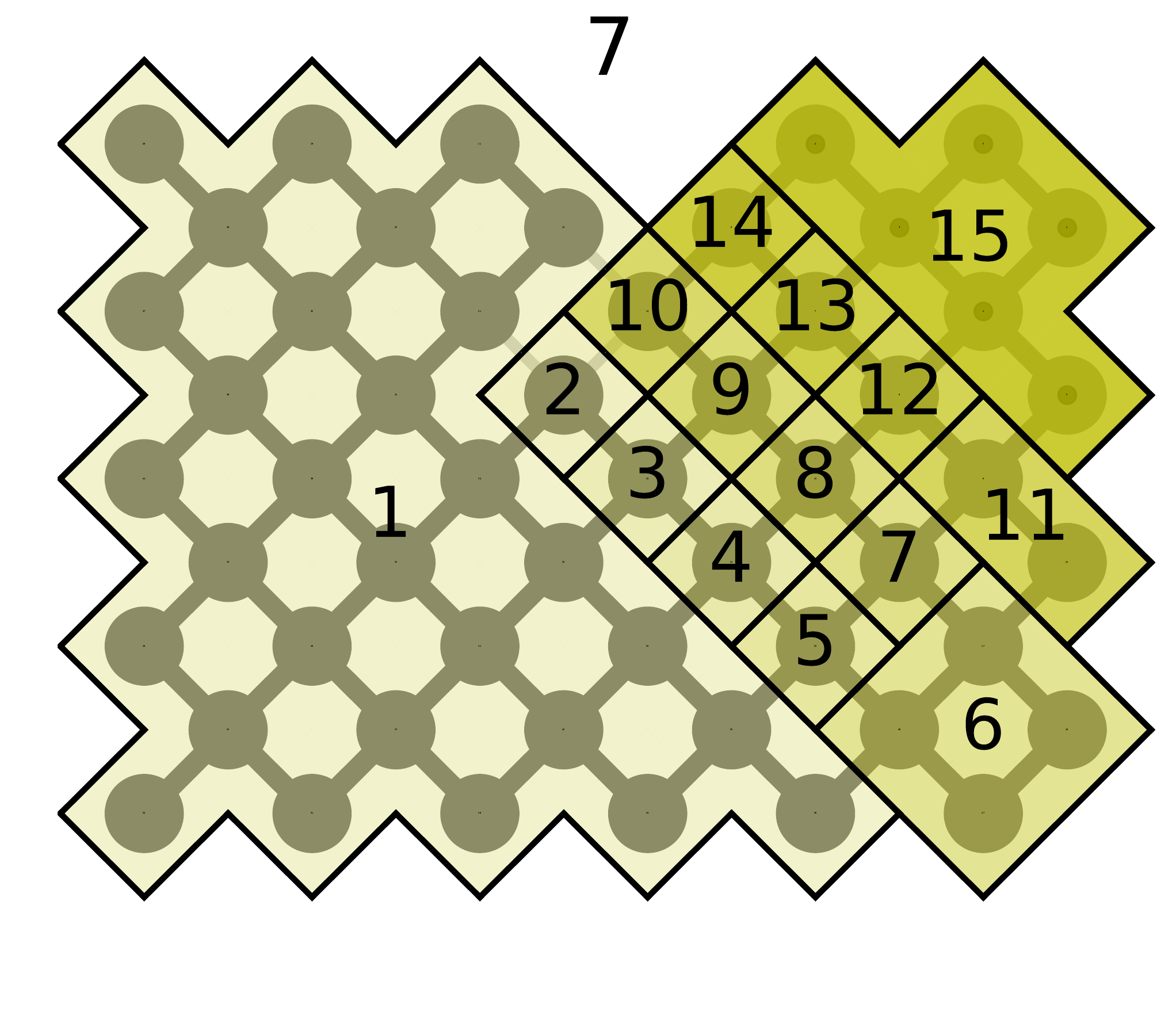}
\includegraphics[width=0.49\columnwidth]{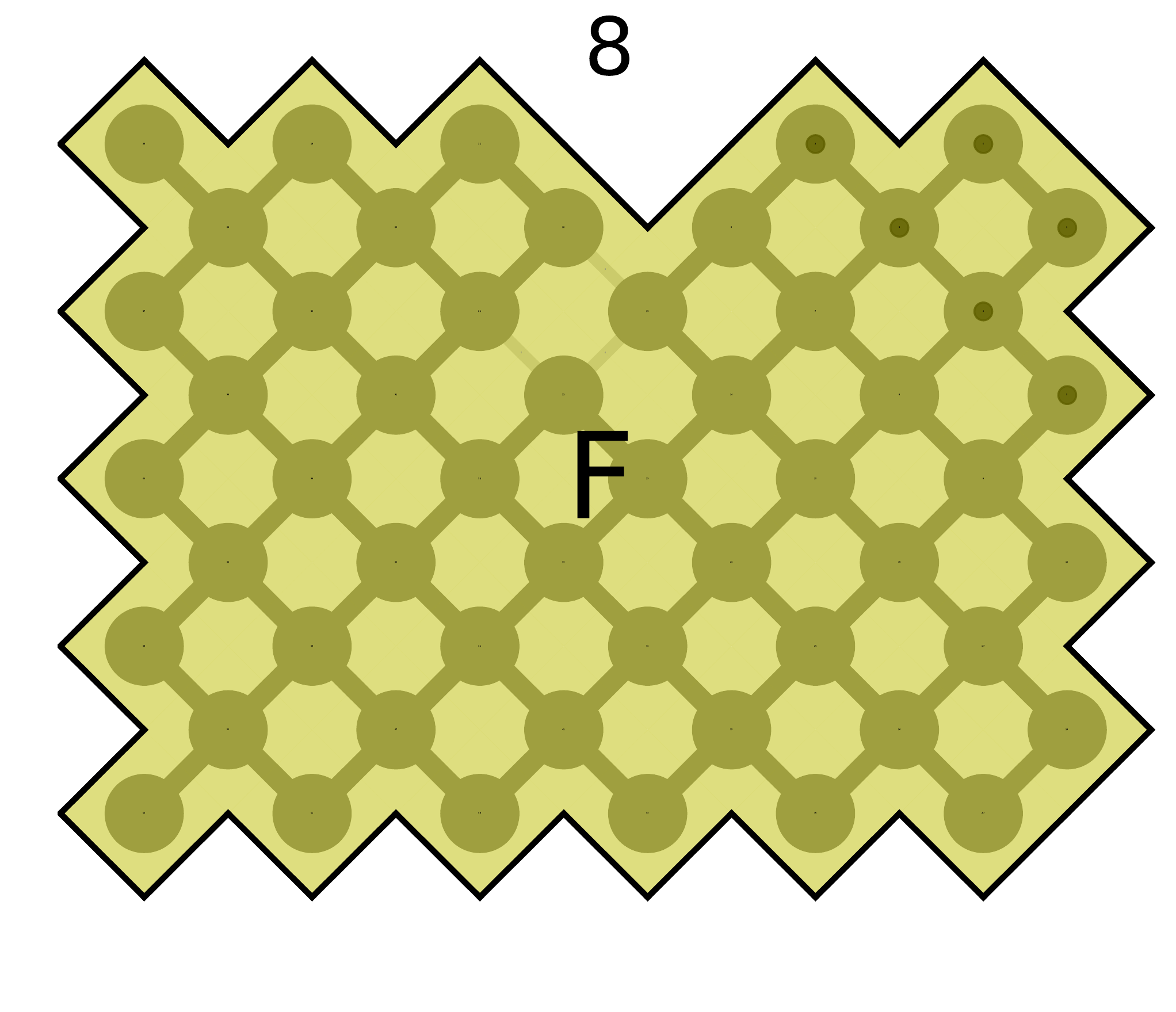}
\caption{\label{fig:qflex_ordering} \textbf{TN contraction ordering for the computation of a batch of amplitudes for the simulation of Sycamore with 12 and 14 cycles.} Dotted qubits are used for fast sampling; the output index is left open. Three indexes are cut, with remaining bond dimensions given in Fig.~\ref{fig:qflex_bond_dimensions}, and all possible cut instances are labelled by variables $\alpha$, $\beta$, and $\gamma$ (panel 1). Tensors $A$, $B$, and $C$ are independent of cut instances, and so are contracted only once (panels 2 and 3) and reused several times. Given a particular instance of $\alpha$ and $\beta$, tensors $D$ (panels 3 and 4) and subsequently $E$ (panels 5 and 6) are contracted; tensor $E$ will be reused in the inner loop. For each instance of $\gamma$ (inner loop), tensor $F$ is contracted (panels 7 and 8), which gives the contribution to the batch of amplitudes (open indexes on $C$ and specified output bits otherwise) from a particular $(\alpha, \beta, \gamma)$ instance (path). The sequence of tensor contractions leading to building a tensor are enumerated, where each tensor is contracted to the one formed previously. For simplicity, the contraction of two single-qubit tensors onto a pair before being contracted with others (\emph{e.g.}, tensor 10 in the yellow sequence of panel 5) is not shown on a separate panel; these pairs of tensors are computed first and are reused for all cut instances.}  
\end{figure*}

qFlex was introduced in Ref.~\cite{villalonga2019flexible} and later adapted to GPU architectures in Ref.~\cite{villalonga2019establishing} to allow efficient computation on Summit, currently the world's Top-1 supercomputer. qFlex is open source and available at \url{https://github.com/ngnrsaa/qflex}.
Given a random quantum circuit, qFlex computes output bitstring amplitudes by adding all the Feynman path contributions via tensor network (TN) contractions \cite{MS08, Boi17}, and so it follows what we call a Feynman approach (FA) to circuit sampling. TN simulators are known to outperform all other methods for circuits with low depth or a large number of qubits (\emph{e.g.}, Ref.~\cite{villalonga2019establishing} successfully simulates 121 qubits at low depth using this technique), as well as for small sample sizes ($N_s$), since simulation cost scales linearly with $N_s$.

TN simulators compute one amplitude (or a few amplitudes; see below) per contraction of the entire network. In order to sample bitstrings for a given circuit, a set of random output bitstrings is chosen before the computation starts. Then, the amplitudes for these bitstrings are computed and either accepted or rejected using
\emph{frugal rejection sampling}~\cite{markov_quantum_2018}. This ensures that the selected subset of bitstrings is indistinguishable from bitstrings sampled from a quantum computer. The cost of the TN simulation is therefore linear in the number of output bitstrings. This makes TN methods more competitive for small sets of output bitstrings.

The optimization of qFlex considers a large number of factors
to achieve the best time-to-solution on current supercomputers, an approach that often
diverges from purely theoretical considerations on the complexity of TN contractions.
More precisely, qFlex implements several features such as:
\begin{itemize}
    \item \textbf{Avoidance of distributed tensor contractions:} by ``cutting'' the TN (slicing some indexes), the contraction of the TN is decomposed into many \emph{paths} that can be contracted locally and independently, therefore avoiding internode communication, which is the main cause for the slowdown of distributed tensor contractions.
    \item \textbf{Contraction orderings for high arithmetic intensity:} TN contraction orderings are chosen so that the expensive part of the computation consists of a small number of tensor contractions with high arithmetic intensity. This lowers the time-to-solution.
    \item \textbf{Highly efficient tensor contractions on GPU:} the back-end TAL-SH library~\cite{lyakh_tal_sh} provides fully asynchronous execution of tensor operations on GPU and fast tensor transposition, allowing out-of-core tensor contractions for instances that exceed GPU memory. This achieves very high efficiency (see Table~\ref{table:qflex_runs}) on high arithmetic intensity contractions.
\end{itemize}

In addition, qFlex implements two techniques in order to lower the cost of the simulation:
\begin{itemize}
    \item \textbf{Noisy simulation:} the cost of a simulation of fidelity $\mathcal{F}<1$ ($\mathcal{F}\approx 5\times10^{-3}$ in practice) is lowered by a factor
    $1/\mathcal{F}$, \emph{i.e.}, is linear in $\mathcal{F}$~\cite{markov_quantum_2018,villalonga2019flexible}.
    \item \textbf{Fast sampling technique:} the overhead in applying the frugal rejection sampling mentioned above is removed by this technique, giving an order of magnitude speedup~\cite{villalonga2019flexible}. This involves the computation of the amplitudes of a few correlated bitstrings (\emph{batch}) per circuit TN contraction.
\end{itemize}

As shown in Table~\ref{table:qflex_runs}, qFlex is successful in simulating Sycamore with 12 cycles on Summit, sampling 1M bitstrings with fidelity close to $0.5\%$ in 1.29 hours. At 14 cycles, we perform a partial simulation and extrapolate the simulation time for the sampling of 1M bitstrings with fidelity close to $0.5\%$ using Summit, giving an estimated 68 days to complete the task. Sampling 3M bitstrings at 14 cycles with fidelity close to $1.0\%$ (average experimentally realized fidelity) would take an estimated 1.1 years to complete. Other estimates for different sample sizes and fidelities can be found in Table~\ref{table:qflex_runs}. At 16 cycles and beyond, however, the enormous amount of Feynman paths required so that the computation does not exceed the 512 GB of RAM of each Summit node makes the computation impractical.\\

The contraction of the TNs involved in the computation of amplitudes from Sycamore using qFlex is preceded by a simplification of the circuits, which allows us to decrease the bond (index) dimension of some of the indexes of the TN. This comes from the realization that ${\rm fSim(\theta=\pi/2, \phi)} = -i\cdot[{\rm R_z (-\pi/2) \otimes R_z(-\pi/2)}]\cdot{\rm cphase(\pi+\phi)}\cdot{\rm SWAP}$ (see Sections~\ref{sec:calib_metro}~and~\ref{subsec:quantum_gates}); note that the ${\rm SWAP}$ gate can be applied either at the beginning or at the end of the sequence. We apply this transformation to all ${\rm fSim}$ gates at the beginning (end) of the circuit that affect qubits that are not affected by any other two-qubit gate before (after) in the circuit. The ${\rm SWAP}$ is then applied to the input (output) qubits and their respective one-qubit gates trivially, and the bond dimension remaining from this gate is 2, corresponding to the ${\rm cphase}$ gate, as opposed to the bond dimension 4 of the original ${\rm fSim}$ gate. Note that in practice this identity is only approximate, since $\theta\approx \pi/2$; we find that transforming all gates described above causes a drop in fidelity to about $95\%$.

After the above simplification is applied, we proceed to cut (slice) some of the indexes of the TN (see Ref.~\cite{villalonga2019flexible} for details). The size of the slice of the index involved in each cut (the effective bond dimension of the index) is variable, and is chosen differently for different number of cycles on the circuit. Cutting indexes decomposes the contraction of the TN into several simpler contractions, whose results are summed after computing them independently on different nodes of the supercomputer.

Fig.~\ref{fig:qflex_bond_dimensions} shows the bond dimensions of the TN corresponding to the circuits with 12 and 14 cycles simulated. We can see the decrease in bond dimension after the ${\rm fSim}$ simplification is applied, as well as the remaining bond dimension on the indexes cut for each case.

Finally, we contract the tensor network corresponding to the computation of a set of amplitudes (for fast sampling) for a particular batch of output bitstrings. The contraction ordering, which is chosen (together with the size and position of the cuts) in order to minimize the time-to-solution of the computation (which involves a careful consideration of the memory resources used and the efficiency achieved on the GPUs) is shown in Fig.~\ref{fig:qflex_ordering}. The computation can be summarized in the following pseudo-code, where $\alpha$, $\beta$, and $\gamma$ are variables that denote the different instances of the cuts:
\begin{verbatim}
# Qubits on C are used for fast sampling.
# size_of_batch amps. per circuit contraction.
size_of_batch = 2^num_qubits(C)

# Placeholder for all amplitudes in the batch.
batch_of_amplitudes = zeros(size_of_batch)

# Start contracting...
contract(A)  # Panel 2
contract(B)  # Panel 2
contract(C)  # Panel 2

# alpha labels instances of 1st cut
for each alpha {

    # beta labels instances of 2nd cut
    for each beta {
        contract(D)  # Panels 3 & 4
        contract(E)  # Panels 5 & 6
        
        # gamma labels instances of 3rd cut
        for each gamma {
            contract(F)  # Panels 7 & 8
            
            # Add contribution from this
            # path (alpha, beta, gamma).
            batch_of_amplitudes += F
        }
    }
}
\end{verbatim}
Dotted qubits on Fig.~\ref{fig:qflex_ordering} denote the region used for fast sampling, where output indexes are left open. The circuit TN contraction leads to the computation of $64$ amplitudes of correlated bitstrings (tensor $F$). Note that computing only a fraction $\mathcal{F}$ of the paths results in amplitudes with a fidelity roughly equal to $\mathcal{F}$. Computing a set of perfect fidelity batches of amplitudes, where the number of batches is smaller than the number of bitstrings to sample also provides a similar fidelity $\mathcal{F}$ in the sampling task, where $\mathcal{F}$ is equal to the ratio of the number of batches to the number of bitstrings in the sample. A hybrid approach (fraction of batches, each only with a fraction of paths), which we use in practice, also provides a similar sampling fidelity. See Refs.~\cite{markov_quantum_2018, villalonga2019flexible} and Section~\ref{subsec:sa_sfa} for more details.

A new feature of qFlex, implemented for this work, is the possibility to perform out-of-core tensor contractions (of tensors that exceed GPU memory) over more than one GPU on the same node. Although the arithmetic intensity requirements to achieve high efficiency are now higher (about an arithmetic intensity of 3000 for an efficiency close to $90\%$ over three GPUs, as opposed to 1000 for a similar efficiency using a single GPU), the fact that a large part of a node is performing a single TN contraction lets us work with larger tensors, which implies reducing the number of cuts, as well as increasing the bond dimension of each cut; this, in turn, achieves better overall time-to-solution for sampling than simulations based on TNs with smaller tensors and with a lower memory footprint during their contraction (which could perhaps show a higher GPU efficiency due to the simultaneous use of each GPU for independent TNs). It is worth noting that the TN contraction ordering presented in Fig.~\ref{fig:qflex_ordering} provides us with the best time-to-solution after considering several possibilities for the simulation of sampling from Sycamore using qFlex for both 12 and 14 cycles. This is generally not the case, since different numbers of cycles generate different TNs, which generally have different contraction schemes for best simulation time-to-solution.\\

Sampling of random circuits on Sycamore is difficult to simulate with TN simulators at 16 cycles and beyond.  Indeed, FA simulators suffer from an exponential scaling of runtime with circuit depth. For qFlex, this is manifested in the large size of the tensors involved in the circuit TN contraction (this size grows exponentially with the number of cycles of the circuit), which require a large number of cuts in order not to exceed the RAM of a computation node, and which in turn generates an impractical number of Feynman paths. For other simulators, such as the one presented in Ref.~\cite{chen2018classical}, the number of projected variables is expected to be so large that the computation time (which increases exponentially with the number of projected variables) on a state-of-the-art supercomputer makes the computation impractical; see Section~\ref{subsec:treewidth} for a detailed analysis. For TN-based simulators that attempt the circuit contraction distributed over several nodes (without cuts)~\cite{guo2019general}, we expect the size of the largest tensor encountered during the TN contraction (which grows exponentially with depth) to exceed the RAM available on any current supercomputer. Not having enough memory for a simulation is the problem that led to developing FA simulators in the first place, for circuits of close to 50 qubits and beyond, for which the Schr\"odinger simulator (see Section~\ref{subsec:sa_simulator}) requires more memory to store the wave function than available. FA simulators give best performance as compared to other methods in situations with a large number of qubits and low depth. For circuits where both the number of qubits and the number of cycles are considered large enough to make the computation expensive, and contribute equally in doing so (formally, each linear dimension of the qubit grid is comparable to the time dimension), like the supremacy circuits considered in this work, we expect SFA of Section~\ref{subsec:sa_sfa} to be the leading approach for sampling from a random circuit, given a large enough sample size ($\sim$~1M in this work); note the linear dependence of the runtime of FA with sample size, which is absent for SFA.

\subsection{Supercomputer Schr\"odinger simulator}
\label{subsec:sa_simulator}

We also performed supercomputer Schr\"odinger simulations in the  J\"ulich Supercomputing Centre.
For a comprehensive description of the universal quantum computer simulators
JUQCS-E and JUQCS-A, see Refs.~\cite{RAED07x} and \cite{RAED19a}.

For a given quantum circuit $U$ designed to generate a random state,
JUQCS-E~\cite{RAED19a} executes $U$ and computes (in double precision floating point) the probability distribution $p_U(j)$
for each output or bitstring $j\in\{0,\ldots,D-1\}$, where $D=2^n$, $n$ denoting the number of qubits.
JUQCS-E can also compute (in double precision floating point) the corresponding distribution function
$P_U(k)=\sum_{j=0}^k p_U(j)$ and sample bitstrings from it.
We denote by ${\cal U}$ the set of $m$ states generated by executing the circuit $U$.
A new feature of JUQCS-E, not documented in Ref.~\onlinecite{RAED19a},
allows the user to specify a set ${\cal Q}$ of $M$ bitstrings
for which JUQCS-E calculates $p_U(j)$ for all $j\in{\cal Q}$ and saves them in a file.

Similarly, for the same circuit $U$, JUQCS-A~\cite{RAED19a} computes (with adaptive two-byte encoding)
the probability distribution $p_A(j)$ for each bitstring $j\in\{0,\ldots,D-1\}$.
Although numerical experiments  with Shor's algorithm for up to 48 qubits indicate that
the results produced by JUQCS-A are sufficiently accurate,
there is, in general, no guarantee that $p_A(j)\approx p_U(j)$.
In this sense, JUQCS-A can be viewed as an approximate simulator of a quantum computing device.

In principle, sampling states with probabilities $p_A(j)$ requires the knowledge of the
distribution function $P_A(k)=\sum_{j=0}^k p_A(j)$.
If $D$ is large, and $p_A(j)\approx {O}(1/D)$, as in the case of random states, computing
$P_A(k)$ requires the sum over $j$ to be performed with sufficiently high precision.
For instance, if $D=2^{39}$, $p_A(j)\approx{O}(10^{-12})$ and
even with double precision arithmetic ($\approx 16$ digits), adding
$D=2^{39}$ small numbers requires some care. Note that in practice, each MPI process
only calculates a partial sum, which helps to reduce the loss of significant digits.
JUQCS-A can compute $P_A(k)$ in double precision and sample bitstrings from it.
We denote by ${\cal A}$ the set of $M$ bitstrings generated by JUQCS-A after executing the circuit $U$.
Activating this feature requires additional memory, effectively reducing the maximum number of qubits that
can be simulated by three.
This reduction of the maximum number of qubits might be avoided as follows.
In the case at hand, we know that all $p_A(j)\approx {O}(1/D)$.
Then, since $p_A(j)$ is known, one might as well sample the states from a uniform distribution,
list the weight $w_A(j)=N p_A(j)$ for each generated state $j$
and use these weights to compute averages.
We do not pursue this possibility here because for the present purpose,
it is essential to be able to compute $p_U(j)$ and therefore,
the maximum number of qubits that can be studied is limited
by the amount of memory that JUQCS-E, not JUQCS-A, needs to perform the simulation.

For an XEB comparison, the quantities of interest are
\begin{eqnarray}
\alpha_{U,U}&\equiv&\log D +\gamma + \sum_{j=0}^{D-1} p_U(j) \log p_U(j),
\label{alpha0}
\\
\alpha_{A,U}&\equiv&\log D +\gamma + \sum_{j=0}^{D-1} p_A(j) \log p_U(j),
\\
\alpha_{A,A}&\equiv&\log D +\gamma + \sum_{j=0}^{D-1} p_A(j) \log p_A(j),
\\
\alpha_{{\cal X},U}&\equiv&\log D +\gamma + \frac{1}{M}\sum_{j\in{\cal X}} \log p_U(j),
\label{alpha1}
\end{eqnarray}
where ${\cal X}$ is one of the four sets ${\cal U}$, ${\cal A}$,
${\cal M}$ (a collection of bitstrings generated by the experiment), or
${\cal C}$ (obtained by generating bistrings distributed uniformly).
If M is sufficiently large ($M=500000$ in the case at hand), we may expect that
$\alpha_{{\cal U},U}\approx \alpha_{U,U}$ and
$\alpha_{{\cal A},U}\approx \alpha_{A,U}$.

In addition to the cross entropies Eqs.~(\ref{alpha0})--(\ref{alpha1}), we also
compute the linear cross entropies
\begin{eqnarray}
\widehat\alpha_{U,U}&\equiv&\sum_{j=0}^{D-1} p_U(j) (D p_U(j) -1),
\label{alpha2}
\\
\widehat\alpha_{A,U}&\equiv& \sum_{j=0}^{D-1} p_A(j) (D p_U(j)-1),
\\
\widehat\alpha_{A,A}&\equiv&\sum_{j=0}^{D-1} p_A(j) (D p_A(j)-1),
\\
\widehat\alpha_{{\cal X},U}&\equiv& \frac{1}{M}\sum_{j\in{\cal X}} (D p_U(j)-1).
\label{alpha3}
\end{eqnarray}

Table~\ref{tab1} presents simulation results
for the $\alpha$'s defined by Eqs.~(\ref{alpha0})--(\ref{alpha1}) and
for the $\widehat\alpha$'s defined by Eqs.~(\ref{alpha2})--(\ref{alpha3}),
obtained by running JUQCS-E and JUQCS-A on the supercomputers at the J\"ulich Supercomputer Centre.
For testing quantum supremacy using these machines, the maximum number of qubits that a
universal quantum computer simulator can handle is 43 (45 on the Sunway TaihuLight at Wuxi China~\cite{RAED19a}).

The fact that in all cases, $\alpha_{U,U}\approx\alpha_{A,A}\approx 1$ supports
the hypothesis that the circuit $U$, executed by either JUQCS-E or JUQCS-A, produces a Porter-Thomas distribution.
The fact that in all cases, $\alpha_{{\cal U},U}\approx1$ supports the
theoretical result
that replacing the sum over all states by the sum over $M=500000$ states
yields an accurate estimate of the former (see Section \ref{sec:xeb_theory}).
Although $\alpha_{A,A}\approx 1$ in all cases, using the sample ${\cal A}$
generated by JUQCS-A to compute $\alpha_{{\cal A},U}$
shows an increasing deviation from one, the deviation becoming larger as the number of qubits increases.
In combination with the observation that $\alpha_{A,A}\approx 1$, this suggests that
JUQCS-A produces a random state, albeit not the same state as JUQCS-E.
Taking into account that JUQCS-A stores the coefficients of each of the basis states
as two single-byte numbers and not as two double precision floating point numbers (as JUQCS-E does), this is hardly a surprise.

From Table~\ref{tab1} it is clear that
the simulation results for  $\alpha_{{\cal X},U}$ and $\widehat\alpha_{{\cal X},U}$
where ${\cal X}={\cal A},{\cal M},{\cal C}$ are consistent.
The full XEB fidelity estimates $\alpha_{{\cal M},U}$ and $\widehat\alpha_{{\cal M},U}$, that is the values computed
with the bitstrings produced by the experiment, are close to the
fidelity estimates of the probabilistic model, patch XEB, and elided
XEB, as seen in Fig.~4(a) of the main text. 

\begin{table*}[htbp]
\caption{
Simulation results for various $\alpha$'s as defined by Eqs.~(\ref{alpha0})--(\ref{alpha1}), obtained
by JUQCS-E and JUQCS-A.
The results for the $\widehat\alpha$'s defined by Eqs.~(\ref{alpha2})--(\ref{alpha3}) are given in parenthesis.
The set of bitstrings ${\cal M}$ has been obtained from experiments.
In the first column, the number in parenthesis is the circuit identification number.
Horizontal lines indicate that data is not available (and would require additional simulation runs to obtain it).
}
\begin{ruledtabular}
\begin{tabular}{ccccccc}
qubits &
$\alpha_{U,U}        $ &
$\alpha_{A,A}        $ &
$\alpha_{{\cal U},U} $ &
$\alpha_{{\cal A},U}\;(\widehat\alpha_{{\cal A},U}) $ &
$\alpha_{{\cal M},U}\;(\widehat\alpha_{{\cal M},U}) $ &
$\alpha_{{\cal C},U}\;(\widehat\alpha_{{\cal C},U}) $ \\
\hline\noalign{\smallskip}
30 & $1.0000$ &  $1.0000$ & $0.9997$ &   $0.8824\;(0.8826)$ &  $0.0708\;(0.0711)$ & $+0.0026\;(+0.0017)$ \\
39(0) & $1.0000$ &  $1.0000$ & $0.9992$ &   $0.4746\;(0.4762)$ &  $0.0281\;(0.0261)$ & $-0.0003\;(-0.0011)$ \\
39(1) & $1.0000$ &  $1.0000$ & $1.0002$ &   $\mbox{-----};(\mbox{-----})$ &  $0.0350\;(0.0362)$ & $\mbox{-----};(\mbox{-----})$ \\
39(2) & $1.0000$ &  $1.0000$ & $0.9996$ &   $\mbox{-----};(\mbox{-----})$ &  $0.0351\;(0.0332)$ & $\mbox{-----};(\mbox{-----})$ \\
39(3) & $1.0000$ &  $1.0000$ & $0.9999$ &   $\mbox{-----};(\mbox{-----})$ &  $0.0375\;(0.0355)$ & $\mbox{-----};(\mbox{-----})$ \\
42(0) & $1.0000$ &  $1.0001$ & $0.9998$ &   $0.4264\;(0.4268)$ &  $0.0287\;(0.0258)$ & $-0.0024\;(-0.0001)$ \\
42(1) & $1.0000$ &  $1.0000$ & $1.0027$ &   $\mbox{-----};(\mbox{-----})$ &  $0.0254\;(0.0273)$ & $\mbox{-----};(\mbox{-----})$ \\
43(0) & $1.0000$ &  $1.0001$ & $1.0013$ &   $0.3807\;(0.3784)$ &  $0.0182\;(0.0177)$ & $-0.0010\;(-0.0003)$ \\
43(1) & $1.0000$ &  $1.0000$ & $\mbox{-----}$ &   $\mbox{-----};(\mbox{-----})$ &  $0.0217\;(0.0204)$ & $\mbox{-----};(\mbox{-----})$ \\
\end{tabular}
\end{ruledtabular}
\label{tab1}
\end{table*}

For reference, in Tables~\ref{SUPER} and \ref{TIMING} we present some technical information about the supercomputer
systems used to perform the simulations reported in this appendix and give some indication of the computer resources
used.

\begin{table*}[htbp]
\caption{Specification of the computer systems at the J\"ulich Supercomputing Centre used to
perform all simulations reported in this appendix.
The row ``maximum \# qubits'' gives the maximum number of qubits $n$ that JUQCS-E (JUQCS-A) can simulate
on a specific computer.
}
\begin{center}
\begin{ruledtabular}
\begin{tabular}{cccccc}
 Supercomputer                       & JURECA-CLUSTER~\cite{JURECA}        & JURECA-BOOSTER~\cite{JURECA}             & JUWELS~\cite{JUWELS}           \\
\hline\noalign{\vskip 4pt}
 CPU                                 & Intel Xeon            & Intel Xeon Phi 7250-F      & Dual Intel Xeon  \\
                                     & E5-2680 v3 Haswell    & Knights Landing            &  Platinum 8168   \\
\hline\noalign{\vskip 4pt}
 Peak performance                    & 1.8 PFlop/s    & 5 PFlop/s              & 10.4 PFlops/s  \\
 Clock frequency                     & 2.5 GHz        & 1.4 GHz                & 2.7 GHz        \\
 Memory/node                         & 128 GB         & 96 GB + 16 GB (MCDRAM) & 96 GB          \\
 \# cores/node                       & $2\times12$    & 64                     & $2\times24$    \\
 \# threads/core used                & 1              & 1                      & 3              \\
 maximum \# nodes used               & 256            & 512                    & 2048           \\
 maximum  \# MPI processes used      & 4096           & 32768                  & 32768          \\
 maximum \# qubits                   & 40 (43)        & 41 (44)                & 43 (46)        \\
\end{tabular}
\end{ruledtabular}
\label{SUPER}
\end{center}
\end{table*}

\begin{table*}[htbp]
\caption{Representative elapsed times and number of MPI processes used to perform simulations
with JUQCS-E and JUQCS-A on the supercomputer indicated.
Note that the elapsed times may fluctuate significantly depending on the load of the machine/network.
}
\begin{center}
\begin{ruledtabular}
\begin{tabular}{ccc|cccc|ccc}
\multicolumn{3}{c|}{} & \multicolumn{4}{c|}{JUQCS-E} & \multicolumn{3}{c}{JUQCS-A} \\
qubits & gates & & Supercomputer & MPI processes & Elapsed time & & Supercomputer & MPI processes & Elapsed time \\
\hline\noalign{\vskip 4pt}
30 & 614 & & BOOSTER &     128 &  0:02:28 & &  CLUSTER &    128  & 0:05:23 \\
39 & 802 & & CLUSTER &    4096 &  0:42:51 & &  CLUSTER &   4096  & 1:38:42 \\
42 & 864 & & JUWELS  &   16384 &  0:51:16 & &  JUWELS  &  8192   & 2:15:48 \\
43 & 886 & & JUWELS  &   32768 &  1:01:53 & &  JUWELS &  32768  &1:32:19 \\
\end{tabular}
\end{ruledtabular}
\label{TIMING}
\end{center}
\end{table*}

\subsection{Simulation of random circuit sampling with a target fidelity}
\label{subsec:sim_target_fidelity}

A classical simulator can leverage the fact that experimental sampling from random circuits occurs at low fidelity $\xebfidelity$ by considering only a small fraction of the Feynman paths (see Secs. \ref{subsec:sa_sfa} and \ref{subsec:fa}) involved in the simulation~\cite{markov_quantum_2018}, which provides speedups of at least a factor of $1/\xebfidelity$. This is done by Schmidt decomposing a few two-qubit gates in the circuit and counting only a fraction of their contributing terms (\emph{paths}). A key assumption here is that the different paths result in orthogonal output states, as was studied in Ref.~\cite{markov_quantum_2018} and later in Ref.~\cite{villalonga2019flexible}. In what follows, we argue that, provided the generation of paths through decomposing gates, the Schmidt decomposition is indeed the optimal approach to achieving the largest speedup, \emph{i.e.}, that the fidelity kept by considering only a fraction of paths is largest when keeping the paths with the largest Schmidt coefficient. This is different from proving the optimality of the Schmidt decomposition of a single gate, since here we refer to the fidelity of the entire output state, and decomposed gates are embedded in a much larger circuit. In addition, we show that, for the two-qubit gates used in this work, the speedup is very close to linear in $\xebfidelity$ (and not much larger), since their Schmidt spectrum is close to flat. We close this section by relating the present discussion to Section~\ref{sec:wedge_formation}, where the formation of simplifiable gate patterns in some two-qubit gate tilings of the circuit is introduced.

In summary, this section provides a method to simulate approximate sampling with a classical computational cost proportional to $\xebfidelity$. Sec.~\ref{sec:complexity} argues, based on complexity theory, that this scaling is optimal. We note that Refs~\cite{kalai2014gaussian,bremner2017achieving,yung2017can} propose an alternative method to approximately sample the output distribution at low fidelity. In essence, this method relies on the observation that, for some noise models, the high weight Fourier components of the noisy output distribution decay exponentially to 0. Then this method proposes to estimate low weight Fourier components with an additive error which is polynomial in the computational cost. Nevertheless, Ref.~\cite{boixo2017fourier} shows that all Fourier components of the output distribution of random circuits are exponentially small, and therefore they can not be estimated in polynomial time with this method. The conclusion is then that the noisy output distribution can be approximated by sampling bitstrings uniformly at random, the distribution for which all Fourier components are 0. This is consistent with Ref.~\cite{boixo2018characterizing} and Secs.~\ref{sec:xeb_theory} and~\ref{sec:error_model}, but it will produce a sample with $\xebfidelity =0$, while the output of the experimental samples at 53 qubits and $m=20$ still has $\xebfidelity \ge 0.1\%$

\subsubsection{Optimality of the Schmidt decomposition for gates embedded in a random circuit}

Consider a two-qubit gate $V_{ab}$ acting on qubits $a$ and $b$. We would like to replace it by a tensor product operator $M_a \otimes N_b$. The final state of the ideal circuit is 
\begin{equation}
| \psi \rangle := U_2 V_{ab} U_1 |0^n \rangle 
\end{equation}
where $U_1 (U_2)$ is a unitary composed by all the gates applied before (after) $V_{ab}$. The final normalized state of the circuit with the replacement by $M_a \otimes N_b$ is 
\begin{equation}
| \phi_{M, N} \rangle := U_2 (M_a \otimes N_b)  U_1 |0^n \rangle / \Vert  U_2 (M_a \otimes N_b)  U_1 |0^n \rangle \Vert.
\end{equation}

We would like to find $M, N$ which maximize the fidelity of the two states, given by 
\begin{equation} \label{overlap}
\langle \psi | \phi_{M, N} \rangle =  \langle  0^n | U_{1}^{\cal y} V_{ab}^{\cal y}  |\beta \rangle / \sqrt{  \langle  \beta | \beta \rangle }   \text{,}
\end{equation}
where
\begin{equation}
| \beta \rangle \equiv  (M_a \otimes N_b)  U_{1} |0^n \rangle 
\end{equation}

As the overlap is invariant if we multiply $(M_a \otimes N_b)$ by a constant, we fix the normalization $\text{tr}[(M_a \otimes N_b)^{\cal y}(M_a \otimes N_b)] = 1$.

We now make the assumption that the circuit is random (or sufficiently scrambling) and that the $V_{ab}$ is a gate placed sufficiently in the middle of the computation that the reduced density matrix of qubits $a$ and $b$ of $U_{1} |0^n \rangle$ shows maximal mixing between the two. In more detail, let 
\begin{equation} \label{boundonHSdistance}
\varepsilon := \left \Vert \text{tr}_{\backslash (a, b)}(U_1 | 0^n \rangle  \langle 0^n | U_1^{\cal y} ) - \frac{I}{4} \right \Vert_2,
\end{equation}
with $\Vert X \Vert_2 := \text{tr} (X^{\cal y}X)^{1/2}$ the Hilbert-Schmidt norm and $\text{tr}_{\backslash (a, b)}$ the partial trace of all qubits except $a$ and $b$. 

Using Eq. (\ref{boundonHSdistance}) and Eq.~(\ref{overlap}), we find

\begin{eqnarray} 
\langle \psi | \phi_{M, N} \rangle &=& \text{tr}( \text{tr}_{\backslash (a, b)}(U_1 | 0^n \rangle  \langle 0^n | U_1^{\cal y} ) V_{ab}^{\cal y} (M_a \otimes N_b))  \\
&=& \frac{1}{4} \text{tr}[ V_{ab}^{\cal y}  (M_a \otimes N_b) ]  \pm \Vert (M_a \otimes N_b) \Vert_2 \Vert V_{ab} \Vert_2 \varepsilon. \nonumber
\end{eqnarray}
As $\Vert (M_a \otimes N_b) \Vert_2 = 1$ and $\Vert V_{ab} \Vert_2 = 2$, we find 
\begin{eqnarray} \label{overlap_epsilon}
\langle \psi | \phi_{M, N} \rangle = \frac{1}{4} \text{tr}[ V_{ab}^{\cal y}  (M_a \otimes N_b) ]  \pm 2 \varepsilon.
\end{eqnarray}

Refs. \cite{nahum2018operator, von2018operator} proved that for a random circuit $U_1$ of depth $D$ in one dimension, $\varepsilon  \leq  (4/5)^D$. In two dimensions we expect $\varepsilon$ to go to zero even faster with depth, so we can ignore the second term of Eq.~(\ref{overlap_epsilon}) for sufficiently large depth.

We now want to find $M_a$, $N_b$ which are optimal for
\begin{equation}  \label{optimization33}
\max_{M_a, N_b : \Vert M_a \Vert_2 = \Vert N_b \Vert_2 } \text{tr}[ V_{ab}^{\cal y}  (M_a \otimes N_b) ] \text{.}
\end{equation}
At this point, we have reduced the problem to finding the optimal decomposition of the gate as a standalone operator.

Consider the operator Schmidt decomposition of $V_{ab}$:
\begin{equation}
V_{ab} = \sum_{i} \lambda_i R_{a, i} \otimes S_{b, i},
\end{equation}
where $R_{a, i}$ ($S_{b, i}$) are orthonormal set of operators in the Hilbert-Schmidt inner product, i.e. $\text{tr}(R_{a, i}^{\cal y} R_{a, j}) = \text{tr}(S_{a, i}^{\cal y} S_{b, j}) =  \delta_{ij}$. The Schmidt singular values $\lambda_1 \geq \lambda_2 \geq \ldots$ are in decreasing order. Then it follows that the solution of Eq.~(\ref{optimization33}) is $\lambda_1$, with optimal solution $M_a = R_{a, 1}$ and $N_b = S_{b, 1}$. Indeed we can write Eq.~(\ref{optimization33}) as 
\begin{equation}
\max_{ |x \rangle, |y \rangle  }   \langle x |  \overline{V} |y \rangle
\end{equation}
where the maximum is over all unit vectors $|x \rangle, |y \rangle$ in $(\mathbb{C}^2)^{\otimes 2}$ and $\overline{V}$ is the matrix 
\begin{equation}
\overline{V} := \sum_i \lambda_i (R_{a, i} \otimes I)|\Phi \rangle \langle \Phi | (S_{b, i}^{\cal y} \otimes I)
\end{equation}
with $| \Phi \rangle = \sum_{i} |i \rangle \otimes |i \rangle$. This can be verified using the fact that any unit vector $| x \rangle$ in $(\mathbb{C}^d)^{\otimes 2}$ can be written as $|x\rangle = (L \otimes I) |\Phi \rangle$ for a matrix $L$ acting on $(\mathbb{C}^d)$ s.t. $\Vert L \Vert_2 = 1$. The result follows by noting that $\lambda_i $ are the singular values of $\overline{V}$. 

The argument above easily generalizes to the problem of finding the optimal operator of Schmidt rank $k$ for replacing the unitary gate. In that case the optimal choice is $\sum_{i=1}^k \lambda_i R_{a, i} \otimes S_{b, i}$.\\

\begin{figure}
    \centering
    \includegraphics[width=8.5cm]{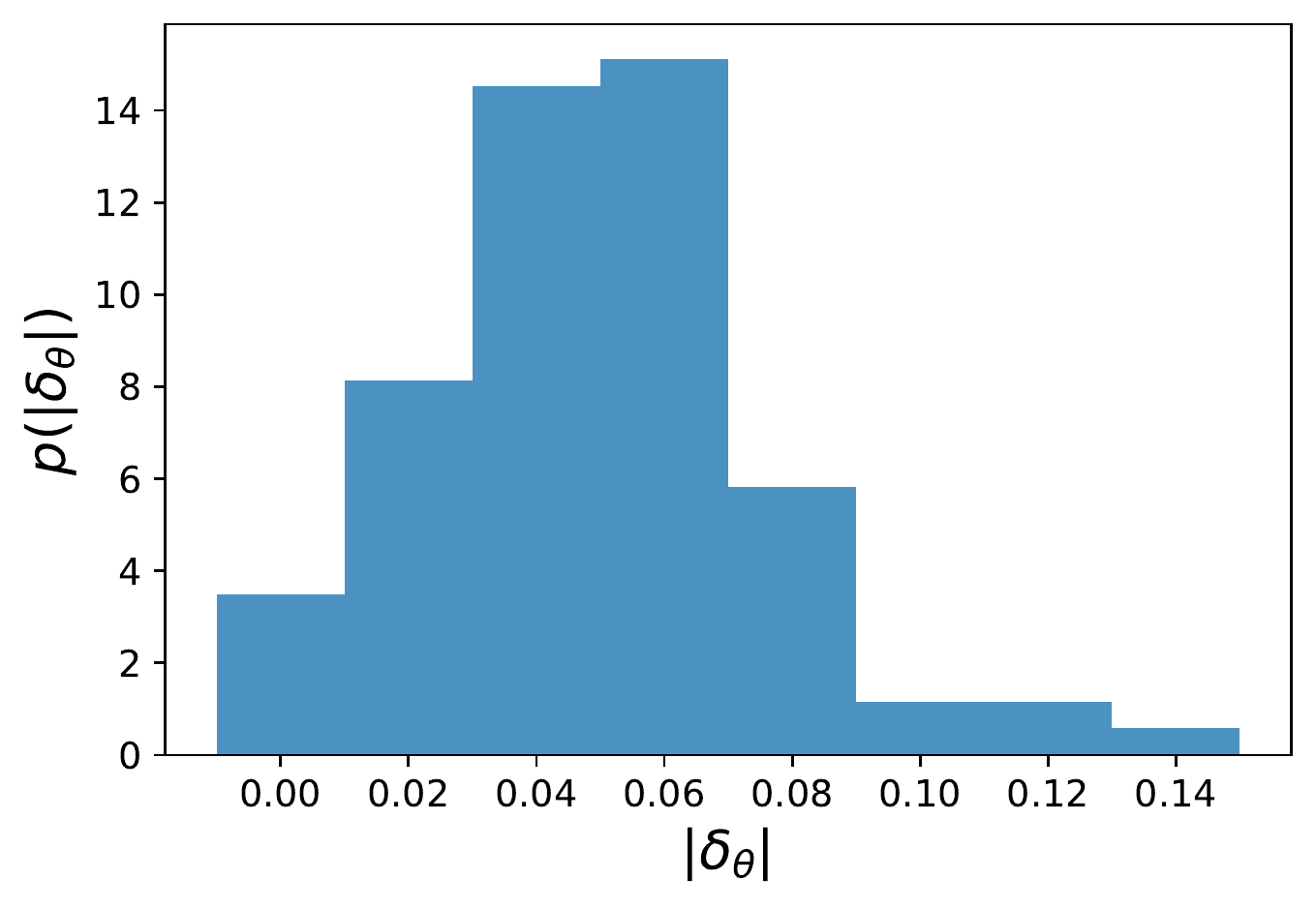}
    \caption{\textbf{Probability distribution of the deviations $|\delta_\theta|$ from $\theta\approx \pi/2$ for $\rm fSim$ gates.} The magnitude of $\delta_\theta$ is directly related to the runtime speedup low fidelity classical sampling can take from exploiting the existence of paths with large Schmidt coefficients. In practice, $|\delta_\theta|\approx 0.05$ radians on average, which imposes a bound of less than an order of magnitude on this potential speedup for the circuits, gates, and simulation techniques considered in this work.}
    \label{fig:p_delta_theta}
\end{figure}

\begin{figure}
    \centering
    \includegraphics[width=8.5cm]{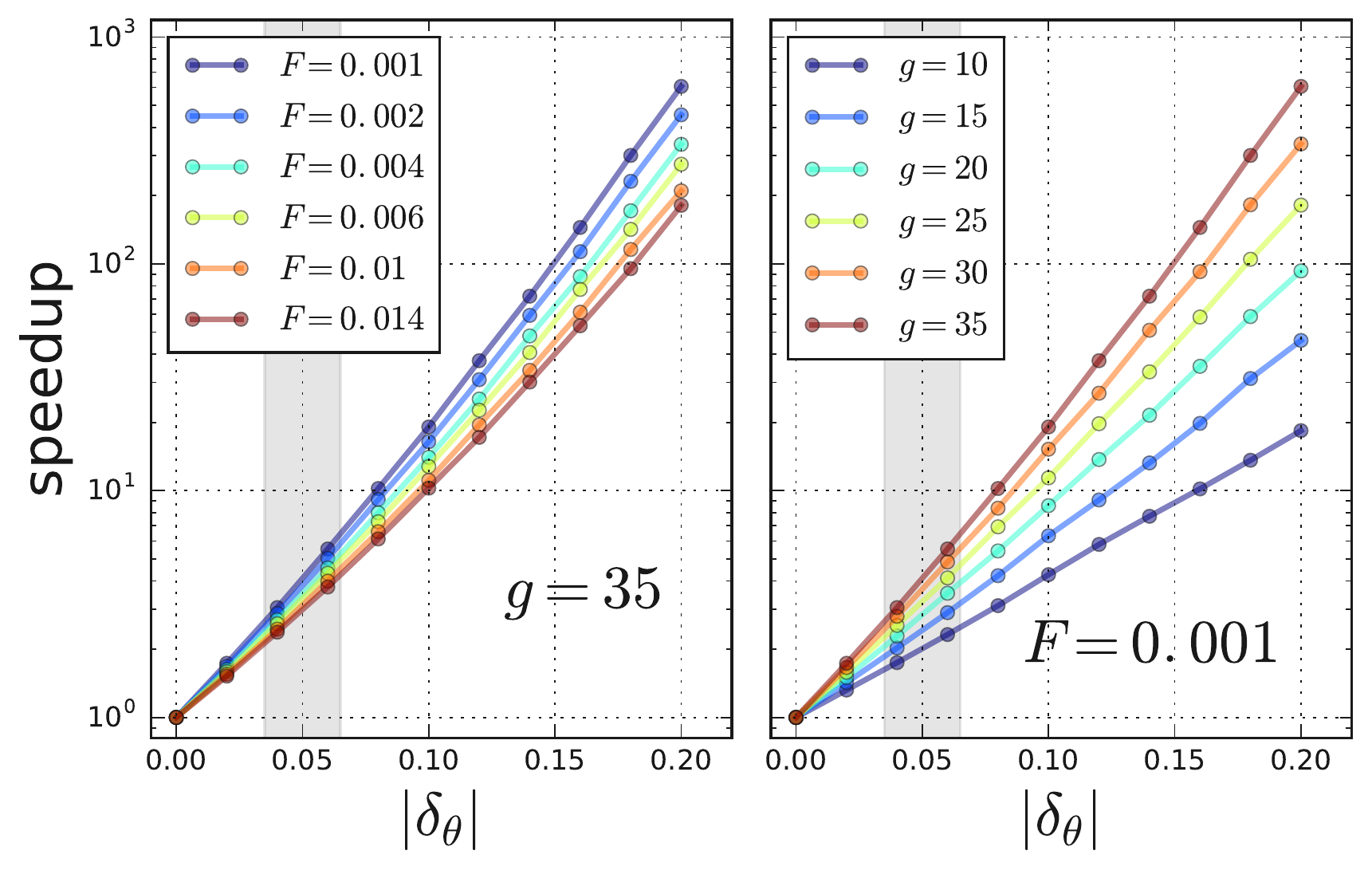}
    \caption{\textbf{Classical speedup given by the imbalance in the Schmidt coefficients of the gates decomposed.} The speedup is computed by comparison with the case where $\theta=\pi/2$ exactly. The classical simulation has a target fidelity $\mathcal{F}$, and $g$ {\rm fSim} gates are decomposed. For simplicity, we assume $\theta = \pi/2 + \delta_\theta$ is the same for all gates, as well as $\phi=\pi/6$. \emph{Left:} speedup at different target fidelities for fixed $g=35$. Note that the speedup decreases with $\mathcal{F}$; this is due to the fact that at very low fidelity, considering a few paths with very high weight might be enough to achieve the target fidelity, while for larger values of $\mathcal{F}$, paths with a smaller weight have to be considered, and so a larger number of them is needed per fractional fidelity increase. \emph{Right:} speedup for fixed fidelity $\mathcal{F}=0.001$ for different values of $g$. As expected, the speedup is greater as $g$ increases, since the weight of the highest contributing paths increases exponentially with $g$. The largest speedup is achieved at large $g$ and small $\mathcal{F}$. For $g=35$ and $\mathcal{F}\gtrapprox 0.001$, we find speedups well below an order of magnitude, given that $|\delta_\theta|\approx 0.05$ radians in practice (shaded area); this case is representative of our simulation of Sycamore with $m=20$ (see Section~\ref{subsec:sa_sfa}) targeting the fidelity measured experimentally.}
    \label{fig:schmidt_speedups}
\end{figure}

\subsubsection{Classical speedup for imbalanced gates}

We now want to analyze the Schmidt spectrum of the two-qubit gates used in this work. The $\rm fSim(\theta, \phi)$ gate is introduced in Section~\ref{subsec:quantum_gates}. This gate, which is presented in matrix form in Eq.~(\ref{eq:fsim}), has the following Schmidt singular values:
\begin{align}
  \label{eq:singular_values}
  \lambda_1 &= \sqrt{1 + 2 \cdot |\cos(\phi/2) \cos \theta| + \cos^2 \theta} \\ 
  \lambda_2 = \lambda_3 &= \sin \theta \\
  \lambda_4 &= \sqrt{1 - 2 \cdot |\cos(\phi/2) \cos \theta| + \cos^2 \theta} \;,
\end{align}
where normalization is chosen so that $\sum_i^4 \lambda_i^2 = 4$. In practice, we have $\theta\approx \pi/2$ and $\phi\approx \pi/6$, and so we obtain $\lambda_i\approx 1$, $\forall i\in \{1, 2, 3, 4\}$, which gives a flat spectrum.

In the case that $\theta=\pi/2\pm\delta_\theta$, the spectrum becomes imbalanced, as expected. When considering the decomposition of a number $g$ of $\rm fSim(\pi/2\pm\delta_\theta, \phi\approx \pi/6)$ gates, the set of weights of all paths is equal to the outer product of all sets of Schmidt coefficients (one per gate). Achieving a fidelity $\xebfidelity>0$ implies (in the optimal case) including the largest contributing paths, and so the advantage one can get from this is upper bounded by the magnitude of the largest weight, which is equal to $\prod^g_{\alpha=1} \lambda^2_{\alpha, max}$, where $\alpha$ labels the gates decomposed and $\lambda_{\alpha, max}$ is the largest Schmidt coefficient for gate $\alpha$. In practice, $|\delta_\theta|$ has values of around 0.05 radians (see Fig.~\ref{fig:p_delta_theta}). The geometric mean of $\lambda_{max}$ is about 1.047, which gives an upper bound of $1.047^{2g}$ to the speedup discussed here. For the largest value of $g$ considered in this work, \emph{i.e.}, the decomposition of $g=35$ gates using the SFA simulator (Section~\ref{subsec:sa_sfa}) on a circuit of $m=20$ cycles, we obtain a value of $1.047^{2\times35}=25.4$. Note that the speedup obtained in practice (as compared to runtimes over circuits with perfectly flat gate Schmidt decompositions) for fidelities of the order of $0.1\%$ and larger is expected to be far smaller than this value, given that one has to consider a large number of paths, from which only an exponentially small number will have a weight close to 25.4.

We can get a better estimate for the speedup achieved in practice, beyond the upper bound of about a factor of 25 that decomposing $g=35$ gates with typical parameters would give. For simplicity, let us assume that all $g$ gates have the same values of $\theta$ and $\phi$. Then the weight of each path arising from this decomposition can be written as $W_{i} = W_{(a, b, c)} = \lambda_1^{2a}\lambda_2^{2b}\lambda_3^{2c}$, where $a+b+c=g$, and that the number of paths for each choice of $(a, b, c)$ is equal to $\#(a, b, c) = \sum_{k=0}^{b} {\rm multinomial}(a, b-k, k, c) = 2^b \times {\rm multinomial}(a, b, c)$. After sorting all $4^g$ weights (and paths) by decreasing value, given a target fidelity, $\mathcal{F}$, one now has to consider the first $S$ paths (\emph{i.e.}, those with the largest weight), up to the point where the sum of their weights $\sum_{i=1}^S \frac{W_i}{4^g}$ matches the target fidelity. The normalization factor $4^g$ guarantees that if one were to consider all paths, the fidelity would be unity, as expected. Compared to the case where we consider a number $\mathcal{F}\times 4^g$ of paths, as for a flat Schmidt spectrum, this provides a speedup equal to $\frac{S}{\mathcal{F}\times 4^g}$. We show the speedup achieved this way in Fig.~\ref{fig:schmidt_speedups}. For the case where we would achieve the largest speedup in the simulations considered in this work, namely the simulation of Sycamore at $m=20$ cycles and a fidelity $\mathcal{F}\approx 0.2\%$ with $g=35$ gates decomposed (see Section~\ref{subsec:cost_estimation}), we estimate that the speedup obtained this way would be well below an order of magnitude, since $|\delta_\theta|$ typically takes values of about $0.05$ radians.

\subsubsection{Verifiable and supremacy circuits}
\label{subsec:verifiable_and_supremacy}

So far we have considered the decomposition of gates one by one, \emph{i.e.}, where the total number of paths is equal to the product of the Schmidt rank of all gates decomposed. However, by fusing gates together in a larger unitary, one can provide some speedup to the classical simulation of the sampling task.

The rationale here comes from the realization that a unitary that involves a number of qubits $q$ cannot have a rank larger than $4^{\min(q_l, q_r)}$ when Schmidt decomposed  over two subsets of qubits of size $q_l$ and $q_r$, with $q_l+q_r=q$. Therefore one might reduce exponentially the number of paths by fusing gates such that the resulting unitary reaches on either side ($l$ or $r$) a number of qubits that is smaller than the product of the ranks of the fused gates to be decomposed. This is at the heart of the formation of \emph{wedges} of Section~\ref{sec:wedge_formation}. These wedges denote particular sequences of consecutive two-qubit gates that only act upon three qubits. Fusing these two-qubit gates together generates 4 paths, as opposed to a naive count of $4^2$ paths if one decomposes each gate separately. Each wedge identified across a circuit cut provides a speedup by a factor of $4$.

In this work, we define two classes of circuits: \emph{verifiable} and \emph{supremacy} circuits. Verifiable circuits present a large number of wedges across the partition used with the SFA simulator (Section~\ref{subsec:sa_sfa}) and are therefore classically simulatable in a reasonable amount of time. These circuits were used to perform full XEB over the entire device up to depth $m=14$ (see Fig.~4a of the main article and Sections~\ref{sec:circuits}~and~ \ref{sec:large_scale_xeb}), which involves perfect fidelity computations. On the other hand, supremacy circuits are designed so that the presence of wedges and similar sequences is mitigated, therefore avoiding the possibility of exploiting this classical speedup.

It is natural to apply the ideas presented here beyond \emph{wedges}. It is also easy to look for similar structures in the circuits algorithmically. This way, we find that for the supremacy circuits there is a small number of such sequences. On the sequence of cycles DCD (see Fig.~\ref{fig:gate_patterns}), three two-qubit gates are applied on qubits 16, 47, and 51 (see Fig.~\ref{fig:qubit_order} for numbering). These three gates can be fused in one. Then, if the two gates between qubits 47 and 51 are decomposed (as is done with the SFA simulations of Section~\ref{subsec:sa_sfa} used in Fig.~4 of the main article), this technique provides a speedup of a factor of 4. The sequence of layouts DCD appears twice for circuits of $m=20$, which provides a total speedup of $4^2=16$ in the simulation of the supremacy circuits. This particular decomposition is currently not implemented, and the estimated timings of Section~\ref{subsec:sa_sfa} and Fig.~4 of the main article do not take it into account.

Beyond this, one has to go to groups of several cycles of the circuit (more than two) in order to identify regions where the fusion of several gates provides any advantage of this kind. In our circuits, the resulting unitaries act upon a large number of qubits, which makes explicitly building the unitary impractical.\

\subsection{Treewidth upper bounds and variable elimination algorithms}
\label{subsec:treewidth}
We explained in Section~\ref{subsec:fa} that the Feynman method to compute individual amplitudes of the output of a quantum circuit can be implemented as a tensor network when quantum gates are interpreted as tensors. All indexes of the tensor network have dimension two because indexes correspond to qubits.
Similarly, Ref.~\cite{Boi17} showed that a quantum circuit can be mapped directly to an undirected graphical model. In the undirected graphical model, vertices or variables correspond to tensor indexes, and cliques correspond to tensors. Individual amplitudes can be computed using a variable elimination algorithm on the undirected graphical model, which is similar to a tensor contraction on a tensor network. The variable elimination algorithm depends on the ordering in which variables are eliminated or contracted. If we define the \emph{contraction width} of an ordering to be the rank of the largest tensor formed along the contraction, the \emph{treewidth} of the undirected graph is equal to the minimum contraction width over all orderings. Therefore, the complexity of a tensor network contraction grows in the optimal case exponentially with the treewidth, and the treewdith can be used to study the complexity of Feynman methods for simulating quantum circuits~\cite{MS08}. Ref.~\cite{Boi17} showed that for diagonal gates the undirected graphical model is simpler, potentially lowering its treewidth, and hence improving the complexity. This simplification is not achievable in the tensor network view without including hyperedges, \emph{i.e.}, edges attached to more than two tensors. Ref.~~\cite{Boi17} also introduced the use of {\tt QuickBB} to find a heuristic contraction ordering~\cite{gogate2004complete}. If allowed to run for long enough, {\tt QuickBB} finds the optimal ordering, together with the treewidth of the graph. However, note that obtaining the treewidth of a graph is an NP-hard problem, and so in practice a suboptimal solution is considered for the simulations described here.\\

Once the width of a contraction is large enough, the largest tensor it generates is beyond the memory resources available. This constraint was overcome in Ref.~\cite{chen2018classical} by \emph{projecting} a subset of $p$ variables or vertices in the undirected graphical model into each possible bistring of 0 and 1 values. This generates $2^p$ similar subgraphs, each of which can be contracted with lower complexity and independently from each other, making the computation embarrassingly parallelizable. Choosing the subset of variables that, after projection, optimally decreases the treewidth of the resulting subgraph is also NP-hard. However, Ref.~\cite{chen2018classical} developed a heuristic approach that works well in practice. The algorithm proceeds as follows:

\begin{enumerate}
    \item Run {\tt QuickBB} for $S$ seconds on the initial graph. This gives a heuristic contraction ordering, as well as an upper bound for the treewidth.
    \item For each variable, estimate the cost of contracting the subgraph after projection. The estimate is done with the ordering inherited from the previous step.
    \item Choose to project the variable which results in the minimum contraction cost. 
    \item Repeat steps 2 and 3 until the cost is within reasonable resources.
    \item Once all variables have been chosen and projected, run {\tt QuickBB} for $S$ seconds on the resulting subgraph to try to improve the contraction ordering inherited from step 1 and lower the contraction cost.
\end{enumerate}

\begin{figure}
    \centering
    \includegraphics[width=1.\columnwidth]{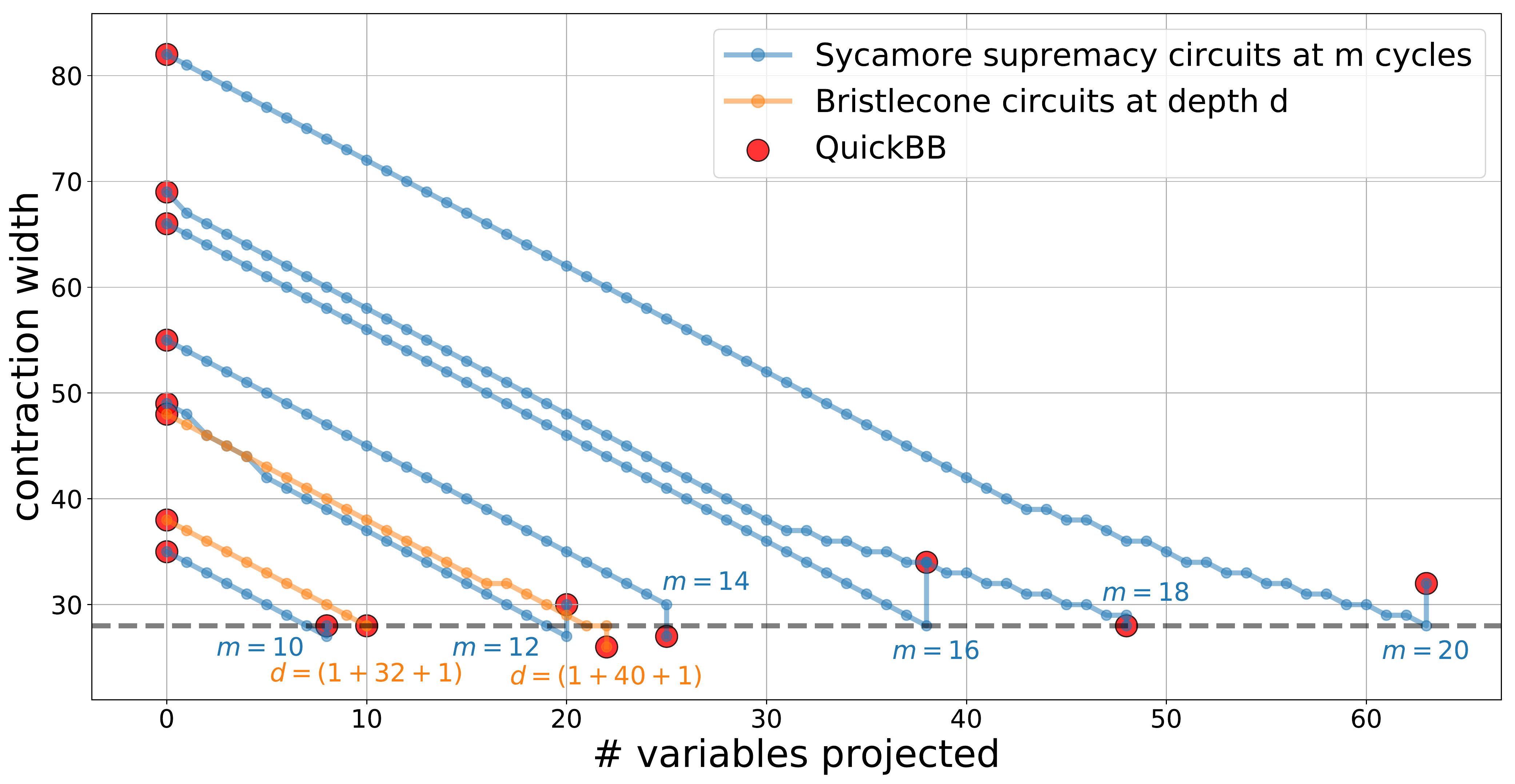}
    \includegraphics[width=1.\columnwidth]{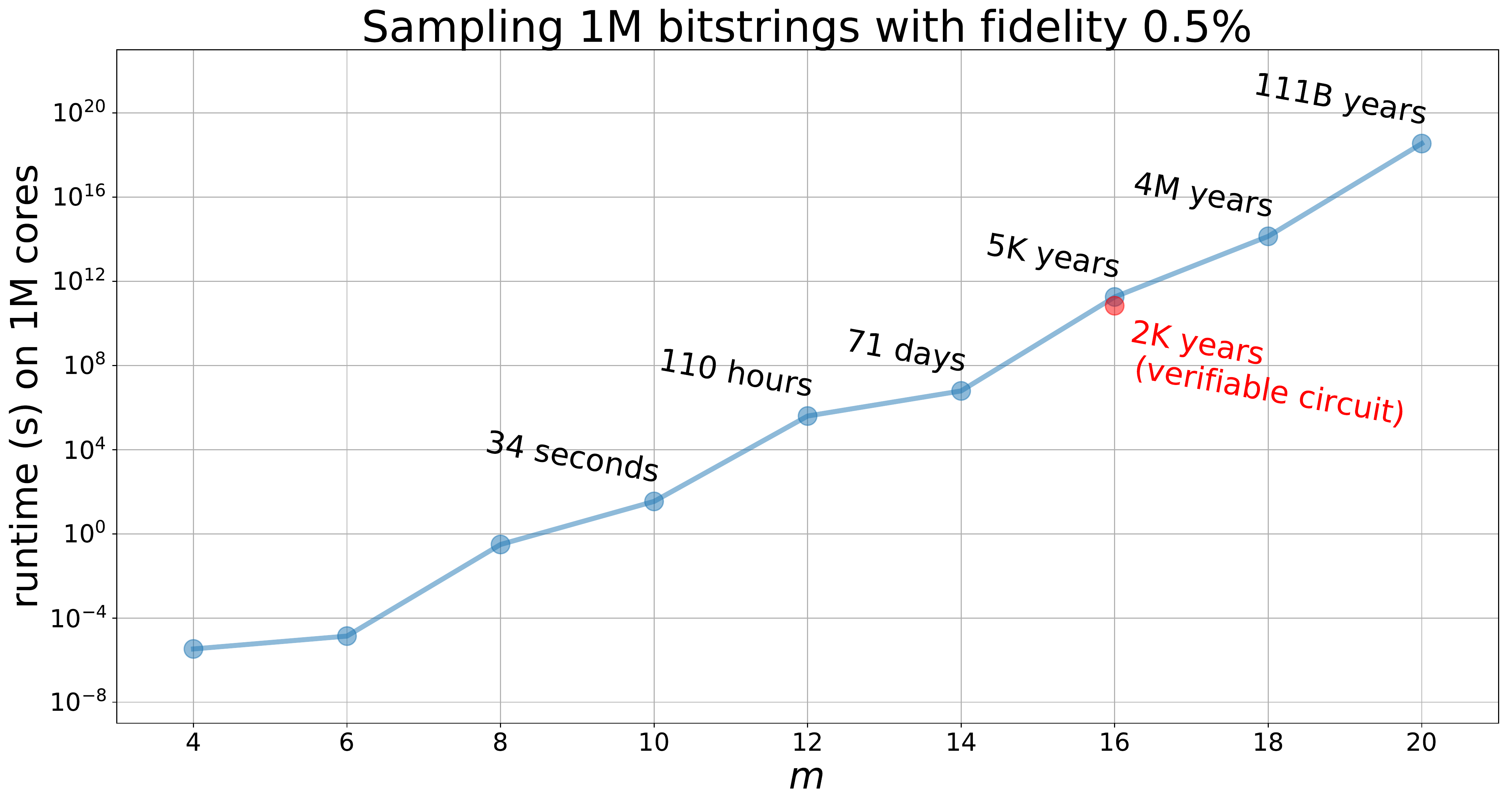}
    \caption{\textbf{Contraction widths and estimated runtimes for classical sampling using the variable elimination algorithm with projected variables of Ref.~\cite{chen2018classical} for Sycamore supremacy circuits.} \emph{Top:} contraction width as a function of the number of variables projected using the algorithm of Ref.~\cite{chen2018classical}. We project enough variables in order to decrease the width to 28 or lower. Note that often the second {\tt QuickBB} run does not decrease the treewidth (and might even increase it), in which case the resulting contraction ordering it is ignored. \emph{Bottom:} estimated runtimes for the classical sampling of 1M bitstrings from the supremacy circuits with fidelity $0.5\%$ using the contraction ordering found by {\tt QuickBB} at the end of the projection procedure shown in the top panel. The red data point shows the estimated runtime for a verifiable circuit; note that the heuristic algorithm analyzed here provides some speedup in this case. Our time estimates assume the use of fast sampling, although it is so far unclear whether this technique can be adapted to the algorithm described here. Failure to do so would result in a slowdown of about an order of magnitude.}
    \label{fig:projection}
\end{figure}

In the top panel of Fig.~\ref{fig:projection} we show the contraction width as a function of the number of variables that are projected for the supremacy circuits used in this paper. In order to decrease the contraction width to 28 or below (a tensor with 28 binary indexes consumes 2 GB of memory using single precision complex numbers), we need to project between 8 and 63 variables, depending on the depth of the circuits. In addition, we report the result of the projection procedure on the Bristlecone circuits considered in Refs.~\cite{villalonga2019flexible,zhang2019alibaba} and available at \url{https://github.com/sboixo/GRCS} for depths (1+32+1) and (1+40+1), since these cases were benchmarked in Ref.~\cite{zhang2019alibaba}. We obtain a contraction width equal to 28 after 10 projections for Bristlecone at depth (1+32+1), and width 26 after 22 projections for Bristlecone at depth (1+40+1), consistent with the results in  Ref.~\cite{zhang2019alibaba}. Even though Ref.~\cite{chen2018classical} uses $S=60$, we run {\tt QuickBB} for 1800 seconds (30 minutes) every time, in order to decrease the contraction width of the Bristlecone simulations to values that match the memory requirements reported in Ref.~\cite{zhang2019alibaba}. Note that Ref.~\cite{zhang2019alibaba} neither reports the value of $S$ used nor the contraction widths found; however, with $S=1800$ we are able to match the scaling of time complexity reported, as is explained below.

To estimate the runtime of the computation of a single amplitude using this algorithm on the circuits presented in this work, we use the following scaling formula:
\begin{align}
    T_{\text{VE}} = C_{\text{VE}}^{-1} \cdot 2^p \cdot (\text{cost after } p \text{ projections}) / n_{\text{cores}}\text{,}
\end{align}
where VE refers to the variable elimination algorithm with projections described in this section, $C_{\text{VE}}$ is a constant factor, $p$ is the number of variables projected, and $n_{\text{cores}}$ is the number of cores used in the computation. The cost of the full contraction of each subgraph is estimated as the sum of $2^\text{rank}$, where the rank refers to the number of variables involved in each individual contraction along the full contraction of the subgraph. We obtain the value of $C_{\text{VE}}$ from the runtimes reported in Ref.~\cite{chen2018classical}, which shows that a single amplitude of Bristlecone at depth (1+32+1) takes 0.43 seconds to compute on 127,512 CPU cores with 10 projected variables, and at depth (1+40+1) it takes 580.7 seconds with 22 projected variables using the same number of cores. We use the benchmark at depth (1+32+1) because it provides the largest value for $C_{\text{VE}}$ (lowest time estimates), which is equal to 52.7 MHz; the benchmark at depth (1+40+1) gives $C_{\text{VE}} = 51.6$ MHz. In order to sample 1M bitstrings from a random circuit with fidelity $0.5\%$, we need to compute 5000 amplitudes. 

We present our estimates for Sycamore supremacy circuits in the bottom panel of Fig.~\ref{fig:projection}. Note that depth (1+40+1) in Refs.~\cite{chen2018classical,zhang2019alibaba} is equivalent to m=20 cycles here because of the denser layout of two-qubit gates. Furthermore, computation times reported previously are for circuit variations less complex than for Sycamore, arising from changes in complexity such as {\rm CZ} vs. {\rm fSim} gates and differing patterns; with this change of gates, depth (1+40+1) in Refs.~\cite{chen2018classical,zhang2019alibaba} is actually equivalent to m=10 cycles here. Finally, note that we present optimistic estimates, since we are assuming that the fast sampling technique discussed in Section~\ref{subsec:fa} is applicable here. To the best of our knowledge, it is not known how to apply this technique for the heuristic variable elimination algorithm discussed here; in the absence of an implementation of this technique, in order to successfully apply rejection sampling we would instead need to compute a few independent amplitudes per sampled bitstring, which would increase the estimated times by about an order of magnitude (see Section~\ref{subsec:fa} and Refs.~\cite{markov_quantum_2018,chen2019quantum} for more details). According to our estimates, sampling from supremacy circuits at $m=16$ and beyond is out of reach for this algorithm. Interestingly, we find some speedup for the simulation of verifiable circuits, as is shown in Fig.~\ref{fig:projection} for $m=16$ (red data point).\\

\begin{figure}
    \centering
    \includegraphics[width=1.\columnwidth]{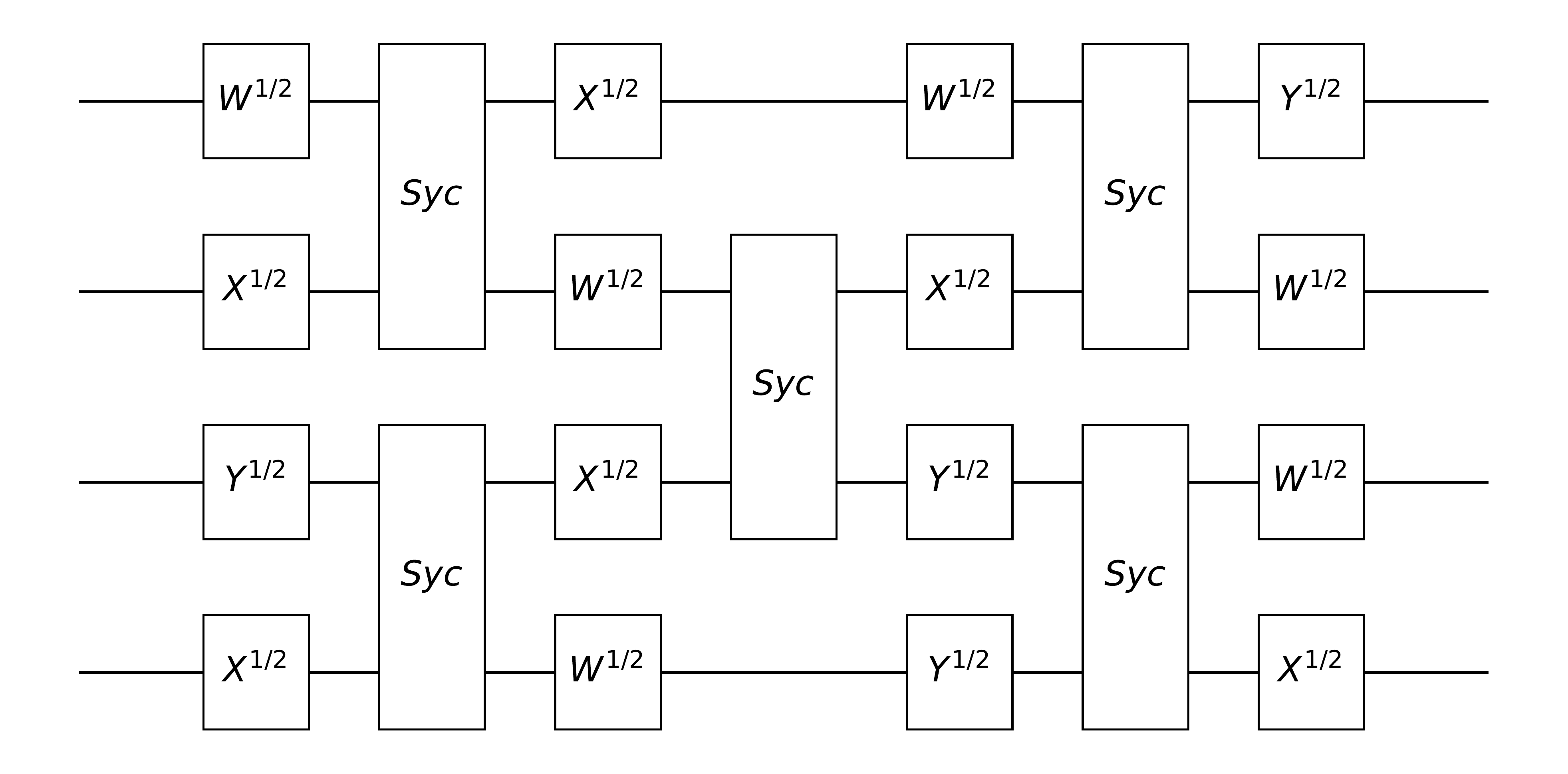}
    \includegraphics[width=0.9\columnwidth]{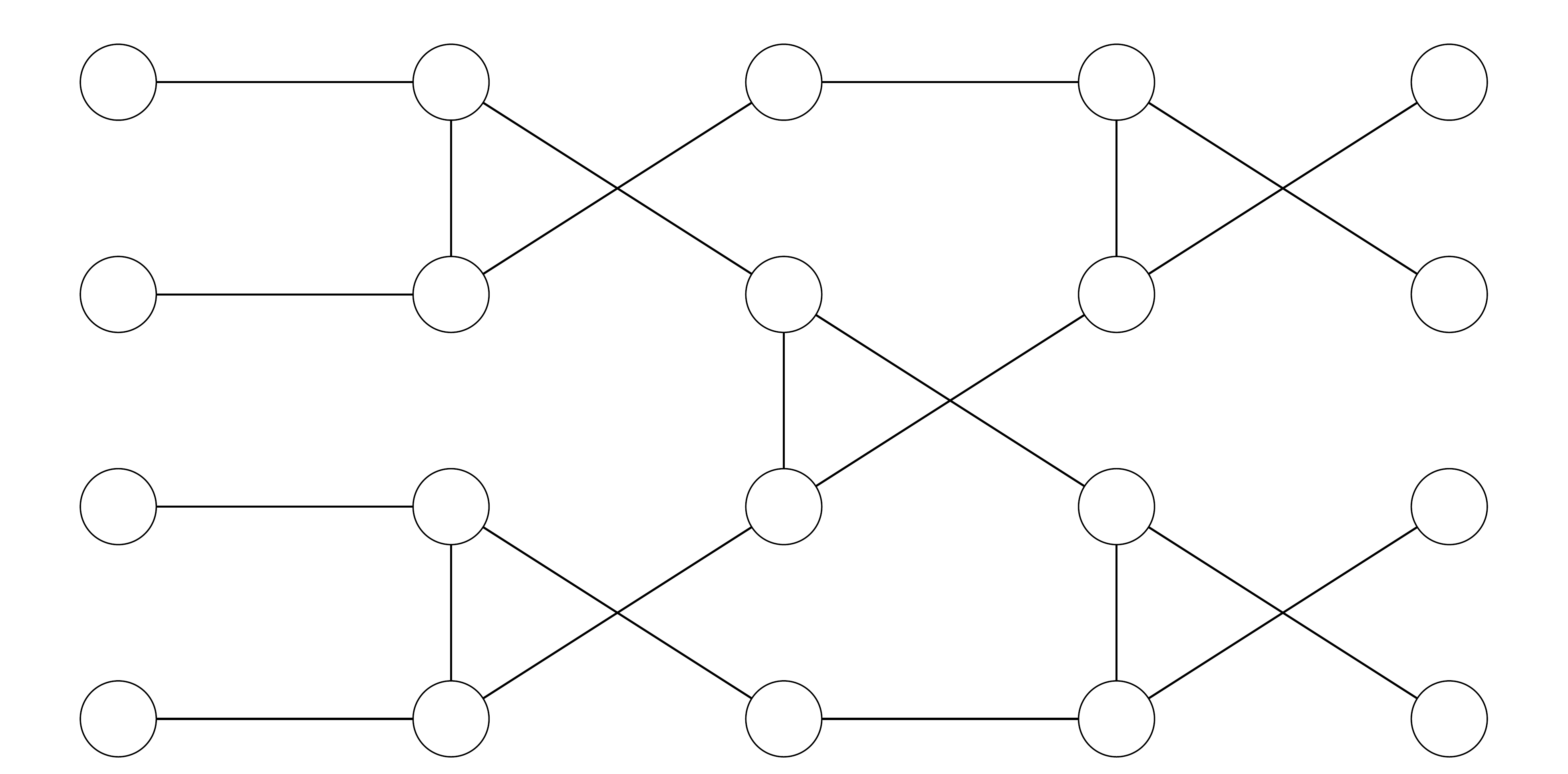}
    \caption{\textbf{Circuit with Sycamore gates (top) and its corresponding undirected graphical model (bottom).} Each non-diagonal single-qubit gate introduces a new vertex or variable. Note that, even though two-qubit gates are generally represented by a clique with four vertices or variables, Sycamore gates can be simplified as a cphase followed by a SWAP. The cphase is represented as an edge between two existing variables. The SWAP, however, provides more complexity to the graph as it swaps the corresponding variables.}
    \label{fig:variable_graph}
\end{figure}

Finally, note that the undirected graphical model derived from the supremacy circuits can take advantage of the structure of the Sycamore gates ({\rm fSim} plus single-qubit ${\rm R_z}$ rotations). Due to the fact that ${\rm fSim(\theta\approx\pi/2, \phi)} \approx -i\cdot[{\rm R_z (-\pi/2) \otimes R_z(-\pi/2)}]\cdot{\rm cphase(\pi+\phi)}\cdot{\rm SWAP}$, the Sycamore gate corresponds to a subgraph of only two variables, which explicitly represents the diagonal {\rm cphase} and the logical {\rm SWAP}. This simplification, used in our estimates, results in an undirected graphical model that is simpler than that one generated by arbitrary two-qubit gates. See Fig.~\ref{fig:variable_graph} for an example.

\subsection{Computational cost estimation for the sampling task}
\label{subsec:cost_estimation}

\begin{figure}
    \centering
    \includegraphics[width=8.5cm]{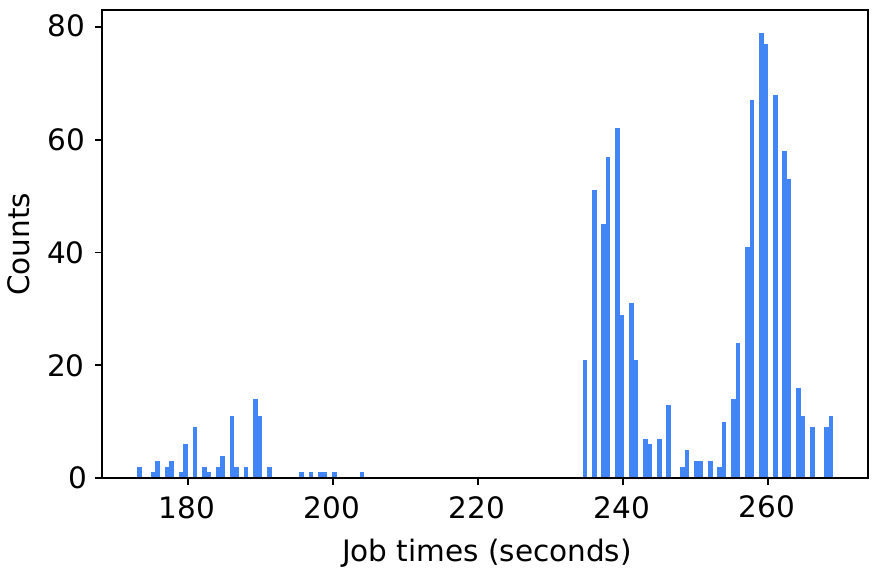}
    \caption{Qsimh execution time for a 53 qubit circuit with 20 cycles for the first 1000 prefix values.  The average job time $\avg{t_\text{prefix}}$ is calculated to be 246 seconds.}
    \label{fig:gcp_simulation_distribution}
\end{figure}
We find that the most efficient simulator for our hardest circuits is the SFA simulator (see Sec. \ref{subsec:sa_sfa}). In order to estimate the computational cost associated with simulating a 53 qubit circuit with 20 cycles, where no gates are elided on the cut, we use a Google cloud cluster composed of 1000 machines with 2 vCPUs and 7.5 GB of RAM each (n1-standard-2).  We use n1-standard-2 because this is the smallest non-custom machine with sufficient RAM for simulating the two halves of the circuit.  In 20 cycles, the circuit contains 35 gates across the cut.  All cross gates have a Schmidt rank of 4 except for the last four gates which can be simplified to $\rm cphase$ with a Schmidt rank of 2.  To obtain a perfect fidelity simulation we would need to simulate all $4^{31} \times 2^{4}$ paths.   We configure qsimh according to Ref.~\cite{markov_quantum_2018} to have a prefix of  30 cross gates, thus requiring $4^{30}$ separate qsimh runs. The first 1000 paths of the required $4^{30}$ were used for timing purposes.  In Figure~\ref{fig:gcp_simulation_distribution} we plot the distribution of simulation times with qsimh consuming two hyperthreads.  The average job time is 246 seconds resulting in a calculated $1.6\times10^{14}$ core hours for a simulation of the circuit with 0.002 fidelity \cite{Note1}. Extrapolated run times for other circuits with 53 qubits are shown in Table \ref{table:qsimhruntimes}. To calculate a total cost for the largest circuit we multiply the Google Cloud preemptible n1-standard-2 price in zone us-central-1 of \$0.02 per hour, 246 seconds average run time, 0.002 target fidelity, and $4^{30}$ qsimh runs.  This results in an estimated cost of 3.1 trillion USD. For perfect fidelity simulations (necessary for XEB), an extrapolation to a fidelity value of 100\% gives a good estimate of the run time. We believe these estimates are a lower bound on costs and simulation time due to the fact that these calculations are likely to compete with each other if they are run on the same nodes.

As a final remark, note that a hypothetical implementation of the decomposition discussed at the end of Section~\ref{subsec:verifiable_and_supremacy} could decrease the computation time presented here by a factor of 16.

\begin{table}
\begin{tabular}{| r | r | r | r | r |}
\hline
qubits, $n$ & cycles, $m$ & total \#paths & fidelity & run time \\
\hline
53 & 12 & $4^{17} 2^4$ & 1.4\% & 2 hours \\
\hline
53 & 14 & $4^{21} 2^4$ & 0.9\% & 2 weeks \\
\hline
53 & 16 & $4^{25} 2^3$ & 0.6\% & 4 years \\
\hline
53 & 18 & $4^{28} 2^3$ & 0.4\% & 175 years \\
\hline
53 & 20 & $4^{31} 2^4$ & 0.2\% & 10000 years \\
\hline
\end{tabular}
\label{table:qsimhruntimes}
\caption{Approximate qsimh run times using one million CPU cores extrapolated from the average simulation run time for 1000 simulation paths on one CPU core.}
\end{table}

\subsection{Understanding the scaling with width and depth of the computational cost of verification}
\label{subsec:scaling}

\subsubsection{Runtime scaling formulas}
\label{subsec:scaling_formulas}

Here we study the scaling of the runtime of the classical computation of exact amplitudes from the output wave function of a circuit with $m$ cycles and $n$ qubits on Sycamore, assuming a supercomputer with 1M cores. This computation is needed in order to perform XEB on the circuits run. We consider two algorithms: a distributed Schr\"odinger algorithm (SA)~\cite{RAED07x,RAED19a} (see Section~\ref{subsec:sa_simulator}) and a hybrid Schr\"odinger-Feynman algorithm (SFA)~\cite{markov_quantum_2018} that splits the circuit in two patches and time evolves each of them for all Feynman paths connecting both patches (see Section~\ref{subsec:sa_sfa}). The latter is embarrassingly parallelizable. Note that these scaling formulas provide rough estimates presented with the intent of building intuition on the scaling of runtimes with the width and depth of the circuits, and that the finite size effects of the circuits can give discrepancies of an order of magnitude or more for the circuit sizes considered in this work.\\

For SA, the runtime is directly proportional to the size of the wave function on $n$ qubits. This is equal to $2^n$. In addition, the runtime is proportional to the number of gates applied, which scales linearly with $n$ and $m$. For this reason, we propose the scaling:
\begin{align}
\label{eq:scaling_sa}
    T_{\text{SA}} = C_{\text{SA}}^{-1} \cdot m n \cdot 2^{n} \text{,}
\end{align}
where the constant $C_{\text{SA}}$ is fit to runtimes observed experimentally when running on a supercomputer, and scaled to 1M cores.
\\

For SFA the runtime is proportional to the number of paths connecting both patches, as well as to the time taken to simulate each pair of patches. When using the \emph{supremacy} two-qubit gate layouts (ABCDCDAB\ldots), each {\rm fSim} gate bridging between the two patches (cross-gates) generates a factor of 4 in the number of paths. The number of cross-gates scales with $\sqrt{n}$ (we assume a two-dimensional grid) and with $m$. The time taken to simulate each patch is proportional to $2^{n/2}$, where $n/2$ estimates the number of qubits per patch, and the exponential dependence comes from a linear scaling of the runtime with the size of the wave function over that patch. The runtime therefore scales as:
\begin{align}
\label{eq:scaling_sfa_supremacy}
    T_{\text{SFA, supremacy}} = C_{\text{SFA}}^{-1} \cdot 2 \cdot 2^{n \over 2} \cdot 4^{B \cdot m \sqrt{n}} \text{,}
\end{align}
where the extra factor of two accounts for the fact that, for every path, two patches have to be simulated. The constant $C_{\text{SFA}}$, with units of frequency, is the effective frequency with which 1M cores simulate paths and is fit from experimentally observed runtime. The constant $B$ accounts for the average number of cross-gates observed per cycle, which depends on the two-dimensional grid considered and on the two-qubit gate layouts used. For Sycamore, with the supremacy layouts, we find 35 cross-gates for $n=53$ and $m=20$, which gives $B=0.24\approx 1/4$.

For SFA, using the \emph{verifiable} two-qubit gate layouts (EFGHEFGH\ldots), the main difference with the supremacy circuits case is the fact that most of the cross-gates can be fused in pairs, forming three-qubit gates we refer to as \emph{wedges} (see Sec.~\ref{sec:wedge_formation} and \ref{subsec:verifiable_and_supremacy}). Each cross-wedge generates only 4 paths, as opposed to the $4^2$ paths the two independent {\rm fSim} gates would have generated. Since every 4 cycles provide 7 cross-gates, and from those 7 gates, 6 are converted into 3 wedges, we count only $4^4$ paths, as opposed to a naive count of $4^7$ for those 4 cycles. In turn, the exponent in the last factor of Eq.~\ref{eq:scaling_sfa_supremacy} is corrected by the fraction $4 \over 7$. This results in:
\begin{align}
\label{eq:scaling_sfa_verifiable}
    T_{\text{SFA, verifiable}} = C_{\text{SFA}}^{-1} \cdot 2 \cdot 2^{n \over 2} \cdot 4^{{4\over 7} B \cdot m \sqrt{n}} \text{.}
\end{align}

\subsubsection{Assumptions and corrections}
\label{subsec:assumptions}

There are several assumptions considered in Section~\ref{subsec:scaling_formulas} and other details that can either (1) contribute to a somewhat large discrepancy between the runtimes predicted by the scaling formulas and the actual runtimes potentially measured experimentally, or (2) be ignored with no significant impact on the accuracy of the predictions. Here we discuss the ones we consider most relevant.\\

Concerning SA, the algorithm is benchmarked in practice on up to 100K cores. Since this is a distributed algorithm, the scaling with number of cores is not ideal and therefore the constant $C_{\text{SA}}$ can only be estimated roughly. We assume perfect scaling in our estimates for runtime on 1M cores, \emph{i.e.}, the runtime on 1M cores is the one on 100K cores divided by 10; this is of course an optimistic estimate, and runtimes should be expected to be larger.

\begin{figure*}[htbp]
    \centering
    \includegraphics[width=7.0in]{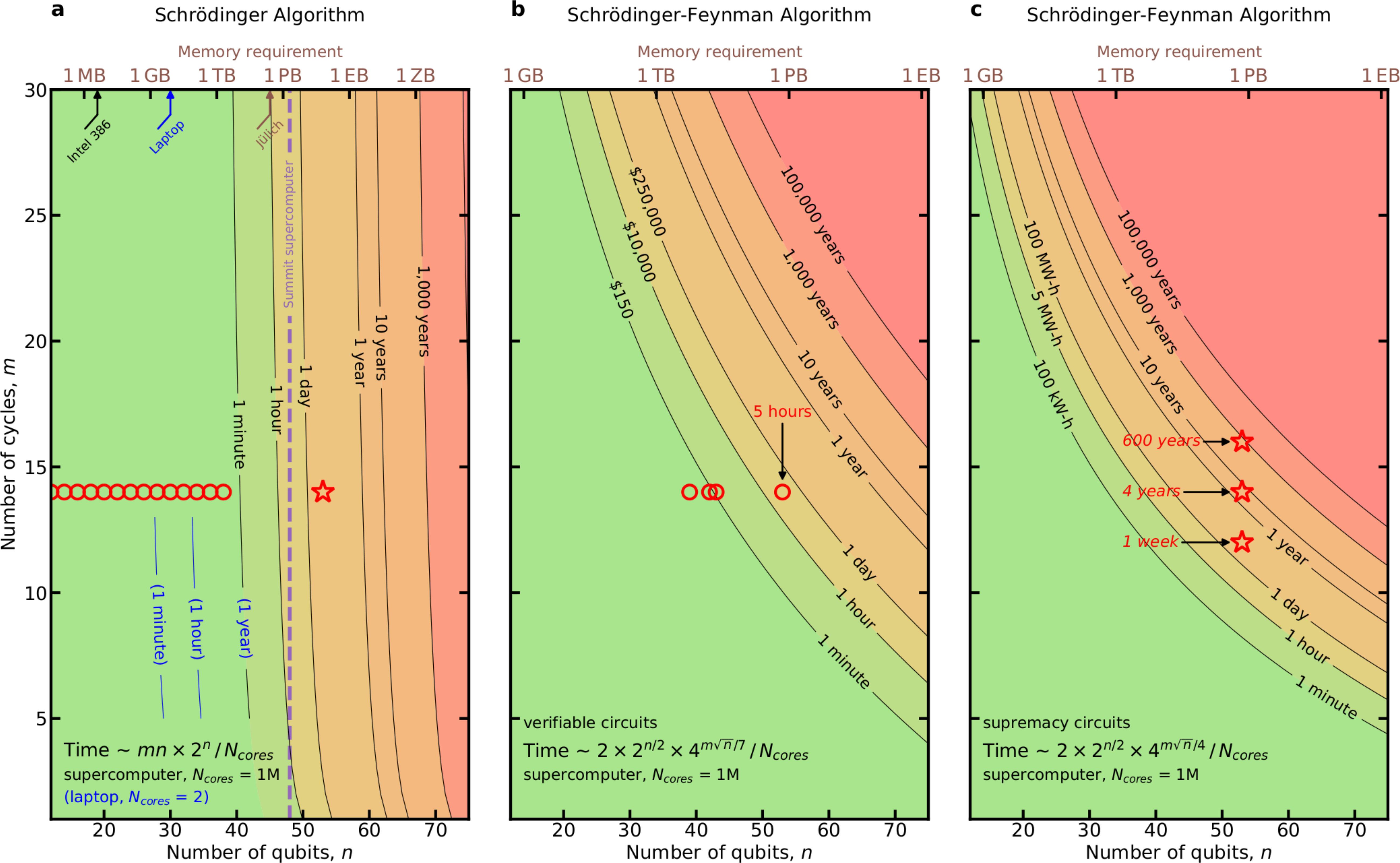}
    \caption{\textbf{Scaling of the computational cost of XEB using SA and SFA.} \textbf{a,} For a Schr\"{o}dinger algorithm, the limitation is RAM size, shown as vertical dashed line for the Summit supercomputer.  Circles indicate full circuits with $n=12$ to 43 qubits that are benchmarked in Fig.~4a of the main paper \cite{arute2019}.  53 qubits would exceed the RAM of any current supercomputer, and is shown as a star.  \textbf{b,} For the hybrid Schr\"{o}dinger-Feynman algorithm, which is more memory efficient, the computation time scales exponentially in depth. XEB on full verifiable circuits was done at depth $m=14$ (circle). \textbf{c,} XEB on full supremacy circuits is out of reach within reasonable time resources for $m=12$, 14, 16 (stars), and beyond. XEB on patch and elided supremacy circuits was done at $m=14$, 16, 18, and 20.}
    \label{fig:scaling}
\end{figure*}

For memory requirement estimates, we assume a 2 byte encoding of complex numbers. Beyond about 49 qubits there is not enough RAM on any existing supercomputer to store the wave function. In those cases, runtimes are given for the unrealistic, hypothetical case that one \emph{can} store the wave function.\\

SFA is embarrassingly parallelizable, and so it does not suffer from non-ideal scaling. However, there are other factors to take into account. First, we have written no explicit dependence of the time to simulate patches of the circuit with $m$; the number of cycles $m$ only plays a role when counting the number of paths to be considered. SFA stores several copies of the state of a patch after its evolution at different depths, iterating over paths over several nested loops. For this reason, most of the time is spent iterating over the inner-most loop, which accounts for the last few gates of the circuit and is similar in cost for all depths. This implies that the amortized time per path is considered approximately equal for all depths and the direct $m$ dependence was correctly ignored.

A factor contributing to the discrepancy between the predicted runtimes of the scaling formulas of Section~\ref{subsec:scaling_formulas} and those expected in practice is due to finite size effects. While these scaling formulas consider the average number of cross-gates encountered per cycle, different cycles have layouts that contribute a few more (or less) gates than others. Since the runtime dependency is exponential in the number of gates, this might cause discrepancies of around an order of magnitude. Furthermore, for verifiable circuits, wedges form over groups of two cycles; this coarse graining exacerbates finite size effects. For the sake of simplicity in the scaling formulas, we do not perform any corrections to include these factors. However, in order to mitigate the propagation of finite size effect errors, we consider different constants $C_{\text{SFA, supremacy}}$ and $C_{\text{SFA, verifiable}}$, that we fit independently.

Finally, we refer to runtimes of our simulations on a hypothetical supercomputer with 1M cores. While this is a realistic size for a Top-5 supercomputer currently, a core-hour can vary significantly between different CPU types. Again, we only intend to provide rough estimates in order to build intuition on the dependence of runtimes with circuit width and depth.

\subsubsection{Fitting constants}
\label{subsec:fitting}

In the case of SA, we fit the constant $C_{\text{SA}}$ with a runtime of 0.1 hours for the simulation with $n=43$ and $m=14$. This runtime is obtained by assuming ideal scaling when extrapolating a runtime of 1 hour on nearly 100K nodes ($2^{15}$ MPI processes, 3 cores per process), as reported in Sec.~\ref{subsec:sa_simulator}. This gives a value of
\begin{align}
\label{eq:fit_sa}
    C_{\text{SA}} = 0.015 \times 10^{6} \text{ GHz.}
\end{align}
\\

For SFA, we consider $B=1/4$ for simplicity. In order to fit $C_{\text{SFA}}$, we consider a runtime of 5 hours and 4 years for the case with $n=53$ and $m=14$ for verifiable and supremacy circuits, respectively (see Fig.~4 of the main text). This gives:
\begin{align}
\label{eq:fit_sfa}
    C_{\text{SFA, verifiable}} = 0.0062 \times 10^{6} \text{ GHz} \nonumber \\
    C_{\text{SFA, supremacy}} = 3.3 \times 10^{6} \text{ GHz.}
\end{align}
As discussed above, these fits provide times estimated for a supercomputer with 1M cores. Contour plots showing the dependency of runtime with $n$ and $m$ are presented in Fig.~\ref{fig:scaling}.

\subsubsection{Memory usage scaling}
\label{subsec:memory}

Let us conclude with a discussion of the memory footprint of both algorithms. For these estimates, we assume a 2-byte encoding of complex numbers, as opposed to 8 bytes (single precision) or 16 bytes (double precision). This results in a lower bound for the memory usage of these two algorithms. These estimates need an extra factor of 4 (8) when using single (double) precision. SA stores the wave function of the state on all qubits. For this reason, it needs $2^n \times 2 = 2^{n+1}$ bytes. SFA simulates the wave function of both halves of the system ($n/2$ qubits) per path, one at a time. This requires $2^{n\over 2} \cdot 2$ bytes per path. In practice, the use of checkpoints implies the need to store more than one wave function per path; for simplicity, and in the same optimistic spirit of other assumptions, we ignore this fact. If 1M cores are used and each path is simulated using a single core, the total memory footprint is estimated to be $10^{6} \times 2^{{n\over 2} + 1}$ bytes. State-of-the-art supercomputers have less than 3~PB of memory.

\subsection{Energy advantage for quantum computing}
\label{sec:energy_advantage}

With the end of Dennard scaling for CMOS circuits, gains in computing energy efficiency have slowed significantly \cite{koomey2016}.  As a result, today's high performance computing centers are usually constrained by available energy supplies rather than hardware costs. For example, the Summit supercomputer at Oak Ridge National Laboratory has a total power capacity of 14 MW available to achieve a design specification of 200 Pflop/s double-precision performance.  We took detailed energy measurements with qFlex running on Summit. The energy consumption grows exponentially with the circuit depth, as illustrated in Table~\ref{table:qflex_runs}.

For a superconducting quantum computer, the two primary sources of energy consumption are: 
\begin{enumerate}
\item \textbf{A dilution refrigerator:} our refrigerator has a direct power consumption of $\sim$10~kW, dominated by the mechanical compressor driving the 3~K cooling stage.  The power required to provide chilled water cooling for the compressor and pumps associated with the refrigerator can be an additional 10~kW or more.
\item \textbf{Supporting electronics:} these include microwave electronics, ADCs, DACs, clocks, classical computers, and oscilloscopes that are directly associated with a quantum processor in the refrigerator.  The average power consumption of supporting electronics was nearly 3~kW for the experiments in this paper.
\end{enumerate}

We estimate the total average power consumption of our apparatus under worst-case conditions for chilled water production to be 26~kW.  This power does not change appreciably between idle and running states of the quantum processor, and it is also independent of the circuit depth.  This means that the energy consumed during the 200~s required to acquire 1M samples in our experiment is $\sim5\times10^6$~J ($\sim1$~kWh).  As compared to the qFlex classical simulation on Summit, we require roughly 7 orders of magnitude less energy to perform the same computation (see Table~\ref{table:qflex_runs}).  Furthermore, the data acquisition time is currently dominated by control hardware communications, leading to a quantum processor duty cycle as low as 2\%.  This means there is significant potential to increase our energy efficiency further.

\section{Complexity-theoretic foundation of the experiment}\label{sec:complexity}

The notion of quantum supremacy was originally introduced by John Preskill~\cite{preskill_2012}. He conceived of it as ``the day when well controlled quantum systems can perform tasks surpassing what can be done in the classical world''. For the purpose of an experimental demonstration we would like to refine the definition.
\\
\newline
\noindent Demonstrating quantum supremacy requires:
\begin{enumerate}
\item A well defined computational task, i.e.~a mathematical specification of a computational problem with a well defined solution. 

Comment: This requirement, standard in computer science, excludes tasks such as ``simulate a glass of water''. However, it would include finding the ground state energy of an H$_{2}$O molecule to a given precision governed by a specific Hamiltonian. Note that a mathematical specification of a computational problem calls for highly accurate control resulting in measurable system fidelity. 

\item Programmable computational device

Comment: Many physics experiments estimate the values of observables to a precision which can not be obtained numerically. But those do not involve a freely programmable computational device and the computational task is often not well defined as required above. Ideally, we would even restrict ourselves to devices that are computationally universal. However, this would exclude proposals to demonstrate quantum supremacy with BosonSampling~\cite{AA11} or IQP circuits~\cite{BM16}.

\item A scaling runtime difference between the quantum and classical computational processes that can be made large enough as a function of problem size so that it becomes impractical for a supercomputer to solve the task using any known classical algorithm. 

Comment: What is impractical for classical computers today may become tractable in ten years. So the quantum supremacy frontier will be moving towards larger and larger problems. But if a task is chosen such that the scaling for the quantum processors is polynomial while for the classical computer it is exponential then this shift will be small. Establishing an exponential separation requires substantial efforts designing and benchmarking classical algorithms~\cite{RAED07x,smelyanskiy2016qhipster,boixo2018characterizing,Boi17,villalonga2019flexible,RAED19a,villalonga2019establishing,chen2018classical,MS08,zhang2019alibaba}, and support from complexity theory arguments~\cite{boixo2018characterizing,aaronson2017complexity,Bou19}. Sampling the output of random quantum circuits is likely to exhibit this scaling separation as a function of the number of qubits for large enough depth. In this context, we note that quantum analog simulations that estimate an observable in the thermodynamic limit typically do not define a problem size parameter. 
\end{enumerate}
The requirements above are satisfied by proposals of quantum supremacy emerging from computer science, such as BosonSampling~\cite{AA11}, IQP circuits~\cite{BM16}, and random circuit sampling~\cite{boixo2018characterizing,neill2018blueprint,aaronson2017complexity,Bou19,movassagh2019cayley}. They are also implicit in the “Extended Church-Turing Thesis”: any “reasonable” model of computation can be efficiently simulated, as a function of problem size, by a Turing machine. 

We note that formal complexity proofs are asymptotic, and therefore assume an arbitrarily large number of qubits. This is only possible with a fault tolerant quantum computer and therefore near term practical demonstrations of quantum supremacy must rely on a careful comparison with highly optimized classical algorithms on state-of-the-art supercomputers. 

So far we have argued for quantum supremacy by comparing the running time of the quantum experiment with the time required for the same task using the best known classical algorithms, running on the most powerful supercomputers currently available. The fastest known algorithm for exact sampling (or for computing transition probabilities) runs in time exponential in the treewidth of the quantum circuit \cite{MS08, Boi17}; for a depth $D$ circuit on a rectangular lattice of sizes $l_x$ and  $l_y$, the treewidth is given by $\min(\min(l_x, l_y)D, l_x l_y)$. For approximate simulation in which one only requires a given global fidelity $F$, the classical cost is reduced linearly in $F$ \cite{markov_quantum_2018}. As classical algorithms and compute power can be improved in the future, the classical cost benchmark is a moving target.  

A complementary approach to back up supremacy claims consists of giving complexity-theoretic arguments for the classical hardness of the problem solved (in our case sampling from the output distribution of a random circuit of a given number of qubits, depth and output fidelity). Previous work gave hardness results for sampling \textit{exactly} from the output distribution of different classes of circuits \cite{AA11, BJS10, HM17, boixo2018characterizing, TD02}. Most relevant to us are Refs. \cite{Bou19, HM18,movassagh2019cayley}, which proved that it is classically intractable (unless the polynomial hierarchy collapses to its third level, which is considered extremely unlikely \cite{AB09}) to sample from the exact probability distribution of outcomes of measurements in random circuits. We note the distribution of circuits considered in \cite{Bou19, HM18,movassagh2019cayley} is different from ours. 

An important clarification is that such results are asymptotic, i.e. they show that, unless the polynomial hierarchy collapses, there are no polynomial-time classical algorithms for sampling from output measurements of certain quantum circuits. But they cannot be used directly to give concrete lower bounds for quantum computations of a fixed number of qubits and depth. Refs. \cite{CNS18, Dal18, MS19} tackled this question using tools from fine-grained complexity, giving several finite size bounds. 

There are also results arguing for the hardness of \textit{approximate} sampling (see e.g. \cite{AA11, BM16, boixo2018characterizing, HM17}), where the task is only to sample from a distribution which is close to the ideal one. As the quantum experiment will never be perfect, this is an important consideration. However those results are weaker than the ones for exact sampling, as the hardness assumptions required have been much less studied (and in fact were introduced with the exact purpose of arguing for quantum supremacy). Another drawback is that the results only apply to the situation where the samples come from a distribution very close to the ideal one (i.e. with high fidelity with the ideal one). This is not the regime in which our experiment operates.

With these challenges in mind, we consider an alternative hardness argument in this section, which will allow us to lower bound the classical simulation cost of noisy quantum circuits by the cost of the ideal one. On one hand, our argument will be more restrictive than previous results in that we will assume a particular noise model for the quantum computer (one, however, which models well the experiment). On the other hand, it will be stronger in two ways: (1) it will apply even to the setting in which the output fidelity of the experimental state with the ideal one can be very small, but still the product of total fidelity with exact computational cost is large; and (2) it will be based on more mainstream complexity assumptions in contrast to the tailor-made conjectures required in e.g. \cite{AA11, BM16, HM17} to handle the case of small adversarial noise. 

\subsection{Error model}

Our error model is the following. We assume that the quantum computer samples from the following output distribution:
\begin{equation} \label{rdistrib}
r_{U, F}(x) := F |\bra{x}U\ket{0}|^2 + (1 - F)/2^n,
\end{equation}
with $U$ the circuit implemented. In words, we assume global depolarizing noise. Ref. \cite{boixo2018characterizing} argues that Eq.~(\ref{rdistrib}) is a good approximation for the output state of random circuits (see Sec. \ref{sec:xeb_theory} and Section III of \cite{boixo2018characterizing}); this form has also been verified experimentally on a small number of qubits. In the experiment, $F$ is in the range $10^{-2}-10^{-3}$. 

We note that while we assume a global white noise model in this section, we do not assume it in the rest of the paper, neither for validating the cross entropy test nor in the comparison with state-of-the-art classical algorithms (and indeed the algorithm considered in Section \ref{sec:classical_sim} samples from an approximate distribution different from the one in Eq.~(\ref{rdistrib})). 

\subsection{Definition of computational problem}

Before stating our result, let us define precisely the computational problem we consider. We start with the ideal version of the problem with no noise:

\vspace{0.2 cm}

\noindent \textbf{Circuit Sampling}: The input is a description of a $n$ qubit quantum circuit $U$, described by a sequence of one- and two-qubit gates. The task of the problem is to sample from the probability distribution of outcomes $p_U(x) := |\langle x | U | 0 \rangle|^2$.

\vspace{0.2 cm}

Circuit sampling is an example of a \textit{sampling problem} \cite{aaronson2014equivalence}. A classical algorithm for circuit sampling can be thought of, without loss of generality, as a function $A$ mapping $m \in \text{poly(n)}$ bits $r = (r_1, \ldots r_m)$ to $n$ bits such that
\begin{equation} \label{samplingfunccyion}
\frac{1}{2^m} | \{ (r_1, \ldots, r_m) \hspace{0.1 cm} \text{s.t.}   \hspace{0.1 cm} A(r_1, \ldots, r_m) = x \}  | = \tilde{p}_U(x), 
\end{equation}
with $\tilde{p}(x)$ an approximation of $p_U(x)$ to $l \in \text{poly(n)}$ bits of precision. So when $r$ is chosen uniformly at random, the output of $A$ are samples from $p$ (up to rounding errors which can be made super-exponentially small).

Assuming the polynomial hierarchy does not collapse, it is known that Circuit Sampling cannot be solved classically efficiently in $n$, meaning any algorithm $A$ satisfying Eq.~(\ref{samplingfunccyion}) must have superpolynomial circuit complexity, for several classes of circuits (such as short depth circuits \cite{TD02}, IQP \cite{BJS10} and Boson Sampling \cite{AA11}). We might also be interested in the average case of circuit sampling (for a restricted class of circuits).

\vspace{0.2 cm}

\noindent \textbf{Random Circuit Sampling}: The input is a set of quantum circuits ${\cal U}$ on $n$ qubits. The task is to sample from $p_U(x) := |\langle x | U | 0 \rangle|^2$ for most circuits $U \in {\cal U}$. 

\vspace{0.2 cm}

Ref. \cite{Bou19} proved that an efficient (in terms of $n$) classical algorithm for this task for random circuits would also collapse the polynomial hierarchy. As every realistic quantum experiment will be somewhat noisy, it is relevant to consider a variant of this task allowing for small deviations from ideal. One possible formulation is the following:

\vspace{0.2 cm}

\noindent \textbf{$\varepsilon$-Approximate Random Circuit Sampling}: The input is a set of quantum circuits ${\cal U}$ on $n$ qubits. The task is to sample for most circuits $U \in {\cal U}$, from any distribution $q_U$ s.t. $d_{\text{VD}}(q_U, p_{U})\leq \varepsilon$, where $d_{\text{VD}}(p, q)$ is the variational-distance between the distributions $p, q$ \cite{Note2} and  $p_U(x) := |\langle x | U | 0 \rangle|^2$.
\vspace{0.2 cm}

Refs. \cite{AA11, BM16, boixo2018characterizing} put forward new complexity-theoretic assumptions about the $\#\text{P}$-hardness of certain problems and proved they imply that several restricted classes of circuits are hard to approximately sample for $\varepsilon$ sufficiently close to zero. However, we cannot use these results here as the $\varepsilon$ we achieve is far from zero. We will resort to the following different variant of approximate circuit sampling.

\vspace{0.2 cm}

\noindent \textbf{Unbiased-Noise $F$-Approximate Random Circuit Sampling}: The input is a set of quantum circuits ${\cal U}$ on $n$ qubits. The task is to sample from the distribution $r_{U, F}$ given by Eq.~(\ref{rdistrib}), for most circuits $U \in {\cal U}$.  

\vspace{0.2 cm}

We note that there are alternatives for defining the computational problem for which supremacy is achieved without having to use sampling problems. These have the advantage that it is possible to verify, for each problem instance, that the task was achieved (whereas while it is in principle possible to verify that one is sampling from the correct distribution by estimating the frequencies of outcomes, this is unfeasible in practice for high entropy distributions with $>2^{50}$ outcomes as the one we consider here). 

One such problem (considered on Refs. \cite{aaronson2017complexity, boixo2018characterizing}) is the following:

\vspace{0.2 cm}

\noindent \textbf{$b$-Heavy Output Generation}: Given as input a number $b > 1$ and a random circuit $U$ on $n$ qubits (drawn at random from a set of circuits ${\cal U}$), generate output strings $x_1, \ldots, x_k$ s.t. \begin{equation}
\frac{1}{k} \sum_{j=1}^k |\langle x_j | U | 0 \rangle|^2 \geq \frac{b}{2^n}     
\end{equation}

\vspace{0.2 cm}

Ref. \cite{aaronson2017complexity} argues for the hardness of this task for every $b > 1$, although here again one has to resort to rather bold complexity-theoretic conjectures. Cross entropy benchmarking allows us to estimate $b$ for a reasonable value of $k$ (though the classical time needed to compute $|\langle x_j | U | 0 \rangle|^2$ still grows very fast), see Sec. \ref{sec:xeb_theory}. In terms of known algorithms, the complexity of solving Heavy Output Generation is equivalent to the complexity of sampling $k$ samples from a noisy distribution corresponding to the same $b$ value. 

The experiment we report in this paper can be interpreted as showing quantum supremacy in solving the $b$-Heavy Output Generation with $b = 1 + F$ and $F$ the fidelity of the output quantum state.

\subsection{Computational hardness of unbiased-noise sampling}

To state our result, we use the complexity class Arthur-Merlin, which is a variant of the class NP and is denoted by AM$[T]$. It is defined as the class of problems for which there is an Arthur-Merlin one-round protocol of the following form: given an instance of a problem in $AM[T]$ (which Arthur would like to decide if it is a YES or NO instance), Arthur first sends random bits to Merlin. Merlin (which is computationally unbounded) then sends back a proof to Arthur. Finally Arthur uses the proof and decides in time $T$ if he accepts. In the YES case, Arthur accepts with probability larger than 2/3. In the NO case, he accepts with probability no larger than 1/3.

\begin{theorem} \label{globaldepolarizing}
Assume there is a classical algorithm running in time $T$ and using $m$ bits of randomness that samples from the distribution $r_{U, F}(x)$ given by Eq.~(\ref{rdistrib}), for a given quantum circuit $U$ on $n$ qubits and $F \geq 0$. Then for every integer $L$, there is an $AM[LT + 2Lm]$ protocol for deciding, given $\lambda > 0$, whether
\begin{equation}
 |\bra{0}U\ket{0}|^2 \geq  \lambda \left(1 + \frac{2}{L}  \right)  +  \frac{2(1-F)}{F L2^n}  
\end{equation}
or 
\begin{equation}
|\bra{0}U\ket{0}|^2 \leq \lambda \left(1 - \frac{2}{L}  \right)  -  \frac{2(1 - F)}{F L2^n}  
\end{equation}
\end{theorem}

Before giving the proof, let us discuss the significance of the result. We are interested in the theorem mostly when $L = c/F$ with $c$ a small constant (say 10). Noting that for a random circuit, with high probability, $|\bra{0}U\ket{0}|^2 \geq  2^{-n}/5$ \cite{HM18}, the theorem states that if we can sample classically in time $T$ from the distribution given in Eq.~(\ref{rdistrib}), then we can calculate a good estimate for $|\bra{0}U\ket{0}|^2$ in time $10T/F$ (with the help from an all-powerful but untrustworthy Merlin). It is unlikely that Merlin can be of any help for this task for random circuits, as estimating $|\bra{0}U\ket{0}|^2$ for random circuits is a $\#\text{P}$-hard problem \cite{Bou19}, and it is believed $\#\text{P}$ is vastly more complex than $AM$ (which is contained on the third level of the polynomial hierarchy \cite{AB09}). Therefore we conclude that global white noise leads to no more than a linear decrease in fidelity in classical simulation time (which is in fact optimal as it is achieved by the method presented in Ref. \cite{markov_quantum_2018}).

Ref. \cite{Aa19} proposed a similar, but more demanding, conjecture about the non-existence of certain AM protocols for estimating transition probabilities of random circuits. This conjecture was applied to show that the output bits of our supremacy experiment can be used to produce certifiable random bits. 

We note Theorem \ref{globaldepolarizing} does not establish a lower bound on the classical computation cost of calculating a transition amplitude with additive error $\delta/{2^{n}}$, for small constant $\delta > 0$. What it does is to show that the sampling problem with unbiased noise is as hard as this task, up to a linear reduction in $F$ in complexity. 

Concerning the hardness of computing $|\langle 0| U | 0 \rangle|^2$ it is known that this problem is $\#\text{P}$ hard for random circuits to additive error $2^{-poly(n)}$ \cite{Bou19}. This implies that there is no subexponential-time algorithms for this task (unless $\#\text{P}$ collapses to P). For finite size bounds, which are more relevant to our experiment, the result of Ref. \cite{CNS18} is the most relevant. It shows that under the Strong Exponential Time Hypothesis (SETH) \cite{calabro2009complexity}, there are quantum circuits on $n$ qubits which require $2^{(1 - o(1))n}$ time for estimating $|\langle 0 | U |0 \rangle|^2$ to additive error $2^{-(n+1)}$ \cite{Note3}. Together with Theorem \ref{globaldepolarizing}, we find there is a quantum circuit $U$ on $n$ qubits for which the distribution $r_{U, F}$ (given by Eq.~(\ref{rdistrib})) cannot be sampled in time $F2^{(1-o(1)n)}$, unless SETH is false. 

It is an open question to show a similar lower bound to the one proved in Ref. \cite{CNS18} for estimating the transition probability of random circuits. Even more relevant for this work, it would be interesting to study if one can show a lower bound of the form $2^{(1 - o(1))\text{treewidth}}$ for a random quantum circuit, under a suitable complexity-theoretic assumption, as the depth of the construction in  \cite{CNS18} is relatively high.  

\subsection{Proof of Theorem \ref{globaldepolarizing}}

The proof will follow along similar lines to previous work \cite{AA11, BM16, HM17}. We will use approximate counting (which can be done in AM) to show that a sampling algorithm for $r_{U, F}$ running in time $T$ implies an AM protocol to
compute $r_{U, F}(0)(1 \pm 1/L)$, with classical verification of order $LT$. Since the noise is unbiased, i.e. $r_{U, F}(0) = F\langle 0 | U | 0 \rangle |^2 + (1-F)/2^n$, we can subtract it and find an AM protocol for estimating $|\langle 0 | U | 0 \rangle |^2$ as stated in the theorem.      

In more detail, suppose there is a classical algorithm for sampling from $r_{U, F}$ given by a function $A$ mapping $m \in \text{poly(n)}$ bits $r = (r_1, \ldots r_m)$ to $n$ bits such that
\begin{multline}  \label{algorithmsamplingr}
\frac{1}{2^m} | \{ (r_1, \ldots, r_m) \hspace{0.1 cm} \text{s.t.}   \hspace{0.1 cm} A(r_1, \ldots, r_m) = x \}  | \\ = r_{U, F}(x). 
\end{multline}

Let $a(r_1, \ldots, r_m)$ be a function which is 1 if $A(r_1, \ldots, r_m) = 0^n$ and zero otherwise. 

We start with the following lemma, showing the existence of $A$ implies an $AM[LT + 2Lm]$ protocol for estimating $r_{U, F}(0)$:

\begin{lemma} \label{AMprotcolforr}
Assume there is an algorithm $A$ given by Eq.~(\ref{algorithmsamplingr}). Then for every $\theta$ and $L$ there is an $AM[LT + 2Lm]$ protocol which determines if (i) $r_{U, F}(0) \geq \theta (1 + 2/L)$ (YES instance) or (ii) $r_{U, F}(0) \leq \theta (1 - 2/L)$ (NO instance).  
\end{lemma}

\textit{Proof:} The protocol is the following:

\begin{enumerate} 

\item For every $t \in [Lm]$, Arthur chooses a function at random $h_{t} \in H_{Lm, t}$ from a family $H_{Lm, t}$ of 2-universal linear hash functions from $\{0, 1\}^{L m}$ to  $\{0, 1 \}^t$ \cite{AB09}. Then he communicates his choice of $(h_{1}, \ldots, h_{Lm})$ to Merlin.

\item Merlin sends an $Lm$-bitstring $w$ to Arthur and an integer $s \in [Lm]$ .

\item Arthur verifies that $h_{s}(w) = 0$ and
\begin{align}
a(w_{1, 1}, \ldots w_{1, m}) \wedge \ldots \wedge a(w_{L, 1}, \ldots w_{L, m}) \nonumber = 0.
\end{align}
He rejects if any of the three equations is not satisfied. Then he checks if $\theta \leq 2^{-m} 20^{1/L} 2^{s/L} (1+2/L)^{-1}$, accepting if it is the case and rejecting otherwise.
\end{enumerate}

The cost to compute $a(w_{1, 1}, \ldots w_{1, m})$ is $T$, and the cost to compute is $h_{s}(w)$ is less than $2Lm$, so the total verification time of the AM protocol is $LT + 2Lm$.

Let us analyze the completeness and soundness of the protocol.

\vspace{0.3 cm}

\noindent \textit{Completeness}: Suppose we have a YES instance, $r_{U, F}(0) \geq \theta (1 + 2/L)$. Let us show that Merlin can send $w$ and $s$ which makes  Arthur accept with high probability. 

Let $M$ be the number of solutions of $a(r_1, \ldots, r_m) = 0$ (i.e. $M = 2^m r_{U, F}(0)$). Then $a(r_{1, 1}, \ldots r_{1, m}) \wedge \ldots \wedge a(r_{L, 1}, \ldots r_{L, m})$ has $M^L$ solutions, $M$ for each copy of the function $a$. As part of the proof Merlin sends $s$ satisfying $20 \geq M^L/2^s \geq 10$ (such a value always exists as $s$ can be an arbitrary integer less than or equal to $Lm$).

Let us apply Lemma \ref{hashfunction} (stated below) with $q = Lm$, $t = s$, $\delta = 1/2$, and $S$ the set of solutions, so $|S| = M^L$. Then indeed $|S|/2^s > 10 > 1/\delta^3$. Therefore, with high probability, the number of solutions of
\begin{equation}
a(x_{1, 1}, \ldots x_{1, m}) \wedge \ldots \wedge a(x_{L, 1}, \ldots x_{L, m})  \wedge h_s(x) 
\end{equation}
is in the interval $[(1/2) M^L/2^s, 2  M^L/2^s]$. Since $(1/2) M^L/2^s \geq 1$, there is a string $w$ s.t. $a(w_{1, 1}, \ldots w_{1, m}) \wedge \ldots \wedge a(w_{L, 1}, \ldots w_{L, m})  \wedge h_s(w) = 0$, which Merlin also sends to Arthur as part of the proof.

Since $M = 2^m r_{U, F}(0) \geq 2^m \theta (1 + 2/L)$ and $M^L/2^s \leq 20$,
\begin{equation}
20 \geq \frac{M^L}{2^s} \geq \frac{2^{Lm}}{2^s} \theta^L \left( 1 + \frac{2}{L} \right)^L,  
\end{equation}
so indeed $\theta \leq 2^{-m} 20^{1/L} 2^{s/L} (1+2/L)^{-1}$ and Arthur will accept with high probability. 

\vspace{0.3 cm}

\noindent \textit{Soundness}:   Suppose we have a NO instance, $r_{U, F}(0) \leq \theta (1 - 2/L)$. Let us show that no matter which witnesses $w, s$ Merlin sends, Arthur will only accept with a small probability. Merlin must send $s$ such that 
\begin{equation}
\theta^L \leq (20) 2^{-Lm}  2^{s} (1+2/L)^{-L},
\end{equation}
otherwise Arthur rejects. By Lemma \ref{hashfunction} (stated below), the number of solutions of
\begin{equation} \label{hashingeqs}
a(x_{1, 1}, \ldots x_{1, m}) \wedge \ldots \wedge a(x_{L, 1}, \ldots x_{L, m})  \wedge h_s(x) 
\end{equation}
will be in the interval $[(1/2) M^L/2^s, 2  M^L/2^s]$, with $M = 2^m r_{U, F}(0) \leq 2^m \theta (1 - 2/L)$. Since 
\begin{eqnarray}
&& 2 M^L/2^s \leq 2 (2^{-s}) 2^{Lm} \theta^L  (1 - 2/L)^{L} \nonumber \\ 
&\leq& 40 (1 - 2/L)^{L} (1 + 2/L)^{-L}  \leq 40 e^{-4} < 1,
\end{eqnarray}
there is no solution to Eq.~(\ref{hashingeqs}) and thus there is no $w$ which will make Arthur accept. This finishes the proof of Lemma \ref{AMprotcolforr}. 

\vspace{0.3 cm}

\noindent \textit{Reduction to AM protocol for $|\langle 0 | U | 0 \rangle|^2$}: Finally let us show how to use Lemma \ref{AMprotcolforr} to build the AM protocol stated in Theorem \ref{globaldepolarizing}. Since $r_{U, F}(0) = F |\bra{0}U\ket{0}|^2 + (1 - F)/2^n$, on one hand:
\begin{equation} \label{alternative1}
|\langle 0 | U | 0 \rangle|^2 \geq   \lambda \left(1 + \frac{2}{L}  \right)  +  \frac{2(1 - F)}{F L2^n}  
\end{equation}
implies that 
\begin{equation}
r_{U, F}(0) \geq   \left( F \lambda + (1 - F)/2^n   \right) \left(1 + \frac{2}{L}  \right).
\end{equation}
On the other hand:
\begin{equation} \label{alternative2}
|\langle 0 | U | 0 \rangle|^2 \leq \lambda \left(1 - \frac{2}{L}  \right)  -  \frac{2(1 - F)}{F L 2^n}  
\end{equation}
implies that 
\begin{equation}
r_{U, F}(0) \leq   \left(  F \lambda + (1 - F)/2^n   \right) \left(1 - \frac{2}{L}  \right).
\end{equation}

Setting $\theta = F \lambda + (1 - F)/2^n$ we see that the AM protocol from before can also be used to decide if Eq.~(\ref{alternative1}) or Eq.~(\ref{alternative2}) hold true. This ends the proof of the theorem. 

\begin{lemma}  \label{hashfunction}
\cite{AB09} For $t \leq q$, let $H_{q, t}$ be a family of pairwise-independent linear hash functions mapping $\{0, 1 \}^q$ to $\{0, 1 \}^t$, and let $\delta > 0$. Let $S \subseteq \{0, 1 \}^n$ be arbitrary with 
$|S| \geq \delta^{-3} 2^t$. Then with probability larger than $9/10$ over the choice of $h \in H_{n, t}$,
\begin{equation}
  (1 - \delta) \frac{|S|}{2^t} \leq |  \{ x \in S | h(x) = 0^t    \}   | \leq(1 + \delta) \frac{|S|}{2^t}  
\end{equation}
Moreover $h(x)$ can be evaluated in time $2n$, for every $h \in H_{n, t}$. 
\end{lemma}

\begin{acknowledgments}
We acknowledge Georg Goerg for consultation on statistical analyses. This research used resources of the Oak Ridge Leadership Computing Facility, which is a DOE Office of Science User Facility supported under Contract DE-AC05-00OR22725.
\end{acknowledgments}

\vspace{1mm}
\textbf{Correspondence and requests for materials}
\small{should be addressed to John M. Martinis~(jmartinis@google.com).}

\vspace{2mm}
$^\dagger$ \small{ Frank Arute$^1$, Kunal Arya$^1$, Ryan Babbush$^1$, Dave Bacon$^1$, Joseph C. Bardin$^{1,2}$, 
\altaffiliation[Also at ]{Department of Electrical and Computer Engineering, University of Massachusetts Amherst, Amherst, MA, 01003-9292}
Rami Barends$^1$, Rupak Biswas$^3$, Sergio Boixo$^1$, Fernando G.S.L. Brandao$^{1,4}$,  
\altaffiliation[Also at ]{Institute for Quantum Information and Matter, Caltech, Pasadena, CA, USA}
David A. Buell$^1$, Brian Burkett$^1$, Yu Chen$^1$, Zijun Chen$^1$, Ben Chiaro$^5$, 
Roberto Collins$^1$, William Courtney$^1$, Andrew Dunsworth$^1$, Edward Farhi$^1$, Brooks Foxen$^{1,5}$, Austin Fowler$^1$, Craig Gidney$^1$, Marissa Giustina$^1$, Rob Graff$^1$, Keith Guerin$^1$, Steve Habegger$^1$, Matthew P. Harrigan$^1$, Michael J. Hartmann$^{1,6}$, Alan Ho$^1$, Markus Hoffmann$^1$, Trent Huang$^1$, Travis S. Humble$^7$, Sergei V. Isakov$^1$, Evan Jeffrey$^1$, Zhang Jiang$^1$, Dvir Kafri$^1$, Kostyantyn Kechedzhi$^1$, Julian Kelly$^1$, Paul V. Klimov$^1$, Sergey Knysh$^1$, Alexander Korotkov$^{1,8}$, Fedor Kostritsa$^1$, David Landhuis$^1$, Mike Lindmark$^1$, Erik Lucero$^1$, Dmitry Lyakh$^{9}$,
Salvatore Mandr\`{a}$^{3,10}$,
Jarrod R. McClean$^1$, Matthew McEwen$^5$,Anthony Megrant$^1$, Xiao Mi$^1$,Kristel Michielsen$^{11,12}$, 
Masoud Mohseni$^1$, Josh Mutus$^1$, Ofer Naaman$^1$, Matthew Neeley$^1$, Charles Neill$^1$, Murphy Yuezhen Niu$^1$, Eric Ostby$^1$, Andre Petukhov$^1$, John C. Platt$^1$, Chris Quintana$^1$, Eleanor G. Rieffel$^3$, Pedram Roushan$^1$, Nicholas C. Rubin$^1$, Daniel Sank$^1$, Kevin J.~Satzinger$^1$, Vadim Smelyanskiy$^1$, Kevin J.~Sung$^{1,13}$, Matthew D. Trevithick$^1$, Amit Vainsencher$^1$, Benjamin Villalonga$^{1,14}$, Theodore White$^1$, Z. Jamie Yao$^1$, Ping Yeh$^1$, Adam Zalcman$^1$, Hartmut Neven$^1$, John M. Martinis$^{1,5}$ 

\noindent
1. Google AI Quantum, Mountain View, CA, USA, \,\,
2. Department of Electrical and Computer Engineering, University of Massachusetts Amherst, Amherst, MA, USA,\,\,
3. Quantum Artificial Intelligence Lab. (QuAIL), NASA Ames Research Center, Moffett Field, USA,\,\,
4. Institute for Quantum Information and Matter, Caltech, Pasadena, CA, USA,\,\,
5. Department of Physics, University of California, Santa Barbara, CA, USA, \,\,
6. Friedrich-Alexander University Erlangen-N\"{u}rnberg (FAU), Department of Physics, Erlangen, Germany,\,\,
7. Quantum Computing Institute, Oak Ridge National Laboratory, Oak Ridge, TN, USA,\,\,
8.  Department of Electrical and Computer Engineering, University of California, Riverside, CA, USA,\,\,
9. Scientific Computing, Oak Ridge Leadership Computing, Oak Ridge National Laboratory, Oak Ridge, TN, USA
10. Stinger Ghaffarian Technologies Inc., Greenbelt, MD, USA, \,\,
11. Institute for Advanced Simulation, J\"{u}lich Supercomputing Centre, Forschungszentrum J\"{u}lich, J\"{u}lich, Germany,\,\,
12. RWTH Aachen University, Aachen, Germany, \,\,
13. Department of Electrical Engineering and Computer Science, University of Michigan, Ann Arbor, MI, USA,\,\,
14. Department of Physics, University of Illinois at Urbana-Champaign, Urbana, IL, USA
}

\vspace{5mm}

\section*{Erratum}
The caption of Figure 4 in the main paper \cite{arute2019} incorrectly states that the error bars in the figure represent both statistical and systematic uncertainty. They represent the statistical uncertainty. See Figure~\ref{fig:stat_and_syst_error} for comparison of both types of uncertainty and discussion in Section~\ref{sec:large_scale_xeb} for details. Note that both types of uncertainty were accounted for in the analysis and all conclusions remain intact.

\bibliographystyle{naturemag}

\providecommand{\noopsort}[1]{}\providecommand{\singleletter}[1]{#1}

\end{document}